\documentclass[11pt]{cernrep}
\usepackage{graphicx}
\usepackage{here}
\usepackage{epsfig}
\usepackage{axodraw}

\begin{document}

\title{ELECTROWEAK PHYSICS}
\author{{\bf Conveners}: S. Haywood, P.R. Hobson, W. Hollik, Z. Kunszt \\
  {\bf Contributing authors}: 
G.~Azuelos,       
U.~Baur,
J.~van der Bij,
D.~Bourilkov,
O.~Brein,
R.~Casalbuoni,
A.~Deandrea,
S.~De Curtis,
D.~De~Florian, 
A.~Denner, 
S.~Dittmaier,
M.~Dittmar,
A.~Dobado,
M.~Dobbs, 
D.~Dominici,
R.~Gatto,
A.~Ghinculov,
F.~Gianotti, 
M.~Grazzini,              
J.B.~Hansen, 
R.~Harper, 
S.~Haywood, 
S.~Heinemeyer,
M.J.~Herrero,
P.R.~Hobson,
W.~Hollik,
M.~Lefebvre, 
M.~Kr\"amer,
Z.~Kunszt,
C.K.~Mackay,
R.~Mazini, 
A.~Miagkov, 
T.~Muller,
D.~Neuberger,
R.~Orr, 
B.~Osculati, 
J.R.~Pel\'aez,              
A.~Pich,                  
D.~Rainwater,
M.~Redi,
S.~Riley,
E.~Ruiz Morales,             
C.~Schappacher,
A.~Signer,
K.~Sliwa, 
H.~Spiesberger,
W.H.~Th\"ummel
D.~Wackeroth,
G.~Weiglein,
D.~Zeppenfeld,
D.~Z\"urcher}               
\institute{~}
\maketitle

\begin{abstract}
In this review, we consider four main topics:
\begin{enumerate}
\item 
The prospects for a significant improvement in the precise measurement of
the electroweak parameters.
\item
NLO QCD description of the production 
$W^+ W^-,  W^\pm Z, Z Z, W^\pm \gamma$ or $Z \gamma$ pairs
with leptonic decays and with  anomalous triple gauge-boson
couplings.
\item
The prospects for significant improvement in the direct measurement
of the non-Abelian gauge-coupling,
with direct limits on triple and quartic anomalous couplings.
\item
Gauge-boson scattering at large centre of mass energy.
\end{enumerate}
\end{abstract}


\section{INTRODUCTION
         \protect\footnote{Section coordinators: 
         W.~Hollik, Z.~Kunszt.}}




\def\beq{\begin{equation}}
\def\eeq{\end{equation}}
\def\beqar{\begin{eqnarray}}
\def\eeqar{\end{eqnarray}}
\def\barr#1{\begin{array}{#1}}
\def\earr{\end{array}}
\def\bfi{\begin{figure}}
\def\efi{\end{figure}}
\def\btab{\begin{table}}
\def\etab{\end{table}}
\def\bce{\begin{center}}
\def\ece{\end{center}}
\def\nn{\nonumber}
\def\disp{\displaystyle}
\def\text{\textstyle}
\def\arraystretch{1.4}

\newcommand{\eV}{\unskip\,\mathrm{eV}}
\newcommand{\GeV}{\unskip\,\mathrm{GeV}}
\newcommand{\MeV}{\unskip\,\mathrm{MeV}}
\newcommand{\TeV}{\unskip\,\mathrm{TeV}}
\newcommand{\fba}{\unskip\,\mathrm{fb}}
\newcommand{\mba}{\unskip\,\mathrm{mb}}
\newcommand{\pba}{\unskip\,\mathrm{pb}}
\newcommand{\nba}{\unskip\,\mathrm{nb}}

\newcommand{\seff}{\sin^2\theta_{\rm eff}}

\def\mathswitchr#1{\relax\ifmmode{\mathrm{#1}}\else$\mathrm{#1}$\fi}
\newcommand{\PB}{\mathswitchr B}
\newcommand{\PW}{\mathswitchr W}
\newcommand{\PZ}{\mathswitchr Z}
\newcommand{\Pg}{\mathswitchr g}
\newcommand{\PH}{\mathswitchr H}
\newcommand{\Pe}{\mathswitchr e}
\newcommand{\Pne}{\mathswitch \nu_{\mathrm{e}}}
\newcommand{\Pane}{\mathswitch \bar\nu_{\mathrm{e}}}
\newcommand{\Pnmu}{\mathswitch \nu_\mu}
\newcommand{\Pd}{\mathswitchr d}
\newcommand{\Pf}{f}
\newcommand{\Ph}{\mathswitchr h}
\newcommand{\Pl}{l}
\newcommand{\Pu}{\mathswitchr u}
\newcommand{\Ps}{\mathswitchr s}
\newcommand{\Pb}{\mathswitchr b}
\newcommand{\Pc}{\mathswitchr c}
\newcommand{\Pt}{\mathswitchr t}
\newcommand{\Pp}{\mathswitchr p}
\newcommand{\Pq}{\mathswitchr q}
\newcommand{\Pep}{\mathswitchr {e^+}}
\newcommand{\Pem}{\mathswitchr {e^-}}
\newcommand{\Pepm}{\mathswitchr {e^\pm}}
\newcommand{\Pmum}{\mathswitchr {\mu^-}}
\newcommand{\PWp}{\mathswitchr {W^+}}
\newcommand{\PWm}{\mathswitchr {W^-}}
\newcommand{\PWpm}{\mathswitchr {W^\pm}}

\def\mathswitch#1{\relax\ifmmode#1\else$#1$\fi}
\newcommand{\MB}{\mathswitch {M_\PB}}
\newcommand{\Mf}{\mathswitch {m_\Pf}}
\newcommand{\Ml}{\mathswitch {m_\Pl}}
\newcommand{\Mq}{\mathswitch {m_\Pq}}
\newcommand{\MV}{\mathswitch {M_\PV}}
\newcommand{\MW}{\mathswitch {M_W}}
\newcommand{\hMW}{\mathswitch {\hat M_\PW}}
\newcommand{\MWpm}{\mathswitch {M_\PWpm}}
\newcommand{\MWO}{\mathswitch {M_\PWO}}
\newcommand{\MA}{\mathswitch {\lambda}}
\newcommand{\MZ}{\mathswitch {M_Z}}
\newcommand{\MH}{\mathswitch {M_\PH}}
\newcommand{\Me}{\mathswitch {m_\Pe}}
\newcommand{\Mmy}{\mathswitch {m_\mu}}
\newcommand{\Mta}{\mathswitch {m_\tau}}
\newcommand{\Md}{\mathswitch {m_\Pd}}
\newcommand{\Mu}{\mathswitch {m_\Pu}}
\newcommand{\Ms}{\mathswitch {m_\Ps}}
\newcommand{\Mc}{\mathswitch {m_\Pc}}
\newcommand{\Mb}{\mathswitch {m_\Pb}}
\newcommand{\Mt}{\mathswitch {m_\Pt}}
\newcommand{\GW}{\mathswitch {\Gamma_W}}

\newcommand{\scrs}{\scriptscriptstyle}
\newcommand{\sw}{\mathswitch {s_{\scrs\PW}}}
\newcommand{\cw}{\mathswitch {c_{\scrs\PW}}}
\newcommand{\swbar}{\mathswitch {\bar s_{\scrs\PW}}}
\newcommand{\swfbar}{\mathswitch {\bar s_{\PW,\Pf}}}
\newcommand{\swqbar}{\mathswitch {\bar s_{\PW,\Pq}}}
\newcommand{\Qf}{\mathswitch {Q_\Pf}}
\newcommand{\Ql}{\mathswitch {Q_\Pl}}
\newcommand{\Qq}{\mathswitch {Q_\Pq}}
\newcommand{\vf}{\mathswitch {v_\Pf}}
\newcommand{\af}{\mathswitch {a_\Pf}}
\newcommand{\gesi}{\mathswitch {g_\Pe}^{\sigma}}
\newcommand{\gem}{\mathswitch {g_\Pe}^-}
\newcommand{\gep}{\mathswitch {g_\Pe}^+}
\newcommand{\GF}{\mathswitch {G_\mu}}

\newcommand{\ri}{{\mathrm{i}}}
\newcommand{\re}{{\mathrm{e}}}
\newcommand{\ieps}{\ri\epsilon}
\newcommand{\rd}{{\mathrm{d}}}
\newcommand{\QCD}{{\mathrm{QCD}}}
\newcommand{\QED}{{\mathrm{QED}}}
\newcommand{\LEP}{{\mathrm{LEP}}}
\newcommand{\SLD}{{\mathrm{SLD}}}
\newcommand{\SM}{{\mathrm{SM}}}
\newcommand{\CM}{{\mathrm{CM}}}
\newcommand{\born}{{\mathrm{Born}}}
\newcommand{\virt}{{\mathrm{virt}}}
\newcommand{\fact}{{\mathrm{fact}}}
\newcommand{\nonfact}{{\mathrm{nonfact}}}
\newcommand{\res}{{\mathrm{res}}}
\newcommand{\PA}{{\mathrm{PA}}}

\def\atn{\mathop{\mathrm{arctan}}\nolimits}
\def\Li{\mathop{\mathrm{Li}_2}\nolimits}
\def\cLi{\mathop{{\cal L}i_2}\nolimits}
\def\Re{\mathop{\mathrm{Re}}\nolimits}
\def\Im{\mathop{\mathrm{Im}}\nolimits}
\def\sgn{\mathop{\mathrm{sgn}}\nolimits}
\def\arc{\mathop{\mathrm{arc}}\nolimits}

\makeatletter

\def\eqnarray{\stepcounter{equation}\let\@currentlabel=\theequation
\global\@eqnswtrue
\global\@eqcnt\z@\tabskip\@centering\let\\=\@eqncr
$$\halign to \displaywidth\bgroup\hskip\@centering
  $\displaystyle\tabskip\z@{##}$\@eqnsel&\global\@eqcnt\@ne
  \hskip 2\arraycolsep \hfil${##}$\hfil
  &\global\@eqcnt\tw@ \hskip 2\arraycolsep $\displaystyle\tabskip\z@{##}$\hfil
   \tabskip\@centering&\llap{##}\tabskip\z@\cr}
\def\appendix{\par
 \setcounter{section}{0} \setcounter{subsection}{0}
 \def\thesection{\Alph{section}}}

\newcommand{\gtrless}
{\;\rlap{\raisebox{+.25em}{$>$}}\raisebox{-.25em}{$<$}\;}
\newcommand{\lsim}
{\;\raisebox{-.3em}{$\stackrel{\displaystyle <}{\sim}$}\;}
\newcommand{\gsim}
{\;\raisebox{-.3em}{$\stackrel{\displaystyle >}{\sim}$}\;}
\def\asymp#1{\;\raisebox{-.4em}{$\widetilde{\scriptstyle #1}$}\;}
\newcommand{\gl}
{\;\raisebox{.25em}{$>$}\hspace{-.75em}\raisebox{-.2em}{$<$}\;}
\def\dsl{\mathpalette\make@slash}
\def\make@slash#1#2{\setbox\z@\hbox{$#1#2$}%
  \hbox to 0pt{\hss$#1/$\hss\kern-\wd0}\box0}

\makeatother

\subsection{Electroweak parameters}
At the LHC, substantial improvement in the 
precise determination of  electroweak parameters,
such as the $W$~boson mass, the top-quark mass
and the electroweak mixing angle, 
will become feasible, as well as an accurate
measurement of the vector-boson self couplings and of the mass
of the Higgs boson. This opens promising
perspectives towards very comprehensive 
and challenging tests of the electroweak theory.

Electroweak precision observables 
provide the basis for important consistency tests of the Standard Model (SM)
or its extensions, in particular the Minimal Supersymmetric Standard
Model (MSSM). 
By comparing
precision data with the predictions of 
specific models, it is possible to derive indirect constraints
on the parameters of the model.
In the case of the
top-quark mass, $m_t$, 
the indirect determination from the precision observables 
in the framework of the SM 
turned out to be in remarkable agreement
with the direct experimental measurement of $m_t$. Since the Higgs boson
mass, $M_H$, enters the predictions for the precision observables only
logarithmically in leading order, the indirect determination of $M_H$
requires very accurate experimental data as well
as high precision of the theoretical predictions.
The uncertainties of the predictions arise from the following sources: 
a) the unknown higher-order
corrections - since the perturbative evaluation is truncated at a certain
order, and b) the parametric uncertainties induced by the
experimental errors of the input parameters.



The most important universal top-quark contribution to the electroweak
precision observables
enters via the $\rho$ parameter, which deviates from unity by a 
loop contribution $\Delta\rho$.
At the one-loop level, the $(t,b)$ doublet yields a term proportional to
$m_t^2$~\cite{Veltman:1977kh}, namely
$\Delta\rho = 3 G_\mu m_t^2/(8\pi^2\sqrt{2})$ 
in the  limit $m_b \to 0$. 
Therefore, it is to be expected that the precision measurement of the 
top-quark mass at the LHC (see Section~\ref{sec:wmass})
will significantly improve the theoretical 
prediction of the $W$ mass, $M_W$ -- 
at present, the experimental error on $m_t$ 
is a limiting factor for the accuracy in the
theoretical predictions of the precision observables.
$M_W$  itself will be measured at the LHC
with a sizably improved accuracy.

The theoretical prediction for $M_W$ is obtained 
from the relation between the vector-boson masses $M_{W,Z}$ 
and the Fermi constant $G_\mu$, which is conventionally written in the form
\begin{equation}
\label{deltar}
M_W^2 \left(1 - \frac{M_W^2}{M_Z^2}\right) =
\frac{\pi \alpha}{\sqrt{2} G_\mu} \frac{1}{1 - \Delta r} \, .
\end{equation}
The quantity $\Delta r=\Delta r(\alpha,M_Z,M_W,m_t,M_H)$, 
first derived in \cite{Sirlin:1980nh,Sirlin:1980nh2} in one-loop order,
summarises the quantum corrections to the vector-boson mass correlation;
it is obtained 
from the calculation of the muon lifetime in the SM
beyond the tree-level approximation.
At one-loop order, $\Delta r$ can be written as
\begin{equation}
\Delta r = \Delta \alpha - \frac{c_W^2}{s_W^2} \Delta\rho + (\Delta
r)_{\mathrm{rem}} .
\end{equation}
$\Delta \alpha$ contains the large logarithmic contributions from
the light fermions, and $\Delta\rho$ the $m_t^2$ dependence;
the non-leading terms are collected in 
$(\Delta r)_{\mathrm{rem}}$ where also the
dependence on $M_H$ enters. 
In Equation~\ref{deltar}, $\Delta r$ is a quantity that accounts also for 
terms of higher order than just one-loop. Moreover,
a partial resummation of large contributions from light
fermions and from the $\rho$ parameter is contained in the expression.
For a discussion see for example the section on the Electroweak Working Group
Report in \cite{Yellowbook95}. Results for $M_W$ that 
were not yet available at the time of the report \cite{Yellowbook95} are the
next-to-leading two-loop terms of ${\cal O}(G_\mu^2 m_t^2
M_Z^2)$~\cite{Degrassi:1996mg,Degrassi:1997ps} 
in an expansion for asymptotically large $m_t$ and the 
result for the Higgs mass dependence of the fermionic two-loop
contributions~\cite{Bauberger:1998ey}. 
Recently, the complete result for the fermionic
two-loop contributions has been obtained~\cite{deltar2lnew}. Furthermore,
the QCD corrections to $\Delta r$ of ${\cal O}(\alpha\alpha_s^2)$ have
been derived~\cite{Chetyrkin:1995js}.

The most recent theoretical prediction~\cite{deltar2lnew} for $M_W$ 
within the SM is displayed in Figure~\ref{fig:MWMHlhc}
as a function of $M_H$.  
To illustrate the comparison between theory and experiment,
the experimental result is included in the figure
for the current uncertainty $\delta M_W = \pm 0.042$~GeV~\cite{cernep9915} 
and the
estimated LHC uncertainty $\delta M_W = \pm 0.015$~GeV 
(see Section~\ref{sec:wmass})
(assuming the same central value).
The uncertainty 
for the current status and for the case where the LHC will have
measured the top-quark mass with much higher accuracy is also displayed, 
in combination with the theoretical uncertainty
from unknown higher-order corrections. 
It is clear 
that both improvements, in $M_W$ and in $m_t$, will lead to a
substantial increase in the significance of Standard Model tests,
with stringent bounds on the Higgs boson mass
to be confronted  with the directly measured value of $M_H$.  

\begin{figure}
\begin{center}
\includegraphics[width=0.6\textwidth,height=0.3\textheight,clip]{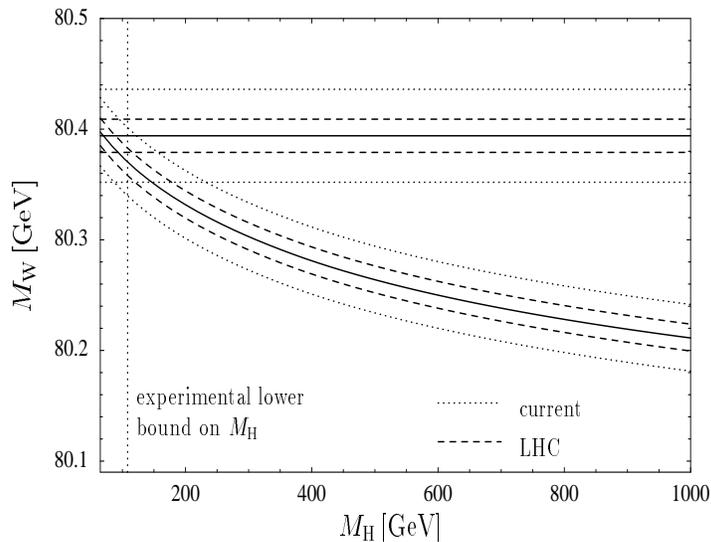}
\end{center}
\caption{The dependence of $M_W$, predicted by means of 
   Equation~\ref{deltar},  
   on $M_H$ is shown for the SM.
   The uncertainty of the predictions corresponds to the present
   and expected parametric uncertainty owing to the top mass, in combination
   with the theoretical uncertainty. The central lines (solid)
   correspond to the present central values of $M_W=80.394$ GeV 
   and $m_t=174.3$ GeV.
\label{fig:MWMHlhc}}
\end{figure}

Besides the $W$ boson mass, the improvement in $m_t$ will also have an
effect on the predictions of the $Z$ pole observables. 
They are conveniently described in terms of 
effective couplings 
\begin{equation}
\label{couplings}
g_V^f = \sqrt{\rho_f}\, (I_3^f - 2 Q_f \sin^2\theta_{\rm eff}^f), \quad
\quad   g_A^f = \sqrt{\rho_f}\,  I_3^f 
\end{equation}
in the neutral-current vertex 
at the $Z$ resonance
for a given fermion species $f$, normalised according to
$J_{\mu}^{\mathrm{NC}} = ( \sqrt{2} G_\mu M_Z^2 )^{1/2}
(g_V^f \gamma_{\mu} - g_A^f \gamma_{\mu} \gamma_5)$.
Besides the overall normalisation factor $\rho_f=1+\Delta\rho +\cdots$,
we mention in particular the
effective mixing angle, which is usually chosen as the on-resonance
mixing angle for the leptons $f=e,\mu,\tau$ in Equation~\ref{couplings} and
denoted as
 $\sin^2 \theta^{\mathrm{lept}}_{\mathrm{eff}}$. 
This quantity also depends sensitively on the top-quark mass,
mainly through $\Delta\rho$.
The theoretical prediction of $\sin^2 \theta^{\mathrm{lept}}_{\mathrm{eff}}$
will definitely be sharpened by the precise measurement of the
top-quark mass;
a sizable improvement concerning the internal consistency test
can be anticipated.
The on-resonance mixing angle for the light quarks 
$\neq b$ is numerically very close
to the leptonic one.  $\sin^2 \theta^{\mathrm{lept}}_{\mathrm{eff}}$
can therefore be measured at the LHC in
the Drell-Yan production of charged-lepton pairs around the $Z$ resonance, via
$q\bar{q} \rightarrow l^+l^-$, where an accuracy of 
$1.4\times 10^{-4}$ on $\sin^2 \theta^{\mathrm{lept}}_{\mathrm{eff}}$
may be feasible (see Section~\ref{sec:drellyan}).


Besides these internal consistency checks of the SM, the 
electroweak precision observables may be useful to
distinguish between different models as candidates for the
electroweak theory.  In Figure~\ref{fig:MWMTlhc}, the
SM prediction of $M_W$ as a function of $m_t$ is compared with the
prediction within the MSSM,
where the MSSM prediction is based on results up to 
${\cal O}(\alpha\alpha_s)$~\cite{Chankowski:1994eu,Djouadi:1998sq}. 
The SM uncertainty arises
from the only unknown parameter, the Higgs boson mass. 
On the other hand, within the
MSSM, the Higgs boson mass is not a free
parameter~\cite{ghkd}, and the uncertainty originates from the unknown SUSY
mass scales. In the small overlap region, the MSSM behaves like the SM,
{\it i.e.}\ all SUSY particles are heavy and decouple from the precision
observables, and the $M_H$ value of the SM stays below 130 GeV, the
upper bound on the lightest MSSM Higgs boson
mass for $m_t = 175$~GeV (see~\cite{Heinemeyer:1999np} and references 
therein). Figure~\ref{fig:MWMTlhc} shows the clear
improvement from the current status to the LHC era, where eventually,
besides direct experimental evidence, a distinction between SM and
MSSM might become feasible.

\begin{figure}
\begin{center}
\includegraphics[width=0.6\textwidth,clip]{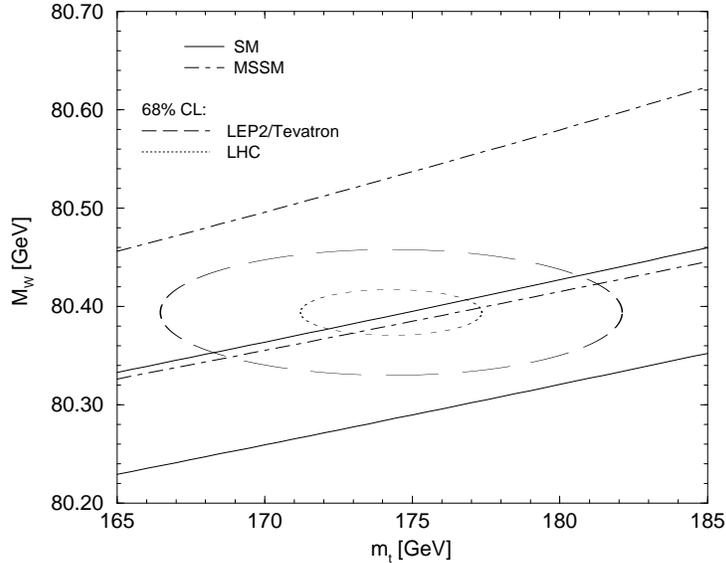}
\end{center}
\caption{The dependence of $M_W$ on $m_t$ is shown for the SM and the
  MSSM. It is compared to the current
  errors and to the errors expected from the LHC.
\label{fig:MWMTlhc}}
\end{figure}

\subsection{Vector-boson pair production and scattering}

At the LHC, the precise measurement of the production
of  $W^+ W^-$,  $W^\pm Z$, $Z Z$, $W^\pm \gamma$ or $Z \gamma$ pairs
is also an  important physics goal. In the simplest studies,
the gauge-bosons will be detected via their leptonic decays.
Already a couple events have been obtained by CDF and D0
for $WW$ and $WZ$ production and D0 has seen about 100 $W\gamma$
and 30 $Z\gamma$ events. The data set at Run II will be about 20 times
larger and about 1000 times larger at the LHC.
For a summary of the experimental situation see 
\cite{diehl-ht,EllisonWudka}.

The production of gauge-boson pairs provide us with the best
test of the non-Abelian gauge-symmetry of the Standard Model
(SM). Deviation from the SM
predictions may come either from the presence of anomalous 
couplings or the production of new heavy particles
and their decays into vector-boson pairs.
If the particle spectrum of the SM has to be enlarged with
new particles (as in the Minimal Supersymmetric Standard Model (MSSM))
with mass values of $\ge 0.5-1 \TeV$, small anomalous
couplings are generated at low energy.  If the Higgs boson
is very heavy, it will  decay mainly into $W^+W^-$ and $ZZ$ pairs.
If the symmetry breaking mechanism is dynamical (technicolor models, 
BESS models),  large anomalous couplings might be generated or new heavy
particles may be produced. In both of these cases, 
vector-boson pair production will 
show deviations from the Standard Model predictions.
At the same time, vector-boson pair production
 gives  the most important background
 for a number of  new physics signals. For example, 
one of the most important physics signal for supersymmetry
at hadron colliders is the production of three charged leptons
and missing transverse momentum. The dominant background for this
process is the production of $W$ plus a $Z$ (real or virtual) or $\gamma$.

The leading order production mechanism of gauge-boson pair
production is $q\bar{q}$ annihilation.
The precise calculation of the cross sections in the QCD
improved parton model have received recently a lot of attention.
The cross sections of the gauge-boson pair production
and its decay into lepton pairs have been
calculated  in next-to-leading order (NLO) accuracy retaining
the full spin correlations of the leptonic decay products.
A significant achievement was that the
 theoretical results in NLO QCD  for the production of 
  $W^+ W^-$,  $W^\pm Z$, $Z Z$, $W^\pm \gamma $ or $Z \gamma$ pairs
could be documented in short
analytic formulae \cite{Dixon:1998py}
allowing for independent numerical implementations.
Subsequently, several 
 so called NLO numerical Monte Carlo programs have been developed
and the complete one loop corrections became available for the
first time 
for    $W^+ W^-$,  $W^\pm Z$, $Z Z$ in~\cite{Campbell:1999ah, Dixon:1999di},
 and for $ W^\pm \gamma,$ or $Z \gamma$ pairs in 
\cite{Deflo-wgam:2000}.
These new  results  have 
 superseded and  confirmed previous NLO results on spin averaged
production gauge-boson pair production   
\cite{ZZOhn, ZZit, WZOhn, WZit, WWOhn, WWit, STN, yOhn},
as well the approximate results
where
spin correlation have been neglected in the virtual corrections
\cite{BHOWZZero, BHOWW, BHOWZ, BHOWg, BHOZg}. 
The agreement between the well documented results in 
\cite{Dixon:1999di} and in
\cite{ZZit, WZit, WWit}
 is within the precise integration error and
the agreement between the results of \cite{Dixon:1999di}  and the recent
programs of 
\cite{BHOWZZero, BHOWW, BHOWZ, BHOWg, BHOZg} 
 is about 3\%.
Therefore, previous experimental simulation studies based on these programs
(see  Section~\ref{sec:bess}) should not be repeated.

 Simple  analytic NLO results exist also for  the anomalous
coupling contributions at NLO accuracy
in \cite{Dixon:1999di, Deflo-wgam:2000}. 
Again, the agreement with previous 
approximate NLO results
\cite{BHOWZZero,BHOWW, BHOWZ, BHOWg, BHOZg}
is also good (see Section~\ref{sec:tgc_nlo}).
Future anomalous coupling studies may like to  use the more
accurate  packages.
At the LHC, contrary to LEP, 
 the phenomenological studies of  anomalous triple gauge-boson 
 coupling constants 
 cannot be treated as constant couplings
 since they lead to
 violation of $SU(2)$ gauge-symmetry and unitarity. 
 The difficulty
comes from truncation of the contribution of an  infinite
series of higher dimensional non-renormalisable gauge-invariant operators. 
In the case of $q\bar{q}$ annihilation to gauge-boson pairs,
a suitable phenomenological approach is the introduction of
form factors for the anomalous couplings (which in principle
are calculable in the true underlying theory).
As long as we do not obtain deviations from the Standard Model,
for practical  purposes, simple dipole form
factors with various cut-off parameters can be used. 
With better data, one can put limits on  the form factor values in 
 small $\sqrt{\hat{s}}$ intervals,
 assuming constant couplings for each interval.
In the case of positive signals, such a  form factor measurement
will  provide us 
  with important
information on the underlying theory (see 
Sections~\ref{sec:ewprecision},~\ref{sec:vbpp} and~\ref{sec:anomtgc}).

 At higher energies,  the higher order
production processes of 
$WW$ and $ZZ$ scattering (the weak boson
are emitted from the incoming quarks)
will become more and more important.
These interactions are the  most sensitive
to the mechanism of the electroweak symmetry breaking.
In particular, if the breaking
of the electroweak symmetry  is due to  
new particles with strong interactions
at the TeV scale, 
enhanced production of longitudinal
gauge-boson pairs will be the most typical 
signal~\cite{technicol, technicol_suss}.
The minimal model to describe this alternative  is obtained
by assuming that the new particles are too heavy
to be produced at LHC and the linear $\sigma$-model Higgs-sector
of the Standard Model is replaced 
 by the non-renormalisable
non-linear $\sigma$-model which can also be considered 
as an effective chiral vector-boson Lagrangian with non-linear 
realisation of the gauge-symmetry \cite{ApBe80, longhitano:81}.
 The question is whether this more phenomenological
approach is  consistent with the precision
data.
In a recent analysis,   a positive answer was obtained~\cite{Bagger:1999te}. 
It has been
 found that due to the screening
of the symmetry breaking sector \cite{veltman},
this alternative still has enough flexibility to be in
perfect agreement  with  the precision data 
up to a cut-off scale of $3\TeV$ (see  
Sections~\ref{sec:anomtgc} and~\ref{sec:vbfs}).
 In the chiral approach,
 the gauge-boson observables are obtained as truncated
series in powers of the external momenta $p^n/(4\pi v)^n$
with $\MW^2\approx gv^2/8$. The approximation is valid
up to energy scales of $E=4\pi v\approx 3\TeV$.
At the LHC, the partonic centre of mass energy can be higher and
the phenomenological implementation is confronted with the problem
of unitarisation~\cite{Chanowitz:1998wi, vanderBij:1999fp, jose}.
 Although  unitarisation
is not unique, the use of the K-matrix formalism~\cite{Chanowitz:1998wi}
or  the ${\cal O}(p^4)$ 
Inverse Amplitude Method~\cite{jose} appear
to give reasonable model independent 
framework to explore the
various possibilities.
When extrapolating to higher energies in particular,
the masses  of  resonances are   rather sensitive
to  the actual value of additional  chiral parameters.
An alternative approach for the phenomenological
formulation of 
the dynamical symmetry breaking consistent
with the precision data  is offered by the BESS model
\cite{Casalbuoni:1996qt} with an
 extended strongly interacting gauge-sector with enhanced global symmetries
and with important decoupling
properties at low energies. The phenomenologically acceptable
technicolor models~\cite{Appelquist:1999dq}
also require  an enhanced global symmetry in the spectrum
  of the theory. In the most pessimistic
parameter ranges, it is rather difficult to
detect the signals of the strong $WW$ and $WZ$ scattering; 
therefore, one has to push the LHC analysis to its limits.
In the future, further clever strategies have to be pursued for this case
(see Section~\ref{sec:vbfs}).

\section{ELECTROWEAK CORRECTIONS TO DRELL-YAN PROCESSES
         \protect\footnote{Section coordinator: 
          W.~Hollik.}}

The basic parton processes for single vector-boson production
are $q\bar{q'} \rightarrow W \rightarrow l \nu_{l}$ and
$q\bar{q} \rightarrow Z \rightarrow l^+l^-$, with  charged
leptons $l$ in the final state. Investigations around the 
$W$ and $Z$ resonance allow a precise measurement of the $W$ mass
and of the electroweak mixing angle from the forward-backward
asymmetry. At high invariant masses of the $l^+l^-$ pair,
deviations from the standard cross section and $A_{\rm FB}$
could indicate scales of new physics, {\it e.g.} associated with
an extra heavy $Z'$ or extra space dimensions.
For the envisaged precision, a discussion of the electroweak 
higher-order contributions is necessary, on top of the QCD corrections.
The electroweak corrections consist of the set of electroweak
loop contributions, including virtual photons, and of the
emission of real photons.

With respect to QCD, the cross sections in this section are
all of lowest order, evaluated with parton distribution functions
at factorisation scales $M_W$ (for $W$ production) and $M_Z$
(for $Z$ production). Hence, the numerical values are not yet
directly the physical ones. They are given here to point out
the structure and the size of the higher-order electroweak
contributions. 
The QCD corrections are considered in the
QCD chapter of this report, where a QCD-related uncertainty
of $\sim$5\% is estimated.   
For illustration, we give the values (in nb) for
$[\sigma(pp\rightarrow W^+) + \sigma(pp\rightarrow W^-)]\cdot
 {BR}(W\rightarrow e\nu)$ and 
$\sigma(pp\rightarrow Z) \cdot {BR}(Z\rightarrow e^+e^-)$ 
in the purely electroweak calculation (EW) and with NNLO QCD
\cite{mrst99}: \\
\begin{center}
$W: \quad   17.9~({\rm EW}) \quad $ and $ \quad 20.3 \pm 1.0~({\rm NNLO}),$ \\
$Z: \quad    1.71~({\rm EW})\quad $ and $ \quad 1.87 \pm 0.09~({\rm NNLO})$.
\end{center}

\subsection{Universal initial-state QED corrections \label{sec:ewqedcor}}

\begin{figure}[htbp] 
\unitlength 1mm
\begin{picture}(160,82)
\put(0,-1){\epsfig{file=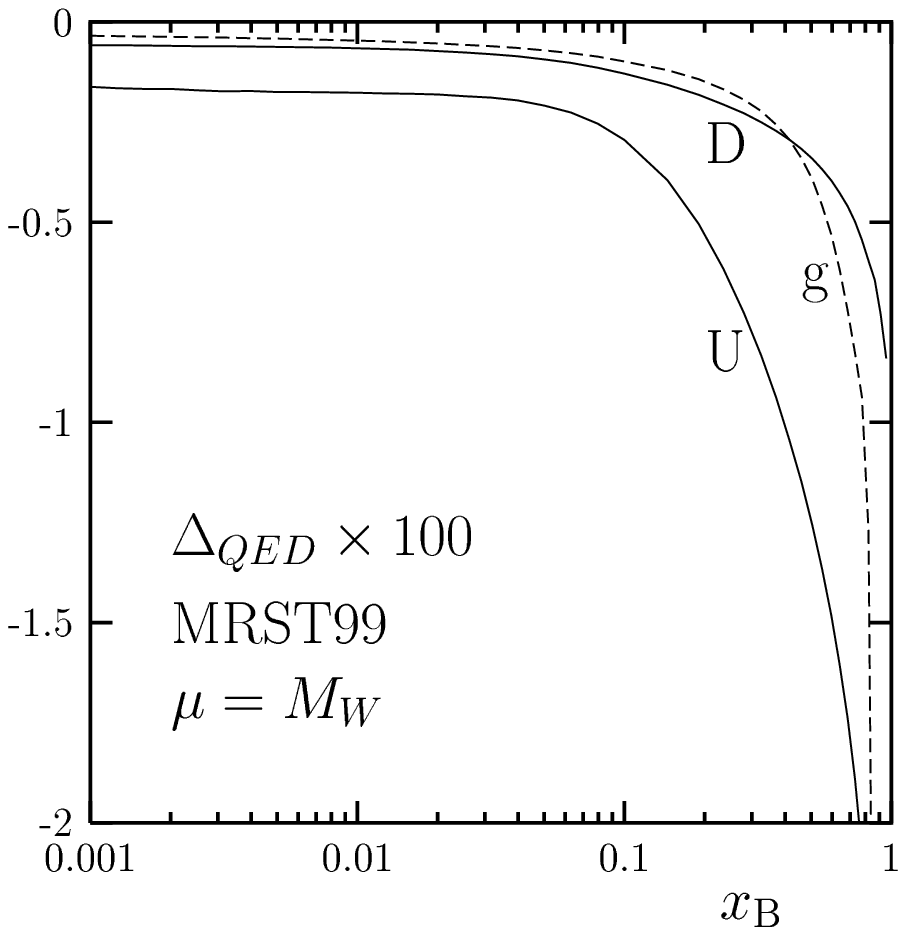,width=8cm}}
\put(40,-1){(a)}
\put(80,-1){\epsfig{file=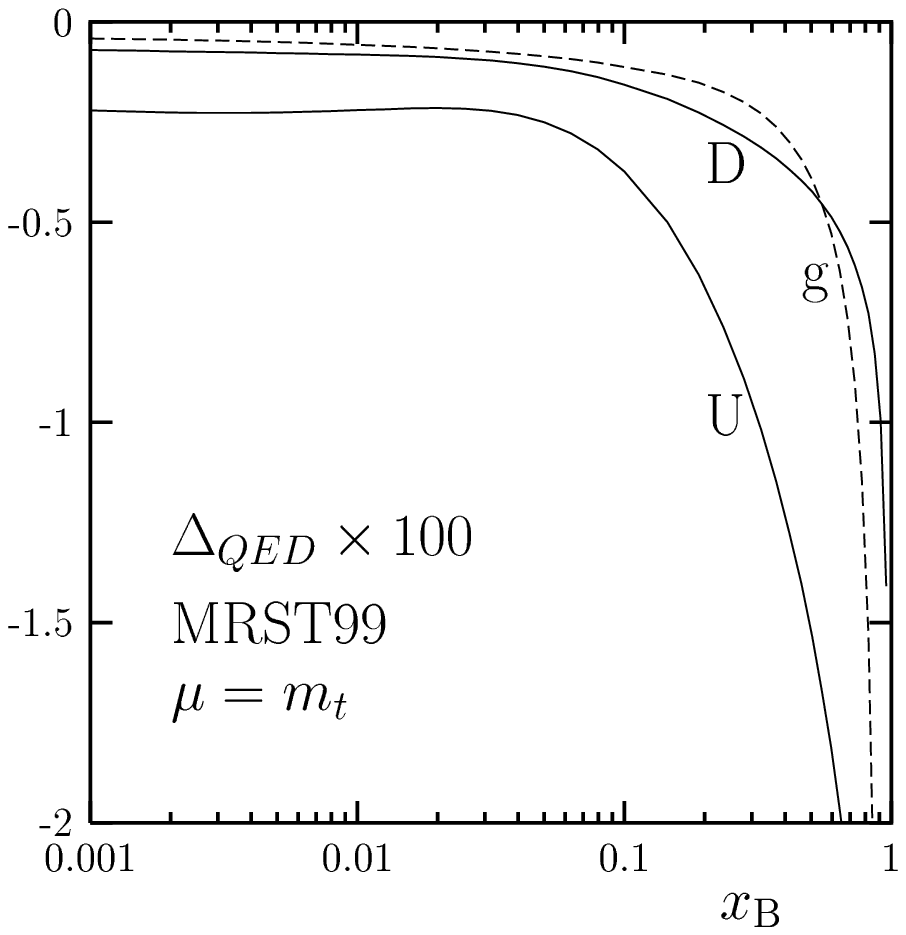,width=8cm}}
\put(120,-1){(b)}
\end{picture}
\caption{QED corrections to the parton distribution functions for {\it
    up}-type quarks, $U(x,\mu^2) = \sum_{gen} \left( u + \bar{u}
  \right)$, {\it down}-type quarks, $D(x,\mu^2) = \sum_{gen} \left( d +
    \bar{d} \right)$ and the gluon $g(x,\mu^2)$ in per cent for the scale
  $\mu = M_W$ (a) and $\mu = m_t$ (b).
}
\label{figqed}
\end{figure}

QED corrections related to the emission of (real or virtual) photons
from quarks contain mass singularities which factorise and therefore can
be absorbed by a redefinition ({\it renormalisation}) of parton
distribution functions \cite{rujula}. This redefinition is well-known in
the calculation of QCD radiative corrections where in complete analogy
to photon radiation, the emission of gluons leads to mass singularities
as well.  By the redefinition, the mass singularities disappear from the
observable cross section and the renormalised distribution functions
become dependent on the factorisation scale $\mu$ which is controlled by
the well-known Gribov-Lipatov-Altarelli-Parisi (GLAP) equations
\cite{ap,AP}. The factorisation scale should be identified with a typical
scale of the process, {\it i.e.}\ a large transverse momentum, or the mass of
a produced particle.

Since mass singularities are universal, {\it i.e.}\ independent of the
process under consideration, the definition of renormalised parton
distributions is also universal. Therefore it is possible to discuss
the bulk of initial-state QED radiative corrections 
in terms of parton distribution
functions. This will be true if there is only one large scale in the
process.

The treatment of mass singularities due to gluonic or photonic radiation
is identical. Photonic corrections can therefore be taken into account
by a straightforward modification \cite{perlt,qedhs} of the standard
GLAP equations which describe gluonic corrections only. The modification
corresponds to the addition of a term of the order of the
electromagnetic fine-structure constant, $\alpha$. Apart from small
non-singular contributions, the resulting modified scale dependence of
parton distribution functions is the only observable effect of 
initial-state QED
corrections in high-energy scattering of hadrons.

The modified evolution equation for the charged parton distribution
functions, $q_f(x,\mu^2)$ for quarks with flavour $f$, can be written as:
\begin{equation}
\begin{array}{ll} \displaystyle
\frac{d}{dt} q_f(x,t) = & \displaystyle \frac{\alpha_s(t)}{2\pi}
\int_x^1 \frac{dz}{z} \left[ P_{q/q} (z,t) q_f( x/z,t)
                            + P_{q/g} (z,t) g ( x/z,t) \right]
\\[1em] & \displaystyle
+ \frac{\alpha(t)}{2\pi}
\int_x^1 \frac{dz}{z} P_{q/q}^{\gamma} (z,t) q_f( x/z,t)
\end{array}
\label{apmod}
\end{equation}
In the leading logarithmic approximation, the splitting functions
$P_{i/j}$ are independent of the scale $t = \ln \mu^2/\Lambda^2$,
and the QED splitting function is given by
\begin{equation}
P_{q/q}^{\gamma}(z) = Q_f^2 \left[ \frac{1+z^2}{(1-z)_+}
                    + \frac{3}{2}\delta(1-z) \right]
                    = \frac{Q_f^2}{C_F} P_{q/q}.
\end{equation}
Since quarks are coupled through the splitting function $P_{q/g}(z) =
\frac{1}{2} \left[ z^2 + (1-z)^2 \right]$ to gluons, the gluon
distribution $g(x,\mu^2)$ is affected by QED corrections as well,
although only indirectly, by terms of the order of ${\cal O}(\alpha
\alpha_s)$.  $\alpha(t)$ is the running electromagnetic fine-structure
constant and $Q_f$ are the fermion charges in units of the positron
charge.

The proper treatment of the mass-singular initial-state QED corrections
would require not only the 
solution of the evolution equations with the QED term,
but also to correct all data that are used to fit the parton distributions
for those QED effects. Apart from a few exceptions, experimental data
have not been corrected for photon emission from quarks. However, one can
illustrate the QED radiative corrections by comparing the modification
of the parton distributions relative to the distribution functions
obtained from the evolution equations without the QED terms, which are
used as an input. 

The solution of the evolution equations corresponds to the resummation
of terms containing factors $\alpha (\alpha_s \ln \mu^2)^n$ with
arbitrary power $n$.  In Figures \ref{figqed}a and \ref{figqed}b, 
we show numerical
results for the corrections $\Delta_{QED}$ to the distribution functions
$U(x,\mu^2)$ ($D(x,\mu^2)$) for the sum of all {\it up}-({\it
  down})-type quarks, and the gluon distribution $g(x,\mu^2)$.  The
figures show the QED corrections in per cent relative to the
distribution functions obtained from the GLAP equations without the QED
term. The input distributions were taken from \cite{mrst99}. One finds
small, negative corrections at the per-mille level for all values of $x$
and $\mu^2$ relevant in the LHC experiments.  Only at large $x \
{\stackrel{>}{\scriptstyle \sim}} \  0.5$ and large $\mu^2 \
{\stackrel{>}{\scriptstyle \sim}} \ 10^3$~GeV$^2$ do the corrections reach
the magnitude of one per cent.  The increase of corrections for $x
\rightarrow 1$ is due to the $\ln (1-x)$ terms appearing in the
evaluation of the ``$+$'' distributions.

The largest corrections are obtained for up-type quarks due to the
larger charge factor $4/9$ as compared to $1/9$ for down-type quarks.
The gluon distribution, being of order ${\cal O}(\alpha \alpha_s)$, is
corrected by less than $0.1 \%$ up to values of $x$ of about 0.2.

The corrections vanish for $\mu^2 \rightarrow \mu^2_0$ since it was
assumed that the input distributions $q_f(x,\mu^2_0)$ and $g(x,\mu^2_0)$
have been extracted from experiment at the reference scale $\mu_0^2$
without subtracting quarkonic QED corrections.

The asymptotic behaviour for $x \rightarrow 0$ can be checked
analytically. The singular behaviour of distributions $\propto
x^{-\eta}$ for $x \rightarrow 0$ remains unchanged by the GLAP
equations if $\eta > 1$. Thus the ${\cal O}(\alpha)$ corrected
distributions have the same power behaviour as the uncorrected ones, the
ratio consequently reaching a constant value for $x \rightarrow 0$. The
valence parts of $U(x)$ and $D(x)$, however, which vanish at $x=0$,
receive positive corrections at small $x$, thus producing the well-known
physical picture: radiation of gluons as well as of photons leads to a
depletion at large $x$ and an enhancement at small $x$, {\it i.e.}\ partons
are shifted to smaller $x$.

Other input distribution functions lead to differences of QED
corrections at the per-mille level which are again irrelevant when
compared with the expected experimental precision of structure-function
measurements.



\subsection{Electroweak corrections to \boldmath $W$ \unboldmath production}

\subsubsection{Physical goals of single $W$ production}

The Drell-Yan-like production of $W$~bosons represents one of the
cleanest processes with a large cross section at the LHC.
This reaction is not only well suited for a precise determination of the
$W$~boson mass $\MW$, it also yields valuable information on the parton
structure of the proton. Specifically, the target accuracy of the order of 
$15\MeV$ \cite{atlas-phystdr2} 
in the $\MW$ measurement exceeds the precision of
roughly $30\MeV$ achieved at LEP2 \cite{lep2repWmass} and Tevatron Run II 
\cite{ai96}, and thus competes with the one of a future $e^+e^-$ collider 
\cite{ac99}. Concerning quark distributions, precise measurements of
rapidity distributions provide information over a wide range in $x$
\cite{mrst99}; a measurement of the $d/u$ ratio would, in particular, be 
complementary to HERA results. The more direct determination of 
parton-parton luminosities instead of single parton distributions is
even more precise \cite{di97}; extracting the corresponding luminosities
from Drell-Yan-like processes allows us to predict related $q\bar q$
processes at the per-cent level.

Owing to the high experimental precision outlined above, the predictions
for the processes $pp \to W \to l\nu_l$ should attain per-cent
accuracy. To this end, radiative corrections have to be included. In the
following some basic features of this processes and recent progress 
\cite{ho97,ba98,ba99,ba00,di00}
on electroweak corrections are summarised; 
a discussion of QCD corrections can be found in the QCD chapter of this report.

\subsubsection{Lowest-order cross section and preliminaries}

We consider the parton process $u\bar d\to \nu_l l^+ (+\gamma)$,
where $u$ and $d$ generically denote the light up- and down-type quarks,
$u=u, c$ and $d=d,s$. The lepton $l$ represents $l=e,\mu,\tau$.
In lowest order, only the Feynman diagram shown in Figure~\ref{fig:Wborndiag}
contributes to the scattering amplitude, 
\bfi
\centerline{\unitlength 1pt
\begin{picture}(120,100)(0,0)
\ArrowLine( 5, 95)(30,50)
\ArrowLine(30, 50)( 5, 5)
\Photon(30,50)(100,50){2}{6}
\Vertex(30,50){1.5}
\Vertex(100,50){1.5}
\ArrowLine(100,50)(125, 95)
\ArrowLine(125, 5)(100, 50)
\put(58,32){$W$}
\put(-8,90){u}
\put(-8, 0){$\bar d$}
\put(130,90){$\nu_l$}
\put(130, 0){$l^+$}
\end{picture}
}
\caption{Lowest-order diagram for $u\bar d\to W^+ \to\nu_l l^+(+\gamma)$.}
\label{fig:Wborndiag}
\efi
and the Born amplitude reads
\beq
{\cal M}_0 = \frac{e^2 V^*_{ud}}{2\sw^2} \,
\left[ \bar v_d\gamma^\mu\omega_-u_u\right] \,
\disp\frac{1}{\hat s-\MW^2+i\MW\GW(\hat s)} \,
\left[ \bar u_{\nu_l}\gamma_\mu\omega_-v_l\right],
\label{eq:m0}
\eeq
with $\hat s$ being the squared centre-of-mass (CM) energy of the parton
system. The notation for the Dirac spinors $\bar v_d$, {\it etc.}, is obvious,
and $\omega_-=\frac{1}{2}(1-\gamma_5)$ is the left-handed chirality 
projector.
The electric unit charge is denoted by $e$, the weak mixing angle is 
fixed by the ratio $\cw^2=1-\sw^2=\MW^2/\MZ^2$ of the $W$ and $Z$~boson
masses $\MW$ and $\MZ$, and $V_{ud}$ is the CKM matrix element for
the $ud$ transition.

Strictly speaking, Equation~(\ref{eq:m0}) already goes beyond lowest order,
since the $W$~boson width $\GW(\hat s)$ results from the Dyson summation of
all insertions of the (imaginary parts of the) $W$~self-energy. Defining
the mass $\MW$ and the width $\GW$ of the $W$~boson in the on-shell scheme
(see {\it e.g.}~\cite{bo86,den93}), the Dyson summation directly leads to a 
{\it running width}, {\it i.e.}\
$\GW(\hat s)|_{\mathrm{run}} = \GW\times(\hat s/\MW^2).$
On the other hand, a description of the resonance by an expansion about
the complex pole in the complex $\hat s$ plane corresponds to a
{\it constant width}, {\it i.e.}\
$\GW(\hat s)|_{\mathrm{const}} = \GW.$
In lowest order these two parametrisations of the resonance region are
fully equivalent, but the corresponding values of the line-shape
parameters $\MW$ and $\GW$ differ in higher orders 
\cite{ho97,ba88,be97}. The numerical difference is given by
$\MW|_{\mathrm{run}}-\MW|_{\mathrm{const}}\approx 26\MeV$ so that it is
necessary to state explicitly which parametrisation is used in a
precision determination of the $W$~boson mass from the $W$~line shape.

The differential lowest-order cross section is easily obtained by
squaring the lowest-order matrix element ${\cal M}_0$ of (\ref{eq:m0}),
\beq
\biggl(\frac{\rd\hat\sigma_0}{\rd\hat\Omega}\biggr) 
= \frac{1}{12} \, \frac{1}{64\pi^2\hat s} \, |{\cal M}_0|^2
= \frac{\alpha^2 |V_{ud}|^2}{192\sw^4 \hat s}
\frac{\hat u^2}{|\hat s-\MW^2+i\MW\GW(\hat s)|^2}, 
\eeq
where $\hat u=(p_u-p_l)^2$ is the squared momentum difference between
the up-type quark and the lepton.
The explicit factor $1/12$ results from the average over the quark
spins and colours, and $\hat\Omega$ is the solid angle of the outgoing
$l^+$ in the parton CM frame.
The electromagnetic coupling $\alpha=e^2/(4\pi)$ can be set to
different values according to different renormalisation schemes.
It can be directly identified with the fine-structure constant $\alpha(0)$
or the running electromagnetic coupling $\alpha(Q^2)$ at a high energy
scale $Q$. 
For instance, it is possible to make use of the value of $\alpha(\MZ^2)$ 
that is obtained by analysing the experimental ratio 
$R=\sigma(e^+e^-\to\mbox{hadrons})/(e^+e^-\to\mu^+\mu^-)$.
These choices are called 
{\it $\alpha(0)$-scheme} and {\it $\alpha(\MZ^2)$-scheme}, respectively,
in the following.
Another value for $\alpha$ can be deduced from the Fermi constant $\GF$,
yielding $\alpha_{\GF}=\sqrt{2}\GF\MW^2\sw^2/\pi$; this choice is
referred to as {\it $\GF$-scheme}. 

\subsubsection{Electroweak corrections}

The electroweak ${\cal O}(\alpha)$ corrections consist of virtual
one-loop corrections and real-photonic brems\-strah\-lung. 
The corrections 
to resonant $W$~production have already been studied in~\cite{ho97,ba98};
detailed discussions of the full calculation, including non-resonant
corrections, can be found in ~\cite{ba00,di00}. 
Since in ${\cal O}(\alpha^2)$ only two-photon bremsstrahlung \cite{ba99}
has been studied so far, the following discussion is restricted to 
${\cal O}(\alpha)$ corrections.

The algebraic structure of the virtual corrections allows for a 
factorisation of the one-loop amplitude ${\cal M}_1$ into the Born 
amplitude ${\cal M}_0$ and a relative correction factor $\delta^{\virt}$. 
Thus, in ${\cal O}(\alpha)$ the correction to the squared amplitude reads
\beq
|{\cal M}_0+{\cal M}_1|^2 = (1+2\Re\{\delta^{\virt}\}) |{\cal M}_0|^2 + \dots.
\eeq
Since only the real part of $\delta^{\virt}$ appears, there is no 
double-counting of the ${\cal O}(\alpha)$ correction that is
already included in ${\cal M}_0$ by the $i\MW\Gamma_\PW$ term.
Moreover, the factorisation trivially avoids potential problems with
gauge-invariance after the introduction of the $W$~decay width in the
resonant terms. Besides the Breit-Wigner factor in $|{\cal M}_0|^2$,
there are logarithmic terms $\ln(\hat s-\MW^2)$ in $\delta^{\virt}$ 
which are singular on resonance. The consistent replacement 
$\ln(\hat s-\MW^2)\to\ln(\hat s-\MW^2+i\Gamma_\PW\MW)$
accounts for a Dyson summation of resonant $W$~propagators in loop diagrams,
without introducing problems with gauge-invariance.

The real corrections are included by adding the lowest-order cross
section for the process $u\bar d\to \nu_l l^+ +\gamma$.
The only non-trivial condition induced by gauge-invariance is
the Ward identity for the external photon, {\it i.e.}\ electromagnetic
current conservation. If the $W$~width is zero, this identity is trivially 
fulfilled. This remains  true even for a constant width, since the
$W$~boson mass appears only in the $W$~propagator denominators, {\it i.e.}\ the
substitution $\MW^2\to\MW^2-i\MW\GW$ is a consistent
reparametrisation of the amplitude in this case.
However, if a running $W$~width is introduced naively, {\it i.e.}\ in the
$W$~propagators only, the Ward identity is violated.
The identity can be restored by taking into account those part of the
fermion-loop correction to the $\gamma WW$ vertex that corresponds to 
the fermion loops in the $W$~self-energy leading to the width in the
propagator \cite{be97,ba95,arg95}. For an external photon, this
modification simply amounts to the multiplication of the $\gamma WW$
vertex by the factor
$f_{\gamma WW}|_{\mathrm{run}} = 1+i\GW/\MW$.

Adding virtual and real corrections, all IR divergences cancel. Mass
singularities of the form $\alpha\ln\Ml$ related to a final-state lepton 
drop out for all observables in which photons within a collinear cone
around the lepton are treated inclusively, in accordance with the KLN 
theorem. As already discussed in 
Section~\ref{sec:ewqedcor}
(see also~\cite{ba98}), mass
singularities to the initial-state quarks are absorbed into renormalised
quark distribution functions.

As long as one is
interested in observables that are dominated by resonant $W$~boson
production, the radiative corrections can be approximated by the
corrections to the production and decay subprocesses to resonant
$W$~bosons. Formally such an approximation can be carried out by a
systematic expansion of all amplitudes about the resonance pole and is,
therefore, called {\it pole approximation} (PA). 
In PA, the virtual correction consists of two parts. 
The first contribution is provided by the (constant) correction factors
to the $Wf\bar f'$ vertex for stable (on-shell) $W$~bosons and is called 
{\it factorisable}. The second contribution, which is called 
{\it non-factorisable}, comprises all remaining resonant corrections. 
It is entirely due to photonic effects and includes, in particular, the 
$\ln(\hat s-\MW^2+i\Gamma_\PW\MW)$ terms. The difference between PA
and the exact result can be estimated by 
$\delta^{\virt}_{\PA}-\delta^{\virt} \sim 
(\alpha/\pi)\ln(\hat s/\MW^2)\ln(\dots)$, 
where $\ln(\dots)$ indicates any logarithmic enhancements.
In principle, also the real corrections can be treated in PA. However, since
a reliable error estimate is not obvious, they are usually
calculated exactly. More details about  PA can be found 
in~\cite{ho97,di00}.

\subsubsection{Numerical results}

The following numerical results have been obtained with the input
parameters of~\cite{di00} and a constant $W$~width; in particular, we
have $\MW=80.35\GeV$ and $\GW=2.08$~GeV.
The QED factorisation is performed in the $\overline{\mbox{MS}}$ scheme 
with $\MW$ being the factorisation scale, and the CTEQ4L \cite{cteq4} 
quark distributions are used in the evaluation of the $pp$ cross
section. For the partonic cross section, the CKM matrix element $V_{ud}$
is set to 1; for the $pp$ cross section a non-trivial CKM matrix is
included in the parton luminosities (see~\cite{di00}).

\begin{figure}[htbp]
\centerline{
\setlength{\unitlength}{1cm}
\begin{picture}(16,7.6)
\put(0,0){\includegraphics{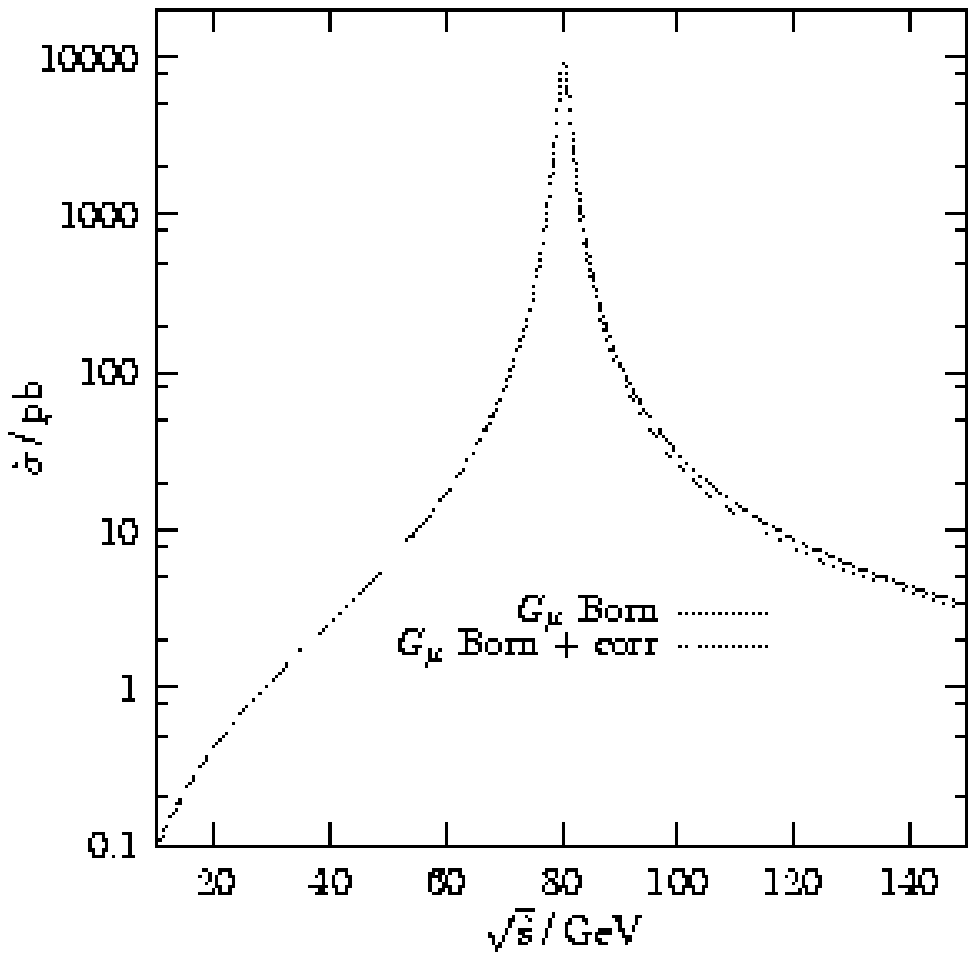}}
\put(8,0){\includegraphics{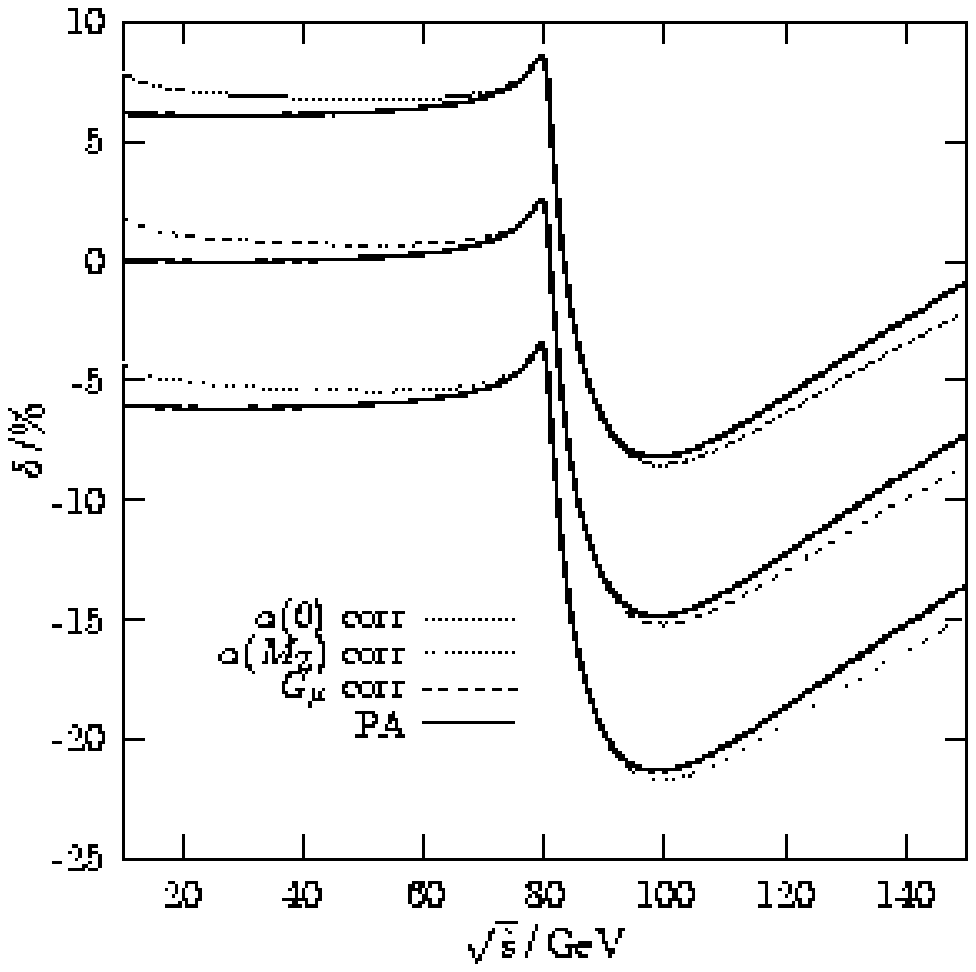}}
\put(1.5,6.5){\small $u\bar d\to\nu_l l^+(+\gamma)$}
\end{picture}
} 
\caption{Total parton cross section $\hat\sigma$ in $\GF$ 
parametrisation and relative corrections $\delta$ for different
parametrisations (results based on \cite{di00}).}
\label{fig:Wparton}
\end{figure}

\begin{table}[htbp]
\begin{center}
\caption{Total lowest-order parton cross section $\hat\sigma_0$ in $\GF$ 
parametrisation and corresponding relative correction $\delta$, 
exact and in PA (results based on \cite{di00}).}
\label{tab:Wparton}
\vskip0.2cm
\begin{tabular}{lccccccc}
\hline
$\sqrt{\hat s}$ (GeV) & 40 & 80 & 120 & 200 & 500 & 1000 & 2000 
\\  \hline
$\hat\sigma_0$ (pb) & 2.646 & 7991.4 & 8.906 & 1.388 & 0.165 & 
0.0396 & 0.00979
\\  
$\delta$ (\%) & 0.7 & 2.42 & $-12.9$ & $-3.3$ & 12 & 19 & 23
\\  
$\delta_\PA$ (\%) & 0.0 & 2.40 & $-12.3$ & $-0.7$ & 18 & 31 & 43
\\  \hline
\end{tabular}
\end{center}
\end{table}

Figure~\ref{fig:Wparton} shows the total partonic cross section $\hat\sigma$
and the
corresponding relative correction $\delta$ for intermediate energies. Note 
that the total cross section and its correction is the same for all
final-state leptons $l=e,\mu,\tau$ in the limit of vanishing
lepton masses. As expected, the $\GF$ parametrisation of the Born cross
section minimises the correction at low energies, since the universal
corrections induced by the running of $\alpha$ and by the $\rho$
parameter are absorbed in the lowest-order cross section. Moreover, the
naive error estimate for the PA taken from above 
turns out to be realistic. The PA 
describes the correction in the resonance region within a few per mille.
Table~\ref{tab:Wparton} contains some results on the partonic cross section 
and its correction up to energies in the TeV range. Far above resonance,
the PA cannot follow the exact correction anymore, since non-resonant
corrections become more and more important. The leading corrections are
due to Sudakov logarithms of the form $\alpha\ln^2(\hat s/\MW^2)$.

\begin{figure}[htbp]
\centerline{
\setlength{\unitlength}{1cm}
\begin{picture}(16,7.6)
\put(0,0){\includegraphics{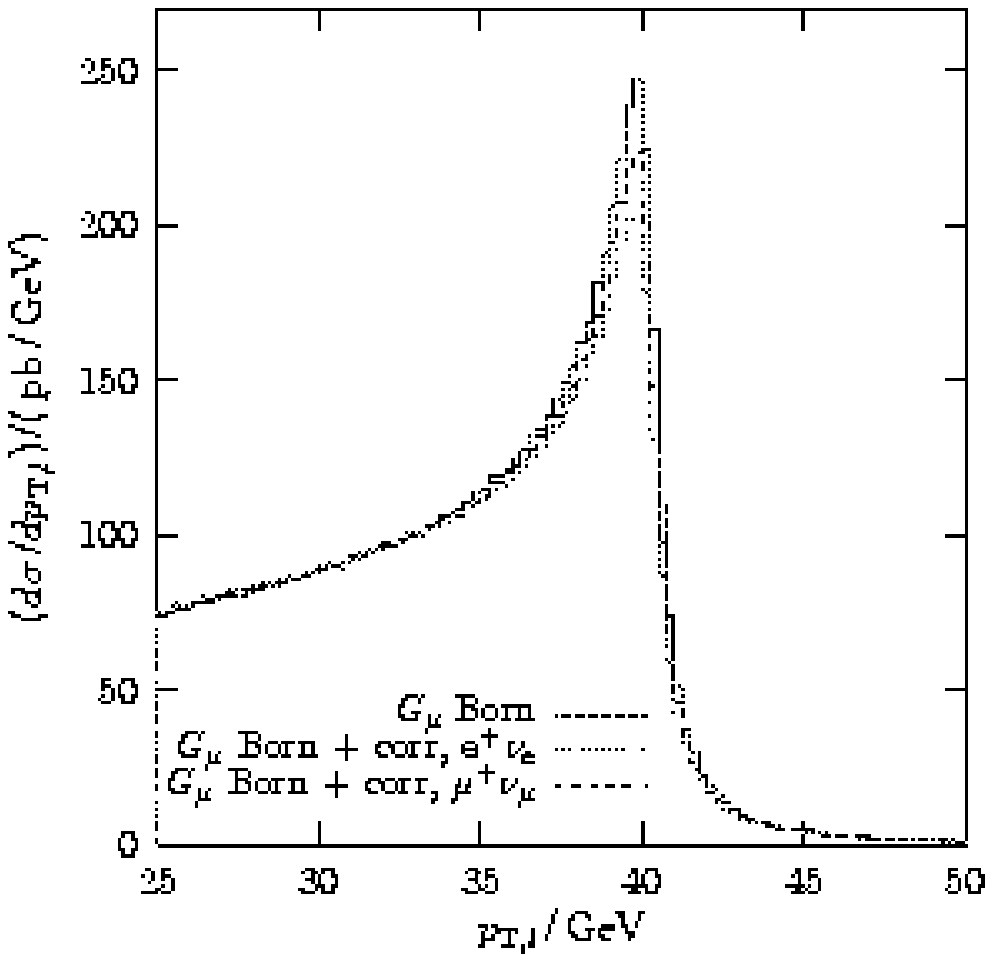}}
\put(8,0){\includegraphics{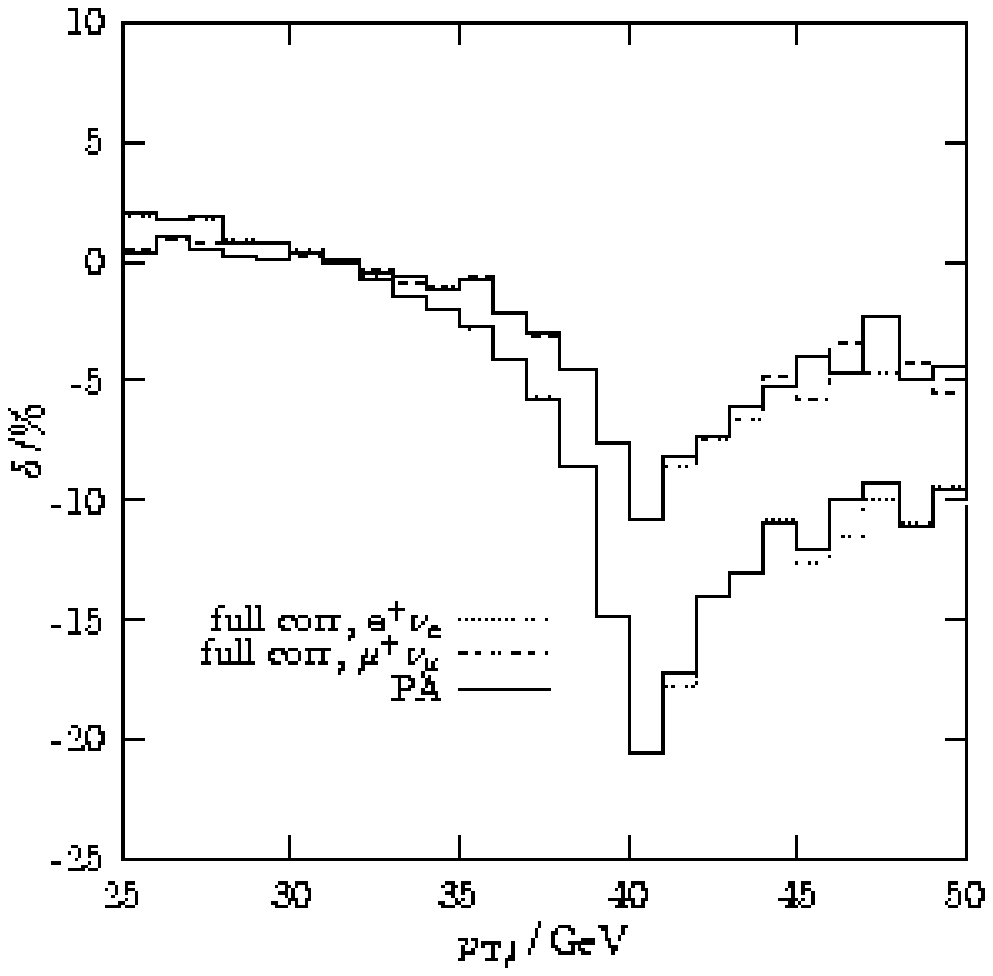}}
\put(1.5,6.6){\small $\Pp\Pp\to\nu_l l^+(+\gamma)$}
\put(1.5,6.0){\small $\sqrt{s}=14\TeV$}
\put(1.5,5.4){\small $p_{\mathrm{T},l},\dsl{p}_{\mathrm{T}}>25\GeV$}
\put(1.5,4.8){\small $|\eta_l|<1.2$}
\end{picture}
} 
\caption{Transverse-momentum distribution 
$(d\sigma/dp_{\mathrm{T},l})$ and relative corrections $\delta$ 
(results based on \cite{di00}).}
\label{fig:Wpp}
\end{figure}

\begin{table}[htbp]
\begin{center}
\caption{Integrated lowest-order $pp$ cross sections $\sigma_0$ 
for different ranges in $p_{\mathrm{T},l}$ and corresponding 
relative corrections $\delta$, 
exact and in PA (results based on \cite{di00}).}
\label{tab:Wpp}
\vskip0.2cm
\begin{tabular}{lcccccc}
\hline
$p_{\mathrm{T},l}$ (GeV) & 
25--$\infty$ & 25--45 & 45--$\infty$ & 50--$\infty$ & 80--$\infty$ & 
200--$\infty$ 
\\  \hline
$\sigma_0$ (pb) & 
1933.3(2) & 1909.9(2) & 23.52(5) & 11.47(2) & 1.682(3) & 0.1014(1)
\\  
$\delta_{e^+\nu_e}$ (\%) &  
$-5.51(5)$ & $-5.45(7)$ & $-11.8(5)$ & $-9.7(4)$ & $-11.7(3)$ & $-17.7(2)$
\\  
$\delta_{e^+\nu_e,\PA}$ (\%) &  
$-5.51(5)$ & $-5.45(7)$ & $-10.9(5)$ & $-8.2(3)$ & $-8.3(3)$ & $-9.0(2)$
\\  
$\delta_{\mu^+\nu_\mu}$ (\%) &  
$-2.98(5)$ & $-2.94(7)$ & $-6.3(6)$ & $-5.7(4)$ & $-8.1(3)$ & $-14.2(3)$
\\  
$\delta_{\mu^+\nu_\mu,\PA}$ (\%) &  
$-2.97(5)$ & $-2.94(7)$  & $-5.7(6)$ & $-4.6(4)$ & $-4.9(3)$ & $-5.6(2)$
\\  \hline
\end{tabular}
\end{center}
\end{table}
Figure~\ref{fig:Wpp} shows the transverse-momentum distribution for the
lepton $l^+$ produced in $pp\to W^+\to\nu_l l^+(+\gamma)$ for the
$pp$ CM energy $\sqrt{s}=14\TeV$ of the LHC. 
The transverse momenta $p_{\mathrm{T}}$ and the lepton pseudorapidity $\eta_l$
are restricted by $p_{\mathrm{T},l},\dsl{p}_{\mathrm{T}}>25\GeV$ and
$|\eta_l|<1.2$.
Since we do not recombine
collinear photons and leptons, the corrections for different leptons do
not coincide, but differ by corrections of the form $\ln(\Ml/\MW)$.
In the total cross section without any cuts these logarithms cancel, 
and the correction
is again universal for all leptons in the massless limit.
Since the $\ln\Ml$ corrections are strongest for electrons, and since
collinear photon emission reduces the momentum of the produced
lepton, the correction $\delta$ for electrons is more negative (positive) 
for large (small) momenta than in the case of the muon.
In particular, Figure~\ref{fig:Wpp} demonstrates the reliability of the
PA for transverse lepton momenta $p_{\mathrm{T},l}\lsim \MW/2$, where
resonant $W$~bosons dominate. However, high $p_{\mathrm{T},l}$ values may
also be interesting in searches for new physics. Table~\ref{tab:Wpp}
shows the contributions to the total cross section divided by different
ranges in $p_{\mathrm{T},l}$. From the above discussion of the parton
cross section it is clear that the PA is not applicable for very large
$p_{\mathrm{T},l}$, where the $W$~boson is far off shell.

The above results underline the importance of electroweak radiative
corrections in a precise description for the $W$~boson cross section
at the LHC. Although the corrections of ${\cal O}(\alpha)$ are well
under control now, there are still some topics to be studied, such as
the impact of realistic detector cuts and photon recombination
procedures or the inclusion of higher-order effects.


The impact of final state photon radiation on $W$ observables strongly
depends on the lepton identification requirements imposed by the
experiment. In addition to the lepton $p_T$, $p\llap/_T$ and
pseudorapidity cuts, one usually imposes requirements on the separation
of the charged lepton and the photon. For muons, the energy of the
photon is required to be less than a critical value, $E^\gamma_c$, in a cone of
radius $R^\mu_c$ around the muon. For electrons, the finite resolution of
the electromagnetic calorimeter makes it difficult to separate electrons 
and photons for small opening angles between the particles. Their four
momentum vectors are therefore recombined if their separation is smaller 
than a critical value $R_c^e$. Finally, uncertainties in the energy and
momentum measurements of the charged lepton and the missing transverse
energy need to be taken into account. They can be simulated by Gaussian
smearing of the particle four-momentum vectors with standard deviation
$\sigma$ which depends on the particle type and the detector.

To illustrate how finite detector resolution affects
the size of the electroweak corrections, we show in
Figure~\ref{fig:ptrw} the ratio of the NLO and lowest-order cross sections 
as a function of the $p_T$ of the electron in 
$pp\to \nu_e e^+(\gamma)$ obtained with the
Monte Carlo generator {\tt WGRAD}~\cite{ba98}.  
The solid histogram
shows the cross section ratio taking only transverse-momentum and
pseudorapidity cuts into account. The dashed histogram displays the result 
obtained when, in addition, the four-momentum vectors are smeared
according to the ATLAS specifications~\cite{atlas-phystdr2},
and electron and
photon momenta are combined if 
$\Delta R(e,\gamma)<0.07$~\cite{atlas-phystdr2}.
Recombining the electron and photon four-momentum
vectors eliminates the mass-singular logarithmic terms of the form
$\alpha\ln m_e$, and strongly reduces the size of the electroweak
corrections. 

  \begin{figure}
\begin{center}
    \includegraphics[width=0.48\textwidth,clip]{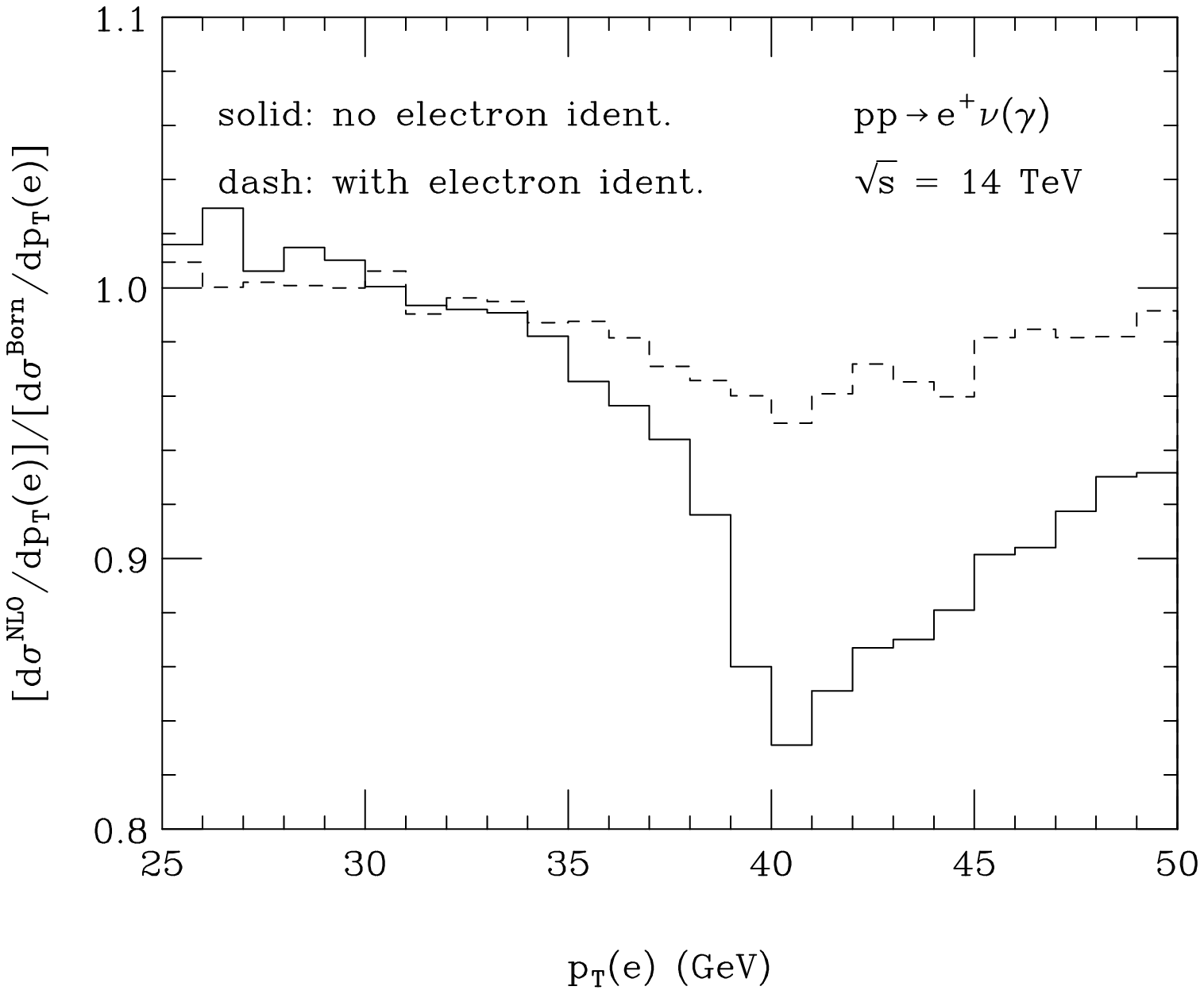}
    \caption{Ratio of the \protect{${\cal O}(\alpha^3)$} and lowest order
             differential cross sections as a function of the 
             transverse momentum of the electron with and without 
             lepton identification requirements
             (results based on~\cite{ba98}). 
             The cuts imposed are described in the text.}
    \label{fig:ptrw}
\end{center}
  \end{figure}

\subsection{Electroweak corrections to \boldmath $Z$ \unboldmath production and
           continuum neutral-current processes}

\subsubsection{QED corrections \label{sec:moreewqedcor}}

The mass-singular universal QED corrections from initial-state
radiation from quarks have already been discussed in 
Section~\ref{sec:ewqedcor}. 
They are part of the quark distribution functions.
The residual QED initial-state corrections,
together with final-state corrections and interference of
initial-final radiation are treated separately by an explicit
diagrammatic computation.   

A complete calculation of the QED ${\cal O}(\alpha)$ radiative
corrections to $p p \to Z,\gamma \to l^+ l^-$ has been carried out 
in~\cite{Baur:1998wa}.  The calculation is based on an explicit
diagrammatic approach. The collinear singularities associated with
initial-state photon radiation are factorised into the parton
distribution functions (see Section~\ref{sec:ewqedcor}). Absorbing the 
initial-state mass singularities into
the pdf's introduces a QED factorisation-scale dependence. The results
presented here are obtained within the QED DIS scheme which is defined
analogously to the QCD DIS factorisation scheme. 
The MRS(A) parton distributions are used, with a factorisation scale $M_Z$.
Due to mass-singular logarithmic terms associated
with photons emitted collinear with one of the final-state leptons, QED
radiative corrections strongly affect the shape of the di-lepton 
invariant mass distribution, the lepton transverse momentum spectrum,
and the forward-backward asymmetry, $A_{\rm FB}$. 

The effect of the QED
corrections on the di-muon invariant mass distribution in the region
$45$~GeV $<m(\mu^+\mu^-)<105$~GeV is shown in
Figure~\ref{fig:mllratio}a where we plot the ratio of the ${\cal
O}(\alpha^3)$ and lowest-order differential cross sections as a function 
of $m(\mu^+\mu^-)$. The lowest-order cross section has been evaluated in 
the effective Born approximation (EBA) which already takes into account those
higher-order corrections which can be absorbed into a redefinition of
the coupling constants and the effective weak mixing angle. More details 
on the EBA can be found in Section~\ref{sec:nqcebd}.
In the region shown in the figure, the cross-section ratio is seen
to vary rapidly. Below the
$Z$ peak, QED corrections significantly enhance the cross section. At
the $Z$ pole, the differential cross section is reduced by about 20\%. 
Photon radiation from one of the leptons lowers the di-lepton invariant
mass. Therefore, events from the $Z$ peak region are shifted towards 
smaller values of $m(\mu^+\mu^-)$, thus reducing the
cross section in and above the peak region, and increasing the rate
below the $Z$ pole. Final-state radiative corrections 
completely dominate over the entire mass range considered. They are
responsible for the strong modification of the 
di-lepton invariant mass distribution. In contrast, initial-state
corrections are uniform and small ($\approx +0.4\%$ in the QED DIS
scheme). 

  \begin{figure}[htbp]
    \includegraphics[width=0.48\textwidth,clip]{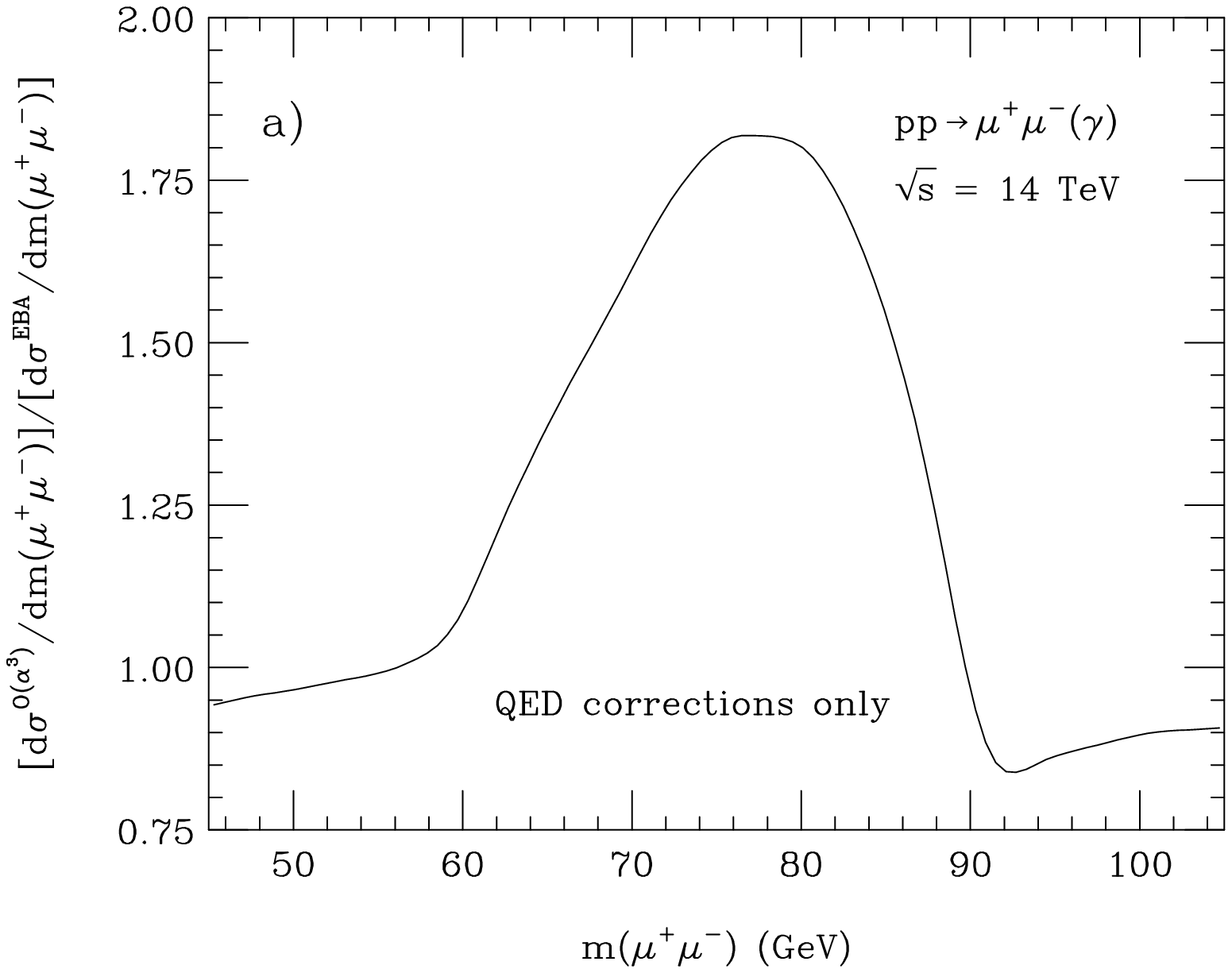}
    \includegraphics[width=0.48\textwidth,clip]{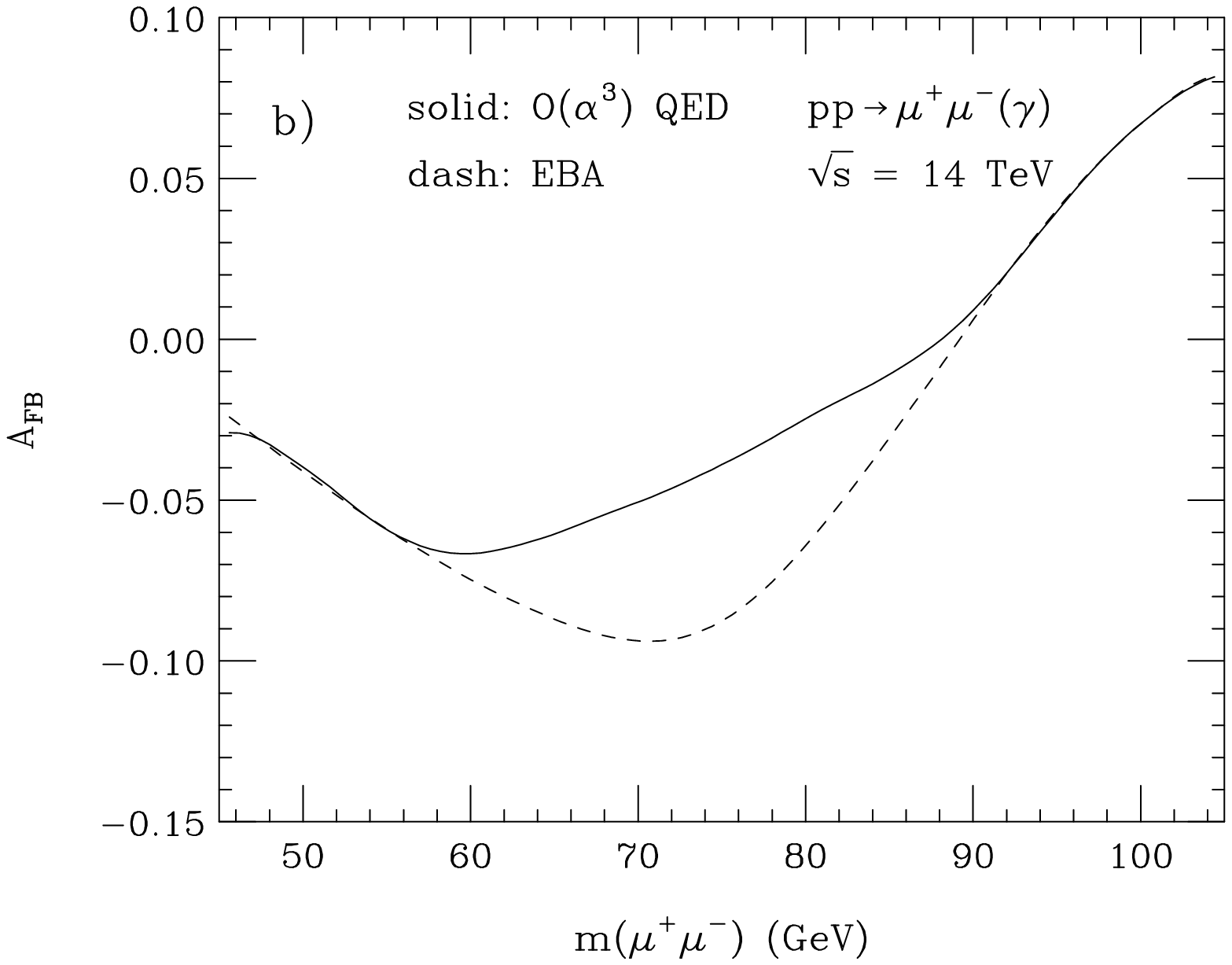}
    \caption{Ratio of the \protect{${\cal O}(\alpha^3)$} and lowest-order
differential cross sections, and the forward-backward asymmetry,
$A_{\rm FB}$, as a function of the $\mu^+\mu^-$ invariant mass. The cuts
imposed are described in the text.}
    \label{fig:mllratio}
  \end{figure}

As pointed out earlier, at the LHC a precise measurement of the
effective mixing angle $\sin^2 \theta_{\rm eff}^{\rm lept}$ using the 
forward-backward asymmetry may be possible. In Figure~\ref{fig:mllratio}b, 
the forward-backward 
asymmetry is shown in the EBA (dashed line), and including QED
corrections (solid line) for $pp\to\mu^+\mu^-(\gamma)$ in the di-muon invariant
mass range from 45~GeV to 105~GeV. Here, $A_{\rm FB}$ is 
defined by~\cite{Baur:1998wa}
\begin{equation}
A_{\rm FB}={F-B\over F+B}
\label{EQ:DEFAFB}
\end{equation}
where
\begin{equation}
F=\int_0^1{{\rm d}\sigma\over {\rm d}\cos\theta^*}\,{\rm d}\cos\theta^*, \qquad
B=\int_{-1}^0{{\rm d}\sigma\over {\rm d}\cos\theta^*}\,{\rm d}\cos\theta^*.
\label{EQ:DEFFB}
\end{equation}
$\cos\theta^*$ is given by
\begin{equation}
\cos\theta^*={|p_z(\mu^+\mu^-)|\over p_z(\mu^+\mu^-)}~{2\over 
m(\mu^+\mu^-)\sqrt{m^2(\mu^+\mu^-)
+p_T^2(\mu^+\mu^-)}}\left [p^+(\mu^-)p^-(\mu^+)-p^-(\mu^-)
p^+(\mu^+)\right ]
\label{EQ:CSTAR}
\end{equation}
in the Collins-Soper frame~\cite{colsop}, with
\begin{equation}
p^\pm={1\over\sqrt{2}}\left (E\pm p_z\right ),
\end{equation}
where $E$ is the energy and $p_z$ is the longitudinal component of the
momentum vector. As expected, the ${\cal O}(\alpha)$ 
QED corrections to $A_{\rm FB}$ are large in the region below the $Z$ peak. 
Since events from the $Z$ peak, where $A_{\rm FB}$ is positive and small,
are shifted towards smaller values of $m(\mu^+\mu^-)$ by photon 
radiation, the forward-backward asymmetry is significantly reduced in 
magnitude by radiative corrections for $55~{\rm 
GeV}<m(\mu^+\mu^-)<90$~GeV. It should be noted that the 
forward-backward asymmetry 
is rather sensitive to the rapidity cuts imposed on
the leptons. More details on $A_{\rm FB}$ and the measurement of the
effective weak mixing angle can be found in Section~\ref{sec:AFB}.

The mass singular terms arising from final-state photon radiation
are proportional to $\alpha\log(\hat s/m_l^2)$, where $m_l$ is the
lepton mass. Thus, the corrections to the $Z$ line shape and $A_{\rm FB}$ for 
electrons in the final state  are considerably larger than those in 
the muon case~\cite{Baur:1998wa}. 

To simulate detector acceptances, 
we have imposed a $p_T(\mu)>20$~GeV and a $|\eta(\mu)|< 3.2$ cut in 
Figure~\ref{fig:mllratio}. Except for the threshold
region, the effects of the lepton acceptance cuts approximately cancel
in the cross section ratio. In a more realistic simulation of how QED
corrections affect observables in Drell-Yan production, lepton and
photon identification requirements need to be taken into account in
addition to the lepton acceptance cuts. Muons are identified in a hadron 
collider detector by hits in the muon chambers. In addition to a hit in 
the muon chambers, one requires that the associated track is
consistent with a minimum ionising particle. This limits the energy of a
photon which traverses the same calorimeter cell as the muon to be 
smaller than a critical value $E^\gamma_c$. For electrons, the finite 
resolution of the electromagnetic calorimeter makes 
it difficult to separate electrons and photons for small opening angles
between their momentum vectors. Therefore, electron and photon four-momentum
vectors are  recombined if their separation in the azimuthal
angle--pseudorapidity plane is smaller than a critical value, $R_c$. This
eliminates the mass-singular terms associated with final-state photon
radiation (KLN theorem) and thus may  reduce significantly the effect QED
corrections have on physical observables in $pp\to
e^+e^-(\gamma)$~\cite{Baur:1998wa}. Specific results sensitively depend on the 
value of $R_c$, which is detector dependent. 

\subsubsection{Non-QED corrections and effective Born description
\label{sec:nqcebd}}

The amplitude for the parton process 
$q(p)+\bar{q}(\bar{p}) \rightarrow l^+(k_+)+ l^-(k_-)$
of quark-antiquark annihilation into charged-lepton pairs
is in lowest order described by photon and $Z$~boson exchange.
In the kinematical variables for the parton system
\beq
 \hat{s} = (k_+ + k_-)^2, \quad
       t = (p-k_-)^2, \quad 
       u = (p-k_+)^2
\eeq
the differential parton cross section can be written as follows
($\theta$ denotes the scattering angle in the parton CMS):
\beq
\label{diffxsec}
     64 \pi^2 \hat{s} \,
    \frac{{\rm d}\hat{\sigma}}{{\rm d}\Omega}\, = \,
     2\,  {\cal A}_0\, \frac{u^2+t^2}{\hat{s}^2}
        + {\cal A}_1\, \frac{u^2-t^2}{\hat{s}^2} \, 
     = \, {\cal A}_0\, (1+\cos^2\theta) \, + \, {\cal A}_1 \, \cos\theta
\eeq
with
\begin{eqnarray}
\label{Zborn}
 {\cal A}_0 & = & Q_q^2 Q_l^2\, e(\hat{s})^4 \, + \, 
           2 v_q v_l Q_q Q_l\, e(\hat{s})^2\,  
           {\rm Re}\, \chi(\hat{s})\, + \,
           (v_q^2+a_q^2)(v_l^2+a_l^2)\, |\chi(\hat{s})|^2 , \nonumber \\
 {\cal A}_1 &= & 4 Q_q Q_l a_q a_l\,  e(\hat{s})^2\, 
          {\rm Re}\, \chi(\hat{s}) \, + \,
          8 v_q a_q v_l a_l\,  |\chi(\hat{s})|^2 \, . 
\end{eqnarray}
This expression is an effective Born approximation, which incorporates
several entries from higher-order calculations:
the effective (running) electromagnetic charge containing the 
photon vacuum polarisation (real part) 
\beq
 e(\hat{s})^2 = \frac{4\pi\alpha}{1-\Delta\alpha(\hat{s})}\,  ;
\eeq
the $Z$ propagator, together with the overall normalisation factor of the
neutral-current couplings in terms of the Fermi constant $G_\mu$,  
\beq
 \chi(\hat{s}) = (G_\mu M_Z^2 \sqrt{2})^2\, 
                 \frac{\hat{s}}{\hat{s}-M_Z^2 + 
                  {\rm i}\hat{s} \Gamma_Z/M_Z } \, ,
\eeq 
containing the $Z$ width as measured from the $Z$ resonance at LEP;
the vector and axial-vector coupling constants  for $f=l, q$    
\beq 
v_f = I_3^f -2 Q_f \sin^2\theta_{\rm eff} , \quad
a_f = I_3^f ,
\eeq
which contain the effective (leptonic) mixing angle at the $Z$ peak,
which is measured at LEP and SLC. Taking $\Gamma_Z$ and $\seff$ from
higher-order calculations, the formulae above yield a good description
in the region around the $Z$ resonance.

From the cross section (\ref{diffxsec}) a   forward-backward
asymmetry for the produced $l^+l^-$ system can be derived,
 which at the parton level is given by
\beq
\label{afb}
 \hat{A}_{\rm FB} = \frac{\hat{\sigma}_{\rm F} - \hat{\sigma}_{\rm B}}
                   {\hat{\sigma}_{\rm F} + \hat{\sigma}_{\rm B}}
            = \frac{3}{8} \frac{{\cal A}_1}{{\cal A}_0} \, .
\eeq
Around the $Z$ peak, this quantity depends sensitively on $\seff$.
Using a parametrisation of the Born-like expressions in 
Equation~\ref{Zborn}, a measurement of $\hat{A}_{\rm FB}$ allows 
a determination of the mixing angle (see Section~\ref{sec:ewprecision}).
Below we give a quantitative evaluation
of the higher-order electroweak effects
in the integrated cross section and in $\hat{A}_{\rm FB}$ to demonstrate
the quality of the approximation around the $Z$ pole and to
point out deviations at higher invariant masses of the lepton pairs.    

Besides the universal and non-universal QED corrections, the following 
IR-finite next-order electroweak terms contribute, which are 
schematically depicted in Figure \ref{zdiagrams}:
self-energy contributions to the photon and $Z$ propagators,
vertex corrections to the 
$\gamma/Z$-$ll$ and $\gamma/Z$-$q\bar{q}$ 
3-point couplings, and box diagrams with two massive~boson exchanges.
Details of the 
treatment of the resonance region at higher order is equivalent to
that in $e^+e^-$ annihilation in fermion pairs and can be found 
{\it e.g.} in \cite{Yellowbook95}.  
Around the $Z$ pole, the box graphs are negligible, but they increase
strongly with the energy and hence contribute sizeably at high invariant
masses of the lepton pair. A description in terms of an effective-Born
cross section far away from the $Z$ pole becomes insufficient for
two reasons: the effective couplings (based on self-energies and
vertex corrections only) are not static but grow as functions of $\hat{s}$,
and the presence of the box contributions, which cannot be
absorbed in effective vector and axial-vector couplings in a Born-like
structure.
\begin{figure}[htb]
\vspace*{0.5cm}
\center{
\begin{tabular}{p{5.2cm}@{\hspace{0.3cm}\hspace{0.3cm}}p{5.2cm}}
\centerline{\epsfig{figure=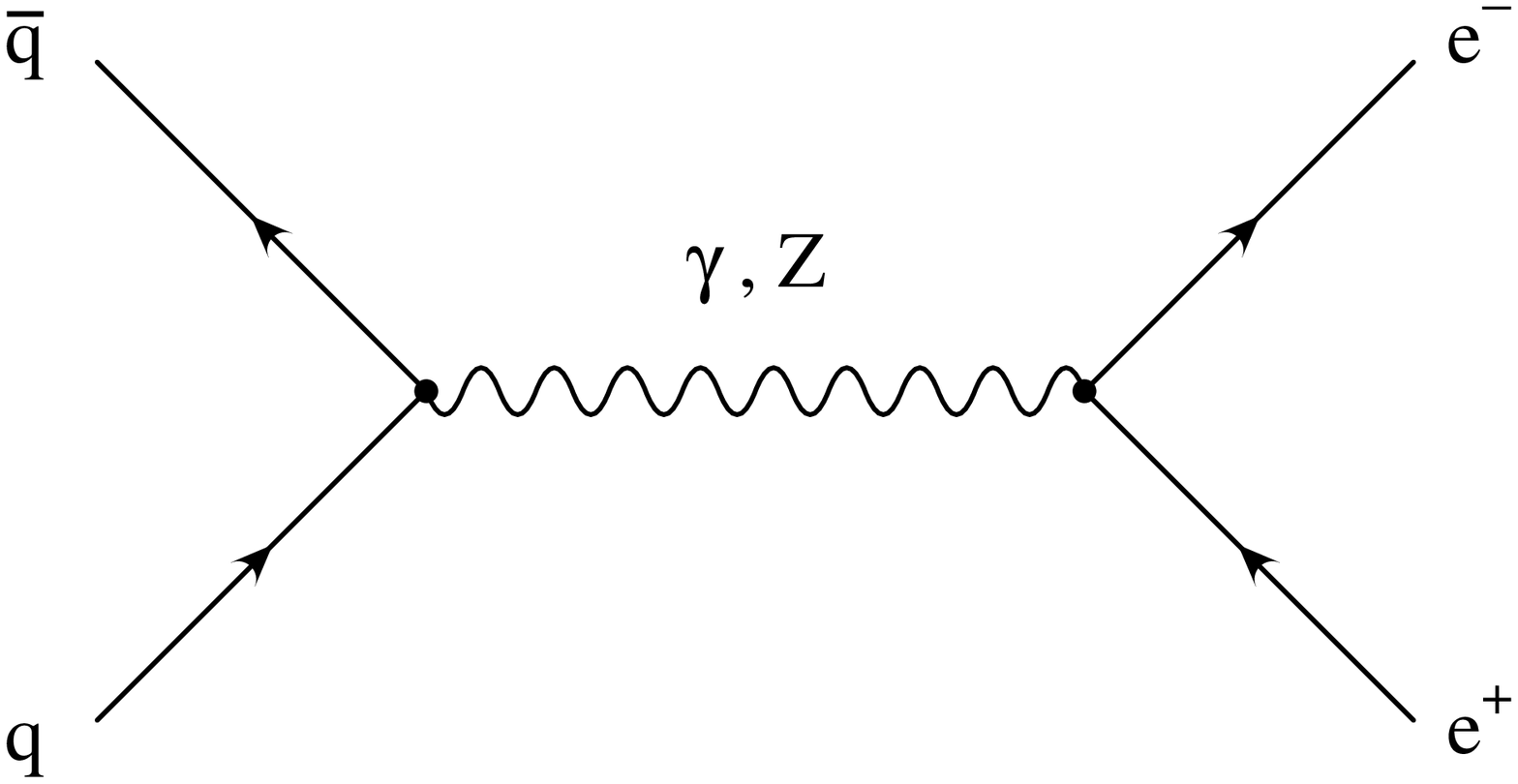,width=4cm}} &
\centerline{\epsfig{figure=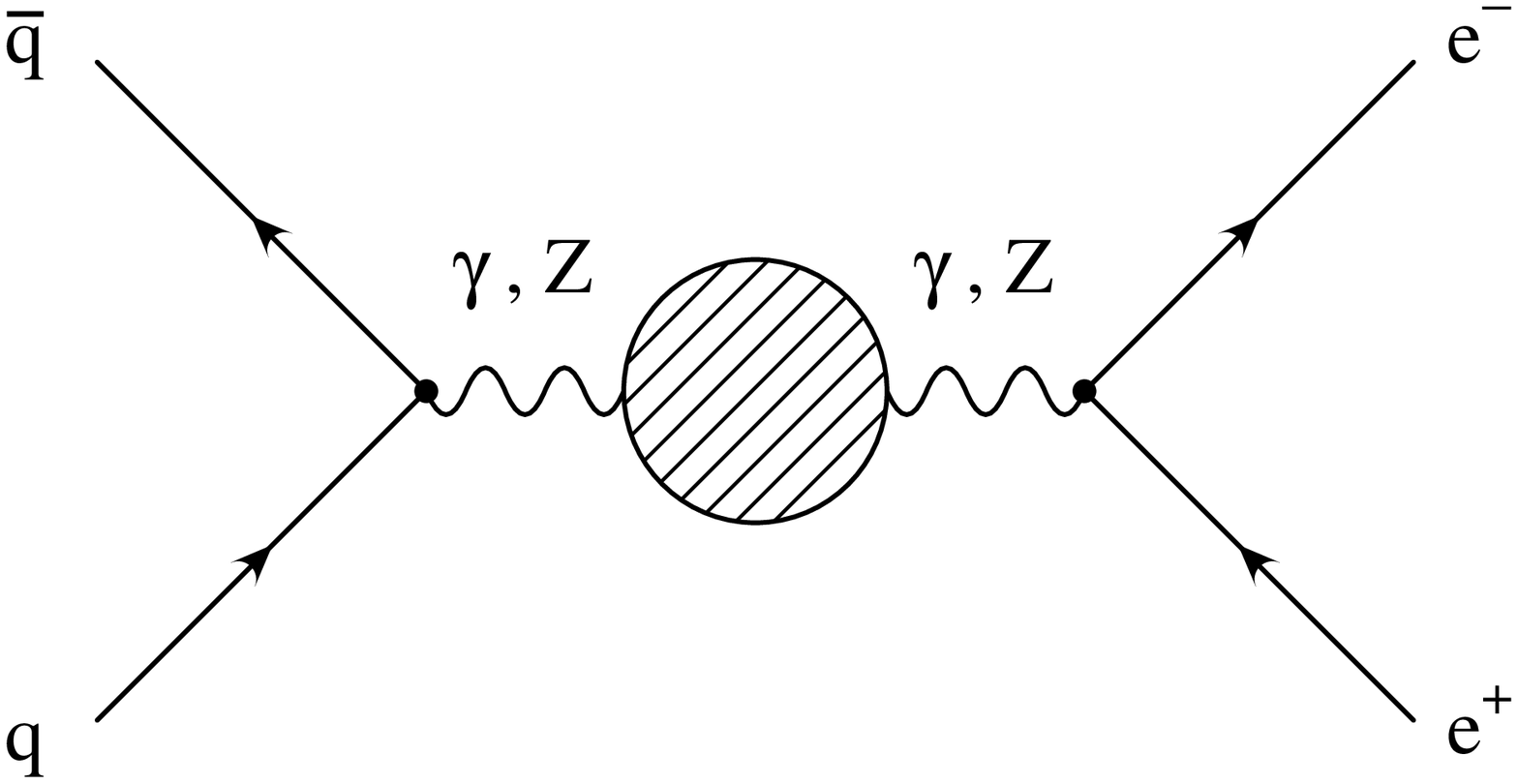,width=4cm}} \\[-0.3cm]
\centerline{\epsfig{figure=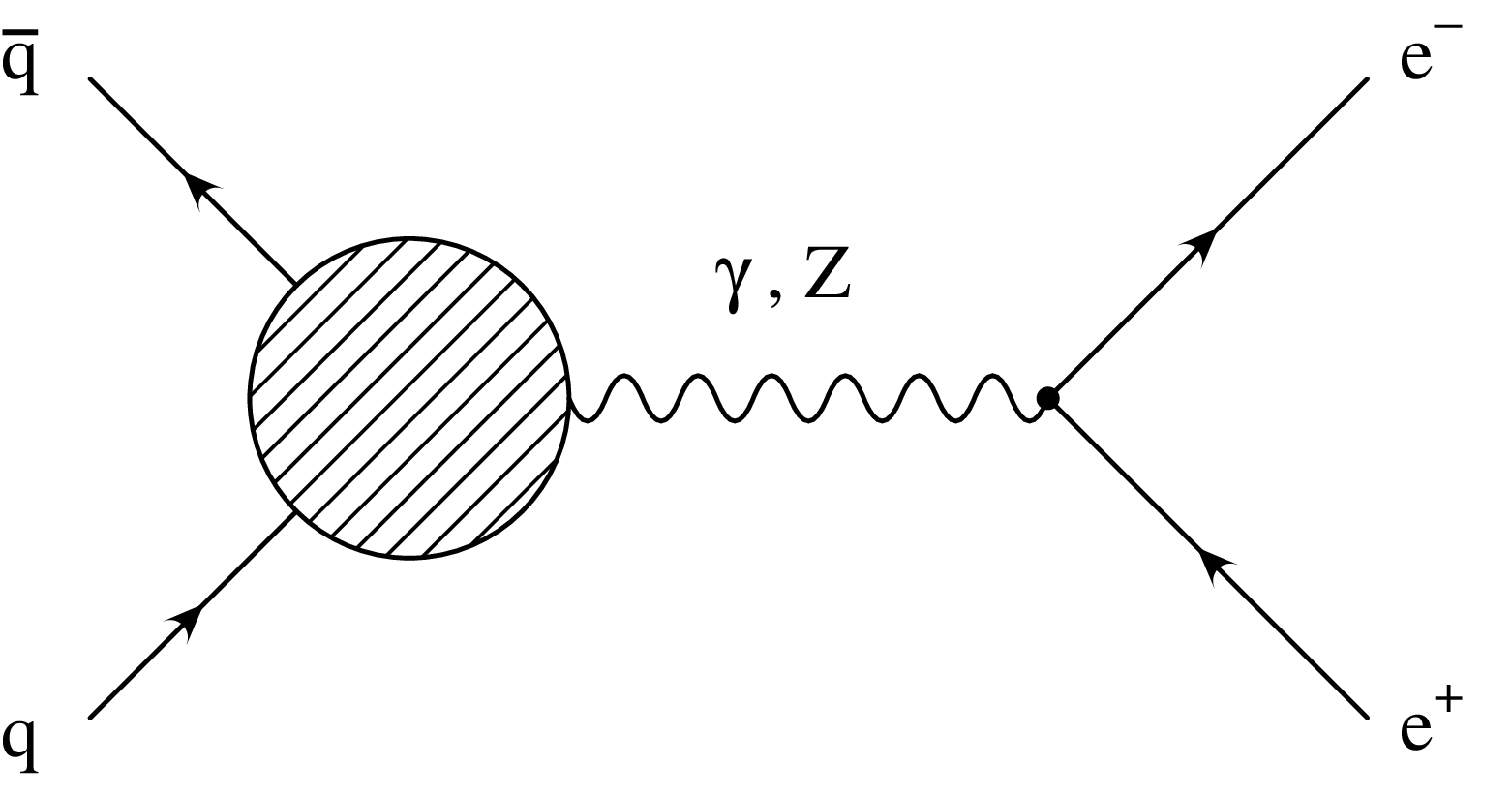,width=4cm}} &
\centerline{\epsfig{figure=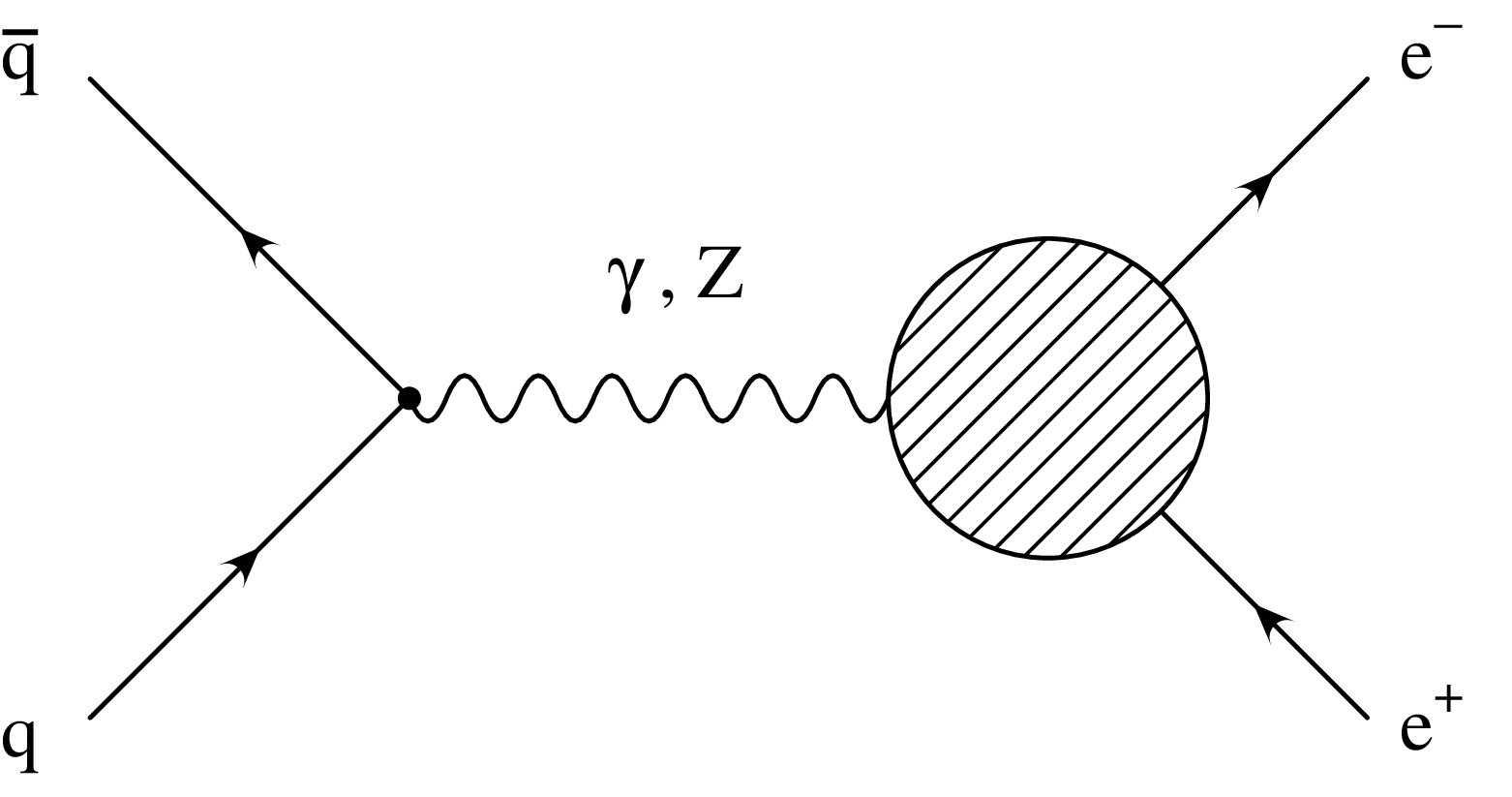,width=4cm}} \\[-0.3cm]
\centerline{\epsfig{figure=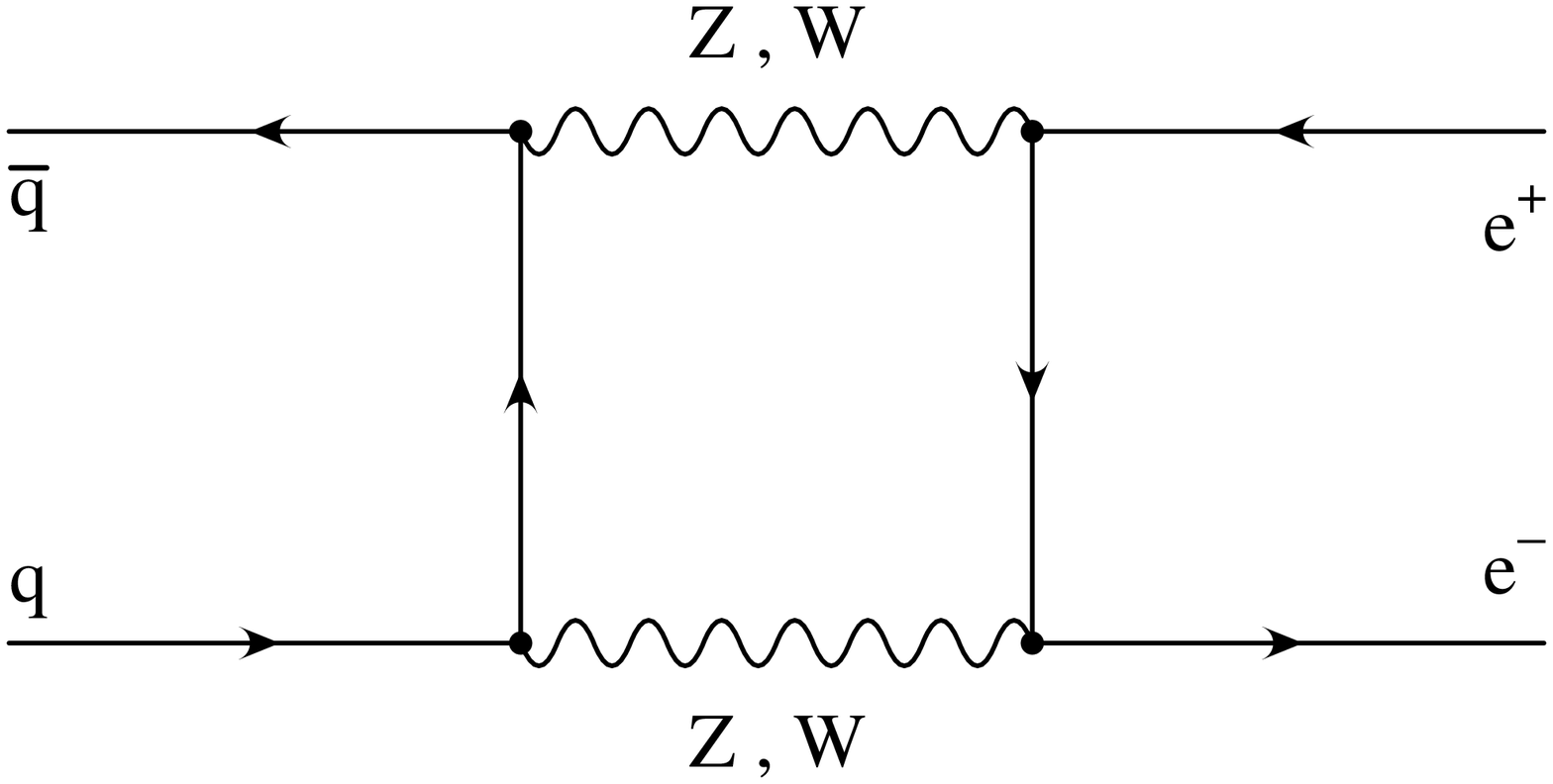,width=4cm}} &
\centerline{\epsfig{figure=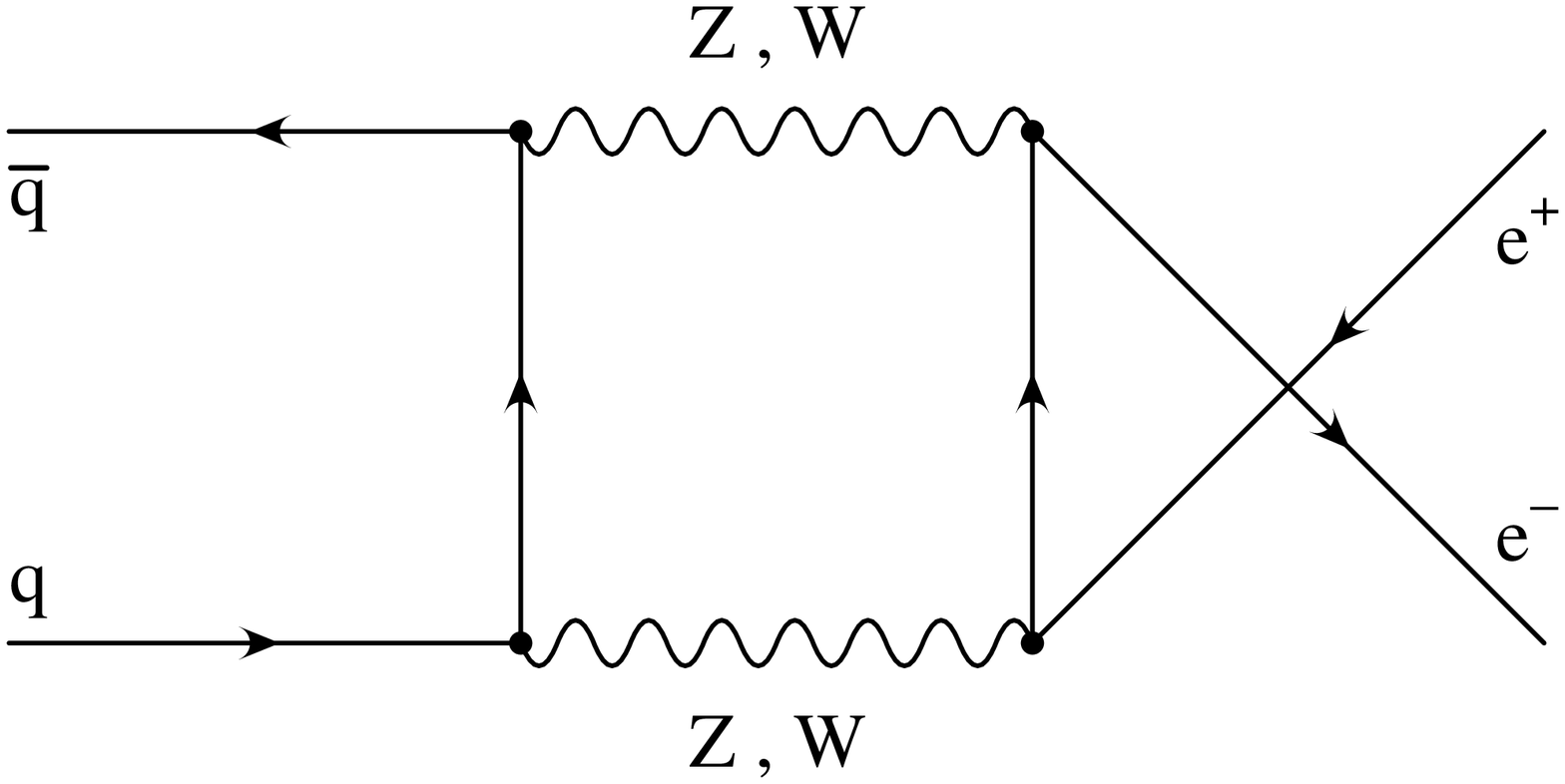,width=4cm}}
\end{tabular}
\vspace*{-1cm}
\caption[]{Born and higher-order electroweak contributions to
         $q\bar{q} \rightarrow e^+ e^-$ in symbolic notation.}
\label{zdiagrams} }
\vspace*{0.5cm}
\end{figure}

In Figures \ref{z123} and \ref{z45}
we compare the integrated cross section $\hat{\sigma}$ 
and the asymmetry $A_{\rm FB}$ at the parton level in the 
approximation corresponding to Equations~\ref{diffxsec} and~\ref{Zborn} with
results obtained by a complete one-loop calculation with proper
treatment of higher-order terms around the $Z$ resonance.
For demonstrational purpose, the effect of the box diagrams is
displayed separately. As one can see, the region where the effective Born
description starts to become unsatisfactory is at rather high   
values of the parton energy. 

\begin{figure}[htbp]
\centerline{\epsfig{figure=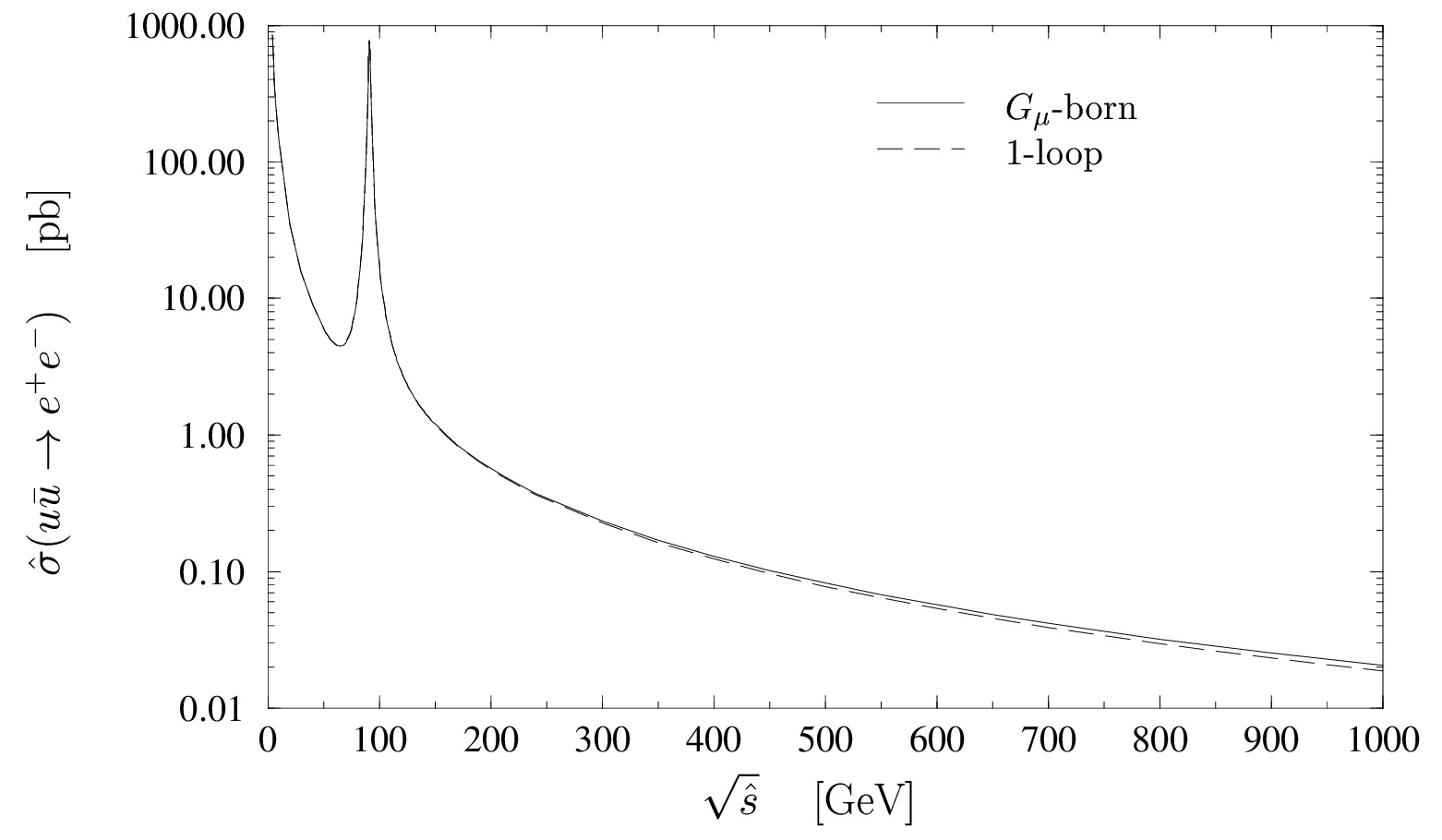,width=10cm}} 
\centerline{\epsfig{figure=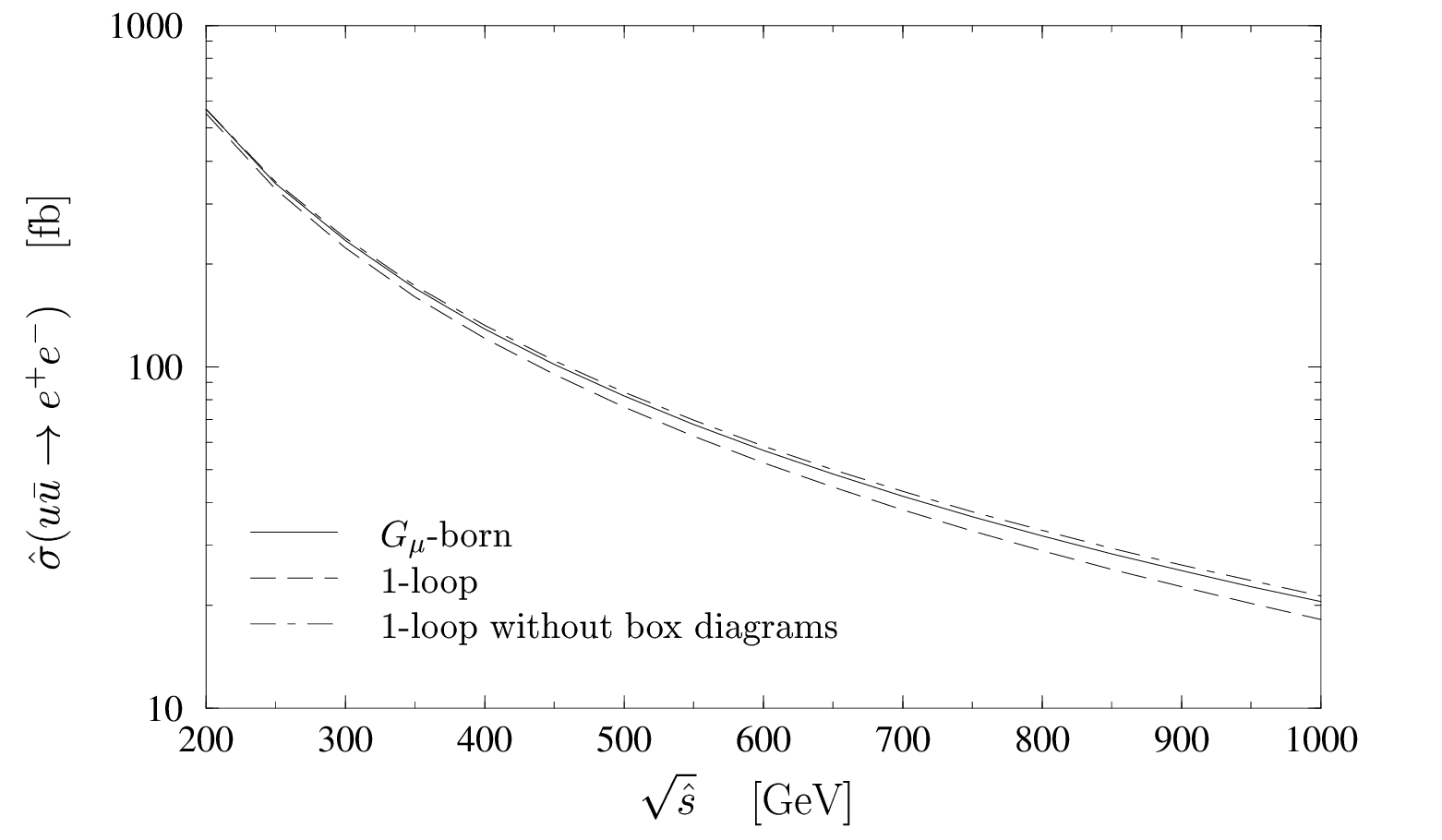,width=10cm}} 
\centerline{\epsfig{figure=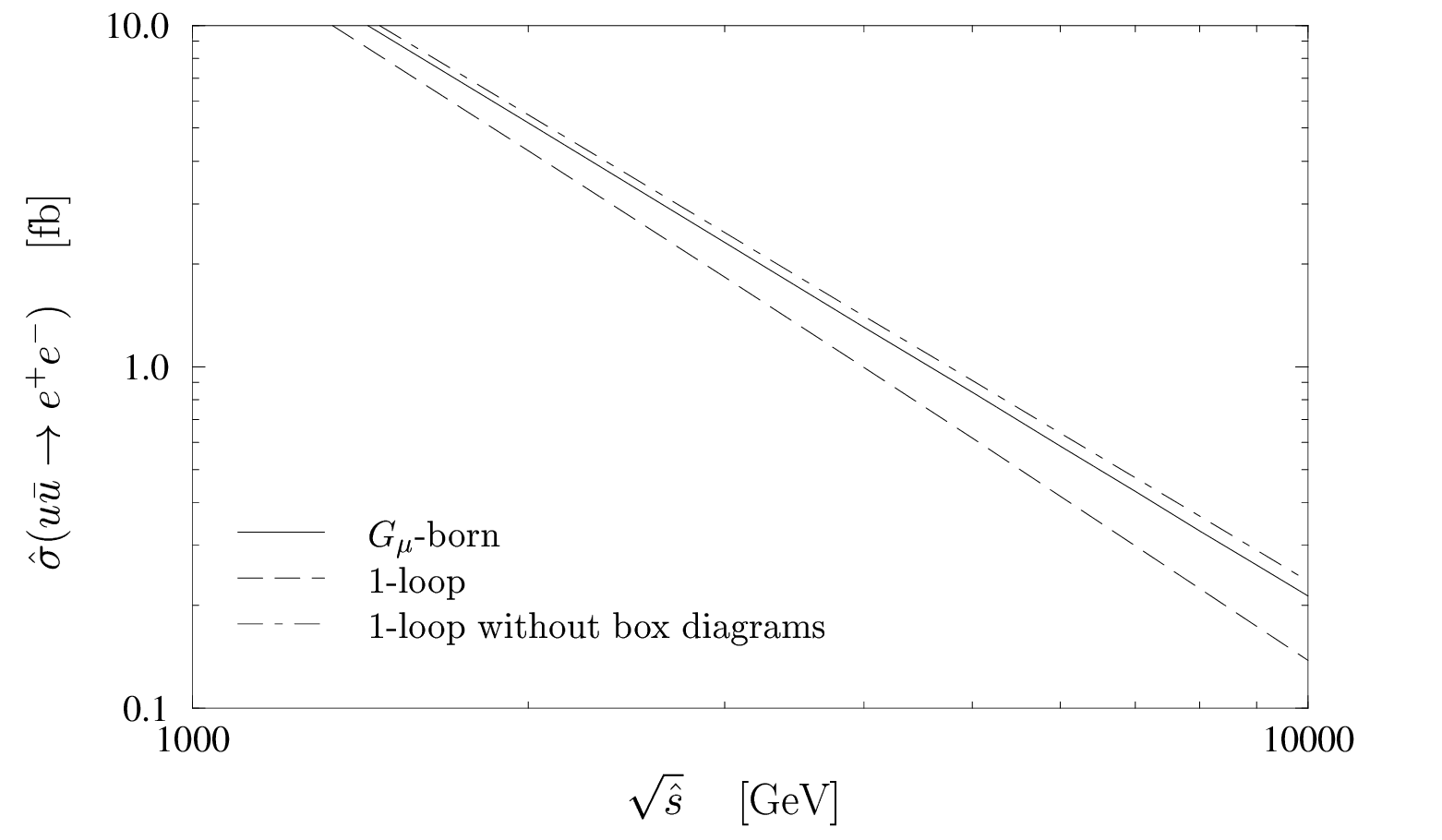,width=10cm}} 
\caption[]{$u\bar{u} \to e^+e^-$. Energy dependence of $\hat{\sigma}$
  in various steps of the approximation. 
  $M_H=100$ GeV and $m_t=174$ GeV.}
\label{z123}
\end{figure}

\begin{figure}[htb]
\centerline{\epsfig{figure=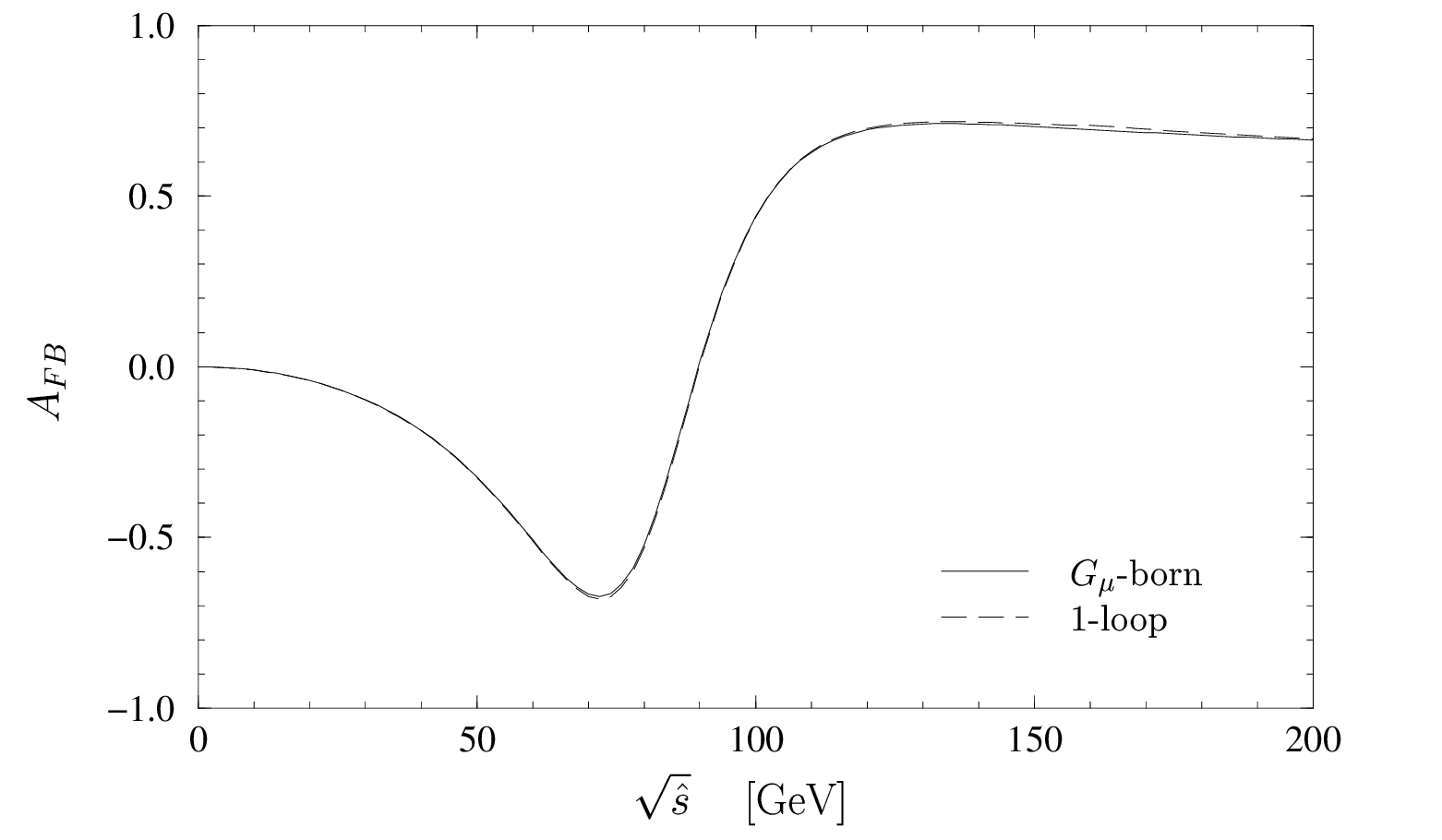,width=10cm}} 
\centerline{\epsfig{figure=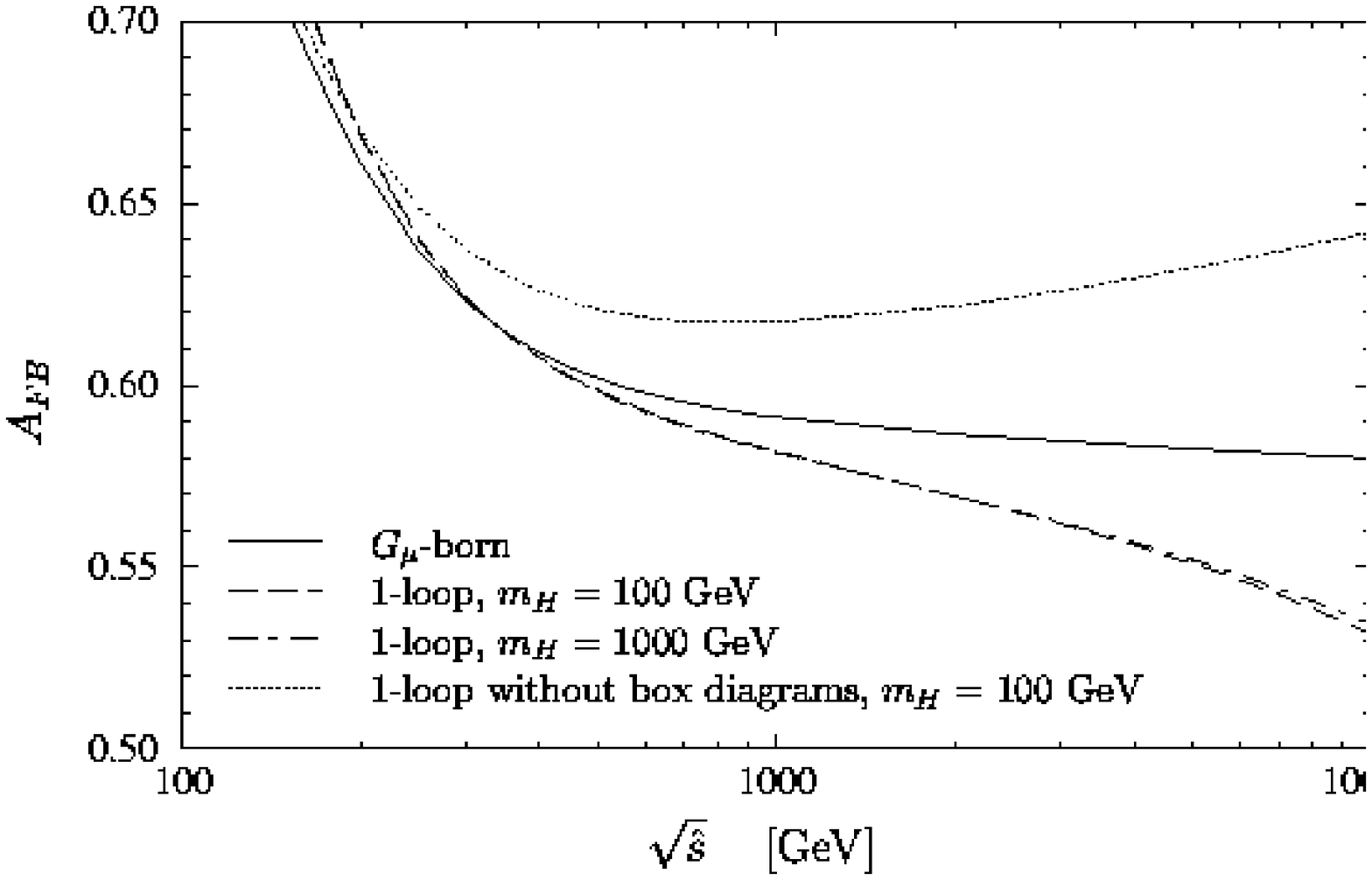,width=10cm}} 
\caption[]{$u\bar{u} \to e^+e^-$. Energy dependence of 
  $\hat{A}_{FB}$ at the parton level, for 
  $m_t=174$ GeV and different values of the Higgs mass,
  in various steps of the approximation.}
\label{z45}
\end{figure}

In order to give an idea of the effects remaining in the
hadronic cross section after convolution with the quark 
distribution functions,  Table~\ref{Zhadronic} contains the
relative deviations of the cross section based on the higher-order
parton results from those based on the Born approximation 
Equation~\ref{Zborn}.
Also listed are the estimated experimental accuracies with which the
cross section in the various bins can be measured. The comparison shows
that at high invariant masses the radiative corrections remain sizeable
and should be taken into account for studies at high $\hat{s}$,
for example in the search for new physics effects originating from a heavy
extra gauge-boson $Z'$.

\begin{table}[htbp]
\begin{center}
\caption[]{Hadronic cross section for $e^+e^-$ pairs with invariant mass
in certain energy ranges. Columns two and three show the predicted 
cross sections in the effective Born approximation and the full one-loop 
calculation. Columns four and five show the relative corrections 
to the effective Born approximation arising from the full 
one-loop calculation as well as the estimated 
experimental errors for the cross section measurements in the
various bins. }
\label{Zhadronic}
\vskip0.2cm
\begin{tabular}[5]{lcccc}
\hline
Energy range  & Born  & Full &  Relative correction & Relative experimental \\
(for $e^+e^-$ pairs)   & cross section & cross section & to Born cross section
 & error \\
(TeV) & (fb) & (fb) & (\%) & (\%) \\
\hline
0.9  - 1.1        &      6.2299     &     5.6524      &    - 9.3       &
3\\
1.1  - 1.5        &      3.5205     &     3.1491      &    -11.0       &
4\\
1.5  - 1.75       &      0.6076     &     0.5317      &    -12.5       &
9.5\\
1.75 - 2.0        &      0.2681     &     0.2314      &    -13.7       &
14\\
2.0  - 2.5        &      0.1886     &     0.1590      &    -15.7       &
17\\
2.5  - 3.0        &      0.04895    &     0.04031     &    -17.7       &
30\\
3.0  - 4.0        &      0.01837    &     0.01464     &    -20.3       &
50\\
\hline
\end{tabular}
\end{center}
\end{table}


\subsubsection{The full electroweak ${\cal O}(\alpha)$ corrections: Monte
Carlo simulations with {\tt ZGRAD2}}

The QED corrections described in Section~\ref{sec:moreewqedcor} 
have been combined with the
weak corrections summarised in the previous section in a new Monte Carlo 
program called {\tt ZGRAD2}~\cite{BBHSW}. 
In Figure~\ref{fig:mll}a we show the ratio of the full ${\cal O}(\alpha^3)$ 
electroweak and the ${\cal O}(\alpha^3)$ QED differential cross sections 
for $pp\to\mu^+\mu^-(\gamma)$ obtained with {\tt ZGRAD2} as a function
of the $\mu^+\mu^-$ 
invariant mass. As in Section~\ref{sec:moreewqedcor}, 
we have imposed a $p_T(\mu)>20$~GeV and 
a $|\eta(\mu)|<3.2$ cut, and used the EBA to evaluate the lowest-order
contribution to the ${\cal O}(\alpha^3)$ QED cross section. Thus, the ratio 
directly displays the effect of the weak box-diagrams and the
energy dependence of the weak coupling form factors. While the additional
weak contributions only change the differential cross section by 0.6\% at
most, they do modify the shape of the $Z$ resonance curve.

Figure~\ref{fig:mll}b compares the effect of the ${\cal O}(\alpha^3)$ QED 
corrections and the full ${\cal O}(\alpha^3)$ electroweak corrections
on the di-muon invariant mass distribution for $m(\mu^+\mu^-)$ values
between 200~GeV and 2~TeV. Due to the presence of logarithms of the form
$\log(\hat s/M_Z^2)$, the weak
corrections become significantly larger than the QED corrections at
large values of $m(\mu^+\mu^-)$, and, eventually, may have to be
resummed~\cite{KPS}. For $m(\mu^+\mu^-)=2$~TeV, the full ${\cal
O}(\alpha^3)$ electroweak corrections are found to reduce the
differential cross section by more than 20\%.

 \begin{center}
  \begin{figure}[htbp]
    \includegraphics[width=0.48\textwidth,clip]{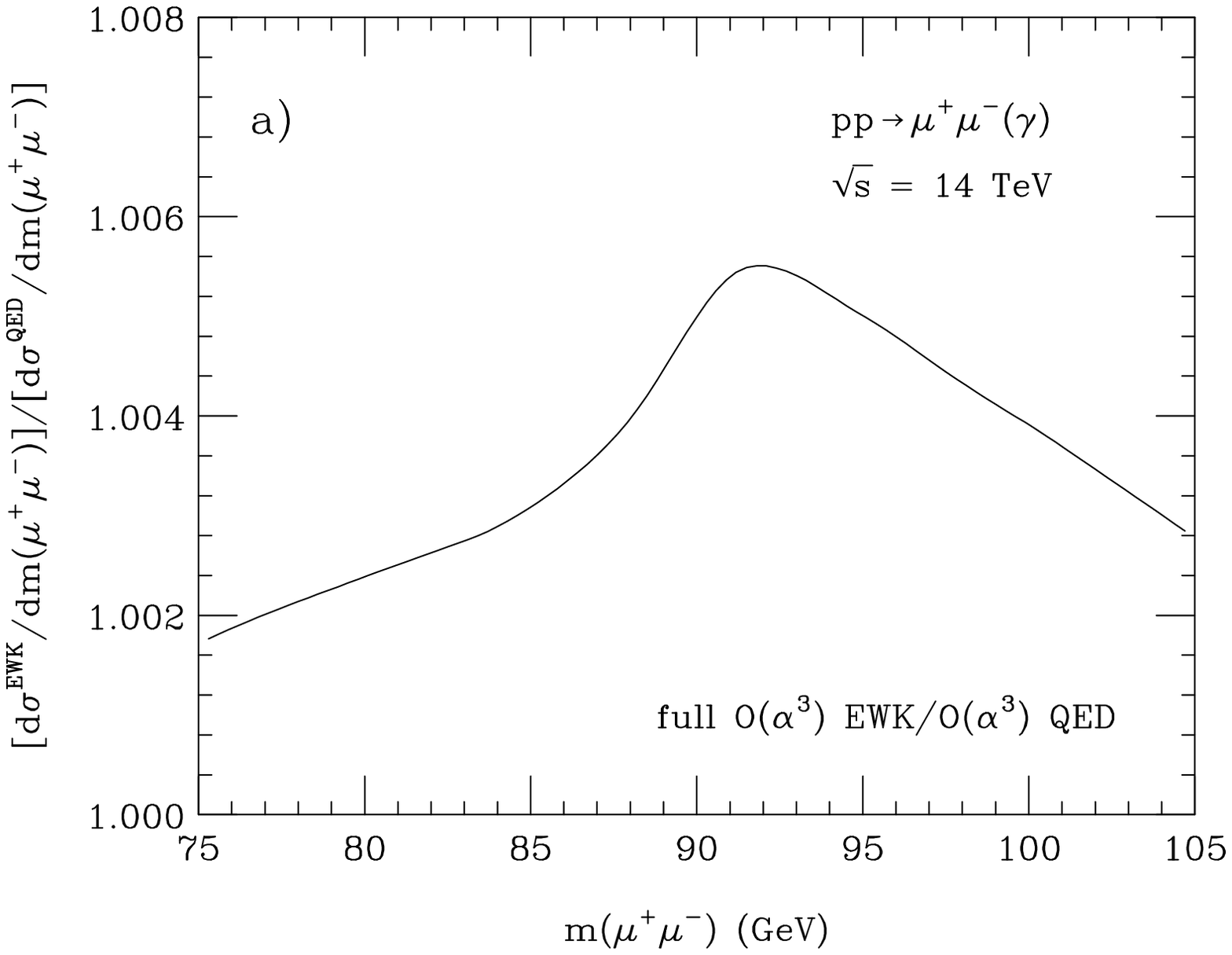}
   \includegraphics[width=0.48\textwidth,clip]{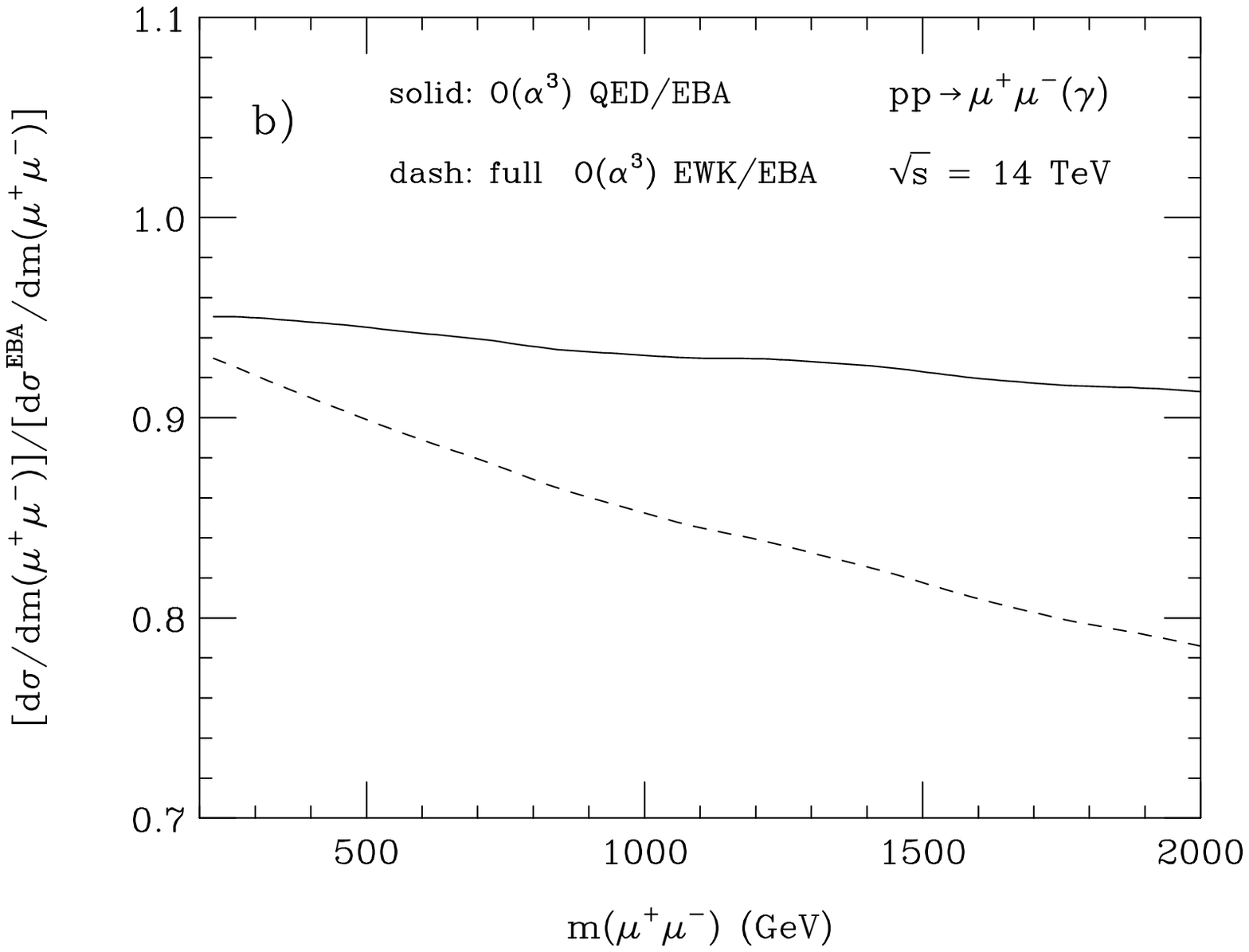}
    \caption{a) Ratio of the full \protect{${\cal O}(\alpha^3)$}
electroweak  and
the \protect{${\cal O}(\alpha^3)$} QED differential cross sections in
the vicinity of the $Z$ pole. b) Differential cross section ratios,
displaying the size of the full \protect{${\cal O}(\alpha^3)$} electroweak and
the \protect{${\cal O}(\alpha^3)$} QED corrections for large 
values of $m(\mu^+\mu^-)$. The cuts imposed are described in the text.}  
  \label{fig:mll}
  \end{figure}
\end{center}

Finally, in Figure~\ref{fig:afb} we show how the ${\cal O}(\alpha^3)$
corrections affect the forward-backward asymmetry 
(see Equations~\ref{EQ:DEFAFB} to~\ref{EQ:CSTAR}). 
Both QED and weak corrections reduce $A_{\rm FB}$, and
their size increases with growing di-muon masses. For
$m(\mu^+\mu^-)=2$~TeV, the weak corrections are about twice as large as 
the QED corrections. Note that the electroweak corrections affect
$A_{\rm FB}$ much less than the lepton pair invariant mass distribution. In
the $Z$ pole region, $75~{\rm GeV}<m(\mu^+\mu^-)<105$~GeV, the weak
corrections change the forward-backward asymmetry by at most $5\times
10^{-4}$. Results qualitatively similar to those shown in
Figures~\ref{fig:mll} and~\ref{fig:afb} are obtained for $pp\to
e^+e^-(\gamma)$. 

\begin{center}
  \begin{figure}[htbp]
    \includegraphics[width=0.48\textwidth,clip]{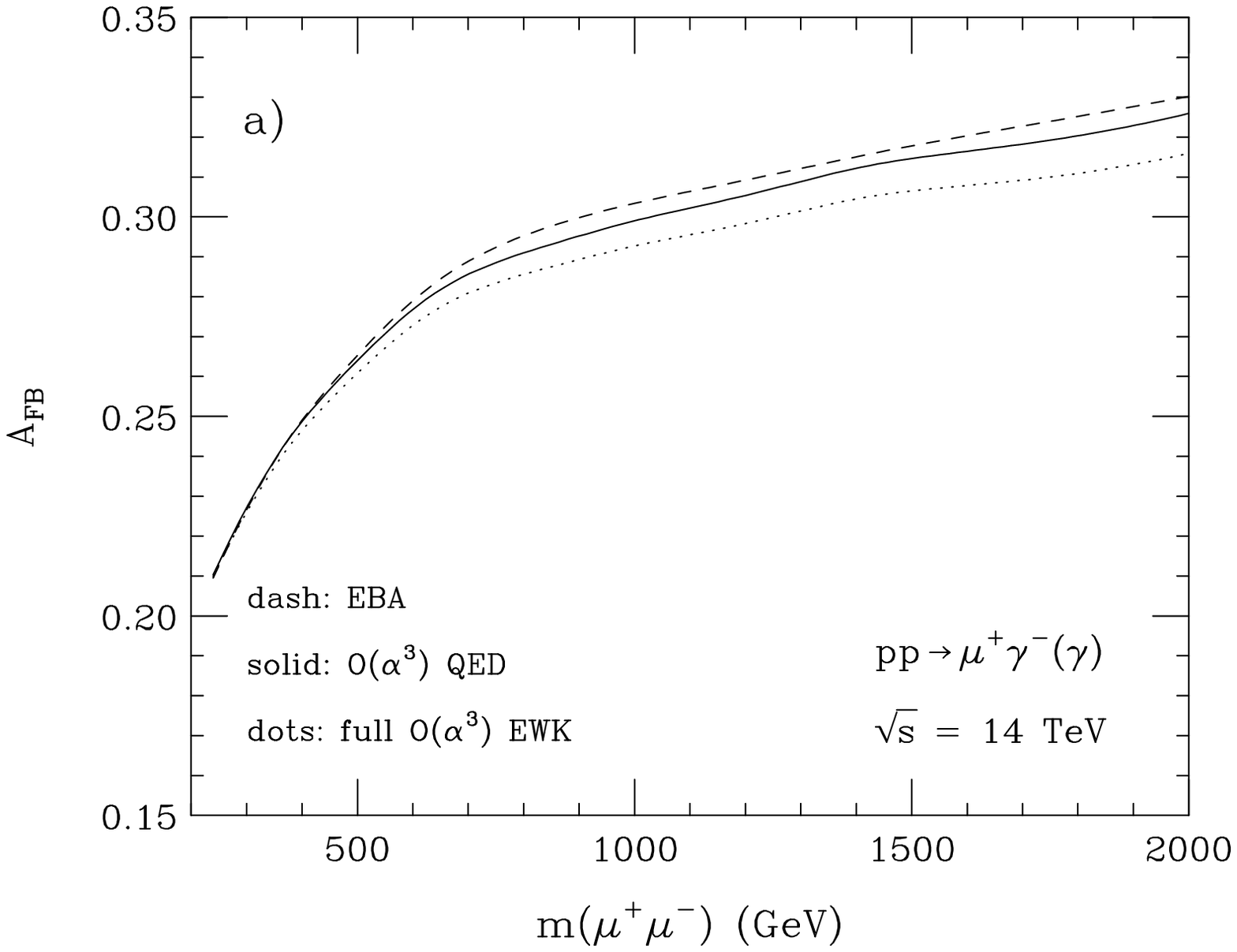}
   \includegraphics[width=0.48\textwidth,clip]{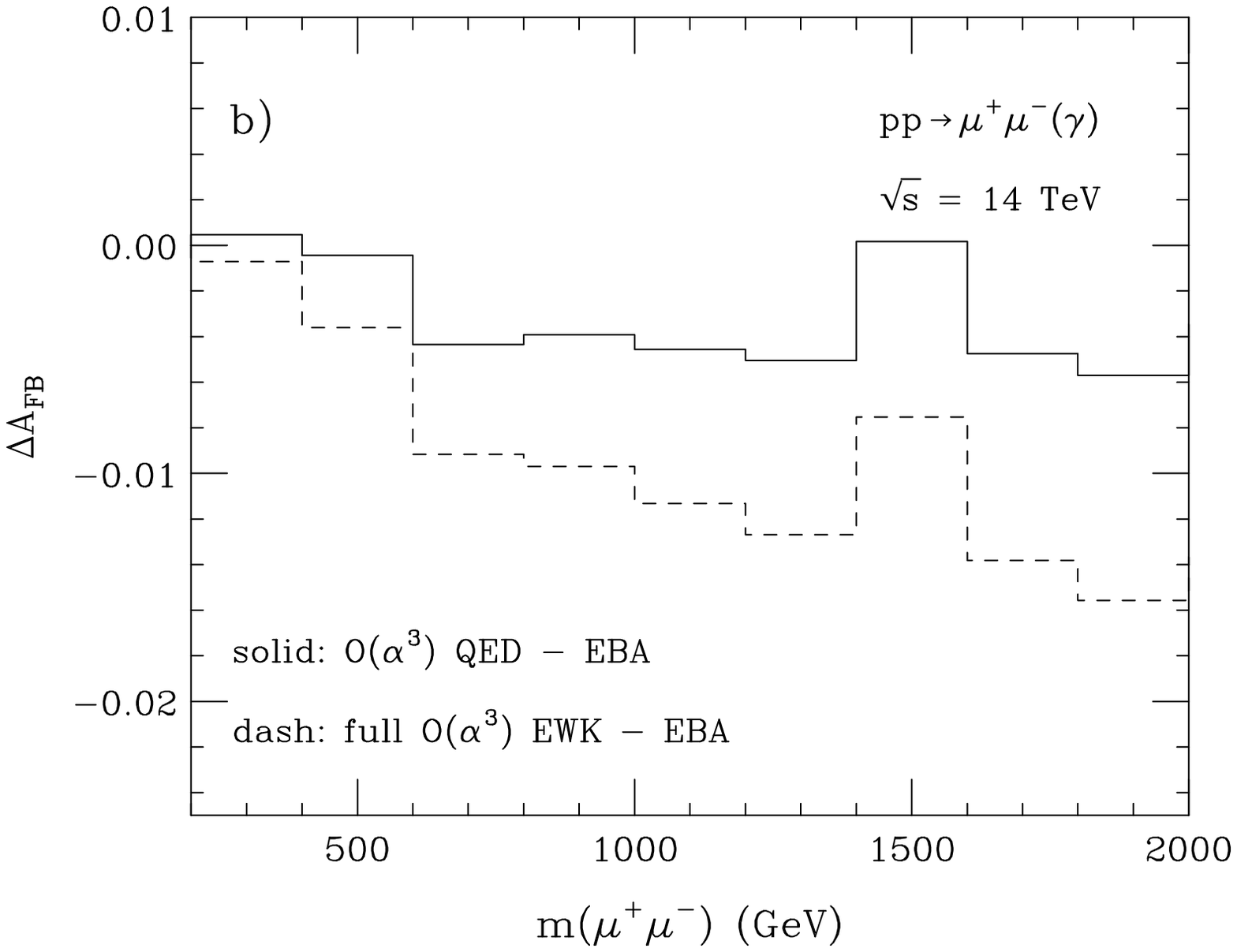}
    \caption{{\tt ZGRAD2} predictions of a) the forward-backward
asymmetry, and b) the change $\Delta A_{\rm FB}$ due to the \protect{${\cal
O}(\alpha^3)$} electroweak and QED corrections. The cuts imposed are
described in the text.  }  
  \label{fig:afb}
  \end{figure}
\end{center}

{\tt ZGRAD2} includes the
complete weak one-loop corrections and the full non-universal QED 
${\cal O}(\alpha)$ corrections. The collinear singularities associated with
initial-state photon radiation are factorised into the parton
distribution functions. However, QED corrections to the evolution of the 
parton distribution functions (see Section~\ref{sec:ewqedcor}) 
are not included in 
{\tt ZGRAD2}. These corrections should be part of a complete global fit 
of the pdf's including all QED effect - this is beyond the scope of the
calculation presented here. None of the current fits to the pdf's 
include QED corrections.
 
\subsection{\boldmath $Z^\prime$ \unboldmath 
indication from new APV data in cesium and searches
             at LHC
}

The weak charge $Q_W$ for a heavy atom is defined in terms of the
number of $u,d$ quarks $N_u= 2Z+N$, $N_d=2N+Z$ in the nucleus $(Z,N)$
and the coefficients $C_{1u,d}$ in the 
parity-violating part of the electron-quark  Hamiltonian, 
\begin{equation}
 {\cal H}_{PV} = - \frac{G_F}{\sqrt{2}}\,
    \bar{e}\gamma_\mu\gamma_5 e \, \left( 
    C_{1u}\, \bar{u} \gamma^\mu u + C_{1d}\, \bar{d} \gamma^\mu d \right) \, ,
\end{equation}
via the relation
\begin{equation}
  Q_W = 2 (N_u C_{1u} + N_d C_{1d}) \, .
\end{equation}
In the SM: $C_{1q} = I_3^q - 2 Q_q \sin^2\theta_W$.

In a recent paper \cite{bennett} a new determination of the weak
charge of atomic cesium has been reported. 
The most precise atomic
parity violating (APV) experiment compares the mixing among $S$ and
$P$ states due to neutral weak interactions to an induced Stark
mixing \cite{wood}. The 1.2\% uncertainty on the previous
measurement of the weak charge $Q_W$ was dominated by the
theoretical calculations on the amount of Stark mixing  and on the
electronic parity violating
  matrix elements. In  \cite{bennett}
the Stark mixing was measured and, incorporating new experimental
data, the uncertainty in the electronic  parity violating matrix
elements was reduced. The new result 
$ Q_W(^{133}_{55}{\rm Cs})=-72.06\pm
(0.28)_{\rm expt}\pm (0.34)_{\rm theor}$ represents a considerable
improvement with respect to the previous determination
\cite{wood,noecker,blundell,dzuba}. 
The discrepancy between the standard model (SM) and
the experimental data is now given by  $ Q_W^{\rm
expt}-Q_W^{SM}=1.18 (1.28)\pm 0.46$ (for $m_t=175$~GeV and
$M_H=100(300)$~GeV). This corresponds to 2.6(2.8) standard 
deviations \cite{apv}, excluding
 the SM at 99\%~CL  and, {\it a fortiori}, all the models
leading to negative additional contributions to $Q_W$, as for example
models with a sequential $Z^\prime$ \cite{apv}. This  deviation
could be explained by assuming the existence of an extra $Z^\prime$
from $E_6$ or $O(10)$ or from $Z^\prime_{LR}$ of left-right (LR)
models \cite{bennett,apv,erlanga}.
The high-energy data at the $Z$ resonance strongly bound the $Z-Z'$
mixing \cite{gross};
 for this reason we will assume zero mixing. In
this case, the new physics contribution to  $Q_W$ is due to the direct exchange of the $Z^\prime$ and is
completely fixed by the $Z^\prime$ parameters,
$
\delta_NQ_W=
16 a_e^\prime [(2Z+N)
v_u^\prime+
(Z+2N)  v_d^\prime]
{M_Z^2}/{M_{Z^\prime}^2}
$, where
$a_f^\prime,
v_f^\prime$ are the couplings  $Z^\prime$ to fermions and, for 
$^{133}_{55}{\rm Cs}$,
$Z=55$ and $N=78$.
 The relevant couplings of
 the $Z^\prime$ to the electron and to the up and down quarks are given in the
Table 1 of \cite{apv}.

\begin{figure}[tbh]
\epsfysize=8truecm
\centerline{\epsffile{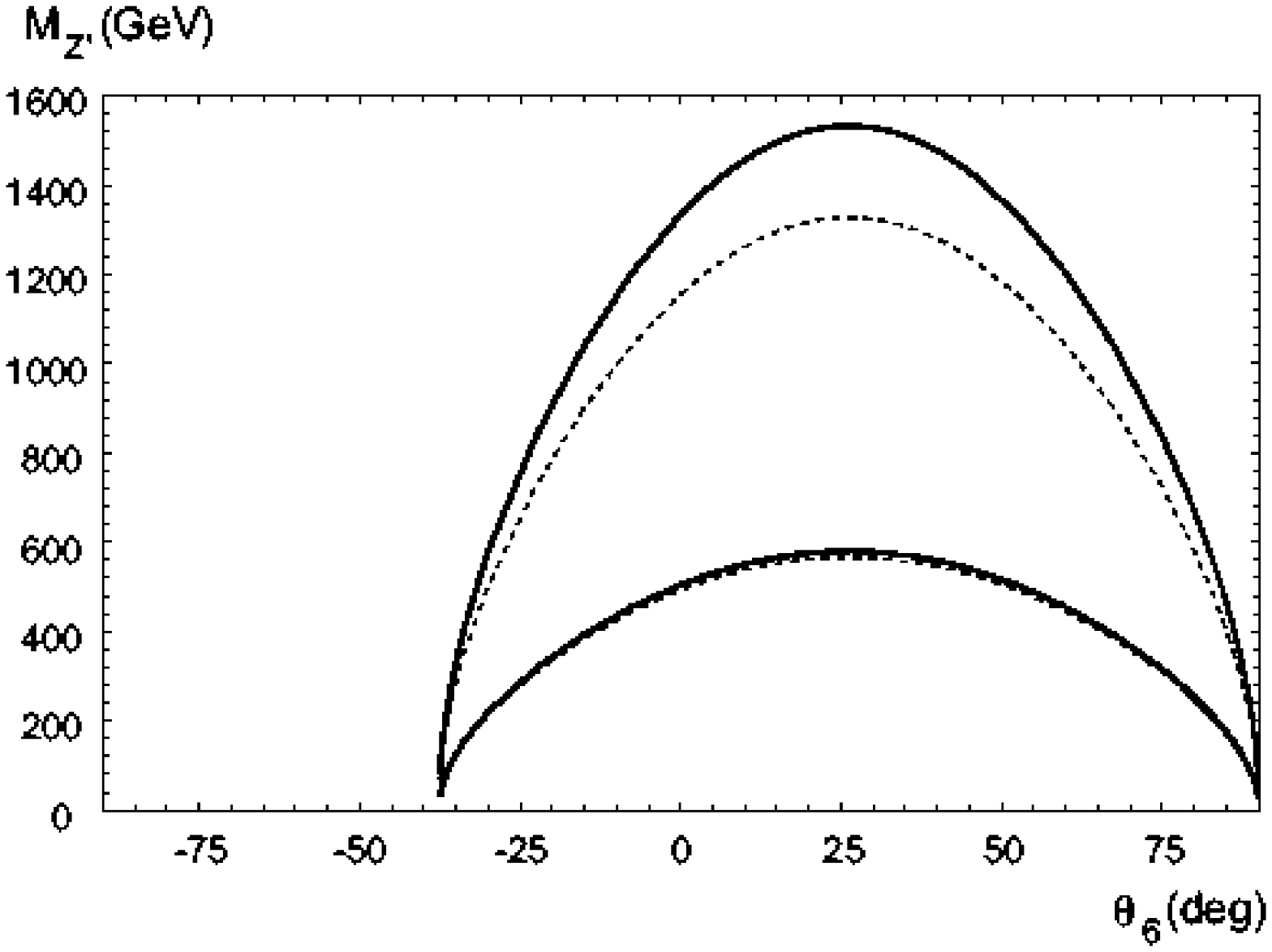}}
\caption[] {The 95\%~CL lower and upper bounds for
$M_{Z^\prime}$ for the extra-U(1) models versus $\theta_6$. The
solid  (dash) line corresponds to $M_H=100(300)$~GeV.}
\label{apvz}
\end{figure}

In the case of the  LR model  considered in \cite{apv},
the extra contribution to the weak charge is
$
\delta_NQ_W=-{M_Z^2}/{M_{Z^\prime}^2}Q_W^{SM}
$.
For this model one has a 95\%~CL lower bound on $M_{Z^\prime_{LR}}$
from the Tevatron \cite{tevatron} given by $M_{Z^\prime_{LR}}\ge
630$~GeV. An LR model could then explain the APV data allowing for a
mass of the $Z^\prime_{LR}$ varying between the intersection from
the 95\%~CL bounds $540\le M_{Z^\prime_{LR}}$~(GeV) $\le 1470$ deriving
from $Q_W$ and the lower bound of $630$~GeV.
In the case
of the extra-U(1) models,  
the CDF experimental lower bounds for the masses vary
according to the values of the angle  $\theta_6$ which parameterises
different extra-U(1) models, but in general they are about 600~GeV at 
95\%~CL
\cite{tevatron}.
 For the particular  models   $\eta$, $\psi$, $\chi$,
 corresponding  to $\theta_6=\arctan{(-\sqrt{5/3})},~\pi/2,~0$,
the 95\%~CL lower bounds are 
 $M_{Z^\prime_\eta}\simeq 620$~GeV,
$M_{Z^\prime_\psi}\simeq 590$~GeV,  $M_{Z^\prime_\chi}\simeq 595$~GeV.
In Figure~\ref{apvz}, the 95\%~CL
bounds on  $M_{Z^\prime}$ from APV are plotted versus $\theta_6$
(the direct lower bounds
from the Tevatron are about $600 $~GeV). We see that  
an extra $Z^\prime$ can explain the discrepancy with
the SM prediction for the $Q_W$ for a wide range of $\theta_6$
angle.  
In  particular,   the models $\eta$ and  $\psi$ are excluded, whereas
the $\chi$ model  is
allowed for $M_{Z^\prime_\chi}$ less than about 1.2~TeV.

In the near  future,  the Tevatron upgrade and LHC can confirm or
disprove this indication coming from $Q_W$. The
 existing bounds for $E_6$ models  from direct searches
at the Tevatron will be upgraded by the future run with $\sqrt{s}=2$~TeV and
1~fb$^{-1}$ to $M_{Z^\prime}\sim 800-900$~GeV and pushed to $\sim 1$~TeV for
10~fb$^{-1}$. The bounds are based on 10 events in the
$e^+e^-+\mu^+
\mu^-$ channels and decays to SM final-states only are assumed \cite{rizzo}.
At the LHC with an integrated luminosity of $100$~fb$^{-1}$, one can explore a mass
range up to $4-4.5$~TeV depending on the $\theta_6$ value.
 Concerning LR models, the 95\%~CL lower limits from the Tevatron run with
$\sqrt{s}=2$~TeV and 1(10)~fb$^{-1}$ are $\sim 900(1000)$~GeV and extend to $\sim
4.5 $~TeV at LHC \cite{rizzo}. Ratios of $Z^\prime$ 
couplings to fermions can be probed
 at LHC, by considering  the forward-backward asymmetries, 
ratios of cross sections in different rapidity bins 
and other observables. For example, for $M_{Z^\prime}=1$~TeV, the LHC can determine the
magnitude of normalised $Z^\prime$ quark and lepton couplings to around $10-20\%$ \cite{rizzo}.
Therefore, if the deviation for the weak charge $Q_W$ with respect to the  SM prediction
is not due to a statistical fluctuation, the new physics described by an extra gauge-boson model
like $Z_\chi^\prime$ can explain the discrepancy and the LHC will be able to verify this possible evidence.

\section{PRECISION MEASUREMENTS
         \protect\footnote{Section coordinator: S.~Haywood}
         \label{sec:ewprecision}}

\subsection{Measurement of the \boldmath
$W$ \unboldmath mass
\label{sec:wmass}}

At the time of the LHC start-up, the $W$ 
mass will be known with a precision of 
about 30~MeV from measurements at LEP2 \cite{altarelli-96} and 
Tevatron  \cite{amidei-96}. The 
motivation to improve on such a precision is discussed briefly below. 
The $W$ mass, which is one of the fundamental parameters of the 
Standard Model, 
is related to other parameters of the theory, {\em i.e.} 
the QED fine structure 
constant $\alpha$, the Fermi constant $G_F$ 
and the Weinberg angle $\sin \theta_W$, through the relation
\begin{equation}
M_W = \sqrt{\frac{\pi \alpha}{G_F \sqrt{2}}} \cdot
      \frac{1}{\sin \theta_W \sqrt{1 - \Delta r}}
\label{eqn:mw}
\end{equation}
where $\Delta r$ 
accounts for the radiative corrections which amount to about 4\%. The 
radiative corrections depend on the top mass as $\sim m_{t}^2$
and on the Higgs mass 
as $\sim \log M_H$. 
Therefore, precise measurements of both the $W$ mass and the top 
mass constrain the mass of the Standard Model Higgs boson or of the $h$ 
boson of the MSSM. This constraint is relatively weak because of the 
logarithmic dependence of the radiative corrections on the Higgs mass.
When it comes to making a comparison of the measurements of $(M_W, m_{t})$
with the SM predictions, it is not very useful if one measurement is much 
more restrictive than the other. To ensure that the two mass determinations
have equal weight in a $\chi^2$ test, 
the precision on the top mass 
and on the 
$W$ mass should be related by the expression
\begin{equation}
\Delta M_W \approx 0.7 \times 10^{-2} \Delta m_{t}
\label{eqn:deltamw}
\end{equation}  
Since the top mass will be measured with an accuracy of about 2~GeV at the LHC 
\cite{atlas-phystdr2}, 
the $W$ mass should be known with a precision of about 15~MeV, 
so that it does not become the dominant error in the test of the 
radiative corrections and in the estimation of the Higgs mass. Such a 
precision is beyond the sensitivity of Tevatron and LEP2. 

A study was performed to assess whether the LHC will be able to measure the 
$W$ mass to about 15~MeV \cite{gianotti-99,richter-98}.
The ATLAS experiment was taken as an 
example, but similar conclusions hold also for CMS. Such a precise 
measurement, which will be performed already in the initial phase at low 
luminosity as will the top mass measurement, 
would constrain the mass of the Higgs 
boson to better than 30\%. When and if the Higgs boson will be found, such 
constraints would provide an important consistency check of the theory, and in 
particular of its scalar sector. Distinguishing between the Standard Model and 
the MSSM might be possible, since the radiative corrections to the $W$ 
mass are 
expected to be a few percent larger in the latter case.

The measurement of the $W$ 
mass at hadron colliders is sensitive to many subtle 
effects which are difficult to predict before the experiments start. However, 
based on the present knowledge of the LHC detector performance and on the 
experience from the Tevatron, it is possible to make a reasonable estimate of 
the total uncertainty and of the main contributions to be expected. In turn, 
this will lead to requirements for the detector performance and the 
theoretical inputs which are needed to achieve the desired precision. This is 
the aim of the study which is described in the next sections. 

\subsubsection{The method}
 
The measurement of the $W$ 
mass at hadron colliders is performed in the leptonic 
channels. Since the longitudinal momentum of the neutrino cannot be measured, 
the transverse mass $m_T^W$ is used. This is calculated using the transverse 
momenta of the neutrino and of the charged lepton, ignoring the longitudinal 
momenta: 
\begin{equation}
m_T^W = \sqrt{2 p_T^l p_T^{\nu} (1 - \cos \Delta \phi)}
\label{eqn:mtw}
\end{equation}
where $l = e, \mu$. 
The lepton transverse momentum $p_T^l$ is measured, whereas the 
transverse momentum of the neutrino $p_T^{\nu}$ 
is obtained from the transverse 
momentum of the lepton and the momentum $\vec{u}$ of the system 
recoiling against 
the $W$ in the transverse plane (hereafter called ``the recoil''):
\begin{equation}
p_T^{\nu} = - | \vec{p_T}^l + \vec{u} |
\label{eqn:ptnu}
\end{equation}
The angle between the lepton and the neutrino in the transverse plane is 
denoted by $\Delta \phi$. 
The distribution of $m_T^W$, and in particular the trailing edge of the 
spectrum, is sensitive to the $W$ mass. Therefore, by fitting the experimental 
distribution of the transverse mass with Monte Carlo samples generated with 
different values of $M_W$, 
it is possible to obtain the mass which best fits the 
data. The trailing edge is smeared by several effects, such as the $W$ 
intrinsic 
width and the detector resolution. This is illustrated in 
Figure~\ref{fig:wmass}, which 
shows the distribution of the $W$ transverse mass as 
obtained at particle level 
(no detector resolution) and by including the energy and momentum resolution 
as implemented in a fast particle-level simulation and reconstruction of the 
ATLAS detector ({\tt ATLFAST}, \cite{richter-98}). 
The smearing due to the finite resolution 
reduces the sharpness of the end-point and therefore the sensitivity to $M_W$. 

\begin{figure}[htbp]
\begin{center}
\includegraphics[width=7cm]{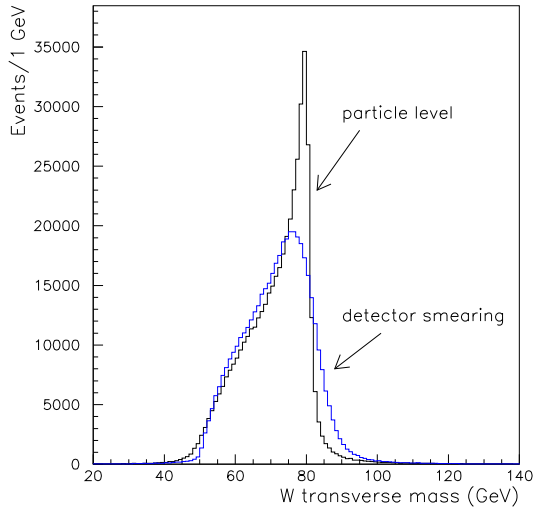}
\end{center}
\caption{Distribution of the $W$ transverse mass as obtained at particle 
level and by including the expected ATLAS detector resolution.}
\label{fig:wmass}
\end{figure}

When running at high luminosity, the pile-up will smear significantly the 
transverse mass distribution, therefore the use of the transverse-mass method 
will probably be limited to the initial phase at low luminosity. Alternative 
methods are mentioned in Section~\ref{wmass-results}. 

\subsubsection{$W$ production and selection \label{wmass-sel}}
 
At the LHC, the cross-section for the process 
$pp \rightarrow W+X$ with $W \rightarrow l\nu$ and $l = e,\mu$ 
is 30 nb. Therefore, about 300 million events are expected to be produced in 
each experiment in one year of operation at low luminosity (integrated 
luminosity 10 fb$^{-1}$). 
Such a cross-section is a factor of ten larger than at 
the Tevatron ($\sqrt{s} =$~1.8~TeV). 

To extract a clean $W$ signal, one should require:
\begin{itemize}
\item
An isolated charged lepton ($e$ or $\mu$) with $p_T >$~25~GeV inside 
the pseudorapidity region devoted to precision physics $|\eta|<2.4$. 
\item
Missing transverse energy $E_T^{\rm miss} >$~25~GeV. 
\item
No jets in the event with $p_T >$~30~GeV.
\item
The recoil should satisfy $|\vec{u}|<$~20~GeV.
\end{itemize}
The last two cuts are applied to reject $W$'s produced with high $p_T$, 
since for 
large $p_T^W$ 
the transverse mass resolution deteriorates and the QCD background 
increases. The acceptance of the above cuts is about 25\%. 
By assuming a lepton 
reconstruction efficiency of 90\% and an identification efficiency of 80\% 
\cite{atlas-phystdr1},
a total selection efficiency of about 20\% should be achieved. 
Therefore, after all cuts about 60 million $W$'s are expected in one year of 
data taking at low luminosity in each experiment, which is a factor of about 
50 larger than the statistics expected from the Tevatron Run 2. 

\subsubsection{Expected uncertainties \label{sec:wmass_uncertain}}

Due to the large event sample, the statistical uncertainty on the $W$ mass 
should be smaller than 2 MeV for an integrated luminosity of 10~fb$^{-1}$. 

Since the $W$ mass is obtained by fitting the experimental distribution of the 
transverse mass with Monte Carlo samples, the systematic uncertainty will come 
mainly from the Monte Carlo modelling of the data, {\em i.e.} 
the physics and the 
detector performance. Uncertainties related to the physics include the 
knowledge of: the $W$ $p_T$ spectrum and angular distribution, the parton 
distribution functions, the $W$ width, 
the radiative decays and the background. 
Uncertainties related to the detector include the knowledge of: the lepton 
energy and momentum scale, the energy and momentum resolution, the detector 
response to the recoil and the effect of the lepton identification cuts. At 
the LHC, as now at the Tevatron, most of these uncertainties will be 
constrained {\em in situ} by using data samples such as $Z \rightarrow ll$ 
decays. The latter 
will be used to determine the lepton energy scale, to measure the detector 
resolution, to model the detector response to the $W$ recoil and the $p_T$ 
spectrum of the $W$, {\em etc.}. 

The advantages of the LHC with respect to the Tevatron experiments are:
\begin{itemize}
\item
The large number of $W$ events mentioned above. 
\item
The large size of the `control samples'. About six million $Z \rightarrow ll$
decays, where $l = e, \mu$, 
are expected in each experiment in one year of data 
taking at low luminosity after all selection cuts. This is a factor of about 
50 larger than the event sample from the Tevatron Run 2. 
\item
ATLAS and CMS are in general more powerful than CDF and D0 
are, in terms of energy resolution, particle identification capability, 
geometrical acceptance and granularity. What may be more important for this 
measurement is the fact that ATLAS and CMS will benefit from extensive 
and detailed simulations and test-beam studies of the 
detector performance, undertaken even before the start of data-taking
\end{itemize}

Nevertheless, the LHC experiments have complex detectors, which will require a 
great deal of study before their behaviour is well understood. 

To evaluate the systematic uncertainty on the $W$ 
mass to be expected in ATLAS, 
$W \rightarrow l\nu$ 
decays were generated with {\tt PYTHIA~5.7} and processed with {\tt ATLFAST}. After 
applying the selection cuts discussed above, a transverse mass spectrum was 
produced for a reference mass value (80.300~GeV). All sources of systematic 
uncertainty affecting the measurement of the $W$ mass from CDF Run~1 
\cite{cdf-wmass,wagner-99}
were then considered as an example\footnote{Similar results have been
obtained by the D0 experiment \cite{wagner-99,d0-wmass}.}. 
Their magnitude was evaluated in 
most cases by extrapolating from the Tevatron results, on the basis of the 
expected ATLAS detector performance. The resulting error on the $W$ mass was 
determined by generating new $W$ samples, each one including one source of 
uncertainty, and by comparing the resulting transverse mass distributions with 
the one obtained for the reference mass. A Kolmogorov test \cite{eadie-71}
was used to evaluate the compatibility between distributions. 

Since the goal is a total error of $\sim 20$~MeV 
per experiment, the individual 
contributions should be much smaller than 10~MeV. A large number of events was 
needed to achieve such a sensitivity. With three million events after all 
cuts, corresponding to twelve million events at the generation level, a 
sensitivity at the level of 8~MeV was obtained. 

The main sources of uncertainty and their impact on the $W$ 
mass measurement are 
discussed one by one in the remainder of this section. The total error and some concluding 
remarks are presented in Section~\ref{wmass-results}.
 
\paragraph{Lepton energy and momentum scale}

This is the dominant source of uncertainty on the measurement of the $W$ mass 
from Tevatron Run~1, where the absolute lepton scale is known with a precision 
of about 0.1\% \cite{cdf-wmass,wagner-99,d0-wmass}.
Most likely, this will be the dominant error 
also at the LHC. In order to measure the $W$ mass with a precision of better 
than 20~MeV, the lepton scale has to be known to 0.02\%. 
The latter is the most 
stringent requirement on the energy and momentum scale from LHC physics. It 
should be noted that a very high precision (0.04\%) must be achieved also by 
the Tevatron experiments in Run~2, in order to measure the $W$ mass to 40~MeV 
\cite{amidei-96}. 
If such a precision will indeed be demonstrated at the Tevatron, it 
would represent a good benchmark for the LHC experiments.

The lepton energy and momentum scale will be calibrated {\em in situ} 
at the LHC by 
using physics samples, which will complement the information coming from the 
hardware calibration, from the magnetic field mapping of solenoids and 
toroids, and from test-beam measurements. The muon scale will be calibrated by 
using mainly $Z \rightarrow \mu\mu$ events, 
and the electromagnetic calorimeter scale will be 
calibrated by using mainly $Z \rightarrow ee$ events or $E/p$ measurements 
for isolated electrons, where $E$ and $p$ are the electron energy 
and momentum as measured in 
the electromagnetic calorimeter and in the inner detector respectively. 
Leptonic decays of other resonances ($\Upsilon$, $J/\psi$) 
should provide additional 
constraints which minimise the extrapolation error to lower masses than the 
$Z$ boson mass. 

Similar methods are used today at the Tevatron, where the uncertainty on the 
absolute lepton scale is dominated by the statistical error due to the limited 
$Z$ data sample. The main advantage of the LHC compared to the Tevatron is 
the large sample of $Z \rightarrow ll$ decays.
The $Z$ boson is close in mass 
to the $W$ boson, therefore the extrapolation error from the point where the 
scale is determined to the point where the measurement is performed is small. 

A preliminary study of the error on the absolute electron scale to be expected 
in ATLAS was performed by using a sample of 500000 $Z \rightarrow ee$ 
decays processed 
through a full {\tt GEANT}-based simulation of the ATLAS detector 
\cite{atlas-phystdr1}. Several 
possible sources of uncertainties were considered: the knowledge of the amount 
of material in the inner detector, which affects the electromagnetic 
calorimeter scale because of photon bremsstrahlung; radiative $Z$ decays, 
which 
distort the reconstructed mass spectrum; the modelling of the underlying event 
and of the pile-up at low and high luminosity. Table~\ref{tab:wmass-ecal}
shows that the 
impact of these uncertainties on the electron scale in the calorimeter can 
most likely be kept below 0.02\%. The most stringent requirement to achieve 
this goal is the knowledge of the material in the inner detector to 1\%, which 
will require scrutiny during construction plus {\em in situ} 
measurements with photon 
conversions and $E/p$ for isolated electrons. More details can be found in 
\cite{atlas-phystdr1}. 

\begin{table}[htb]
\begin{center}
\caption{Expected contributions to the uncertainty on the electron energy 
scale of the ATLAS electromagnetic calorimeter, as determined using a 
fully-simulated sample of $Z \rightarrow ee$ decays 
(from \cite{atlas-phystdr1}). }
\label{tab:wmass-ecal}
\vskip0.2cm
\begin{tabular}{lcc}
\hline
Source & Requirement & Uncertainty on scale \\
\hline
Material in inner detector &    Known to 1\%           &  $<$ 0.01\% \\
Radiative decays           &    Known to 10\%          &  $<$ 0.01\% \\
Underlying event           &    Calibrate and subtract &  $\ll$ 0.03\% \\
Pile-up at low luminosity  &    Calibrate and subtract &  $\ll$ 0.01\% \\
Pile-up at high luminosity &    Calibrate and subtract &  $\ll$ 0.01\% \\
\hline
\end{tabular}
\end{center}
\end{table}

Several experimental constraints will be needed to achieve a 0.02\% 
uncertainty 
on the inner detector muon scale: the solenoidal magnetic field in the inner 
cavity must be known locally to better than 0.1\%, the alignment must be 
understood locally to $\sim$~1~$\mu$m in the bending plane, {\em etc.}. 
A detailed discussion 
on how to meet these goals can be found in \cite{atlas-phystdr1,haywood-99}.

The scale calibration of the external muon spectrometer depends on the 
knowledge of the magnetic field, on the chamber alignment and on the 
knowledge of the muon energy losses in the calorimeters. The latter must be 
understood to a precision of 0.25\% 
in order to achieve the goal uncertainty of 
0.02\% on the absolute scale. A preliminary study based on a full {\tt GEANT} 
simulation of the ATLAS detector demonstrated that with a sample of only 10000 
$Z \rightarrow \mu\mu$ decays a scale uncertainty of 0.1\% 
should be attained in the muon 
spectrometer. More details can be found in \cite{atlas-phystdr1,aleksa-99}.

In conclusion, to achieve the needed precision on the lepton scale, several 
experimental constraints will have to be satisfied. In addition, cross-checks 
and combined fits between different sub-detectors (inner detector and 
electromagnetic calorimeter for the electron scale, inner detector and muon 
system for the muon scale) will be needed. Indeed, only in an over-constrained 
situation will it be possible to disentangle the various contributions to the 
detector response, and therefore to derive a reliable systematic error. 

\paragraph{Lepton energy and momentum resolution}

To keep the uncertainty on the $W$ mass from the lepton resolution to less 
than 
10~MeV, the energy resolution of the electromagnetic calorimeter and the 
momentum resolution of the inner detector and muon system have to be known 
with a precision of better than 1.5\%. 

The lepton energy and momentum resolutions will be determined at the LHC by 
using information from test-beam data and from Monte Carlo simulations of the 
detector, as well as {\em in situ} measurements of the $Z$ width in 
$Z \rightarrow ll$ final 
states. The $E/p$ distribution for electrons from $W$ decays 
provides an additional tool. These methods are used presently at the Tevatron. 
As an example, the statistical error on the momentum resolution obtained by 
CDF in Run~1A is 10\%, 
whereas the systematic error is only 1\% and is dominated 
by the uncertainty on the radiative decays of the $Z$ \cite{cdf-wmass}. 
Since the ATLAS 
performance in terms of momentum resolution is expected to be similar to that 
of CDF in the momentum range relevant to $W$ production and decays, and since 
the statistical error at the LHC will be negligible, a total error of much 
less than 1.5\% should be achieved. This uncertainty might further be 
decreased 
if improved theoretical calculations of radiative $Z$ decays will become 
available. 

\paragraph{Recoil modelling}

The transverse momentum of the system recoiling against the $W$, together with 
the lepton transverse momentum, is used to determine the $p_T$ of the neutrino 
(see Equation~\ref{eqn:ptnu}). 
The recoil is mainly composed of soft hadrons from the 
underlying event, for which neither the physics nor the detector response are 
known with enough accuracy. Therefore, in order to get a reliable recoil 
distribution in the Monte Carlo, information from data is used at the 
Tevatron. By exploiting the 
similar production mechanisms of $W$ and $Z$ bosons, in each Monte Carlo event 
with a given $p_T^W$ (determined from the truth information)
the recoil is replaced by the recoil measured in the data for 
$Z$ events characterised by a $p_T^Z$ (measured by the leptons) 
similar to $p_T^W$. The resulting error on the $W$ 
mass from CDF Run~1A is 60~MeV per channel, and is dominated by the limited 
statistics of $Z$ data. The result obtained from Run~1B (about 30~MeV) shows 
that this uncertainty scales with $\sqrt{N}$, where $N$ 
is the number of events. 
Extrapolating to the LHC data sample, an error of smaller than 10~MeV per 
channel should be achieved. It should be noted that the recoil includes the 
contribution of the pile-up expected at low luminosity (two minimum-bias 
events per bunch crossing on average). 

\paragraph{\boldmath $W$ $p_T$ \unboldmath spectrum}
 
The modelling of $p_T^W$ 
in the Monte Carlo is affected by both theoretical and 
experimental uncertainties. Theoretical uncertainties arise from the 
difficulty in predicting the non-perturbative regime of soft-gluon emission, 
as well as from missing higher-order QCD corrections.
Experimental uncertainties are mainly related to the difficulty of simulating 
the detector response to low-energy particles. 

Therefore, the method used at the Tevatron to obtain a reliable estimate of 
$p_T^W$ consists of measuring the $p_T$ distribution of the $Z$ boson from 
$Z \rightarrow ll$
events in the data, exploiting the fact that both gauge-bosons have similar 
$p_T$ 
distributions, and using the theoretical prediction for the ratio 
$p_T^W/p_T^Z$ (in 
this ratio several uncertainties cancel) to convert the measured $p_T^Z$
into $p_T^W$. 
The resulting error on the $W$ mass obtained by CDF is 20~MeV, 
dominated by the limited $Z$ statistics. 

At the LHC, the average transverse momentum of the $W$ ($Z$) 
is 12~GeV (14~GeV), 
as given by {\tt PYTHIA~5.7}. Over the range $p_T$ ($W$,$Z$) $<$~20~GeV, 
both gauge-bosons 
have $p_T$ spectra which agree to within $\pm$10\%. By assuming a negligible 
statistical error on the knowledge of $p_T^Z$, which will be measured with 
high-statistics data samples, and by using the $p_T^Z$ spectrum instead of 
the $p_T^W$ 
distribution, an error on the $W$ mass of about 10~MeV 
per channel was obtained 
without any further tuning. Although the leading-order parton shower approach 
of {\tt PYTHIA} is only an approximation to reality, this result is encouraging. 
Furthermore, improved theoretical calculations for the ratio of the $W$ and 
$Z$ $p_T$ 
distributions should become available at the time of the LHC, so that the 
final uncertainty will most likely be smaller than 10~MeV.
 
\paragraph{Parton distribution functions}

Parton momentum distributions inside the protons determine the $W$ 
longitudinal 
momentum, and therefore affect the transverse mass distribution through lepton 
acceptance effects. At the Tevatron, parton distribution functions (pdf),
in particular the $u/d$ ratio, are 
constrained by measuring the forward-backward charge asymmetry
of the $W$ rapidity distribution.
Such an asymmetry, which is typical of $p\bar{p}$ 
collisions, is not present in $pp$ 
collisions and therefore cannot be used at the LHC. However, it has been shown 
\cite{di97} 
that pdf can be constrained to a few percent at the LHC by using 
mainly the pseudorapidity distributions of leptons produced in $W$ and $Z$ 
decays. 
The resulting uncertainty on the $W$ mass should be smaller than 10~MeV. 

\paragraph{\boldmath $W$ \unboldmath width}

At hadron colliders, the $W$ width can be obtained from the measurement of 
$R$, 
the ratio between the rate of leptonically decaying $W$'s and leptonically 
decaying $Z$'s: 
\begin{equation}
R = \frac{\sigma_W}{\sigma_Z} \times 
\frac{BR(W \rightarrow l\nu)}{BR(Z \rightarrow ll)}
\label{eqn:rratio}
\end{equation}
where the $Z$ branching ratio ($BR$) is obtained from 
LEP measurements, and the 
ratio between the $W$ and the $Z$ cross-sections is obtained from theory. By 
measuring $R$, the leptonic branching ratio of the $W$ 
can be extracted from the 
above formula, and therefore $\Gamma_W$ can be deduced assuming Standard Model 
couplings for $W \rightarrow l\nu$. 
The precision achievable with this method is limited by 
the theoretical knowledge of the ratio of the $W$ to the $Z$ cross-sections. 
Another method consists of fitting the high-mass tails of the transverse mass 
distribution, which are sensitive to the $W$ width. 

By using these methods, the $W$ width was measured with a precision of 
about 60~MeV by CDF in Run~1, 
which translates into an error of 10~MeV per channel on 
the $W$ mass measurement. 

In Run 2, the $W$ width should be measured with a precision of 30~MeV 
\cite{amidei-96},
which contributes an error of 7~MeV per channel on the $W$ mass. 
This is however 
a conservative estimate for the LHC, where the $W$ 
width should be measured with 
higher precision than at Tevatron by using the high-mass tails of the 
transverse mass distribution. The measurement of $R$, on the other hand, in 
addition to being model-dependent would require very precise theoretical 
inputs. It should be noted that one could also use the value of the $W$ width 
predicted by the Standard Model. 

\paragraph{Radiative decays}

Radiative $W \rightarrow l\nu\gamma$ decays produce a 
shift in the reconstructed transverse mass, 
which must be precisely modelled in the Monte Carlo. Uncertainties arise from 
missing higher-order corrections, which translate into an error of 20~MeV on 
the $W$ mass as measured by CDF in Run~1. 
Improved theoretical calculations have 
become recently available \cite{baur-97}. 
Furthermore, the excellent granularity of 
the ATLAS electromagnetic calorimeter, and the large statistics of radiative 
$Z$ decays, should provide useful additional information. 
Therefore, a $W$ mass error 
of 10~MeV per channel was assumed in this study. This is a conservative 
estimate, since the D0 error from Run~1 is smaller than 10~MeV 
\cite{wagner-99}. 

\paragraph{Background}
 
Backgrounds distort the $W$ 
transverse mass distribution, contributing mainly to 
the low-mass region. Therefore, uncertainties on the background normalisation 
and shape translate into an error on the $W$ mass. 
This error is at the level of 
5~MeV (25~MeV) in the electron (muon) channel for the measurement performed by 
CDF in Run~1, where the background is known with a precision of about 10\%. 

A study was made of the main backgrounds to $W \rightarrow l\nu$ 
final states to be expected 
in ATLAS. The contribution from $W \rightarrow \tau\nu$ 
decays should be of order 1.3\% in both
the electron and the muon channel. The background from $Z \rightarrow ee$ 
decays to the $W \rightarrow e\nu$ 
channel is expected to be negligible, whereas the contribution of 
$Z \rightarrow \mu\mu$
decays to the $W \rightarrow \mu\nu$ 
channel should amount to 4\%. The difference between these 
two channels is due to the fact that the calorimetry coverage extends up to 
$|\eta| \sim 5$, whereas the coverage of the muon spectrometer is limited to 
$|\eta| < 2.7$. 
Therefore, muons from $Z$ decays which are produced with $|\eta| > 2.7$ escape 
detection and thus give rise to a relatively large missing transverse 
momentum. On the other hand, electrons from $Z$ decays produced with 
$|\eta| > 2.4$
are not efficiently identified, because of the absence of tracking devices and 
of fine-grained calorimetry, however their energy can be measured up to 
$|\eta| \sim 5$. 
Therefore these events do not pass the $E_T^{miss}$ cut described in 
Section~\ref{wmass-sel}.
Finally, $t\bar{t}$ 
production and QCD processes are expected to give negligible contributions.

In order to limit the error on the $W$ mass to less than 10~MeV, 
the background 
to the electron channel should be known with a precision of 30\%, which is 
easily achievable, and the background to the muon channel should be known with 
a precision of 7\%. The latter could be monitored by using  
$Z \rightarrow ee$ decays. 

\subsubsection{Results \label{wmass-results}}

The expected contributions to the uncertainty on the $W$ mass measurement, of 
which some are discussed in the previous sections, are summarised in 
Table~\ref{tab:wmass-errors}. 
For comparison, the errors obtained by CDF in Run~1A (integrated 
luminosity $\sim$~20~pb$^{-1}$) and Run~1B 
(integrated luminosity $\sim$~90~pb$^{-1}$) are also 
shown separately. The evolution of the uncertainty between Run~1A and Run~1B 
shows the effect of the increased statistics and of the improved knowledge of 
the detector performance and of the physics, and provides a solid basis for 
the LHC results presented here. 

With an integrated luminosity of 10~fb$^{-1}$, 
which should be collected in one 
year of LHC operation, and by considering only one lepton species ($e$ or 
$\mu$), a 
total uncertainty of smaller than 25~MeV should be achieved by each LHC 
experiment. By combining both lepton channels, which should also provide 
useful cross-checks since some of the systematic uncertainties are different 
for the electron and the muon sample, and taking into account common 
uncertainties, the total error should decrease to less than 20~MeV per 
experiment. Finally, the total LHC uncertainty could be reduced to about 
15~MeV 
by combining ATLAS and CMS together. Such a precision would allow the LHC 
to compete with the expected precision at a Next Linear Collider 
\cite{accomando-97}. 

The most serious experimental challenge in this measurement is the 
determination of the lepton absolute energy and momentum scale to 0.02\%. All 
other uncertainties are expected to be of the order of (or smaller than) 
10~MeV. However, to achieve such a goal, improved theoretical calculations of 
radiative decays, of the $W$ and $Z$ $p_T$ spectra, and of higher-order QCD 
corrections will be needed.
 
The results presented here have to be considered as preliminary and far from 
being complete. It may be possible that, by applying stronger selection cuts, 
for instance on the maximum transverse momentum of the $W$, the systematic 
uncertainties may be reduced further. Moreover, two alternative methods to 
measure the $W$ mass can be envisaged. 
The first one uses the $p_T$ distribution of 
the charged lepton in the final state. Such a distribution features a Jacobian 
peak at $p_T^l \sim M_W/2$ 
and has the advantage of being affected very little by the 
pile-up, therefore it could be used at high luminosity. However, the lepton 
momentum is very sensitive to the $p_T$ of the $W$ 
boson, whereas the transverse 
mass is not, and hence a very precise theoretical knowledge of the $W$ $p_T$ 
spectrum would be needed to use this method. Another possibility is to use the 
ratio of the transverse masses of the $W$ and $Z$ bosons 
\cite{rajagopalan-96}. The $Z$ transverse 
mass can be reconstructed by using the $p_T$ of one of the charged leptons, 
while the second lepton is treated like a neutrino whose $p_T$ is measured by 
the first lepton and the recoil. By shifting the $m_T^Z$ 
distribution until it 
fits the $m_T^W$ 
distribution, it is possible to obtain a scaling factor between 
the $W$ and the $Z$ 
masses and therefore the $W$ mass. The advantage of this method 
is that common systematic uncertainties cancel in the ratio. The main 
disadvantage is the loss of a factor of ten in statistics, since the 
$Z \rightarrow ll$
sample is a factor of ten smaller than the $W \rightarrow l\nu$ sample 
(and only events near to the Jacobian peak contribute significantly to the mass
determination). 
Furthermore, 
differences in the production mechanism between the $W$ and the $Z$ ($p_T$, 
angular 
distribution, {\em etc.}), 
and possible biases coming from the $Z$ selection cuts, 
will give rise to a non-negligible systematic error. 

\begin{table}
\begin{center}
\caption{Expected contributions to the uncertainty on the $W$ mass 
measurement in ATLAS for each lepton family and for an integrated luminosity 
of 10~fb$^{-1}$ (fourth column). The corresponding uncertainties of the CDF 
measurement in the electron channel, as obtained in Run~1A \cite{cdf-wmass} 
and Run~1B \cite{wagner-99}, 
are also shown for comparison (second and third column). }
\label{tab:wmass-errors}
\vskip0.2cm
\begin{tabular}{lccc}
\hline
Source & $\Delta M_W$ (CDF Run 1A) &  $\Delta M_W$ (CDF Run 1B) & 
$\Delta M_W$ (ATLAS)  \\
  &  (MeV) & (MeV) & (MeV) \\
\hline
Statistics                      & 145   & 65    & $<$ 2   \\
$E-p$ scale                     & 120   & 75    &  15  \\
Energy resolution               &  80   & 25    &   5  \\ 
Recoil model                    &  60   & 33    &   5  \\
Lepton identification           &  25   & $-$   &   5  \\
$p_T^W$                         &  45   & 20    &   5  \\ 
Parton distribution functions   &  50   & 15    &  10 \\ 
$W$ width                       &  20   & 10    &   7  \\
Radiative decays                &  20   & 20    & $<$ 10   \\
Background                      &  10   &  5    &   5  \\
\hline
TOTAL                           & 230   &113    & $<$ 25   \\
\hline
\end{tabular}
\end{center}
\end{table}

The final measurement will require using all the methods discussed above, in 
order to cross-check the systematic uncertainties and to achieve the highest 
precision. 

\subsubsection{Conclusions}

Preliminary studies indicate that measuring the $W$ mass at the LHC with a 
precision of about 15~MeV should be possible, although very challenging. The 
biggest single advantage of the LHC is the large statistics, which will result 
in 
small statistical errors and good control of the systematics. To 
achieve such unprecedented precision, improved theoretical calculations in 
many areas will be needed ({\em e.g.} radiative decays, pdf's, $p_T^W$), 
and many stringent experimental requirements will 
have to be satisfied.

\subsection{Drell-Yan production of lepton pairs
\label{sec:drellyan}}


\def\a34{\cos\alpha_{34}}
\def\as{\alpha_{s}}
\def\bz{\cos\chi_{BZ}}
\def\ca{C_{A}}
\def\cf{C_{F}}
\def\ee{\mathrm{e^{+}e^{-}}}
\def\AFB{\mathrm{\rm A_{FB}\;}}
\def\AFBS{\mathrm{\rm A^{s}_{FB}\;}}
\def\COST{\mathrm{\rm \cos\theta\;}}
\def\CP{\mathrm{\rm \cal P \;}}
\def\TO{\mathrm{\longrightarrow\ }}
\def\EE{\mathrm{\rm \;e^+e^-\;}}
\def\EEG{\mathrm{\rm \;e^+e^-(\gamma)\;}}
\def\LLG{\mathrm{\rm \;l^+l^-(\gamma)\;}}
\def\QQG{\mathrm{\rm \;q\bar q(\gamma)\;}}
\def\G{\mathrm{\rm \;\gamma\;}}
\def\GG{\mathrm{\rm \;\gamma\gamma\;}}
\def\MM{\mathrm{\;\mu^+\mu^-\;}}
\def\TT{\mathrm{\;\tau^+\tau^-\;}}
\def\l{\mathrm{\lambda}}
\def\L{\mathrm{\Lambda}}
\def\SP{\mathrm{s^{\prime}}}
\def\SC{\mathrm{\stackrel{\sim}{c}}}
\def\ST{\mathrm{\stackrel{\sim}{t}}}
\def\SU{\stackrel{\sim}{u}}
\def\SD{\stackrel{\sim}{d}}
\def\SQ{\mathrm{\stackrel{\sim}{q}}}
\def\SN{\mathrm{\stackrel{\sim}{\nu}}}
\def\SNE{\mathrm{\stackrel{\sim}{\nu}_{e}}}
\def\SNM{\mathrm{\stackrel{\sim}{\nu}_{\mu}}}
\def\SNT{\mathrm{\stackrel{\sim}{\nu}_{\tau}}}
\def\STIL{\stackrel{\sim}{S}}
\def\etal{{\it et~al.}}
\def\g{\mathrm{g}}
\def\mij{m_{ij}}
\def\mtot{M_{tot}}
\def\nf{N_{F}}
\def\nr{\cos\theta_{NR}^{\star}}
\def\qq{\mathrm{q\bar{q}}}
\def\uu{\mathrm{u\bar{u}}}
\def\dd{\mathrm{d\bar{d}}}
\def\rs{\sqrt{s}}
\def\tf{T_{F}}
\def\tr{T_{R}}
\def\xi{x_{i}}
\def\yc{y_{cut}}
\def\yij{y_{ij}}
\def\Rtau{\Gamma(\tau\longrightarrow e\nu\bar\nu)/\Gamma(\tau\longrightarrow\mu\nu\bar\nu)}
\def\Rtaumu{\Gamma(\tau\longrightarrow e\nu\bar\nu)/\Gamma(\mu\longrightarrow e\nu\bar\nu)}
\def\wangle{\sin^{2}\theta_\mathrm{eff}^\mathrm{lept}(M^2_Z)}

\subsubsection{Introduction}

\paragraph{Parton level:}

In the Standard Model (SM), the production of lepton pairs in
hadron-hadron collisions (the Drell-Yan process)
is described by $s$-channel exchange of photons or $Z$ bosons.
The parton cross section in the centre-of-mass system has the form:
\begin{equation}
\frac{d \hat{\sigma}}{d \Omega} = \frac{\alpha^2}{4s} [A_0(1+\cos^2\theta)+A_1\cos\theta]
\label{llnewph}
\end{equation}
where $\hat{\sigma} = \frac{4 \pi \alpha^2}{3s} A_0$ and
$A_{FB} = \frac{3}{8}\frac{A_1}{A_0}$ give the total cross section and the forward-backward
asymmetry, respectively. 
The terms $A_0$ and $A_1$ are fully determined by
the electroweak couplings of the initial- and final-state fermions.
At the $Z$ peak, the $Z$ exchange dominates while the interference term
is vanishing. At higher energies, both photon and $Z$ exchange contribute
and the large value of the forward-backward asymmetry arises from 
the interference between the neutral currents.

Fermion-pair production above the $Z$ pole is a
rich search field for new phenomena at present and future high-energy 
colliders~\cite{Bourilkov:1998}. 
The differential cross section is given by
\begin{equation}
\frac{d \hat{\sigma}}{d \Omega} \sim |\gamma_s+Z_s +  \mathrm{New\;Physics\;?!}|^2
\label{llxsec}
\end{equation}
where many proposed types of new physics can lead to observable
effects by adding new amplitudes or through their interference
with the neutral currents of the SM.

\paragraph{At hadron colliders:}

The parton cross sections are folded with the
parton distribution functions (pdf's):
\begin{equation}
\frac{\mbox{d}^2\sigma}{\mbox{d}M_{ll}\mbox{d}y} 
(pp\rightarrow l_1 l_2) \ \sim \ \sum_{ij} 
\left(f_{i/p}(x_1)  f_{j/p}(x_2) +(i \leftrightarrow j)\right)\,  \hat{\sigma}  
\label{llpdf1}
\end{equation}
where
$\hat{\sigma}$ is the cross section for the partonic subprocess $ij\rightarrow l_1 l_2$,
$M_{ll}=\sqrt{\tau s} = \sqrt{\hat{s}}$ and $y$ are the invariant mass and 
rapidity of the lepton pair, 
$x_1=\sqrt{\tau}e^y$ and $x_2=\sqrt{\tau}e^{-y}$ are the parton momentum fractions, and
$f_{i/p(\bar{p})}(x_i)$ is the probability to find a parton $i$ with momentum
fraction $x_i$ in the (anti)proton.
\begin{eqnarray}
\sigma_{F\pm B}(y,M) & = & [\int_{0}^{1} \pm \int_{-1}^{0}]\sigma_{ll} d(\cos \theta^{*}) \\
A_{FB} (y,M) & = & \frac{\sigma_{F-B}(y,M)}{\sigma_{F+B}(y,M)}
\label{llpdf2}
\end{eqnarray}
The total cross section and the forward-backward asymmetry are
functions of observables which are well measured experimentally:
the invariant mass and the rapidity of the final state lepton-pair.
For a pair of partons ($x_1 \geq x_2$), there are four combinations of
quarks initiating Drell-Yan production:
$u\bar u, \bar u u, d\bar d, \bar d d$.
In $pp$ collisions, the antiquarks come always from the sea while the quarks
can have valence or sea origin. The $x$-range probed depends on the mass
and rapidity of the lepton-pair as shown in Table~\ref{tab:xmasrap}.
Going to higher rapidities increases the difference between
$x_1$ and $x_2$ and hence the probability that the first quark is
a valence one.

\begin{table}[htb]
  \begin{center}
\caption{$x_1$ and $x_2$ for different masses and rapidities.}
\label{tab:xmasrap}
\vskip0.2cm
\begin{tabular}{l|ccc|ccc|ccc}
\hline
$M$ (GeV) & \multicolumn{3}{|c|}{91.2} & \multicolumn{3}{|c|}{200} 
& \multicolumn{3}{|c}{1000} \\
  $y$      & 0 & 2 & 4 & 0 & 2 & 4 & 0 & 2 & 4 \\
\hline
$x_1$ &   0.0065      &   0.0481    &    0.3557 &   0.0143      &   0.1056    &    0.7800   &
 0.0714      &   0.5278    &       -        \\
$x_2$ &   0.0065      &   0.0009    &    0.0001    &   0.0143      &   0.0019    &    0.0003  &
  0.0714      &   0.0097    &       -   \\
\hline
    \end{tabular}
  \end{center}
\end{table}

\subsubsection{Event rates}

The expected numbers of events for the Tevatron Run~2 (TEV2) and the LHC 
are shown in Table~\ref{tab:nevent} and Figure~\ref{fig:lhctevev}.
The estimation is based on simulations with {\tt PYTHIA~5.7}
\cite{Bourilkov:2000}
by applying the following cuts:
\begin{enumerate}
\item
For LHC: both leptons $|\eta| < 2.5$; 
for TEV2: one lepton $|\eta| < 1$, the other $|\eta| < 2.5$.
\item
For both leptons, $p_T > 20$ GeV.
\end{enumerate}
\vspace{0.2cm}

The data sample can be divided into three classes:\\
{Events near the $Z$ pole:}
\begin{itemize}
\item There will be a huge 
sample of $Z$ events at the LHC. These will allow study of the 
interplay between $\wangle$ and the pdf's.
\end{itemize}
{High mass pairs (110-400 GeV):}
\begin{itemize}
\item LEP2 will study this region up to 200 GeV. 
\item TEV2 will collect a sizeable sample of events in this region.
\item LHC will be able to do precision studies.
\end{itemize}
{Very high mass pairs (400-4000 GeV):}
\begin{itemize}
\item TEV2 will have a first glance.
\item LHC will collect a sizeable sample for
   tests of the SM at the highest momentum transfers ($Q^2$) and for
   searches of new phenomena at the TeV scale.
\end{itemize}

\begin{table}[htb]
  \begin{center}
\caption{{\tt PYTHIA} estimate: expected number of events for one
experiment in the $e^+e^-$ and $\MM$ channels.
For LEP2 and CDF the observed number of events is shown.}
\label{tab:nevent}
\vskip0.2cm
\begin{tabular}{lcccc}
\hline
Pair Mass        &    LEP2       & CDF           & TEV2         & LHC           \\
      & 600 pb$^{-1}$ & 110 pb$^{-1}$ & 10 fb$^{-1}$ & 100 fb$^{-1}$ \\
      & SM / Data     &  Data         & {\tt PYTHIA}       & {\tt PYTHIA}        \\
\hline                                                     
$Z$ pole      &      -        &       -       &$\sim1.5 \times 10^6$&$\sim 134 \times 10^6$\\
$>$ 110 GeV &   12500       & 148 ($>$ 150 GeV) &    46000     & $2.6 \times 10^6$    \\
$>$ 400 GeV &      -        &       1       &      250     &  33000       \\
\hline
    \end{tabular}
  \end{center}
\end{table}
\begin{center}
  \begin{figure}
    \includegraphics[width=0.80\textwidth,height=0.39\textheight]{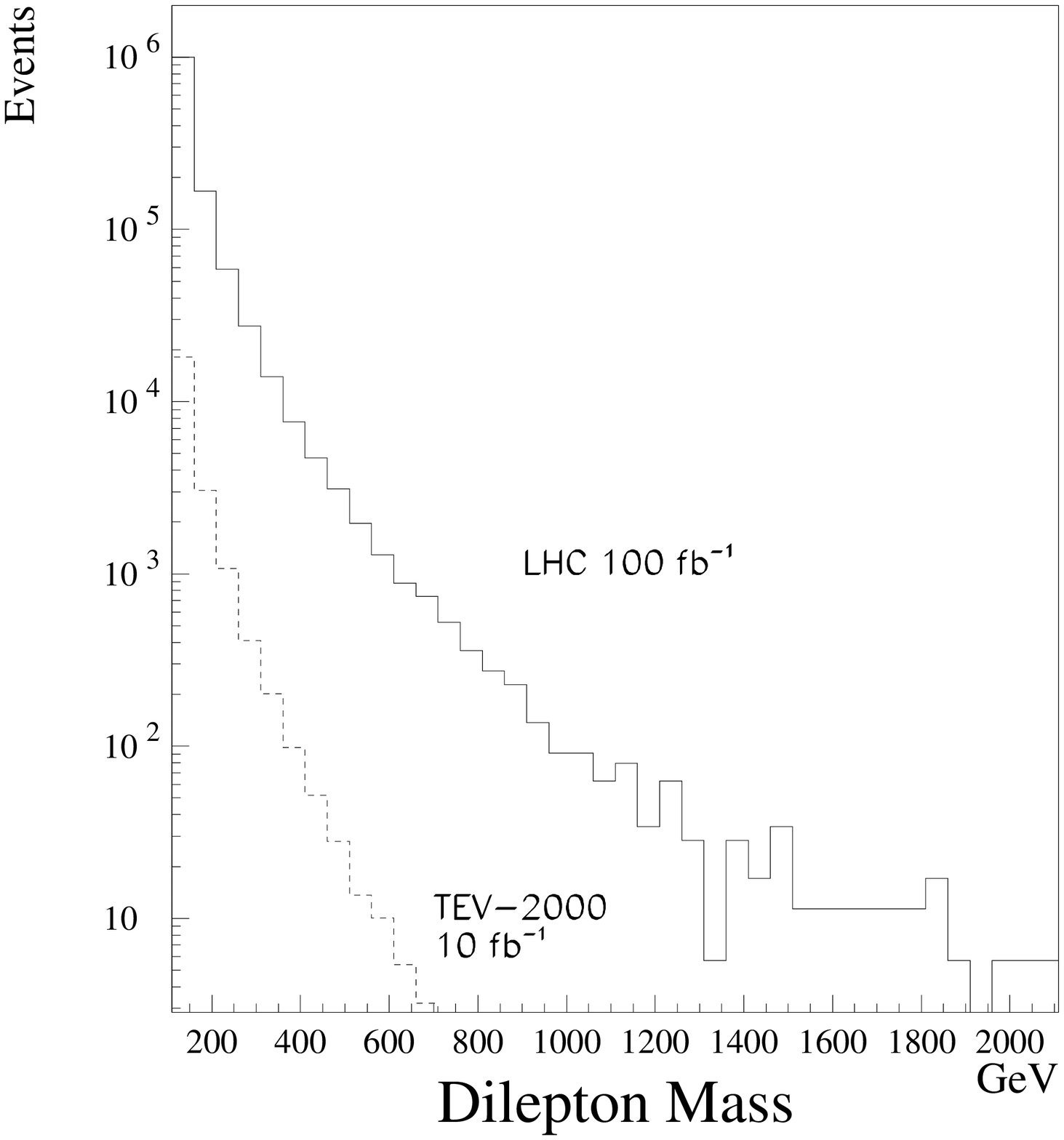}
    \caption{Expected number of events for TEV-2000 and LHC
     in one channel/experiment as a function of the dilepton mass.}
    \label{fig:lhctevev}
  \end{figure}
\end{center}
\vspace{-5mm}

\subsubsection{Measurements of $\sigma$ and $A_{FB}$}

The experimental signature for Drell-Yan events is distinctive:
a pair of well isolated leptons with opposite charge.
This should be straight forward for the ATLAS and CMS detectors
to identify.
The backgrounds are low:
$W^+W^-$, $\TT$, $c\bar{c}$, $b\bar{b}$, $t\bar{t}$;
fakes, cosmics {\it etc.}.
If the need arises, they can be further suppressed by acoplanarity and
isolation cuts. The selection cuts used in this study have already been
described in the section on simulations.

An important ingredient in the cross section measurement is
the precise determination of the luminosity. A promising possibility is
to go directly to the parton luminosity~\cite{di97} by 
using the  $W^{\pm}$ ($Z$) production of single (pair) leptons:
\begin{itemize}
 \item Constrain the pdf's.
 \item Measure directly the parton-parton luminosity.
\end{itemize}
In this way, the 
systematic error on $\sigma_{\mathrm{DY}}^{\mathrm{high} \ Q^2}$ relative
to $\sigma_Z$ can be reduced to $\sim$~1\%.

In order to measure the forward-backward asymmetry, it is necessary to tag
the directions of the incoming quark and antiquark.
At the Tevatron, the $p\bar{p}$ collisions provide a natural label for
the valence (anti)quark. In contrast at the LHC, the $pp$ initial state
is symmetric. 
But in the reaction $q\bar{q} \TO l^+l^-$ only $q$ can be a
valence quark, carrying on average a higher momentum compared
to the sea antiquarks. 
Therefore at the LHC, $A_{FB}$ will be signed according to the sign
of the rapidity of the lepton pair $y(ll)$.
Consequently, $A_{FB}$ increases as a function of $y(ll)$ 
\cite{Rosner:1987,Dittmar:1996} (see Figure~\ref{fig:AFBvsy}).

A precise measurement of $\sigma$ and $A_{FB}$ at large $\hat s$
requires good knowledge of the different types of 
electroweak radiative corrections to the DY process:
vertex, propagator, EW boxes.
A complete one-loop parton cross section calculation has been 
performed~\cite{BBHSW}.
The size of these corrections after folding with
the pdf's and the expected experimental
precision on the cross section measurement are compared in
Figure~\ref{fig:radcor}. The LHC experiments
can probe these corrections up to $\sim$~2 TeV.

\begin{center}
  \begin{figure}[htbp]
\resizebox{0.85\textwidth}{9.4cm}{\includegraphics{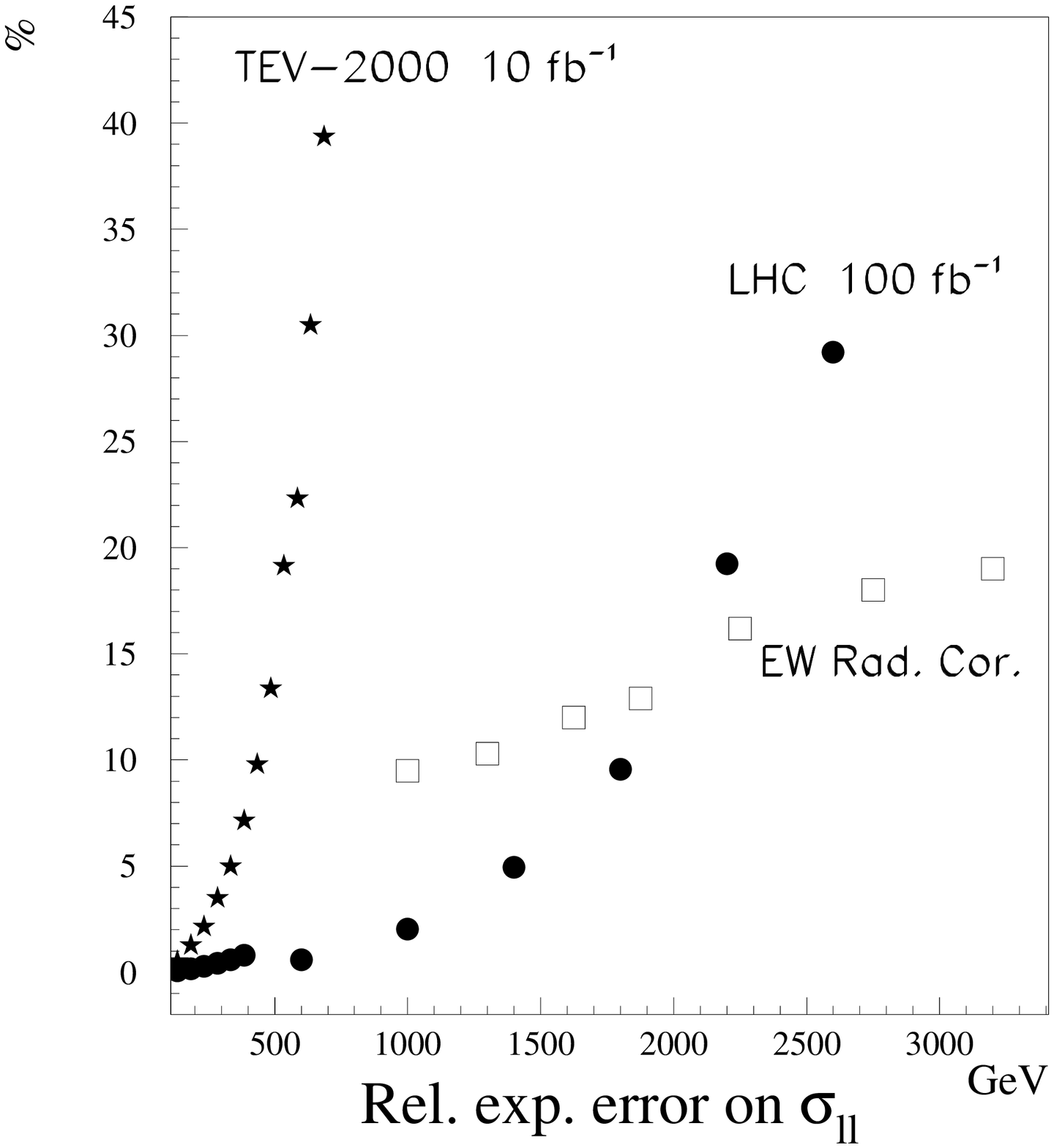}}
    \caption{Size of the electroweak radiative corrections and the 
     expected relative experimental precision on the cross section 
     measurement for $e^+e^-$ and $\MM$ from one experiment in \% 
     as a function of the dilepton mass.}
    \label{fig:radcor}
  \end{figure}
\end{center}

\subsubsection{Determination of $\wangle$ \label{sec:AFB}}

A very precise determination of $\wangle$ will constrain the Higgs
mass or, if the Higgs boson is discovered, will check the consistency
of the SM~\cite{Fischer:1995}. 
The latest result of the LEP Electroweak Working Group 
from the summer of 1999 is:
\begin{equation}
\rm \wangle = 0.23151 \pm 0.00017
\label{wangle}
\end{equation}

\paragraph{Event selection}

A careful study~\cite{Sliwa:2000} 
of the precision which can be obtained from the
$Z \rightarrow ee$ decay by ATLAS and 
CMS has been made using {\tt PYTHIA~5.7} and {\tt JETSET~7.2}.
Background processes from 
$pp \to 2$~jets and $pp \to t\bar{t} \to e^+e^-$ have been included.
In the regions of precision measurements ($|\eta| \leq 2.5$), the 
precision which can be obtained from $Z \rightarrow \mu\mu$ decays 
should be comparable to that from the electron channel.
In addition, the detectors have calorimetry extending to
$|\eta| \sim 5$ and hence, if it is possible to tag very forward
electrons, albeit 
with significantly lower quality, it may be possible to improve dramatically
the measurement of $\wangle$.

The following cuts were made:
\vspace{0.2cm}
\begin{enumerate}
\item
$p_T^\mathrm{electron}> 20$ GeV/$c$
\item
$85.2$ GeV/$c^2 < M(e^+e^-) < 97.2$ GeV/$c^2$
\end{enumerate}
\vspace{0.2cm}
In all cases, one electron was required in the precision calorimetry
$|\eta| \leq 2.5$. Efficiencies after typical electron identification
cuts were taken from detailed studies reported in 
\cite{atlas-phystdr1}. These are typically around 70\%, with 
corresponding jet rejections of $>10^4$ (there was no advantage for
this measurement of larger rejection factors).
For the second electron, the possibility for it to be identified in
the forward calorimetry $2.5 < |\eta| \leq 4.9$ was considered.
In this region, there is no magnetic tracking.
An electron identification efficiency of 50\% was assumed with a
corresponding jet rejection of $\rho$.
Extending the pseudorapidity coverage for the second electron increases the
range of lepton pair rapidity from 
$|y(e^+e^-)| \leq \sim 2$ to $|y(e^+e^-)| \leq \sim 3$.
Figure~\ref{fig:AFBvsy} shows how the asymmetry varies as a function 
of $|y(e^+e^-)|$.

\begin{figure}[htbp]
\begin{center}
\includegraphics[width=7cm]{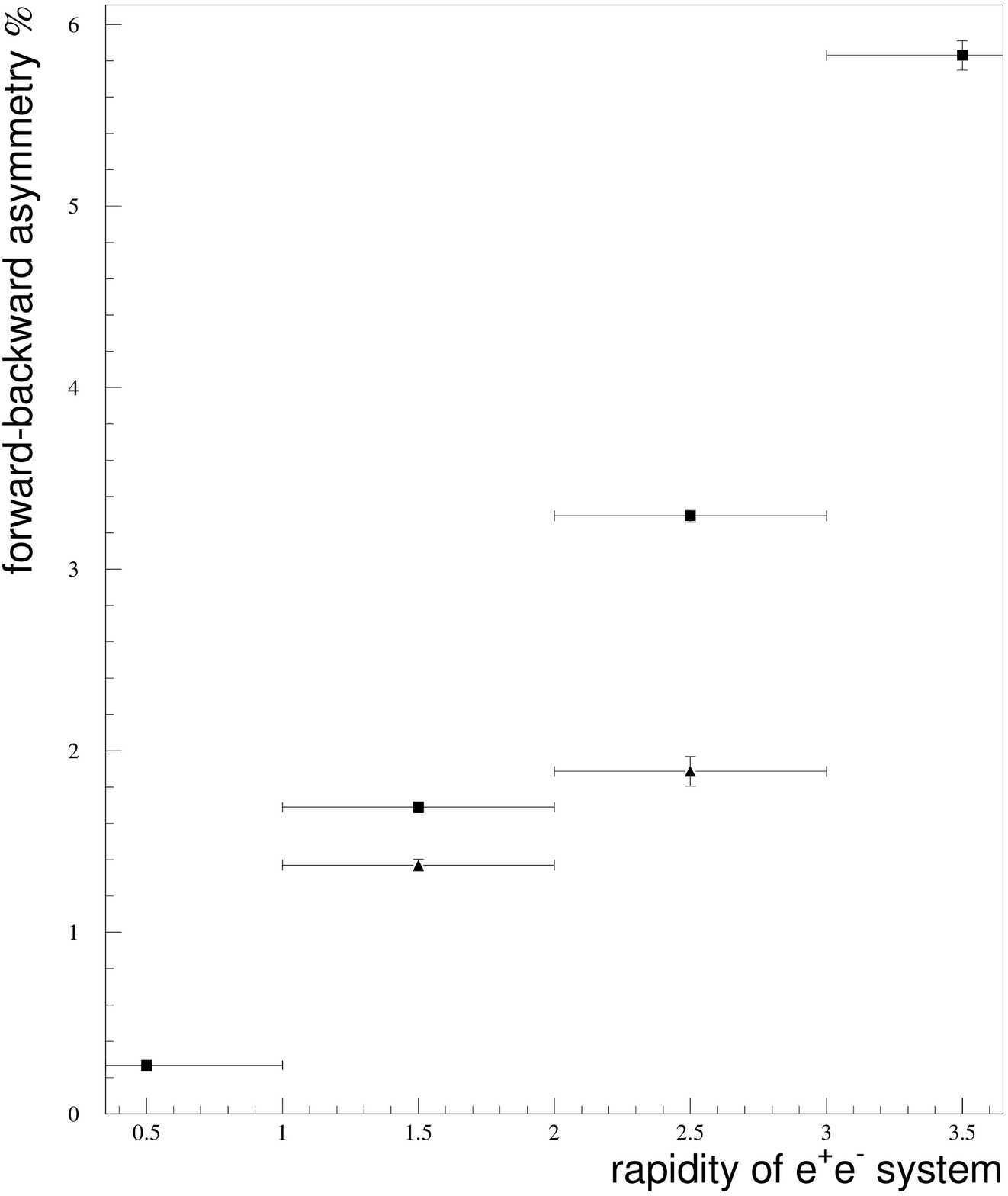}
\end{center}
\caption{Forward-backward asymmetry vs rapidity for $e^+e^-$ pairs from $Z$ 
decays satisfying the selection cuts described in Section~\ref{sec:AFB}. 
The asymmetry is shown where both electrons have $|\eta| < 2.5$ (triangles)
and where one electron is allowed to have  $|\eta| < 4.9$ (squares).
The results are the same for both sets of cuts in the first bin.
\label{fig:AFBvsy}}
\end{figure}

\paragraph{Statistical sensitivity}

The sensitivity of $A_{FB}$ to $\wangle$ can be parametrised
as follows:
\begin{eqnarray}\label{wanrcor}
A_{FB} & = & b (a-\wangle)\\ 
a^{O(\alpha^3)} & = & a^{Born} + \Delta a^{QED} + \Delta a^{QCD} \nonumber \\
b^{O(\alpha^3)} & = & b^{Born} + \Delta b^{QED} + \Delta b^{QCD} \nonumber
\end{eqnarray}
Values of $a$ and $b$ were calculated in~\cite{Baur:1998wa} and have 
been re-evaluated by Baur corresponding to the above cuts - 
see Table~\ref{tab:abparams}.

\begin{table}[hbt]
\begin{center}
\caption{Parameters $a$ and $b$ in Equation~\protect{\ref{wanrcor}}.} 
\label{tab:abparams}
\vskip0.2cm
\begin{tabular}{lcccccccc}
\hline
Cuts    & $a^{Born}$ & $\Delta a^{QED}$ & $\Delta a^{QCD}$ &$a^{O(\alpha^3)}$ &
          $b^{Born}$ & $\Delta b^{QED}$ & $\Delta b^{QCD}$ &$b^{O(\alpha^3)}$ \\ 
\hline
$|\eta| < 2.5$ both $e^\pm$&
        .2481 & .0025 & -.0026 & .2480 & 0.48 & -0.01 & -0.16 & 0.31 \\ 
\hline
$|\eta| < 2.5$ both $e^\pm$&
               &        &         &        &      &       &       &      \\ 
$|y(e^+e^-)| > 1.0$&
          .2503 & -.0009 & -.0069 & .2425 & 0.74 & 0.05 & -0.03 & 0.76 \\      
\hline
$|\eta| < 2.5$ one $e^\pm$&
               &        &         &        &      &       &       &      \\ 
$|\eta| < 4.9$ the other $e^\pm$  &
        .2483 & -.0005 & -.0015 & .2463 & 1.18 & 0.15 & -0.10 & 1.23 \\ 
\hline
$|\eta| < 2.5$ one $e^\pm$&
               &        &         &        &      &       &       &      \\ 
$|\eta| < 4.9$ the other $e^\pm$  &
               &        &         &        &      &       &       &      \\ 
$|y(e^+e^-)| > 1.0$&
        .2486 & .0011 & -.0028 & .2469 & 1.66 & 0.01 & -0.04 & 1.63 \\ 
\hline
\end{tabular}
\end{center}
\end{table}
 
A summary of the statistical errors which can be obtained 
with 100~fb$^{-1}$ are indicated in
Table~\ref{tab:stwprecision}.
With the best rejection factors shown in the table, the effect of the 
background is negligible.
If no jet rejection is possible in the forward calorimetry, the 
statistical precisions which can be obtained on $\wangle$
are $3.4 \times 10^{-4}$ and $4.1 \times 10^{-4}$ for no $y$ cut
and $|y(e^+e^-)| > 1.0$ respectively.
While the sensitivity to $\wangle$ is increased by cutting on 
$|y(e^+e^-)|$ (see Table~\ref{tab:abparams}), the gain is
reduced by the loss of acceptance and increased significance of the 
background when the forward calorimetry is used. 
It is probable that greater sensitivity could be obtained by fitting 
$A_{FB}$ as a function of $|y(e^+e^-)|$.

\begin{table}[htb]
\begin{center}
\caption{Statistical 
precision which can be obtained on $\wangle$ from measurements of 
$A_{FB}$ in $Z \rightarrow ee$ from one LHC experiment with
100~fb$^{-1}$.
Results are given for different jet rejection factors $\rho$ for the
forward calorimetry $2.5 < |\eta| < 4.9$.}
\label{tab:stwprecision}
\vskip0.2cm
\begin{tabular}{lcccc}
\hline
Cuts    & $\rho$ & $A_{FB}$ (\%) & $\Delta A_{FB}$ (\%) & $\Delta \wangle$ \\ 
\hline
$|\eta| < 2.5$ both $e^\pm$ & - & 0.774 & 0.020 & $6.6 \times 10^{-4}$ \\ 
\hline
$|\eta| < 2.5$ both $e^\pm$ &  & &  \\
$|y(e^+e^-)| > 1.0$ & - &        1.66 & 0.030 & $4.0\times 10^{-4}$ \\
\hline
$|\eta| < 2.5$ one $e^\pm$ & $10^{4}$ & 2.02 & 0.017 & 1.4$\times 10^{-4}$\\ 
\cline{2-5}
$|\eta| < 4.9$ the other $e^\pm$ & $10^{2}$ & 1.98 & 0.018 & 1.4$\times 10^{-4}$\\ 
\cline{2-5}
                                 & $10^{1}$ & 1.68 & 0.021 & 1.7$\times 10^{-4}$ \\
\hline
$|\eta| < 2.5$ one $e^\pm$ & $10^{4}$ & 3.04 & 0.022 & 1.35$\times 10^{-4}$\\ 
\cline{2-5}
$|\eta| < 4.9$ the other $e^\pm$ & $10^{2}$ & 2.94 & 0.023 & 1.41$\times 10^{-4}$\\ 
\cline{2-5}
$|y(e^+e^-)| > 1.0$ & $10^{1}$ & 2.31 & 0.030 & 1.83$\times 10^{-4}$ \\
\hline
\end{tabular}
\end{center}
\end{table}

From Table~\ref{tab:stwprecision}, it can be seen that for a 
single lepton species from one LHC experiment, using 
leptons measured in $|\eta| < 2.5$, a statistical precision of
$4.0 \times 10^{-4}$ on $\wangle$ could be obtained.
With the combination of electrons and muons in two experiments,
$2.0 \times 10^{-4}$ could be obtained.

The table shows that for moderate jet rejection ($\geq \sim10^2$)
in the forward calorimetry, 
a statistical precision of $1.4 \times 10^{-4}$ could be reached
by a single experiment using just the electron channel 
(cannot include the muons).
Even a poor rejection $\sim10$, would provide a useful measurement.
While no studies with full detector simulation have been done,
its seems likely that both the ATLAS and CMS
forward calorimetry will be able to 
provide useful electron identification because of moderate 
longitudinal and transverse segmentation.
Combining both experiments 
will permit a further $\sqrt{2}$ reduction in the statistical 
uncertainty.

\paragraph{Systematic uncertainties}
 
In order to be able 
to exploit the possibility of measuring $\wangle$ with such
high precision, the systematic errors have to be comparably small. 
Quick estimates indicate that the following factors are the most 
important ones:
\begin{enumerate}
\item
{\it pdf's:} affect both the lepton acceptance
as well as the results of radiative correction 
calculations.
\item
{\it Lepton acceptance and reconstruction 
efficiency as a function of lepton rapidity:}
while there is some cancellation in the determination of the asymmetry,
the product will need to be known to better than 0.1\%.
CDF \cite{cdfafb} has shown that it is possible to achieve a precision of 
about 1\%, with the largest contribution being due to the uncertainty in 
the pdf's.
\item
{\it Effects of higher order QCD (and electroweak) corrections:}
can be estimated by varying the errors on the parameters $a$ and $b$.
\item
{\it Mass Scale:}
$A_{FB}$ varies as a function of the invariant mass of the lepton pair.
Since the measured asymmetry corresponds to an integration over the $Z$
resonance, it is important to understand the mass scale.
It is expected that this will be known to $\sim 0.02$\% 
(see \ref{sec:wmass_uncertain}) by direct comparison of the $Z$ peak with
the measured LEP parameters.
\end{enumerate}

The most important systematic contribution is that coming from the 
uncertainties in the pdf's.
A study using several ``modern'' pdf's (MRST, CTEQ3 and CTEQ4) 
gave agreement between the 
resulting values of $A_{FB}$ within the 1\% statistical errors of the study
($5\times10^5$ events were generated for each pdf set).
This uncertainty must be reduced by a factor of 10 if it is to be smaller than
the expected statistical precision on $A_{FB}$ shown in 
Table~\ref{tab:stwprecision}.
It remains to be seen whether (a) the differences arising from 
the various pdf's
will shrink with increased statistical sensitivity of the study
and (b) whether the current pdf's actually describe the measured data
sufficiently well (since the pfd's are fitted to common data, 
variations are not necessarily indicative of the actual uncertainties).
New measurements from the Tevatron (and ultimately the LHC itself) 
will improve 
the understanding of the pdf's, but it is unclear at this stage whether 
this will be sufficient.
It may be possible to fit simultaneously $\wangle$ and the pdf's, or 
alternatively, it may be necessary to reverse the strategy and use 
the measurement of $A_{FB}$ combined with existing measurements of 
$\wangle$ to constrain the pdf's.

\subsubsection{Search for new phenomena}

\paragraph{Contact interactions}
Contact interactions offer a
general framework for a new interaction with coupling $g$ and
typical energy scale  $\L \gg \sqrt{\hat s}$. 
At LEP2, the current limits~\cite{Bourilkov:1998,OPAL:1999} for
quark-lepton compositeness at 95\% CL 
vary between 3 and 8~TeV, depending on the model. 
At the LHC scales up to 25-30 TeV are reachable,
as illustrated in Figure~\ref{fig:ci-zpr}.

\paragraph{Search for resonances}
The other extreme is the search for resonances like $Z'$ or $\SN$,
which produce peaks in the mass distributions.
A neutral heavy gauge-boson $Z'$ is characterised by
its mass $m_{Z'}$, by its couplings and by its mixing angle $\theta_M$
with the standard $Z$ boson.
If $\theta_M=0$ and the $Z'$ has SM couplings, the current limit is
$m_{Z'} > 1050$~GeV~\cite{ALEPH:1999}.
For other coupling scenarios the lower limits are model dependent 
and typically of the order of several hundred GeV.
Resonances with masses up to $\sim$~4-5 TeV can be probed at LHC,
as shown in Figure~\ref{fig:ci-zpr}.

\paragraph{R-parity violation}
In SUSY theories with R-parity violation, 
it is possible to couple sleptons to pairs of SM leptons or quarks
through new independent Yukawa couplings 
(9 $\l$ couplings for the slepton-lepton sector and 27 $\l^{'}$ 
couplings for the slepton-quark sector).
This makes the resonance formation of  single scalar neutrino $\SN$ 
in $d\bar{d}$ scattering possible.
It can be observed through the decay of the $\SN$
to lepton pairs, if a suitable combination of
two couplings (e.g. $\l'_{311}\l_{131}$) is present~\cite{Kalinowski:1997}.
The K-factor for slepton production is not calculated yet,
leading to an uncertainty $\sim$ 10\% in the estimate of 
the $\l \l'$ sensitivity.

\paragraph{Low-scale gravity}
An exciting possibility is the search for low-scale gravity  effects
in theories with extra spatial dimensions,
leading to virtual graviton exchange.
The best limits at LEP2 come from combined analysis of Bhabha
scattering~\cite{Bourilkov:1999}:\\
\begin{center}
$\Lambda_T = 1.412 (1.077)$\ TeV  for  $\lambda = +1 (-1)$ at 95\%  CL
\end{center}

In the Drell-Yan process there is an unique contribution from $s$-channel
graviton exchange~\cite{Hewett:1998}, which modifies the form of the
differential cross section and gives a distinct signature:
\begin{equation}
gg \TO l^+l^-
\label{dygrav1}
\end{equation}
\begin{equation}
\frac{d\; \sigma}{d\; \COST} = \frac{\lambda^2\hat s^3}{64\pi M_s^8} (1-\cos^{4}\theta)
\label{dygrav2}
\end{equation}
The large parton luminosity for gluons at LHC may also compensate the
$M_s^{-8}$ suppression. Scales up to $\sim$~5~TeV can be probed with
luminosity 100 fb$^{-1}$.

\begin{center}
  \begin{figure}
  \begin{tabular}{cc}
\resizebox{0.52\textwidth}{9.0cm}{\includegraphics{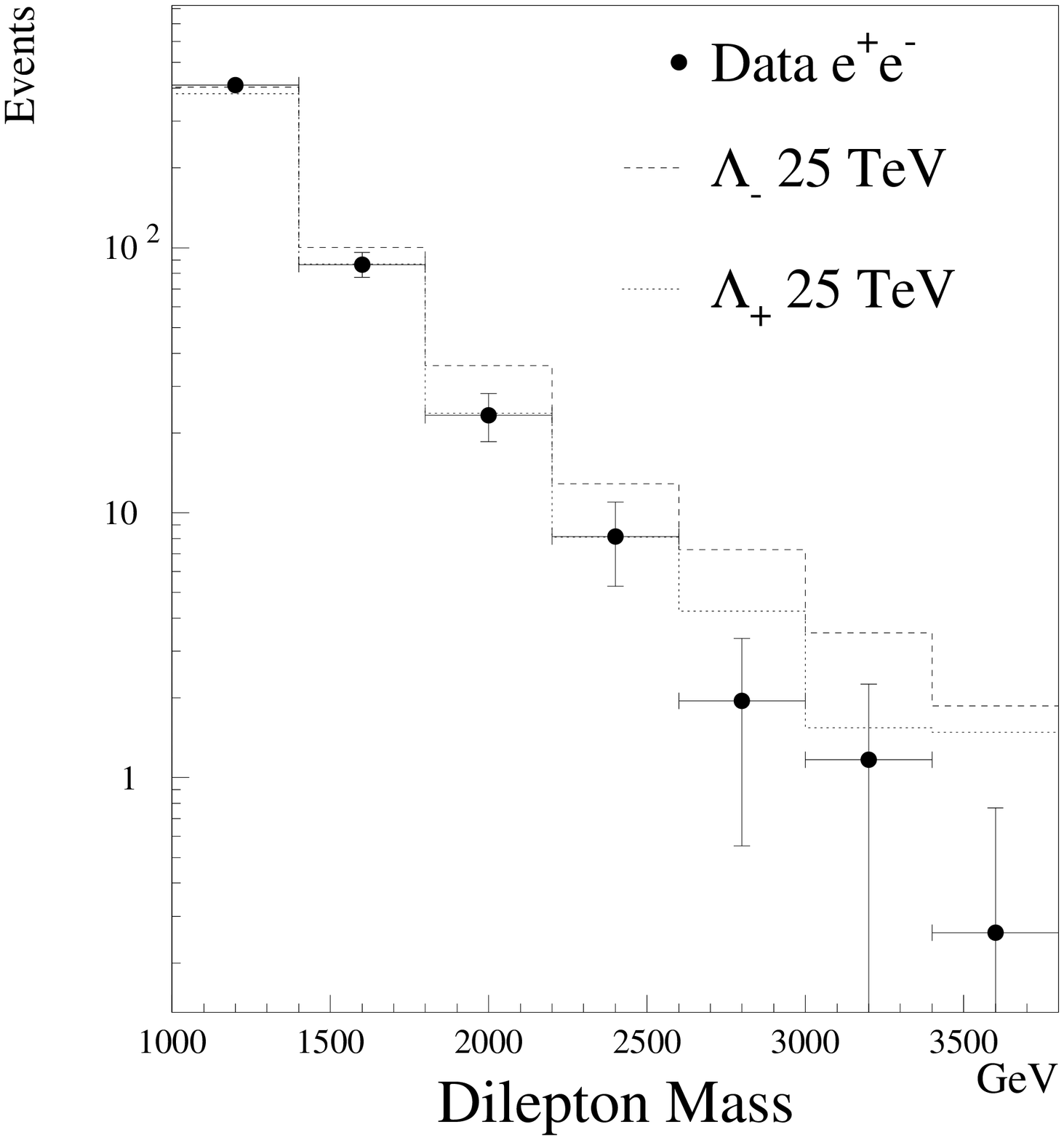}} &
\resizebox{0.47\textwidth}{8.2cm}{\includegraphics{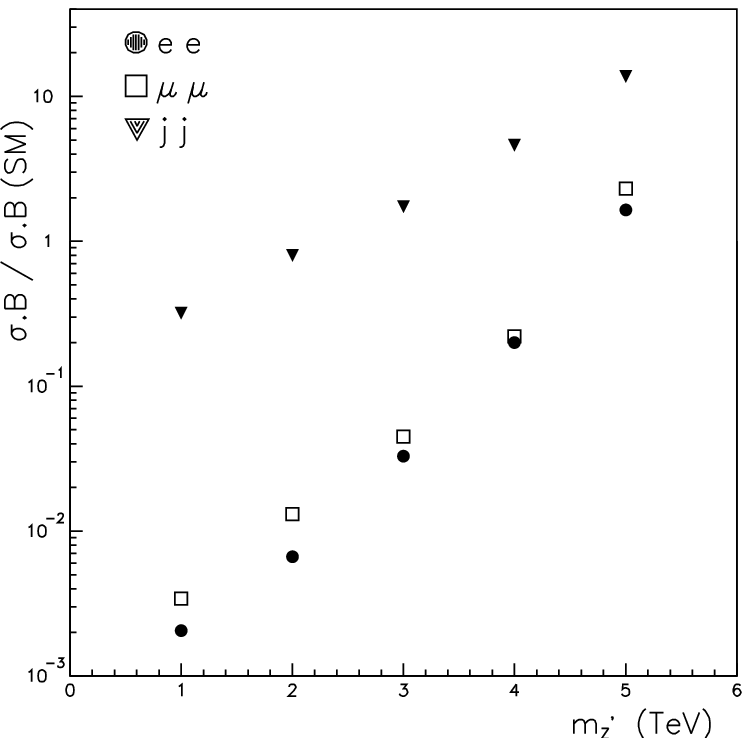}}
  \end{tabular}
    \caption{Left: Contact interaction sensitivity (CMS study).
             Error bars show SM spectrum; histograms show effect of 
             contact term with $\Lambda = 25$~TeV, the sign corresponds to
             the sign of the amplitude.
             Right: $Z'$ mass reach for 100~fb$^{-1}$
             (ATLAS study)~\cite{atlas-phystdr2}.}
    \label{fig:ci-zpr}
  \end{figure}
\end{center}

\subsubsection{Summary}
The main results of this study are:
\begin{itemize}
 \item A competitive measurement of $\wangle$ is hard due to the
       central acceptance of the experiments and the difficulty of controlling
       the pdf's (parton distribution functions)
       with the required precision. However, a detector with 
       extended forward acceptance for one of the leptons offers 
       the possibility to measure $\wangle$ with a statistical precision 
       of $1.4 \times 10^{-4}$.
 \item The total cross-section can be measured with systematic error
       $\frac{\sigma_\mathrm{DY}^{\mathrm{high} \ Q^2}}{\sigma_Z}$ \mbox{$<$ 1\%.}
 \item A non-zero forward-backward asymmetry $A_{FB}$ can be measured up to 2 TeV
       with statistical precision $> 3 \ \sigma$. 
 \item The Drell-Yan process can probe electroweak radiative corrections
       up to 1.5 TeV with statistical precision at the $2\ \sigma$ level
       as a function of $Q^2$.
 \item The high energy and luminosity of LHC offers a rich search field
       at the TeV scale in the Drell-Yan channel: contact interactions,
       resonance formation ($Z'$, scalar neutrinos),
       low-scale gravity, {\it etc.}.
\end{itemize}
\vspace{0.2cm}

Further studies will refine the following points:
\begin{itemize}
 \item The effect of higher order QED corrections (initial- and final-state radiation
  and their interference).
 \item The effect of experimental cuts on the electroweak corrections.
 \item The careful separation of the $\sigma_{u\bar u}$ and 
       $\sigma_{d\bar d}$ contributions.
\end{itemize}

\subsection{Tau physics
}

The $\tau$ lepton is a member of the third generation which decays
into particles belonging to the first and second ones. Thus, $\tau$
physics could provide some clues to the puzzle of the recurring
families of leptons and quarks. One na\"{\i}vely expects the heavier
fermions to be more sensitive to whatever dynamics is responsible for
the fermion-mass generation. 
 The pure leptonic or semileptonic character of $\tau$  decays
provides a clean laboratory to test the structure of the weak
currents  and the universality of their couplings to the gauge-bosons.

 The last few years have witnessed a substantial change in our
knowledge of the $\tau$ properties \cite{tau98,lp99}. The large
(and clean) data samples collected by the most recent experiments
have improved considerably the statistical accuracy and, moreover,
have brought a new level of systematic understanding.

A high-energy hadron collider does not provide a very good
environment to perform precision $\tau$ physics.
Nevertheless, there are a few topics where LHC could contribute
in a relevant and unique way. Moreover, since the $\tau$ is the
heaviest known lepton, it can play a very important role in
searches for new particles (for example, as in Section~\ref{sec:vvhtt}).

\subsubsection{Charged-current universality}

\begin{table}[htb]
\begin{center}
\caption{Present constraints on charged-current lepton universality
\protect{\cite{lp99}}.}
\vskip0.2cm
\begin{tabular}{lccc}
\hline
 & $|g_\mu/g_e|$  & $|g_\tau/g_\mu|$ & $|g_\tau/g_e|$ \\
\hline
$B_{\tau\to\mu}/B_{\tau\to e}$   & $1.0009\pm 0.0022$ & --- & --- 
\\
$B_{\tau\to e}\,\,\tau_\mu/\tau_\tau$ & ---  & $0.9993\pm 0.0023$ &
--- 
\\
$B_{\tau\to\mu}\,\,\tau_\mu/\tau_\tau$ & ---  & ---  & $1.0002\pm
0.0023$
\\
$B_{\pi\to e}/B_{\pi\to\mu}$  & $1.0017\pm 0.0015$  &  --- & --- 
\\
$\Gamma_{\tau\to\pi}/\Gamma_{\pi\to\mu}$ & ---  & $1.005\pm 0.005$ &
--- 
\\
$\Gamma_{\tau\to K}/\Gamma_{K\to\mu}$ & ---  & $0.981\pm 0.018$ & --- 
\\
$B_{W\to l}/B_{W\to l'}$ \  ($p\bar p$) & $0.98\pm 0.03$ &  --- &
$0.987\pm 0.025$
\\
$B_{W\to l}/B_{W\to l'}$ \ (LEP2) & $1.002\pm 0.016$ & 
$1.008\pm 0.019$ & $1.010\pm 0.019$
\\ \hline
\end{tabular}
\label{tab:universality}
\end{center}
\end{table}

Table~\ref{tab:universality}  shows the present
experimental tests on the universality of the
leptonic charged-current couplings. The leptonic
$\tau$ branching ratios are already known with a quite impressive
precision of $0.3\% $; this translates into a test of $g_\mu/g_e$
universality at the 0.22\% level. 
However, in order to test the ratios $g_\tau/g_\mu$ and $g_\tau/g_e$,
one needs precise measurements of the $\tau$
mass and lifetime, in addition. At present, these quantities are known with a
precision of  0.016\% ($m_\tau= 1777.05\,{}^{+0.29}_{-0.26}$ MeV)
and 0.34\% ($\tau_\tau = 290.77\pm 0.99$ fs),
respectively \cite{lp99}, which
leads to a sensitivity of 0.23\% for the three
$g_l/g_{l'}$ ratios.

Future high-luminosity $e^+e^-$
colliders running near the $\tau^+\tau^-$ production threshold
could perform more precise measurements of the leptonic $\tau$
branching fractions and the $\tau$ mass. However, one needs a
high-energy machine to measure the $\tau$ lifetime. 
Clearly, the future tests of lepton universality will be limited by
the $\tau_\tau$ accuracy.
It is not clear whether the
$B$-factories would be able to improve the present
$\tau_\tau$ measurement in a significant way. Thus, it is important
to know how well $\tau_\tau$ can be determined at LHC. 

A less precise but more direct test on the universality of the
leptonic $W^\pm$ couplings is provided by the comparison of the
different $W^+ \to l^+\nu_l$ branching fractions. 
LEP2 has already achieved a better sensitivity than the Tevatron
collider, and a further improvement is expected when the full
LEP2 statistics will be available. It is an open question whether
LHC could be competitive at this level ($\sim 1\% $) of precision.

\subsubsection{Tau lifetime
}

The current world average for the $\tau$ lifetime is 
$290.8 \pm 1.0$~fs ($c\tau = 87 \mu$m) \cite{lp99}.
Improvements in this measurement would be welcome in order to 
give better tests of the Standard Model, in particular lepton 
universality and electroweak calculations. In this section, the 
results of a preliminary study to examine the LHC potential are 
given.

In LEP experiments, $\tau$ pairs are produced back-to-back with 
well defined momenta - this will not be the case at the LHC.
The first feature allows valuable correlations to be made between the 
two $\tau$ decays, while the second provides the boost required
to obtain proper lifetime estimates. At the LHC, 
$Z \rightarrow \tau\tau$ events will be triggered by requiring
one $\tau$ to decay to an electron or muon, while the 
lifetime is estimated from the other $\tau$ which is required
to decay to three charged particles. 

\paragraph{Tau reconstruction}
 
A study was made using fully simulated events in the ATLAS detector
(see \cite{atlas-phystdr1} for more details).
When the $Z$ has some transverse momentum, 
the momenta of the $\tau$'s can be deduced by projecting the 
recoil momentum vector measured by the calorimetry along the 
lines of flight of the two 
$\tau$'s (determined from the direction of the lepton and
the hadronic jet, respectively).
Due to resolution effects, this procedure works best 
when the $\tau$'s are not back-to-back. 
The following cuts were made:
\begin{itemize}
\item
The lepton should have $p_T > 24$~GeV, $|\eta| < 2.5$.
\item
The identified hadronic jet should contain three charged tracks and 
satisfy $E_T > 30$~GeV, $|\eta| < 2.5$.
\item
Transverse mass of lepton and missing energy should be $< 50$~GeV.
\item
The angle $\Delta\phi$ between the $\tau$'s in the transverse plane should
satisfy: $1.8 < \Delta\phi < 2.7$ or $3.6 < \Delta\phi < 4.5$.
\item
The invariant mass of the $\tau$ pair should satisfy: 
$60 < m_{\tau\tau} < 120$~GeV.
\end{itemize}
These cuts result in an efficiency of 1.5\%.
For these events, the $\tau$ momenta could be estimated with a 
resolution of 15\%.

A vertex was formed from the charged tracks in the hadronic jet.
It was required that the vertex should be within 2~cm of the 
interaction point and the invariant mass of the particles should
be between 0.4 and 1.78~GeV.
The efficiency for this was 70\% and resulted in a 
resolution on the vertex 
position in the transverse plane of $490 \mu$m, corresponding to a 
resolution on the proper decay length of $17 \mu$m.

\paragraph{Lifetime estimate}

The statistical resolution on the proper decay length from 
the combination of the vertexing and the estimate of the tau momentum 
is of the order of $21~\mu$m (corresponding to 55~fs). 
A simple Monte Carlo study was made to estimate the 
statistical uncertainty on the $\tau$ lifetime ($\tau_{\tau}$) 
which could be 
achieved with $N$ hadronic $\tau$ decays.
Since the resolution of the lifetime for a single event (55~fs)
is a fair bit smaller than the $\tau$ lifetime (291~fs),
the statistical error which can be obtained is dominated by the
number of events: 
$\sigma(\tau_{\tau}) \approx \tau_{\tau}/\sqrt{N}$.

At the LHC, the cross section for $Z \rightarrow \tau\tau$ 
will be 1.5~nb, with a 
branching ratio of 11\% for a lepton and a three-prong hadronic decay. 
The reconstruction and selection described above results in an 
efficiency of 0.54\%. If 30~fb$^{-1}$ were collected in a low luminosity 
run, then 26,000 reconstructed $\tau$'s could be used, leading to a 
statistical error on the lifetime of 1.8~fs.
To make this competitive would require increased efficiency for 
selecting the $\tau$ decays - this is probably a low luminosity 
measurement and so cannot benefit from the statistics of a
high luminosity run.

Increasing the efficiency may not be simple, since the cuts were designed
to control the background.
$W+$jet events will be removed by the mass cuts, 
and apart from a small amount of gluon splitting to heavy flavour, 
the jets should not contain significant lifetime information, hence 
this background should not be a problem. The $B$ lifetime is a factor 
of five larger than that of the $\tau$, and hence more care will be required 
with $b\bar{b}$ events. 

Concerning systematic errors coming from the determination of the 
decay length in the silicon tracking, the average radial position 
of the detectors in the vertexing layer will need to be understood to 
better than $10~\mu$m. This will be challenging but studies suggest this 
may be feasible \cite{haywood-99}.
It should be possible to control the systematics on the measurement of 
recoil momentum of the $Z$ by comparison with $Z \rightarrow ee$ or 
$Z \rightarrow \mu\mu$ events, where the recoil can be measured accurately
by the leptons.

\subparagraph{The use of \boldmath $W \rightarrow \tau\nu$}

It may be possible to use the decays $W \rightarrow \tau\nu$
which have a higher cross section than $Z \rightarrow \tau\tau$.
In ATLAS, such events could be triggered by a special $\tau$-jet and missing
$E_T$ trigger \cite{atlas-phystdr1}.
Information about the $\tau$ momentum can be deduced
by comparing the energy and direction of the hadronic jet with the
direction of the $\tau$ and using the $\tau$ mass constraint, where
the $\tau$ direction can be determined from the reconstructed decay
vertex.
In principle, 
it is possible to solve completely for the $\tau$ momentum, although 
resolution effects on the vertex position and complications arising from
$\pi^0$'s in the hadronic jet mean that sometimes solutions are not physical.
Alternatively, an approximate estimator can be formed 
which does not employ the mass constraint~\cite{bib:e653}.
This uses the $\tau$-jet energy, mass and $p_T$ relative
to the $\tau$ direction - all three quantities
being determined from the charged tracks alone.
This is more robust but its behaviour is sensitive to the selection cuts.
It is yet to be proved that a $W \rightarrow \tau\nu$ signal can be 
identified with sufficient efficiency above the huge QCD (and in particular,
$b\bar{b}$) background.

\subsubsection{Rare decays}

Owing to the huge backgrounds,
it will not be possible to make a general search for rare
decay modes of the $\tau$. However, the lepton-number violating
decay $\tau^-\to\mu^+\mu^-\mu^-$ has a clean signature, which is
well suited for the LHC detectors. The present experimental bound
\cite{cleoLV} is
$$
BR(\tau^-\to\mu^+\mu^-\mu^-) < 1.9 \times 10^{-6}
\quad (90\%\;\mbox{\rm CL}) 
$$
This limit reflects the size of  the existing $\tau$ data samples. LHC will
produce a huge statistics, several orders of magnitude larger
than the present one.
The achievable limit will then be set by systematics and backgrounds,
which need to be properly estimated.
A sensitivity at the level of $10^{-8}$ does not seem out of reach.
This could open a very interesting window into new physics phenomena,
since many extensions of the Standard Model framework can lead to
signals in the $10^{-6}$ to $10^{-8}$ range.

Although more difficult to detect, other lepton-number
violating decays  such as $\tau\to\mu\mu e$,$\mu e e$,$eee$,
$\mu\gamma$ are worth studying.

\section{VECTOR-BOSON PAIR PRODUCTION
         \protect\footnote{Section coordinator: Z.~Kunszt}
         \label{sec:vbpp}}

\subsection{\boldmath $W^+W^-, W^{\pm}Z, ZZ$ \unboldmath production}

\subsubsection{Recent numerical implementations}

As already  is noted in the introduction,
 for the description of $W^+W^-, W^{\pm}Z, ZZ$ production
with their subsequent decays into lepton pairs
two new numerical parton-level Monte Carlo programs have  recently become
available \cite{Campbell:1999ah}(MCFM), \cite{Dixon:1999di}(DKS).
These packages consider the production of four leptons  in the double 
resonance approximation with complete 
${\cal O}( \alpha_s) $ corrections.
They can be used to compute any infra-red safe quantity with
arbitrary experimental cuts on the leptonic decay products. 
These packages have already been used for updating and cross-checking
previous results. The DKS program is available in fortran90 and
fortran77 versions and  includes  anomalous triple gauge-boson
couplings. The MCFM program is more complete in the
sense that single resonance background diagrams are also added
and finite width effects are included in some approximation
which respects gauge-invariance. However, it does not include anomalous 
triple
gauge-boson  couplings.
The results of the MCFM and DKS programs
agree with each other within the
 integration error of  $\le 0.5\%$. Similar agreement
is found with the spin averaged cross section indicated 
in~\cite{ZZit, WZit, WWit}.
In the past  years the majority  of 
the experimental studies used    the programs  described in 
\cite{BHOWZZero, BHOWZ, BHOWW} (BHO).
 A recent comparison between
the DKS and BHO programs finds  agreement at the level of 1\% for
$WZ$ production and 3-4\% for $WW$ production (further details see 
Section~\ref{sec:tgc_nlo}).
This  confirms  the assumptions of~\cite{BHOWZZero}
that the  spin correlations effects coming from virtual corrections
are small.
Note that 
recently a new ${\cal O}( \alpha_s$) package has been
written also for  $W\gamma$ and
$Z\gamma$ production with  anomalous couplings \cite{Deflo-wgam:2000}
 and for the
first time the complete one loop QCD corrections are available
also for these processes (see Section~\ref{WgNLOprod}).

\subsubsection{Input parameters and bench mark cross sections}

In using these packages,  one should be careful with
input parameters.
The QCD input is standard: the latest
next-to-leading order parton number densities have to be used 
with the corresponding
running coupling constant at some physical scale
 defined in terms of the kinematics of the outgoing particles.

The  helicity amplitudes coded  into these  programs   are calculated in 
${\cal O}(\alpha_s)$ but they are   leading
order in the  electroweak theory. However, the one loop electroweak  
radiative corrections are not completely negligible.
The dominant corrections are given by   light fermion
loops and  large custodial symmetry violating contributions of the
top quark. Fortunately, they are universal and can be taken into account
in the spirit of the ``improved Born approximation'' ~\cite{IBA1,IBA2}.
Universality means that their contributions can  modify only 
 the leading order relation between $M_Z$, $M_W$ and $\sin^2\theta_W$ which
  can be taken into account with the use of 
 the  effective coupling 
\beq
   \sin^2\theta_W \equiv { \pi \alpha(M_Z) \over \sqrt{2} G_F M_W^2 } \,,
\eeq
where $G_F = 1.16639 \times 10^{-5}$~GeV$^{-2}$ is the Fermi constant
and $\alpha(\mu)$ is the running QED coupling.  With the values of
the gauge-bosons masses
of $M_Z=91.187$~GeV and
$M_W=80.41$~GeV, one obtains 
 $\alpha=\alpha(M_Z)=1/128$ 
and $\sin^2\theta_W=0.230$.
Ignoring this correlation  
may lead to  about 5-6\% discrepancy in the cross section values.
The remaining electroweak  corrections are 
estimated to be less than  2\%  as long the parton sub-energy
is below $0.5 -1\TeV$.
However, above the $1\,\TeV$ scale double logarithmic
corrections of ${\cal O}(
\alpha_W\log^2 \hat{s}/M_W^2)$  become  non-negligible. The origin of these
large contributions is 
the incomplete cancellation of 
the soft singularities of massless gauge-boson emission (the Bloch-Nordsieck 
theorem is not valid for non-Abelian theories~\cite{Ciafaloni:2000df}). Since 
the physical cross section  decreases strongly  with the
 increase  of the invariant mass of the gauge-boson pairs, 
 these corrections are not important at the LHC.
The validity of the improved Born approximation and the presence  of 
the double logarithmic corrections has been tested 
for $W$ pair production at LEP2  where the full next-to-leading order
corrections are available \cite{IBA1,IBA2}.

Additional  electroweak input parameters are  the
matrix elements of the CKM mixing matrix.
In the light quark sector, one should use the best experimental 
values  \cite{PDGcaso}. 
In the case of the heavy quark contributions, the calculation
is approximate since the 
 ${\cal O}(\alpha_s)$ helicity amplitudes  have been calculated assuming
massless quarks 
\cite{Dixon:1998py}. This  assumption is clearly not valid for the top
contributions.
 $WW$ pair production receives  
 contributions from diagrams with 
the $t$-channel exchange of the top quark (with
$|V_{td}|=|V_{ts}|=0$ and $|V_{tb}|=1$).
However, it is  
suppressed due to  the large top mass  and small $b$-quark  parton densities; 
 therefore,  it is  reasonable to use
 $|V_{tb}|=0$.
The contribution of the subprocess
$b \bar{b} \to W^+ W^-$ (treating the top as massless)
is of the order of 2\% for the LHC
\cite{Dixon:1998py} giving an upper limit on the theoretical
ambiguity coming from this source.
In the case of $W^\pm Z$ production, one can neglect
the subprocess $ b g \to W^- Z t $. It is present at
next-to-leading order but  again it is strongly suppressed by the large top
quark mass, as well as the small $b$-quark distribution function.
For the numerical results presented here,  values
 $|V_{ud}|=|V_{cs}|=0.975; \ |V_{us}|=|V_{cd}|=0.222$ and
 $|V_{ub}|=|V_{cb}|=|V_{td}|=|V_{ts}|=|V_{tb}|=0$ are used.   
We present  cross-section values without including the
branching ratios.
To get  event signals, they
have to be multiplied   with the leptonic branching
ratios of the vector-bosons. We use 
$$ BR(Z\to e^+e^-\ \ {\rm or}\ \  \mu^+\mu^-  )
 =  3.37\%\quad
 BR(Z\to\sum_{i=e,\mu,\tau}\nu_i\bar{\nu_i})
=20.1\%$$
$$BR(W^+\to e^+\nu_e\ \ {\rm or}\ \  \mu^+\nu_{\mu}^-  )
   =10.8\% 
$$
These ratios implicitly incorporate QCD corrections to 
the hadronic decay widths of the $W$ and $Z$.

Most of the results are obtained with some ``standard cuts'' 
defined as follows:   a transverse momentum cut of $p_T>20$~GeV 
and pseudorapidity cut of $|\eta| \le 2.5$ is applied for all charged 
leptons and  
$p_T^{\rm miss}\ge 20\GeV$
 is required for $WZ$ production
while $p_T^{\rm miss}\ge 25\GeV$ for $W$ pair production.
We use two different parton distributions, MRST 
\cite{mrst} with $M_W=80.41~\GeV$ and CTEQ(4M) \cite{cteq4}
with $M_W=80.33~\GeV$ which we refer to simply as
MRST and CTEQ. $\alpha_s(M_Z) = 0.1175$  is used for MRST 
and $\alpha_s(M_Z) = 0.116$ is used for CTEQ. 
In all computations, we set the renormalisation  and factorisation
scales equal to each other.

In Table~\ref{tab:XSlhc}, we present the total cross section values
for  the various
processes at the LHC, for the MRST and CTEQ parton
distributions.  We tabulated the results
for $\sigma^{\rm tot}$ (the cross sections without any cuts applied)
as well as $\sigma^{\rm cut}$ (the cross sections with the
standard cuts defined above).  
 The cross section values are given for the scale
\beq\label{scalemw}
 \mu =
(M_{V_1} + M_{V_2})/2,
\eeq
 where $M_{V_i}$ are the masses of the two
produced vector bosons.

\begin{table}[htbp]
\begin{center}
\caption[dummy]{\small Cross sections in ${\rm pb}$ for $p p$ collisions at
$\sqrt{s} = 14$ TeV. The statistical errors are $\pm 1$ on the last
digit.  \label{tab:XSlhc}}
\vskip0.2cm
\begin{tabular}{lcccccccc} \hline
 & \multicolumn{2}{c}{$ZZ$} & \multicolumn{2}{c}{$W^+W^-$} 
 & \multicolumn{2}{c}{$W^-Z$} & \multicolumn{2}{c}{$W^+Z$} \\ 
 & LO & NLO & LO & NLO & LO & NLO & LO & NLO \\ \hline  
$\sigma^{\rm tot}$(MRST) 
  & 11.6 & 15.5 & 78.7 & 117 & 11.2 & 19.3 & 17.8 & 30.6 \\ \hline
$\sigma^{\rm tot}$(CTEQ) 
  & 11.8 & 15.8 & 81.3 & 120 & 11.4 & 19.6 & 18.6 & 31.9 \\ \hline
$\sigma^{\rm cut}$(MRST) 
  & 4.07 & 5.47 & 25.0 & 40.18 & 3.49 & 6.58 & 5.20 & 9.68 \\ \hline
$\sigma^{\rm cut}$(CTEQ) 
  & 4.09 & 5.51 & 25.6 & 42.0 & 3.59 & 6.72 & 5.32 & 9.83 \\ \hline 
\end{tabular}
\end{center}
\end{table}

In previous publications~\cite{ZZit, WZit,WWit, BHOWZZero,
BHOWZ, BHOWW,Campbell:1999ah,  Dixon:1999di}
a number of phenomenologically interesting questions have been
considered. Here 
  we restrict ourselves to recall two interesting and typical
features: the scale dependence of the radiative
corrections for $WW$ production and radiation zeros for
$WZ$ production.

\subsubsection{Scale dependence}

The one-loop corrections to the total cross sections 
 are of the order 50\% of the leading order term and they 
can be much larger for the kinematical range of
 larger transverse momenta or invariant
mass of the vector-boson pair. 
For differential distributions where $p_T$ is
not integrated out completely, the scale choice  
\beq\label{scalept} 
\mu^2 = \mu_{\rm st}^2 \equiv 
\frac{1}{2}(p_T^2(V_1)+p_T^2(V_2)+M_{V_1}^2+M_{V_2}^2)
\label{must}
\eeq 
appears to be appropriate. For the total cross section,
the difference between the two
 scale choices expressed in Equations~\ref{scalemw} and~\ref{scalept} 
is very small  since
it is dominated by low-$p_T$ vector-bosons. However, for more exclusive
quantities, the differences can be substantial. 
At the LHC, the huge one-loop
corrections  in the tails of the distributions 
 are dominated by the bremsstrahlung
contributions; therefore 
it is natural to  consider the cross sections
 with and without the  jet veto (that is, with or without
the cut $E_T^{\rm jet}<40$~GeV).

\begin{figure}[htbp]
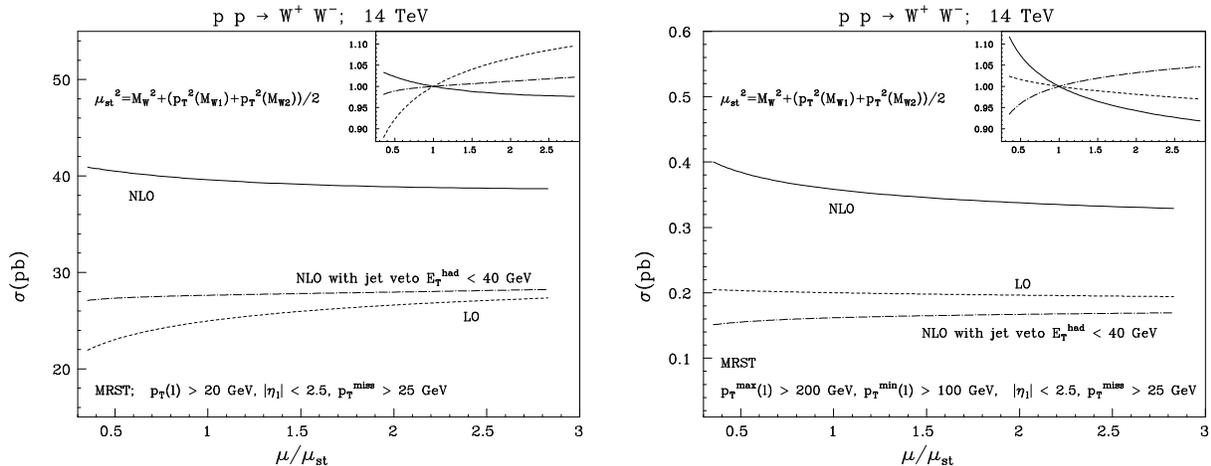

\begin{center}
   \epsfig{figure=XSlhc1.ps,width=0.48\textwidth,clip=}
   \hfill
   \epsfig{figure=XSlhc2.ps,width=0.48\textwidth,clip=} 
\caption{\label{fig:XSlhc} Scale dependence of 
   the cross section for $W$-pair production at the LHC with
   standard cuts. The scale is given in units of $\mu_{\rm st}$ as
   defined in Equation~\ref{must}. We show the LO, NLO and NLO with jet veto
   curves without additional cuts (left) and with an additional cut
   $p_T^{\rm max}(l)>200$~GeV and $p_T^{\rm min}(l)>100$~GeV
   (right). The insets show the  curves normalised to 1 at
   $\mu=\mu_{\rm st}$.  }
\end{center}
\end{figure}                                                              

In Figure~\ref{fig:XSlhc}, the scale dependence of $\sigma^{\rm cut}$ is shown
for standard cuts, with a jet veto and with stronger cuts 
on the  transverse momenta of the charged leptons. 
We can see that  the corrections are large
and increase with the additional cuts applied.
The scale dependence at LO is reduced at NLO and it is reduced further
when a jet veto is applied.
In particular, the size of the correction is strongly reduced when
applying the jet veto - an  important feature for  background
studies.

\subsubsection{Approximate radiation zeros in $WZ$ production}

In leading order,  the angular
distribution of $WZ$ production
 exhibits  an approximate radiation zero for
$\cos\theta = (g_1 + g_2)/(g_1 - g_2)$~\cite{BHOWZ}
where $g_1,g_2$ denote the $Z$ boson couplings to the left handed
up and down quarks, respectively.
Since the precise flight direction of the $W$ boson is not known (due to 
the uncertainty in the longitudinal momentum carried by the neutrino) 
it is convenient  
 to plot a distribution in the (true) rapidity difference
between the $Z$ boson and the charged lepton coming from the
decay of the $W$: $\Delta y_{Zl} \equiv y_Z - y_l$.
This quantity is similar to the rapidity difference 
$\Delta y_{WZ} \equiv | y_W - y_Z |$ studied in~\cite{WZit}, but
uses only the observable charged-lepton variables.  It is the direct
analogue of the variable $y_\gamma-y_{l^+}$ considered in~\cite{BEL}
for the case of $W\gamma$ production.  It is possible 
to determine $\cos\theta$ in the $W\gamma$ or $WZ$ rest frame, 
by solving for the neutrino longitudinal momentum using the 
$W$ mass as a constraint, up to a two-fold discrete ambiguity for each 
event~\cite{nulong, bib:gs, bib:s85}.  However, it has been found~\cite{BEL} 
that the
ambiguity degrades the radiation zero - at least if each solution is 
given a weight of 50\% - so that the rapidity difference 
$y_\gamma-y_{l^+}$ is more discriminating than $\cos\theta$.
As one can see from Figure~\ref{fig:RadZFig}, there is a residual dip in the
$\Delta y_{Zl}$ distribution, even at order $\alpha_s$. This dip
can  be enhanced easily by requiring a minimal energy for the decay
lepton from the $W$ and by cutting on the rapidity 
of the $Z$ boson. In Figure~\ref{fig:RadZFig}, we have chosen
$E(l)>100$~GeV with and without $y_Z<0$. 
 Note that the latter two curves are scaled up by a factor of 5.
 At the  LHC, for the first time,
 we shall have enough statistics to test
experimentally for the presence of approximate radiation zeros.

\begin{figure}[htb]
\centerline{ \epsfig{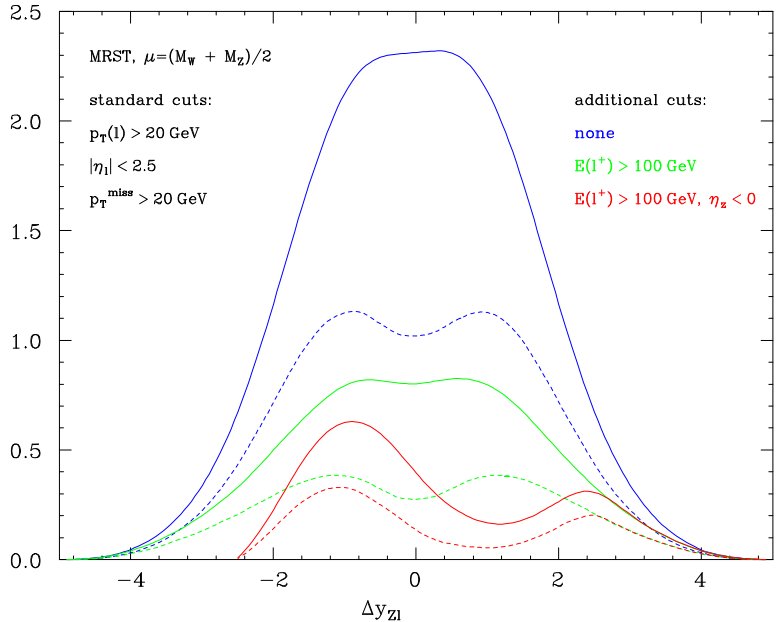} }
   \caption{\label{fig:RadZFig} $WZ$ production 
   followed by leptonic decays of both the $W$ and $Z$ bosons. We plot
   the distribution, in picobarns, in the rapidity difference between
   the $Z$ and the charged lepton $l$ from the decay of the $W$,
   $\Delta y_{Zl} \equiv y_Z - y_l$.  Leptonic branching ratios
   are not included and the scale has been set to $\mu =
   (M_W+M_Z)/2$. The basic cuts used are $p_T(l) > 20$~GeV and
   $|\eta(l)| < 2$ for all three charged leptons, and a missing
   transverse momentum cut of $p_T^{\rm miss} > 20$ GeV.
    We plot the
   $\Delta y_{Zl}$ distribution with these cuts (blue, upper pair), with
    an additional cut on the $W$ decay lepton, $E(l) >
   100$~GeV (green, middle pair) and with a further cut on the 
   rapidity of the $Z$ boson $y_Z<0$ (red, lower pair);  
    the latter curves have been scaled up by a factor of 5.
   The dashed curves are Born-level results; the solid curves include
   the ${\cal O}(\alpha_s)$ corrections.
  }
\end{figure}                                                              

New physics contributions can
 modify the self-interactions of vector-bosons, in
particular the triple gauge-boson vertices.  If  new physics occurs 
at an energy scale well above that being probed experimentally, it
can be integrated out, and the result expressed as a set of 
anomalous (non-Standard Model) interaction vertices.
(The physics of anomalous coupling
 will be considered in  detail in Section~\ref{sec:anomtgc} 
and our standard notation for
the anomalous triple gauge-boson couplings is  given there.)
It is  interesting to know what is the effect of  the 
anomalous $W^+W^-Z$ 
couplings on the approximate radiation zero of  $WZ$ 
production~\cite{bib:ZW88}.
In Figure~\ref{fig:RadZacFig}, the $\Delta y_{Zl}$ distribution is plotted for
two different sets of anomalous couplings at vanishing $q^2$
$
(\Delta g_1=-0.013,\   \lambda^Z=0.02,\ 
 \Delta \kappa^Z=-0.028)\ \ $ and$\ \  (
   \Delta g_1=0.065,\ \lambda^Z=0.04, \ 
\Delta\kappa^Z=0.071)$.
For the  $q^2$ dependence we assumed dipole form factors
of the generic form
\beq\label{dipformf}
 \hat{a}(q^2)=\frac{a}{\left(1+\frac{q^2}{\Lambda^2}\right)^2} 
\eeq
with  $\Lambda=2\TeV$.
As one can see in Figure~\ref{fig:RadZacFig},
the contributions of anomalous
couplings   have the tendency to 
make the dip less pronounced.

\begin{figure}[htbp]
\centerline{ \epsfig{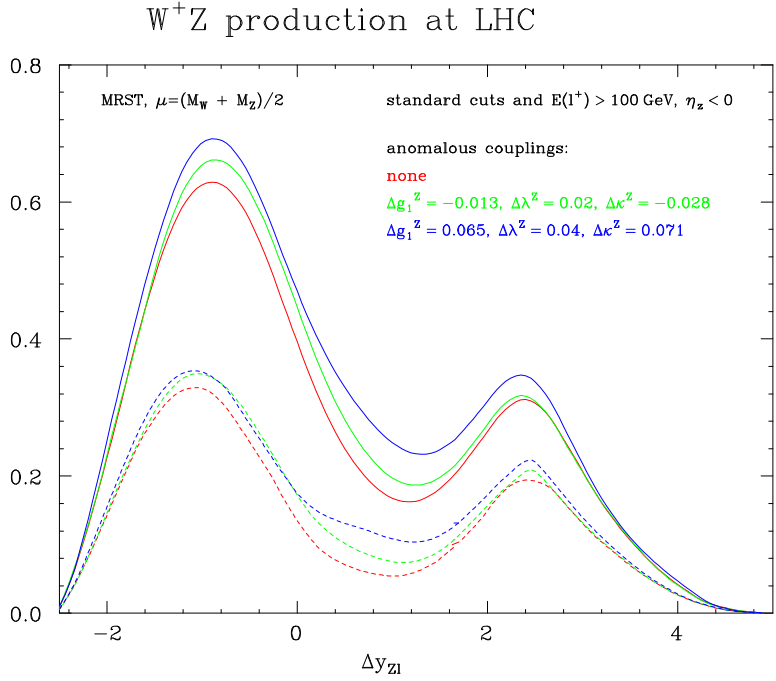} }
   \caption{ \label{fig:RadZacFig} $WZ$ production 
   followed by leptonic decays of both the $W$ and $Z$ bosons. We plot
   the NLO  distribution, in picobarns, in the rapidity difference between
   the $Z$ and the charged lepton $l$ from the decay of the $W$:
   $\Delta y_{Zl} \equiv y_Z - y_l$.  Leptonic branching ratios
   are not included and the scale has been set to $\mu =
   (M_W+M_Z)/2$. The standard  cuts  $p_T(l) > 20$ GeV,
   $|\eta(l)| < 2.5$ for all three charged leptons and  a missing
   transverse momentum cut of $p_T^{\rm miss} > 20$ GeV are applied.
    We plot the
   $\Delta y_{Zl}$ distribution  without anomalous
   couplings (red, lower pair) and with two sets of anomalous couplings
$(\Delta g_1=-0.013,\  \lambda^Z=0.02,\ 
 \Delta \kappa^Z=-0.028)$ (green, middle pair)
 and 
$(\Delta g_1=0.065,\  \lambda^Z=0.04, \ 
\Delta\kappa^Z=0.071)$ (blue, upper pair). The $q^2$
dependence of the couplings is given by 
the dipole form of Equation~\ref{dipformf}
with $\Lambda=2\TeV$.
   Also we plot the same quantities supplementing  the standard cuts 
   with the additional cut on the the $W$ decay lepton, $E(l) > 100$~GeV
   and with the rapidity cut $y_Z<0$; 
   the latter curves have been scaled up by a factor of 5.
   The dashed curves are Born-level results; the solid curves include
   the ${\cal O}(\alpha_s)$ corrections.
    }
\end{figure}                                                              

\subsubsection{Future improvements}

The present state of art of the description of gauge-boson 
pair production is not completely satisfactory yet.
Of the various issues, there are three
which require further theoretical studies.
First, the double resonant approximation is expected to be correct
only up to a few percent accuracy - it is important to
go beyond this approximation. A first attempt has been made  by
Campbell and Ellis~\cite{Campbell:1999ah}
 where, as  already mentioned above, 
the  singly-resonant diagrams have also been included.
These additions  are obviously
relevant in the off-resonant regions.
The inclusion of finite width effect 
is not completely straightforward  because of    
possible  conflict with gauge-invariance.
This issue requires further theoretical study.
Secondly, we  need NLO results also for the semi-leptonic channels
 when
one of the gauge-bosons decays hadronically. 
This requires the inclusion of the contributions of   diagrams describing
the  gluonic corrections to the final-state quarks.
Thirdly, fixed order perturbative QCD description is not applicable for the
description of  the low-$p_T$ behaviour of the gauge-boson
pair. The technique for the resummation of the low-$p_T$
contributions is well known and it can be 
 applied  also to the
case of gauge-boson pair production.
For example, one calculation for the $ZZ$ has been carried out~\cite{bib:hmo}.

\subsubsection{Comparison with {\tt PYTHIA}}

\begin{figure}[htbp]
 \begin{minipage}[htb]{0.47\linewidth}
   \centering
\epsfig{file=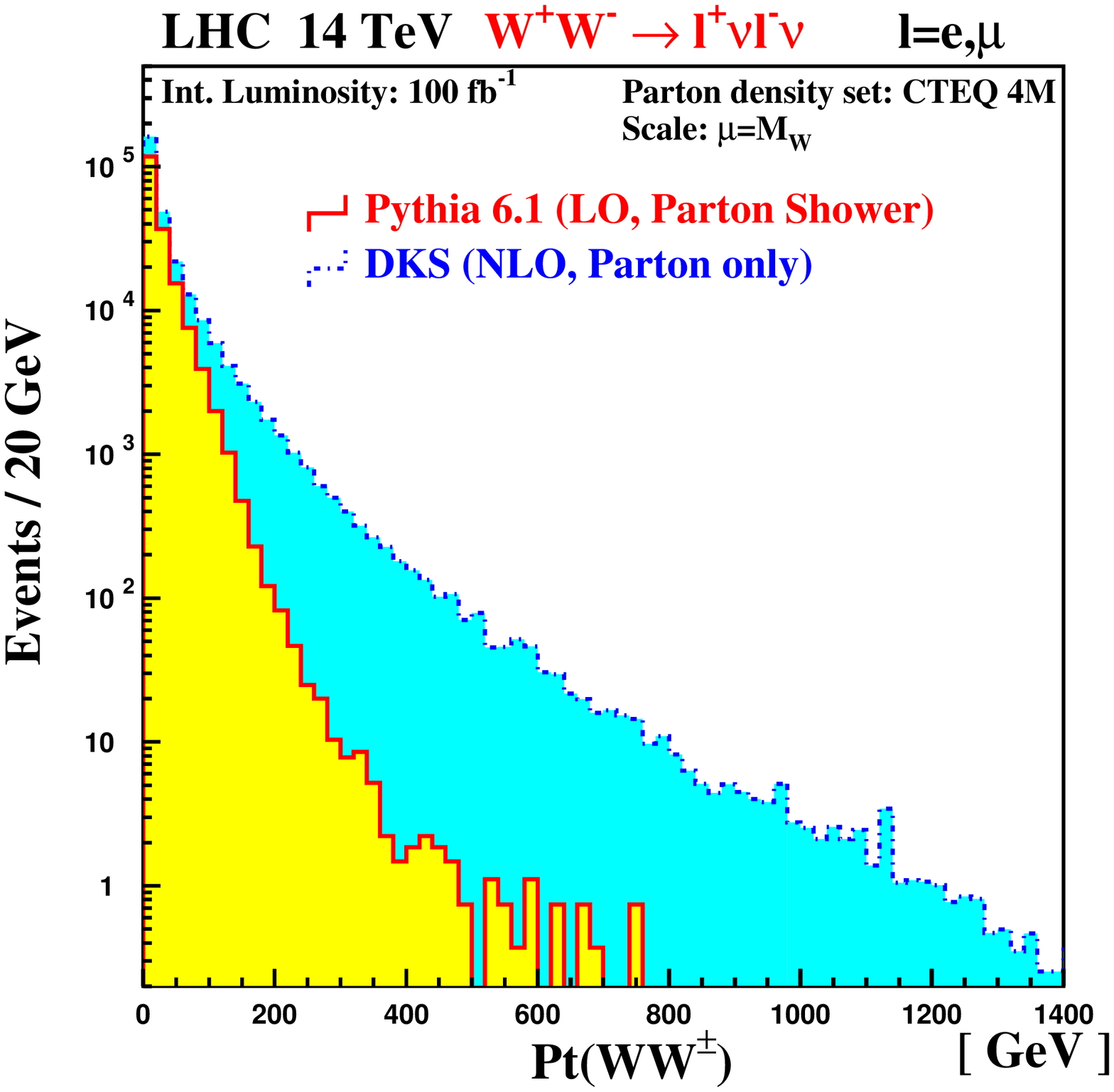,
height=7.5 cm,width=7.5 cm}
   \caption{Transverse momentum of the $WW$ bosons pairs simulated with 
{\tt PYTHIA} and DKS Monte Carlo generators and using the CTEQ~4M 
structure function.} \label{ptww}
 \end{minipage}%
 \begin{minipage}[htb]{0.06\linewidth}
   \hspace{0.3cm}
 \end{minipage}%
 \begin{minipage}[htb]{0.47\linewidth} 
   \centering
 \epsfig{file=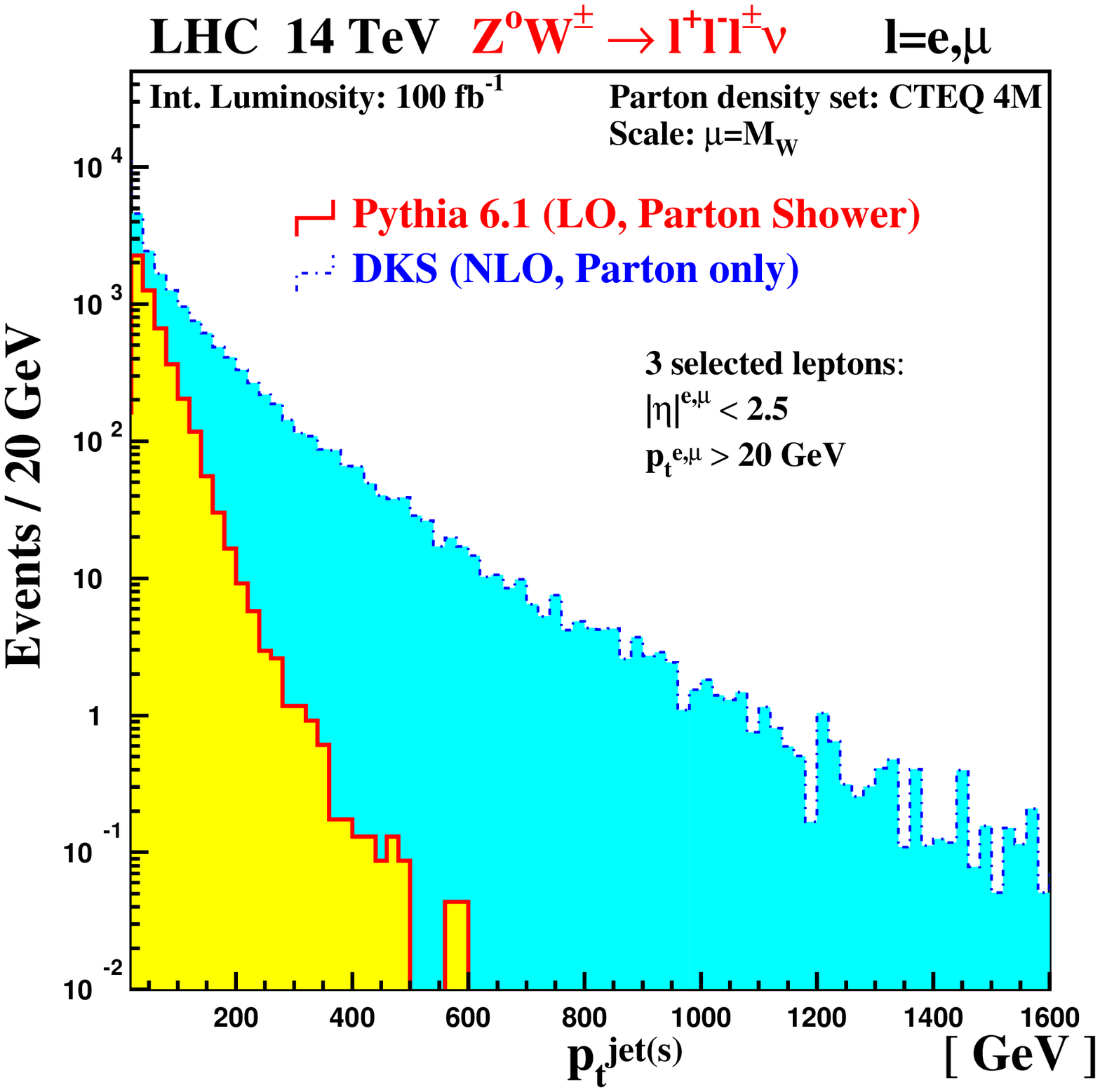,
height=7.5 cm,width=7.5 cm}
   \caption{Transverse momentum of the jets in the case of the $WZ$ production. The 3 leptons fall within the detector acceptance.} \label{wz_ptjet}
 \end{minipage}
\end{figure}

\begin{figure}[htbp]
 \begin{minipage}{0.47\linewidth}
   \centering
\epsfig{file=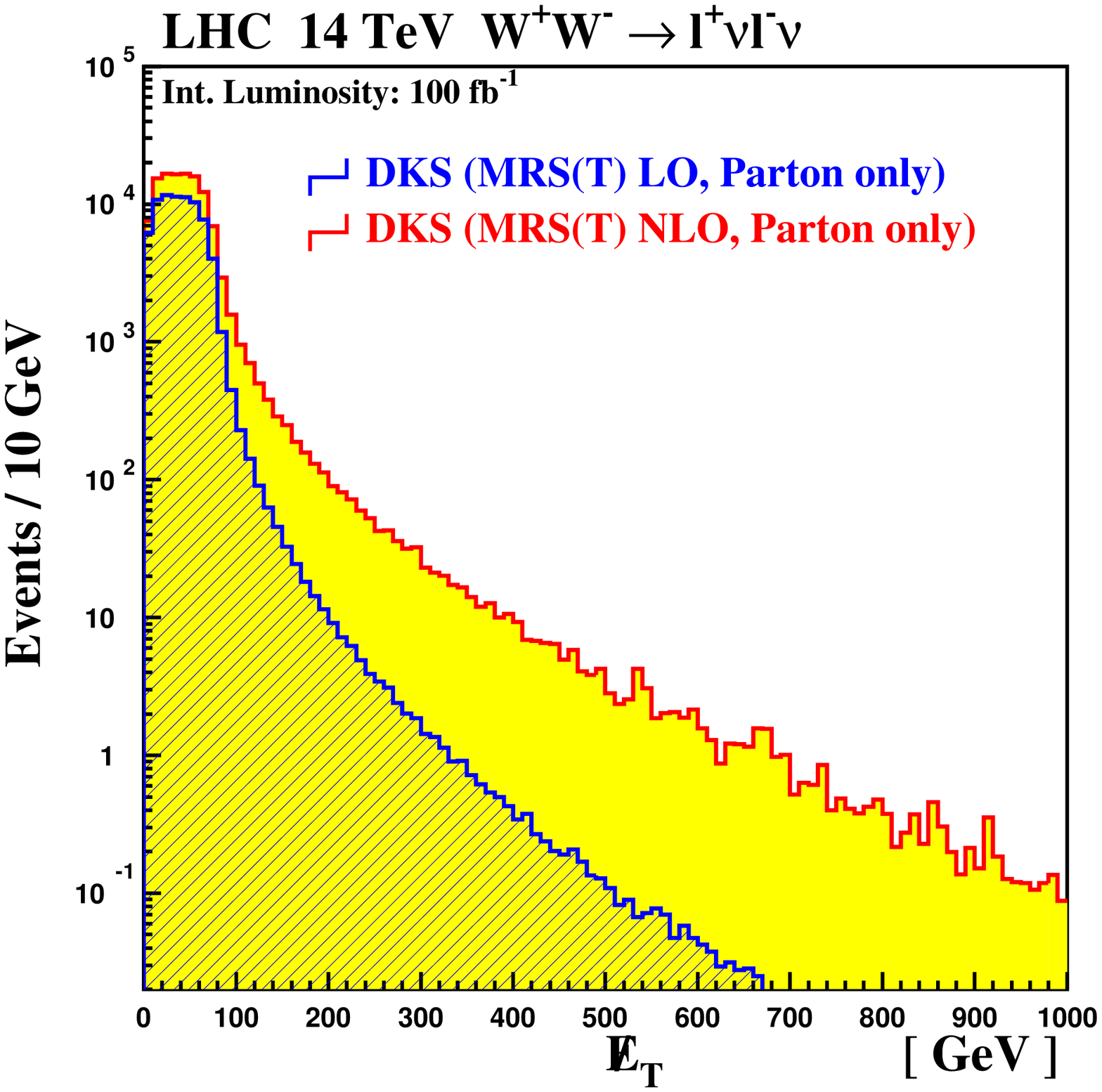,
height=7.5 cm,width=7.5 cm}
   \caption{Missing transverse energy in the $WW$ production. The events are obtained by running the DKS generator with and without including the NLO corrections.} \label{miss_et}
 \end{minipage}%
 \begin{minipage}{0.06\linewidth}
   \hspace{0.3cm}
 \end{minipage}%
 \begin{minipage}{0.47\linewidth} 
   \centering
 \epsfig{file=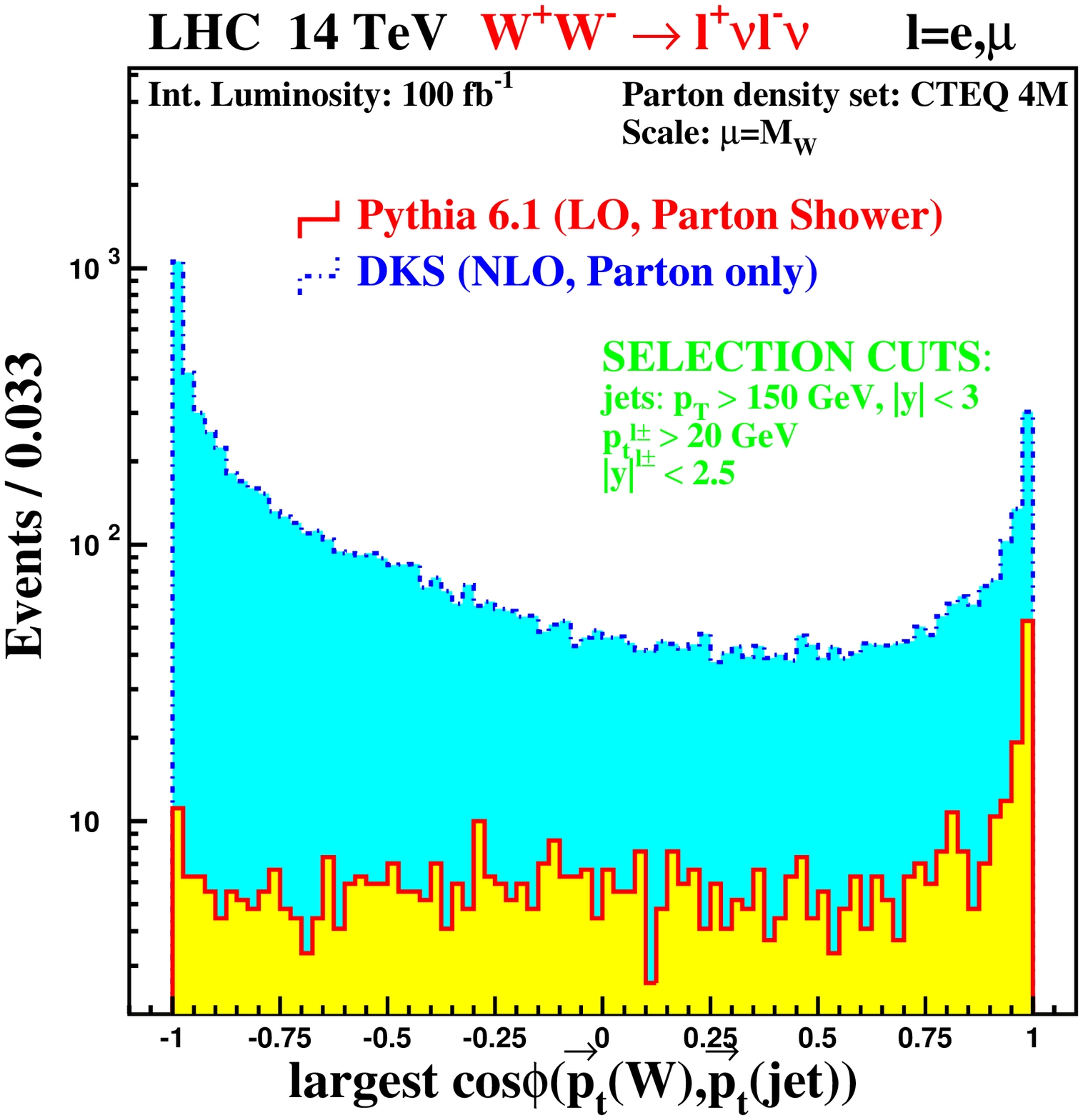,
height=7.5 cm,width=7.5 cm}
   \caption{Smallest angle between one of the $W$'s  and the jet in 
 $WW$ pair production. The two leptons are required to be within the 
detector acceptance and the jet to have a $p_T$ larger 
than 150~GeV.} \label{ang_wjet}
 \end{minipage}
\end{figure}

\begin{figure}[htbp]
\centerline{ \epsfig{figure=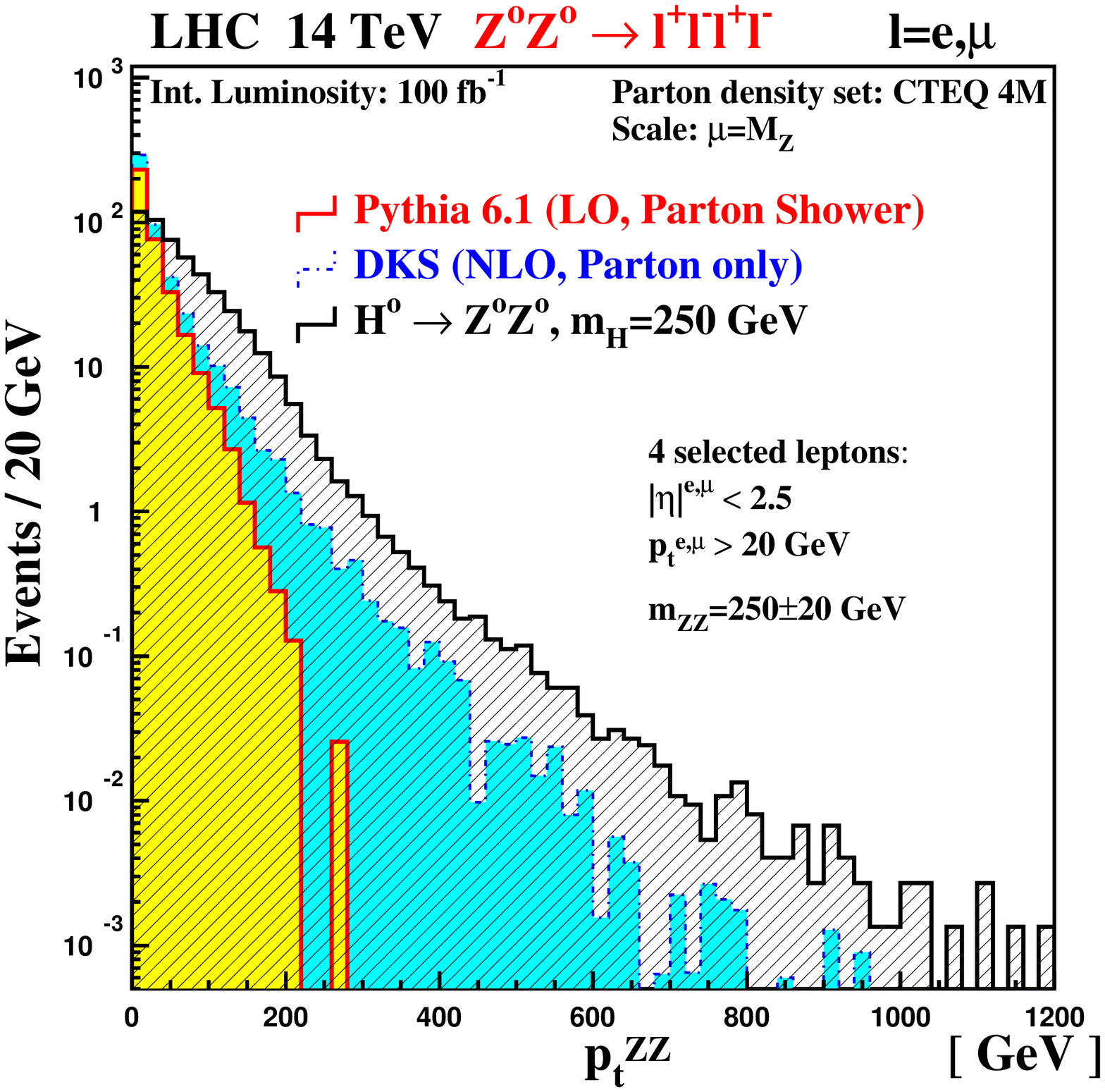,width=0.64\textwidth,clip=} }
   \caption{ \label{pt_h_zz} 
Transverse momentum of $ZZ$ pairs originating 
from a Higgs ($M_H = 250$ GeV), where the two leptons fall into the detector 
 acceptance and the $M_{ZZ}$ is consistent with the
 Higgs mass. The non-resonant $ZZ$ background is simulated
 with (DKS) and without ({\tt PYTHIA}) NLO corrections.
  }
\end{figure}                                                              

In most of the studies carried out so far for the LHC, 
where the production of vector boson pairs played
 an important role, the usual Monte Carlo simulation tool
 has been {\tt PYTHIA}~\cite{pythia} based on  LO matrix elements 
\cite{gunkun} with parton shower.
 In particular, it is expected that    for some optimisation cuts,  
 where  the large corrections
 provided by NLO diagrams (for example by choosing high-$p_T(V)$
 or high-$M_{VV}$ regions) its predictions are not acceptable.
 By making comparison between the predictions of {\tt PYTHIA} and the
 the DKS parton level NLO Monte Carlo~\cite{Dixon:1999di},  
 we investigate here how accurate does {\tt PYTHIA} simulate the di-boson 
cross sections at the LHC, especially in some kinematic regions. 
We relate our analysis to the special case of the CMS detector~\cite{CMSTP}.

In all results presented in this analysis, we assume that the vector-bosons 
always decay leptonically.
 We use the CTEQ(4M) parton distribution~\cite{cteq4}
 in both Monte Carlos and the cross section values are for the scale 
$\mu = (M_{V_1} + M_{V_2})/2$, where $M_{V_i}$ are the masses of the two 
produced 
vector-bosons.
If the DKS Monte Carlo is run at Born-level,
 we obtain very good agreement
 with the total cross sections given by {\tt PYTHIA}.

Figure \ref{ptww} shows the transverse momentum of the $WW$ pairs. 
The comparison between {\tt PYTHIA} and DKS indicates the large
 difference in cross section observables at high-$p_T^{WW}$ values.
 This is related to the fact that at NLO, the sub-processes 
$qg\rightarrow V_1V_2q$ have to be taken into account \cite{WWit,BHOWW}. 
This is also reported in Table~\ref{xsections}. The leptons
 are selected following the CMS criteria, where a $p_T$ larger
 than 20 GeV and a pseudorapidity $|\eta|<$~2.5 are required. 
Jets are selected by: $p_T>$~20~GeV and $|\eta|<$~3. The K-factor
 increases then from 1.5 for the total cross sections up to values of about 60
 if the jets are required to have a $p_T$ larger than 150 GeV.
 The same effect is shown in figure \ref{wz_ptjet} for the $WZ$
 production, where the $p_T$ of the jets is shown  (the jet balances the 
$p_T^{VV}$). For this process the K-factors at large $p_T$-values are even larger than in the $WW$ case (as shown in the table). 
The transverse momentum of the di-boson system (or of the jet(s))
 are not the only variables affected by large NLO corrections. Other variables can show significant differences within their distributions: for example the lepton $p_T$, the invariant mass of the lepton pair $M_{ll}$, the missing transverse energy $\not{\!E}_T$ (as shown in Figure~\ref{miss_et}), the maximal transverse momentum of the two charged leptons $p_T^{max}$, the lepton pseudorapidities $\eta^l$, their difference $\Delta\eta^l=\eta^{l^-}-\eta^{l^+}$, the angle between leptons $cos\theta_{ll}$, the transverse angle between leptons $cos\phi_{ll}$ and so on.

Therefore, it is extremely important to take into account the possible influence of NLO corrections for the vector-boson production at the LHC energy. Every time one is performing an optimisation of  signal selection, 
one should 
be aware of the possible deviations due to the use of a
 LO generator like {\tt PYTHIA}. This is especially true for complicated cuts, where it is difficult to judge whether the effects are large or not. An example is shown for $WW$ events in Figure~\ref{ang_wjet}, where the smallest angle between one of the $W$'s and the jet is shown for events with a high-$p_T$ jet. Not only is the cross section clearly smaller in {\tt PYTHIA} 
but also the shape of the distribution is quite different, changing the result of a possible cut. Another good example is the Higgs search through 
the decay channel $H\rightarrow ZZ \rightarrow 4l$ (see Figure~\ref{pt_h_zz}).
The idea of using $p_T$-cuts to improve the signal-to-background ratio may
 not be as effective as one would expect from using only 
{\tt PYTHIA}. The figure shows indeed that, if the NLO corrections are included, the $p_T$ distribution of the non-resonant background follows 
much more closely  those of the signal, reducing the gain considerably.

\begin{table}[htb]
 \begin{center}
    \caption{Cross sections in pb for $pp$ collision at
 $\sqrt{s}$=14 TeV. The leptons are selected by requiring a $p_T$ larger than 20 GeV and a pseudorapidity $|\eta|<$2.5. The jets should have a $p_T>$20 GeV and $|\eta|<$3.}
    \label{xsections}
\vskip 0.2cm
\begin{tabular}{lcccccc} \hline
&&Selected &Jet &\multicolumn{3}{c}{$p_T^{jet}$ selection (in GeV):} \\
(pb)&$\sigma^{tot}\times BR$& leptons &veto & 20-150 & 150-400 & $>$400  \\\hline
$\sigma_{{\tt PYTHIA}}^{W^+W^-\rightarrow l^+\nu l^-\bar{\nu}} $&3.704&1.704&1.125&0.568&2$\times 10^{-3}$&2.8$\times 10^{-4}$\\
$\sigma_{DKS,\, LO}^{W^+W^-\rightarrow l^+\nu l^-\bar{\nu}} $&3.79&1.71&-&-&-&-\\
$\sigma_{DKS,\, NLO}^{W^+W^-\rightarrow l^+\nu l^-\bar{\nu}} $&5.56&2.58&1.49&0.942&0.135&1.69$\times 10^{-2}$\\
K-factor &1.5& 1.54&1.32&1.66&67&$\sim$60\\\hline
$\sigma_{{\tt PYTHIA}}^{W^{\pm}Z\rightarrow l^+\nu l^+l^-} $&4.35$\times 10^{-1}$&1.45$\times 10^{-1}$&9.47$\times 10^{-2}$&4.91$\times 10^{-2}$&9.33$\times 10^{-4}$&6.5$\times 10^{-6}$\\
$\sigma_{DKS,\, LO}^{W^{\pm}Z\rightarrow l^+\nu l^+l^-} $&4.34$\times 10^{-1}$&1.48$\times 10^{-1}$&-&-&-&-\\
$\sigma_{DKS,\, NLO}^{W^{\pm}Z\rightarrow l^+\nu l^+l^-} $&7.42$\times 10^{-1}$&2.77$\times 10^{-1}$&1.31$\times 10^{-1}$&1.27$\times 10^{-1}$&2.8$\times 10^{-2}$&4.63$\times 10^{-3}$\\
K-factor &1.71&1.91&1.39&2.3&30&$\sim$700\\\hline
$\sigma_{{\tt PYTHIA}}^{ZZ\rightarrow l^+l^-l^+l^-} $&5.13$\times 10^{-2}$&1.79$\times 10^{-2}$&1.15$\times 10^{-2}$&6.26$\times 10^{-3}$&1.33$\times 10^{-4}$&1.5$\times 10^{-6}$\\
$\sigma_{DKS,\, LO}^{ZZ\rightarrow l^+l^-l^+l^-} $&5.31$\times 10^{-2}$&1.84$\times 10^{-2}$&-&-&-&-\\
$\sigma_{DKS,\, NLO}^{ZZ\rightarrow l^+l^-l^+l^-} $&7.07$\times 10^{-2}$&2.55$\times 10^{-2}$&1.58$\times 10^{-2}$&8.79$\times 10^{-3}$&8.23$\times 10^{-4}$&7.78$\times 10^{-5}$\\
K-factor &1.38&1.42&1.38&1.4&6&$\sim$50\\\hline
\end{tabular}
\end{center}
\end{table}

\subsection{\boldmath $W\gamma$ \unboldmath and \boldmath  
$Z\gamma$ \unboldmath  production at NLO \label{WgNLOprod}}

In this section, we present order $\alpha_s$ results for $W\gamma$ and
$Z\gamma$ production at the LHC, including the {\it full} leptonic
correlations and anomalous couplings in the narrow-width approximation
\cite{wgamma-zgamma}. Previous analyses
\cite{BHOWg,OhnemusZ95,BHOZg} included decay correlations only in
the bremsstrahlung amplitudes implementing, as an approximation, the
finite part of the {\it spin-summed} one-loop amplitudes.

To perform the calculation, we use the helicity amplitudes presented
in~\cite{Dixon:1998py}. The amplitudes relevant for the inclusion of
anomalous couplings are given in \cite{wgamma-zgamma}.  In order to
cancel analytically the soft and collinear singularities coming from
the bremsstrahlung and one loop parts, we have used the version of the
subtraction method presented in \cite{FKS}.  
Therefore, the amplitudes are
implemented into a numerical Monte Carlo style program which
allows calculation of any infrared-safe physical quantity with arbitrary
cuts.
 
The results presented in this section correspond to $pp$ scattering at
$\sqrt{s}=14$ TeV using the following cuts: a transverse momentum cut
of $p_T^{l}>25$ GeV for the charged leptons is imposed and the
pseudorapidity is limited to $|\eta|<2.4$ for all detected particles. The
photon transverse momentum cut is $p_T^{\gamma}>50 (100)$~GeV for
$W\gamma$ ($Z\gamma$) production.  For the $W\gamma$ case, we require a
minimum missing transverse momentum carried by the neutrinos $p_T^{\rm
miss}>50$~GeV. Additionally, charged leptons and the photons must be
separated in the pseudorapidity-azimuthal angle by $\Delta R_{l\gamma} =
\sqrt{ (\eta_\gamma -\eta_l)^2 + (\phi_\gamma -\phi_l)^2}> 0.7$.
In order to suppress the contribution from the off-resonant diagrams,
we require the transverse mass $M_T>90$~GeV for $W\gamma$ production
and the invariant mass of the $ll\gamma$ system
$M_{ll\gamma}> 100$~GeV for the $Z\gamma$ case.

Finally, in order to suppress the contribution from the fragmentation
of partons into photons, computed only to LO accuracy, the photons are
required to be isolated from hadrons: the transverse hadronic momentum
in a cone of size $R_0=0.7$ around the photon should be smaller than a
fraction of the transverse momentum of the photon
\begin{equation}
\sum_{\Delta R<R_0} p_T^{\rm had}< 0.15\, p_T^{\gamma}
\end{equation}
This completes the definition of the ``standard'' cuts.

In the  results presented here, the branching ratios of the vector-bosons into
leptons are not included.  For both the LO and NLO results, we use the
latest set of parton distributions of MRST(cor01) \cite{mrst} and the
two loop expression for the strong coupling constant. For the
fragmentation component, we use the fragmentation functions 
from~\cite{GRVfrag}.

The ``standard'' scale for both the factorisation and renormalisation
scales is
\begin{equation}
\mu^2=\mu^2_{\rm st}\equiv M_V^2 +\frac{1}{2} \left[
(p_T^V)^2 + (p_T^\gamma)^2
\right] .
\label{eqn:must} 
\end{equation}

The masses of the vector-bosons have been set to $M_Z=91.187$ GeV and
$M_W=80.41$ GeV and the following values have been used for the
Cabibbo-Kobayashi-Maskawa (CKM) matrix elements:
$|V_{ud}|=|V_{cs}|=0.975$ and $|V_{us}|=|V_{cd}|=0.222$. We do not
include any QED or electroweak corrections but choose the coupling
constants $\alpha$ and $\sin^2\theta_W$ in the spirit of the ``improved
Born approximation'' \cite{IBA1,IBA2}, with
$\sin^2\theta_W=0.230$. Notice that the observable is order
$\alpha^2$; within the same spirit, we use the running
$\alpha=\alpha(M_Z)=1/128$ for the coupling between the vector-boson
and the quarks (to  take into account effectively the EW corrections)
whereas we keep $\alpha=1/137$ for the photon coupling.  It is worth
noticing that this modification results already in more than a 6\% 
change in the normalisation of the cross section with respect to the
standard approach of using both running coupling constants.

\subsubsection{Results at NLO}

For future checks, and for an estimate of the number of events to be
observed at the LHC, some benchmark total cross section numbers are
presented in Table~\ref{tab:XStev}. The first ones were obtained by
imposing only the cut on the transverse momentum of the photon
$p_T^\gamma>50(100)$ GeV for $W\gamma$ ($Z\gamma$) production. 
The importance of the NLO corrections,
as well as the size of the fragmentation contribution before applying
the isolation cut prescription, can be seen from the table.
Furthermore, we also include the
result for the total cross section obtained after the implementation
of the standard cuts.

\begin{table}
\begin{center}
\vskip0.2cm
\begin{tabular}{lccc} \hline 
$\sigma$ (pb) & LO$^*$ & Frag. & NLO  \\ \hline 
$ W^+\gamma \,\,\, (p_T^\gamma >50$ GeV) 
  & 4.79  & 3.02 & 13.89  \\ 
$ W^-\gamma \,\,\, (p_T^\gamma >50$ GeV)
  & 3.08 & 3.55 & 10.15  \\ 
$ Z\gamma \,\,\, (p_T^\gamma >100$ GeV)
  & 1.29 & 0.412 & 2.37 \\ 
$ W^+\gamma $ (std. cuts) 
  &0.436   & 0.094 & 1.71  \\ 
$ W^-\gamma$ (std. cuts) 
  & 0.310 & 0.095 & 1.20  \\ 
$ Z\gamma$ (std. cuts)
  & 0.524 & 0.041 & 0.877\\ 
 \hline
\end{tabular}
\end{center}
\caption[dummy]{\small Cross sections for $pp$ collisions at
  $\sqrt{s}=14 $ TeV. The statistical errors are $\pm$1 within the last digit.
  LO$^*$ corresponds to the direct component only.
\label{tab:XStev}}
\end{table}

In what follows, we will estimate the theoretical uncertainty of the
results by analysing the changes on different distributions when
varying the scale by a factor of two in both directions
$\frac{\mu_{\rm st}}{2} < \mu < 2 \mu_{\rm st}$.

In Figure~\ref{fig:scaleWY}, we show the scale dependence of the $p_T$
distribution of the photon in $W^+\gamma$ production with the standard
cuts (upper curves) and also with the additional requirement of a
jet-veto. As can be observed, the scale dependence is still large 
($\pm$ 10\%) but is considerably reduced when the jet-veto is
applied. The situation is similar to what has been observed in the
case of $WW$ production \cite{Dixon:1999di} and is caused by the suppression
of the contribution from the $qg$ initial state appearing for the
first time at NLO. Since this initial state dominates the
cross section, the NLO result behaves
effectively like a LO one, as far as the scale dependence is concerned.

In the inset plot, we present the ratio between the NLO and LO results
(with the standard scale), which remains larger than 3 and increases
with the photon transverse momentum.  This clearly shows that the LO
calculation is not even sufficient for an understanding of the shape
of the distribution, since the NLO effect goes beyond a simple
normalisation. As is well known \cite{yOhn}, the relevance
of the NLO corrections for this process is mainly due to the breaking
of the radiation amplitude zero appearing at LO and to the large $qg$
initial state parton luminosity at the LHC. It is worth mentioning
that the scale dependence of the LO result turns out to be very
small. This is an artificial effect and illustrates that a small scale
dependence is by no means a guarantee for small NLO corrections.
Furthermore, we present the ratio of the NLO jet-veto and the LO
result. As expected, this ratio is closer to 1, again due to the fact
that most of the contributions coming from the new subprocesses
appearing at NLO are suppressed by the jet-veto.

\begin{figure}
\begin{center}
\includegraphics[width=7cm]{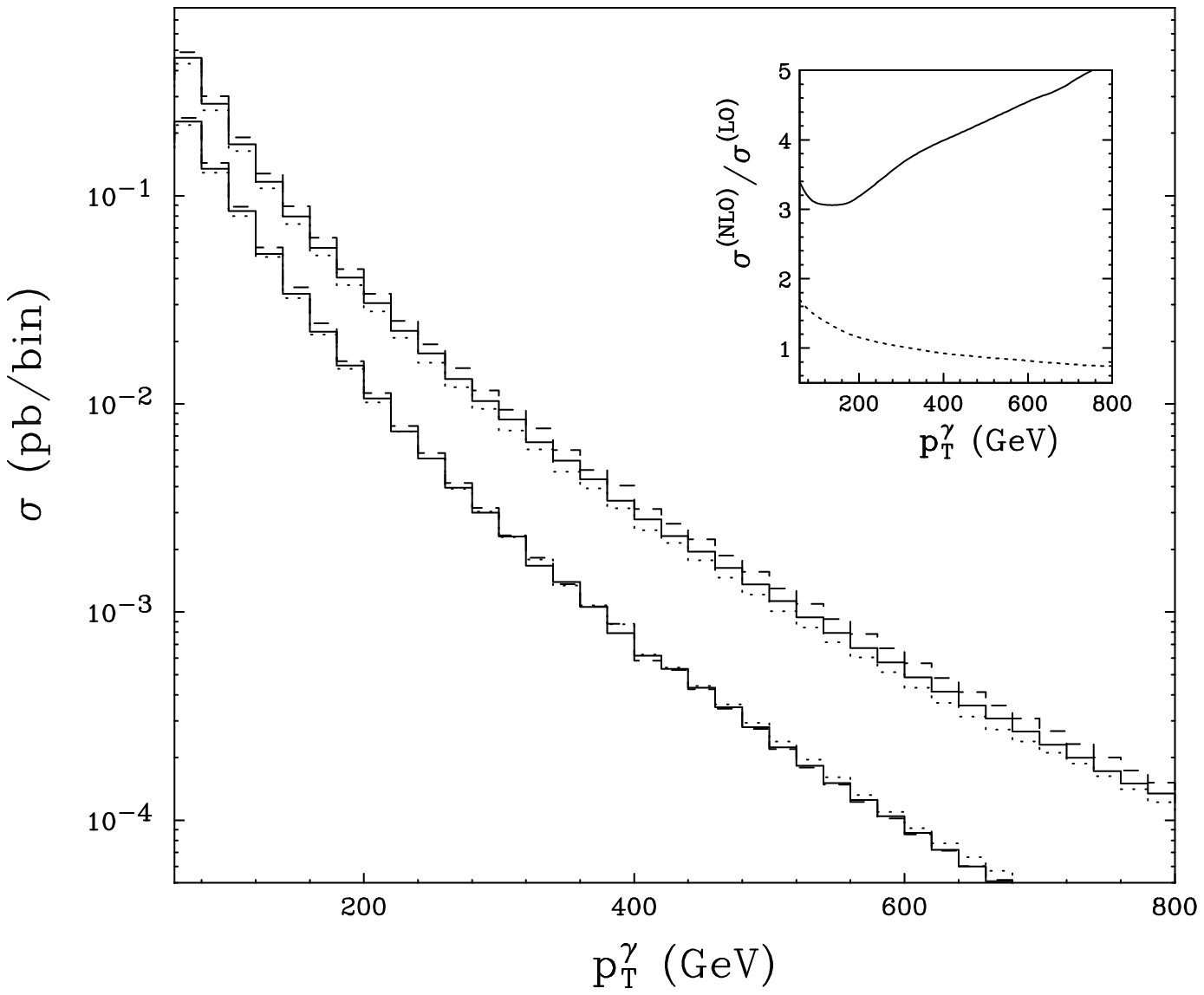}
\end{center}
   \caption{ \label{fig:scaleWY} Scale dependence of
   $\sigma^{NLO}$ without (upper curves) and with (lower curves)
   jet-veto. The scale has been varied according to $\frac{\mu_{\rm
   st}}{2} ({\rm dashes}) < \mu < 2 \mu_{\rm st} ({\rm dots})$. The
   inset plot shows the ratio $\sigma^{NLO}/\sigma^{LO}$, again
   without (solid) and with (dots) jet-veto.  }
\end{figure}                                                              

In Figure~\ref{fig:scaleZY}, we study the lepton correlation in the
azimuthal angle for $Z\gamma$ production
$\Delta\phi_{ll}=|\phi_{l^-}-\phi_{l^+}|$.  Notice that
this observable can be studied at NLO since the spin correlations
between the leptons are fully taken into account in the implementation
of the one-loop corrections. In this case, we observe that the NLO
corrections are rather sizeable and increase the cross section by
$50\%$ for small $\Delta\phi_{ll}$.  The region
$\Delta\phi_{ll}>2$ (with the standard cuts) is kinematically
forbidden unless a jet with a high transverse momentum is produced;
therefore, the cross section vanishes at LO and it is strongly
suppressed for the NLO calculation with jet-veto. In this region, the
full NLO calculation is effectively LO and its scale dependence
becomes larger, as expected.

Because there is no radiation amplitude zero appearing at LO for
$Z\gamma$ production, the NLO corrections are under better control in
the kinematical region where the LO cross section does not
vanish. Nevertheless, for large transverse momentum, the $qg$ initial
state again dominates the NLO contribution and the corrections
increase considerably.

\begin{figure}
\begin{center}
\includegraphics[width=7cm]{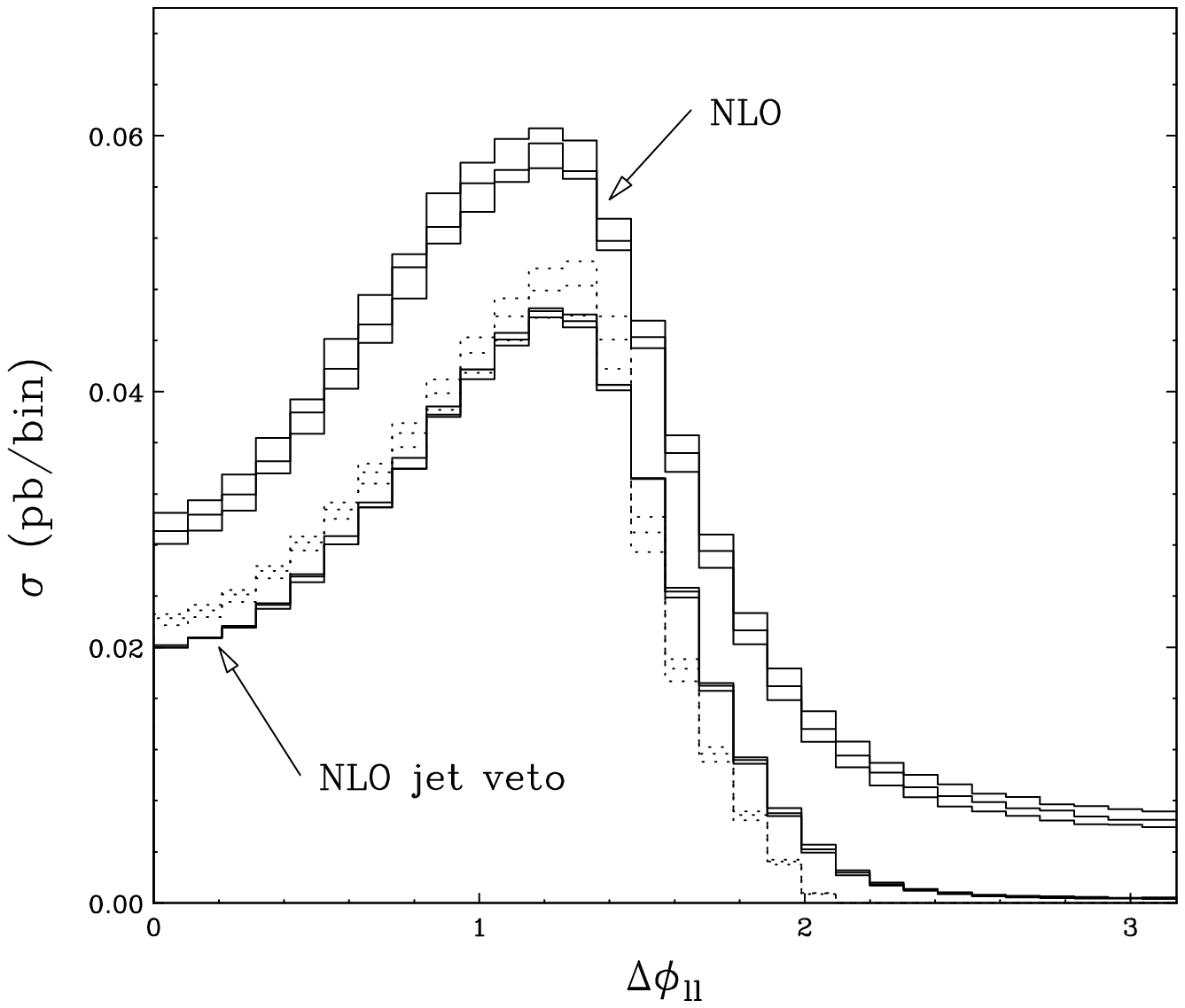}
\end{center}
    \caption{ \label{fig:scaleZY} Scale dependence of
  $\sigma^{NLO}$ without jet-veto (upper solid curves),
   $\sigma^{NLO}$ with jet-veto (lower solid curves) and
   $\sigma^{LO}$ (dotted curves). The scale has been varied
   according to $\frac{\mu_{\rm st}}{2} < \mu < 2 \mu_{\rm st}$.   }
\end{figure}                                                              

\subsubsection{Anomalous couplings without form factors}

The study of triple vector-boson couplings is motivated by the hope
that some physics beyond the Standard Model leads to a modification of
these couplings which eventually could be detected.  In order to
quantify the effects of the new physics, an effective Lagrangian is
introduced which contains all Lorentz invariant terms, in principle.
The new terms spoil the gauge-cancellation in the high energy limit
and, therefore, will lead to violation of unitarity for increasing
partonic centre of mass energy $\hat{s}$. Usually, in an analysis of
anomalous couplings from experimental data in hadronic collisions, this
problem is circumvented by supplementing the anomalous couplings
$\alpha_{\rm AC} $ with form factors. A common choice for the form
factor is
\begin{equation}
\alpha_{\rm AC} \to \frac{\alpha_{\rm
    AC}}{(1+\frac{\hat{s}}{\Lambda^2})^n}
\label{formfac}
\end{equation}
where $n$ has to be large enough to ensure unitarity and $\Lambda$ is
interpreted as the scale for new physics. Obviously, this procedure is
rather {\it ad hoc} and introduces some arbitrariness. Therefore, it would be
very convenient to avoid it in an analysis of anomalous couplings at
hadron colliders.  This would bring these analyses more into line with
those at $e^+e^-$ colliders.  In order to do so, one should analyse the
data at fixed values of $\hat{s}$, as it is done at LEP. This results
in limits for the anomalous parameters which are a function of
$\hat{s}$.

Clearly, it is possible to do such analysis for the production of
$Z\gamma$ when both leptons are detected~\cite{GLR99}, since the
partonic centre of mass energy can be reconstructed from the
kinematics of the final state particles and therefore the cross section can
 be measured for different bins of fixed $\hat{s}$.

The situation is more complicated for $W\gamma$ production since the
neutrino is not observed. Nevertheless, by identifying the transverse
momentum of the neutrino with the missing transverse momentum, and
assuming the $W$ boson to be on shell, it is possible to reconstruct
the neutrino kinematics (particularly the longitudinal momentum) with
a two-fold ambiguity. In the case of the Tevatron, since it is a
$p\bar{p}$ collider, it is possible to choose the ``correct'' neutrino
kinematics 73\% of the times by selecting the maximum (minimum) of the
two reconstructed values for the longitudinal momentum of the neutrino
for $W^+\gamma$($W^-\gamma$).

This is not true at the LHC where, due to the symmetry of the
colliding beams, both reconstructed kinematics have equal chances to
be correct.  Fortunately, in the case of anomalous couplings, we are
interested in a efficient way to reconstruct the $\hat{s}$ rather than
the full kinematics. Again there are two possible values of
$\hat{s}$. It turns out that there is a simple method to choose the
``correct'' one 66\% of the times at the LHC (73\% of the times at
Tevatron) by selecting the minimum $\hat{s}$, $\hat{s}_{\rm min}$, of the two
reconstructed values (for both $W^+\gamma$ and $W^-\gamma$).
Furthermore, we checked that the selected value $\hat{s}_{\rm min}$
differs in almost 90\% of the events by less than 10\% from the exact
value $\hat{s}$. This is likely to be enough precision, since the data
will be collected in sizeable bins of $\hat{s}$ and the
anomalous parameters are not expected to change very rapidly with the
energy in any case.

\begin{figure}
\begin{center}
   \raisebox{0.5cm}{\scriptsize $\sqrt{\hat{s}}$}
   \hspace*{2cm}
   \raisebox{-0.5cm}{\scriptsize $\sqrt{\hat{s}_{\rm min}}$}
   \hspace*{2.8cm}
   \raisebox{2.7cm}{\scriptsize $\sigma$}
   \hspace*{-6.7cm}
\includegraphics[width=7cm]{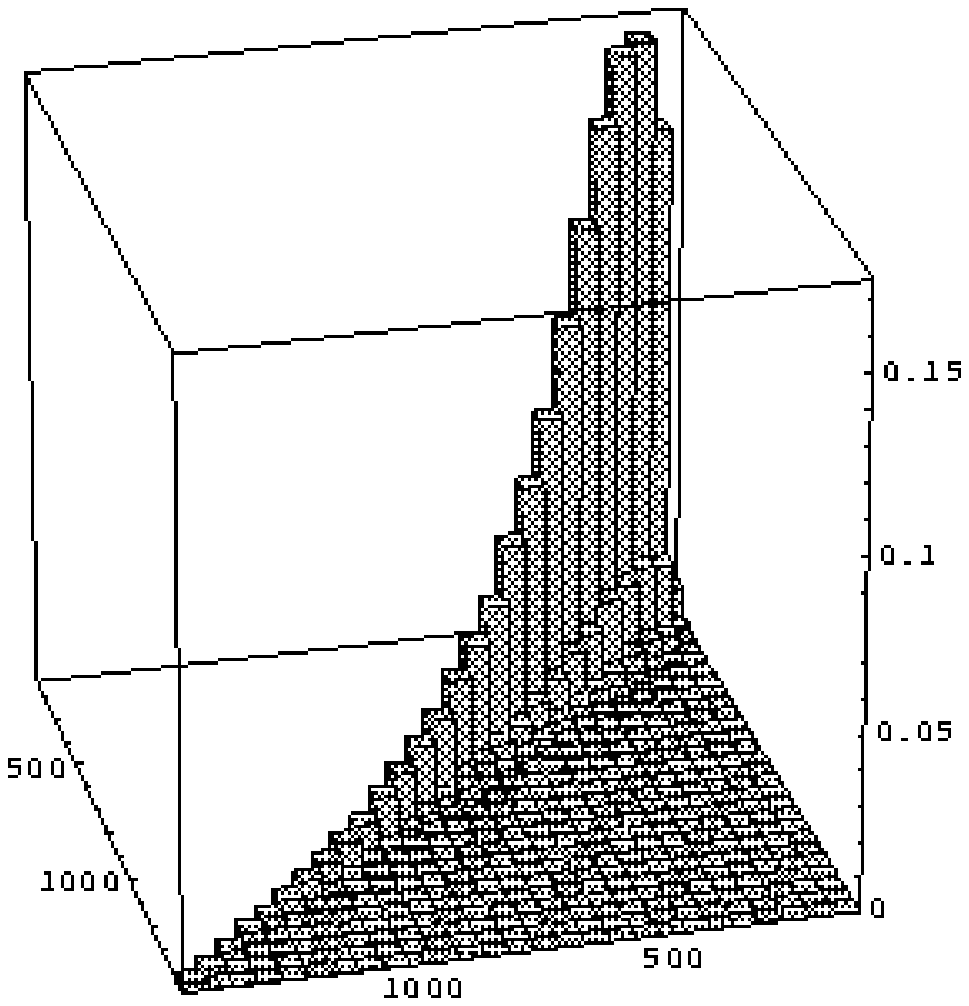}
   \raisebox{0.5cm}{\scriptsize $\sqrt{\hat{s}}$}
   \hspace*{2cm}
   \raisebox{-0.5cm}{\scriptsize $\sqrt{\hat{s}_{\rm min}}$}
   \hspace*{2.8cm}
   \raisebox{2.7cm}{\scriptsize $\sigma$}
   \hspace*{-6.7cm}
\includegraphics[width=7cm]{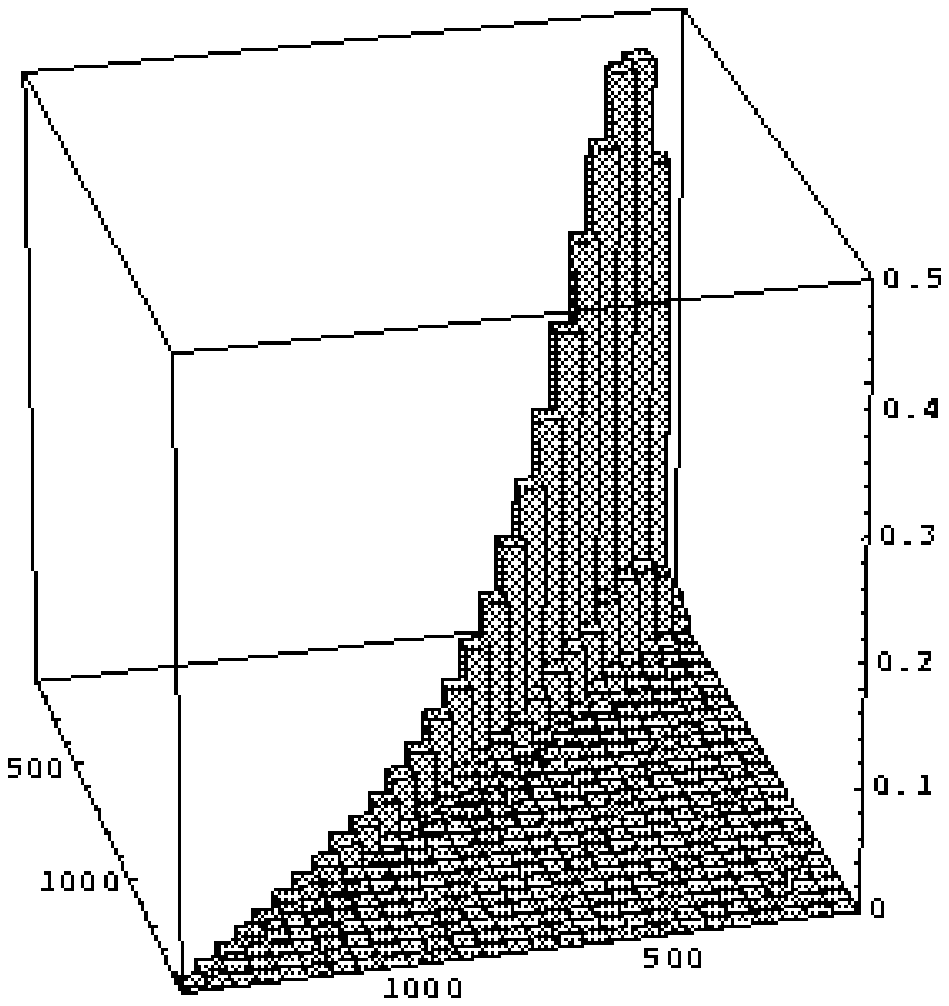}
\end{center}
   \caption{ \label{fig:ShatWY} 
   The cross section for $W^+\gamma$ production (in pb/bin) as as
   function of $\sqrt{\hat{s}}$ and $\sqrt{\hat{s}_{\rm min}}$ (in
   GeV) in order to illustrate the steep fall of $\sigma$ for
   increasing $|\sqrt{\hat{s}}-\sqrt{\hat{s}_{\rm min}}|$. The left
   plot corresponds to the Standard Model, whereas the right plot
   includes anomalous couplings (see text).}
\end{figure}                                                              

To quantify the advantage of the method, we show in
Figure~\ref{fig:ShatWY} the correlations of $\sqrt{\hat{s}_{\rm min}}$
with $\sqrt{\hat{s}}$. The left plot corresponds to the case of pure
Standard Model, whereas the right plot presents results for (already
experimentally ruled out) huge values of anomalous couplings $\Delta
\kappa=0.8$ and $\lambda=0.2$ with an ordinary form factor ($n=2$,
$\Lambda=1$ TeV).

The cross section drops very rapidly for increasing $\sqrt{\hat{s}}-
\sqrt{\hat{s}_{\rm min}}$.  This correlation clearly holds in the
particularly interesting large $\sqrt{\hat{s}}$ region and for both
Standard Model and anomalous contribution.

As a result of this investigation, we conclude that even in the case of
$W\gamma$ production, reliable bounds for anomalous couplings as a
function of $\hat{s}$ (using $\hat{s}_{\rm min}$) can be obtained.
Such a procedure would certainly allow a comparison of various bounds
from different experiments.

\section{ANOMALOUS VECTOR-BOSON COUPLINGS
         \protect\footnote{Section coordinators: 
         P.R.~Hobson, W.~Hollik}
         \label{sec:anomtgc}}

\newcommand{\nc}{\newcommand}
\nc{\bea}{\begin{eqnarray}}
\nc{\eea}{\end{eqnarray}}
\nc{\ba}{\begin{array}}
\nc{\ea}{\end{array}}
\nc{\bpi}{\begin{picture}}
\nc{\epi}{\end{picture}}
\nc{\scs}{\scriptstyle}
\nc{\sss}{\scriptscriptstyle}
\nc{\sst}{\scriptstyle}
\nc{\ts}{\textstyle}
\nc{\ds}{\displaystyle}
\nc{\sctn}[1]{\section{\hspace{-18pt}.\ #1}}
\nc{\subsctn}[1]{\subsection{\hspace{-16pt}.\ #1}}
\nc{\subsubsctn}[1]{\subsubsection{\hspace{-14pt}.\ #1}}

\nc{\al}{\alpha}
\nc{\be}{\beta}
\nc{\ga}{\gamma}
\nc{\Ga}{\Gamma}
\nc{\de}{\delta}
\nc{\De}{\Delta}
\nc{\ep}{\epsilon}
\nc{\ve}{\varepsilon}
\nc{\eb}{\bar{\eta}}
\nc{\et}{\eta}
\nc{\ka}{\kappa}
\nc{\la}{\lambda}
\nc{\La}{\Lambda}
\nc{\Th}{\Theta}
\nc{\ze}{\zeta}
\nc{\p}{\partial}

\nc{\oocw}{\cw^{-1}}
\nc{\cwcw}{\cw^2}
\nc{\cwcwcw}{\cw^3}
\nc{\cwcwcwcw}{\cw^4}
\nc{\oocwcw}{\cw^{-2}}
\nc{\oosw}{\sw^{-1}}
\nc{\swsw}{\sw^2}
\nc{\swswsw}{\sw^3}
\nc{\swswswsw}{\sw^4}
\nc{\tw}{t_\th}
\nc{\twtw}{\tw^2}

\nc{\mb}[1]{\mbox{#1}}
\nc{\unit}{\mb{\bf\large 1}}
\nc{\Tr}{\mb{Tr}}
\nc{\Bb}{\mb{\boldmath $\ds B$}}
\nc{\Wb}{\mb{\boldmath $\ds W$}}
\nc{\Xb}{\mb{\boldmath $\ds X$}}
\nc{\Fb}{\mb{\boldmath $\ds F$}}
\nc{\Tb}{\mb{\boldmath $\ds T$}}
\nc{\Vb}{\mb{\boldmath $\ds V$}}
\nc{\mw}{m_{\sss W0}}
\nc{\mz}{m_{\sss Z0}}
\nc{\half}{{\ts\frac{1}{2}}}
\nc{\ihalf}{{\ts\frac{i}{2}}}
\nc{\dg}{\dagger}
\nc{\Lag}{{\cal L}}
\nc{\Lgf}{\Lag_{\mbox{\scriptsize\em gf}}}
\nc{\Lgh}{\Lag_{\mbox{\scriptsize\em gh}}}
\nc{\Lgfgh}{\Lag_{\mbox{\scriptsize\em gf,gh}}}
\nc{\Lwhc}{\Lag_{\mbox{\scriptsize\em Whc}}}
\nc{\Lbhc}{\Lag_{\mbox{\scriptsize\em Bhc}}}
\nc{\Lvhc}{\Lag_{\mbox{\scriptsize\em Vhc}}}
\nc{\Luhc}{\Lag_{\mbox{\scriptsize\em Uhc}}}
\nc{\od}{{\cal O}}
\nc{\mubar}{\bar{\mu}}
\nc{\Laeff}{\La_{\rm eff}}

\nc{\ltap}{\;\raisebox{-.4ex}{\rlap{$\sim$}}\raisebox{.4ex}{$<$}\;}
\nc{\gtap}{\;\raisebox{-.4ex}{\rlap{$\sim$}}\raisebox{.4ex}{$>$}\;}

The principle of gauge-invariance is used as the basis for the Standard Model.
The non-Abelian gauge-group structure of the theory of electroweak 
interactions predicts very specific couplings between the electroweak 
gauge-bosons. Measurements of these triple gauge-boson couplings (TGCs)  
of the $W$, $Z$ and $\gamma$ gauge-bosons therefore provide powerful 
tests of the Standard Model.

In the most general Lorentz invariant parametrisation, the three 
gauge-boson vertices, $WW\gamma$ and $WWZ$, can be described by fourteen 
independent couplings~\cite{bib:hagiwara}, seven for each vertex. The possible 
four quadruple gauge-boson vertices: $\gamma\gamma WW$, $Z\gamma WW$, 
$ZZWW$ and $WWWW$ require 
36, 54, 81 and 81 couplings, respectively for a general description. 
Assuming electromagnetic gauge-invariance, C- and P-conservation, the set 
of 14 couplings for the three gauge-boson vertices is reduced to 5: 
$g_1^Z$, $\kappa_{\gamma}$, $\kappa_Z$, $\lambda_{\gamma}$ and 
$\lambda_Z$ \cite{bib:baur88}, where their Standard Model values
are equal to $g_1^Z = \kappa_{\gamma} = \kappa_Z = 1$ and 
$\lambda_{\gamma} = \lambda_Z = 0$ at tree level.

The TGCs related to the $WW\gamma$ vertex determine properties of the $W$, 
such as its magnetic dipole moment $\mu_W$ and electric quadrupole moment 
$q_W$:
\begin{equation}
\mu_W = \frac{e}{2M_W} (g_1^Z + \kappa_{\gamma} + \lambda_{\gamma})
\end{equation}
\begin{equation}
q_W = \frac{e}{M_W^2} (\kappa_{\gamma} - \lambda_{\gamma})
\end{equation}

In the following, the anomalous TGCs are denoted by $\Delta g_1^Z$, 
$\Delta \kappa_{\gamma}$, $\Delta \kappa_Z$, $\lambda_{\gamma}$ and 
$\lambda_Z$, 
where the $\Delta$ denotes the deviations of the respective quantity from 
its Standard Model value.

\subsection{Introduction}

The Standard Model is well established by the experiments at LEP and the
Tevatron. Any deviations of the Standard Model can therefore be
introduced only with care. Changes to the Standard Model come with different
forms of severity. In order to see at what level anomalous vector-boson 
couplings can be reasonably discussed, one has to consider these cases
separately. Changes to the gauge-structure of the theory, that do not
violate the renormalisability of the theory, {\it i.e.} the introduction of
extra fermions or possible extensions of the gauge-group are the least severe.
They will typically generate small corrections to vector-boson couplings
via loop effects. In this case also, radiative effects will be generated
at lower energies. For the LHC, the important thing in this case is not to 
measure the anomalous couplings precisely, but to look for the extra particles.
However, this is beyond the scope of this chapter.
In the other case,
a more fundamental role is expected for the anomalous couplings, implying
strong interactions. In this case, one has to ask oneself whether one should
study a model with or without a fundamental Higgs boson. 

Simply removing the
Higgs boson from the Standard Model is a relatively mild change. The model
becomes non-renormalisable, but the radiative effects grow only logarithmically
with the cut-off at the one-loop level.
The question is whether this scenario is ruled out by the
LEP1 precision data. The LEP1 data appear to be in agreement with the Standard
Model, preferring a low Higgs mass. One is sensitive to the Higgs mass
in three parameters, labelled $S$, $T$, $U$ 
or $\epsilon_1, \epsilon_2, \epsilon_3$.
These receive corrections of  the form $g^2 (\log(M_H/M_W) + constant)$,
where the constants are of order one. The logarithmic enhancement is universal
and would also appear in models without a Higgs as $\log(\Lambda)$, where
$\Lambda$ is the cut-off at which new interactions should appear. Only when 
one can determine the three different constants independently, can one say 
that one has established
the Standard Model. At present, the data do not provide sufficient 
precision to do this.

A much more severe change to the Standard Model is the introduction of 
vector-boson couplings not of the gauge-interaction type. 
These new couplings violate renormalisability much more
severely than simply removing the Higgs boson. Typically, quadratically
and quartically divergent corrections would appear to physical observables.
Therefore, it is questionable as to whether 
one should study models with a fundamental
Higgs boson, but with extra anomalous vector-boson couplings. It is hard to
imagine a form of dynamics that could do this. If the vector-bosons 
become strongly interacting, the Higgs probably would exist at most in an
``effective'' way. Therefore, the most natural way is to study anomalous
vector-boson couplings in models without a fundamental Higgs. Actually
when one removes the Higgs boson, 
the Standard Model becomes a gauged non-linear
sigma-model. 
It is well known that the nonlinear sigma-model describes 
low-energy pion
physics. The ``pions'' correspond to the longitudinal degrees of freedom
of the vector-bosons and $f_{\pi}$ corresponds to the vacuum expectation value
of the Higgs field. 
Within this description,  the Standard Model corresponds to the 
lowest-order term
quadratic in the momenta, anomalous couplings to higher derivative terms.
The systematic expansion in terms of momenta is known as chiral perturbation
theory and is extensively used in meson physics. 

Writing down the most general non-linear chiral Lagrangian containing
up to four derivatives gives rise to a large number of terms, which are too
general to be studied
effectively. One therefore has to look for dynamical principles
that can limit the number of terms. Of particular importance are approximate
symmetry principles. In the first place one, expects CP-violation to be
small. We limit ourselves therefore to CP-preserving terms. 
In order to see what this means in practice, it is
advantageous to describe the couplings in a manifestly gauge-invariant
way, using the St\"uckelberg formalism \cite{stuc,longhitano:81}. 
One needs the following definitions:

\begin{equation} F_{\mu \nu}= \frac { i \tau_i}{2} 
                           (\partial_{\mu}W^i_{\nu} - \partial_{\nu}
W^i_{\mu} + g \epsilon^{ijk}W^j_{\mu} W^k_{\nu} ) \end{equation}
is the $SU(2)$ field strength with the $SU(2)$ gauge-coupling $g$;
\begin{equation} D_{\mu} U = \partial_{\mu} U + 
     \frac {{ i}g}{2} \tau_i W^i_{\mu} U
 + {i}g \tan \theta_W \,U \tau_3 B_{\mu} \end{equation}
is the gauge-covariant derivative of the $SU(2)$-valued field $U$, which
describes the longitudinal degrees of freedom of the vector fields
in a gauge-invariant way;
\begin{equation} 
    B_{\mu \nu} = \partial_{\mu} B_{\nu} - \partial_{\nu} B_{\mu} 
\end{equation}
is the hypercharge field strength. In addition,
\begin{eqnarray}
V_{\mu} & = & (D_{\mu} U) U^{\dagger} / g \, , \\  
T & = & U \tau_3 U^{\dagger} / g
\end{eqnarray}
are auxiliary quantities having simple transformation properties.
Excluding CP violation, the non-standard three and four vector-boson couplings
are described in this formalism by the following set of operators:
\begin{eqnarray}
  {\cal L}_1 & = &{\rm Tr} ( F_{\mu \nu} [V_{\mu},V_{\nu}]) \\
{\cal L}_2 & = &  i \frac {B_{\mu \nu}}{2} {\rm Tr} (T [V_{\mu},V_{\nu}]) \\
{\cal L}_3 & = & {\rm Tr} ( T F_{\mu \nu} ) {\rm Tr} (T [V_{\mu},V_{\nu}]) \\
{\cal L}_4 & = & ( {\rm Tr} [V_{\mu} V_{\nu} ]) ^2 \\
{\cal L}_5 & = & ( {\rm Tr} [V_{\mu} V_{\mu} ]) ^2 \\
{\cal L}_6 & = & {\rm Tr}(V_{\mu} V_{\nu}) {\rm Tr}(T V_{\mu}) {\rm Tr}(T V_{\nu}) \\
{\cal L}_7 & = & {\rm Tr}(V_{\mu} V_{\mu}) ({\rm Tr} [T V_{\nu}]) ^2 \\
{\cal L}_8 & = & \frac {1}{2} [ ({\rm Tr} [T V_{\mu}]) ({\rm Tr} [T V_{\nu}])]^2
\end{eqnarray}
In the unitary  gauge $U=1$, one has (with $c_W=\cos\theta_W$, 
$s_W=\sin\theta_W$)
\begin{eqnarray} {\cal L}_1  & = &  {i}  
     ( c_W Z_{\mu \nu} + s_W F_{\mu \nu} ) W_{\mu}^+ 
    W_{\nu}^- + Z_{\nu}/c_W ( W_{\mu \nu}^+ W_{\mu}^- 
   - W_{\mu \nu}^- W_{\mu}^+)    \\
        &  & \quad  +\,\mbox{gauge-induced four boson vertices}, \nonumber\\
         {\cal L}_2  & = &   { i} 
          (c_W F_{\mu \nu} - 
         s_W Z_{\mu \nu}) W_{\mu}^+ W_{\nu}^-  \, ,  \\
         {\cal L}_3 & = &   { i} 
        ( c_W Z_{\mu \nu} + 
         s_W F_{\mu \nu}) W_{\mu}^+ W_{\nu}^- \, .   
\end{eqnarray}      
where
$Z_{\mu \nu}= \partial_{\mu} Z_{\nu} - \partial_{\nu} Z_{\mu}$
and 
$W^{+,-}_{\mu \nu}= \partial_{\mu} W^{+,-}_{\nu}
 - \partial_{\nu} W^{+,-}_{\mu}$.
The Standard Model without a Higgs corresponds to
\beq
\label{lagew}
\ts\Lag_{EW}=
 \half {\rm Tr}(F_{\mu\nu} F^{\mu\nu})
-\frac{1}{4} B_{\mu\nu} B^{\mu\nu}
+\frac{g^2v^2}{4} {\rm Tr}(V_\mu V^\mu) \, .
\eeq

\subsection{Dynamical constraints}

The list given in the previous section
contains terms that give rise to vertices with minimally three
or four vector-bosons. Already with the present data a number of constraints
and/or consistency conditions can be put on the vertices. 
The most important of these come from the limits on the breaking of the
so-called custodial symmetry. If the hypercharge is put to zero, the
effective Lagrangian has a larger symmetry than $SU_{L}(2) \times U_Y(1)$, {\it i.e.}
it has the symmetry $SU_{L}(2) \times SU_{R}(2)$. The $SU_{R}(2)$ invariance
is a global invariance. Within the Standard Model this invariance is an
invariance of the Higgs potential, but not of the full Lagrangian. 
It is ultimately this invariance that is responsible for the fact that the
$\rho$~parameter, which is the ratio of charged to neutral current strength,
is equal to one at the tree level. Some terms in the Lagrangian, {\it i.e.} the
ones containing the hypercharge field explicitly or the terms with $T$,
that project out the third isospin component violate this symmetry explicitly.
These terms, when inserted in a loop graph, give rise to quartically divergent
contributions to the $\rho$~parameter. Given the measurements, this means that
the coefficients of these terms must be extremely small. It is therefore 
reasonable
to limit oneself to a Lagrangian, where hypercharge appears only indirectly
via a minimal coupling, so without explicit $T$. 
This assumption means physically that the ultimate dynamics that is
responsible for the strong interactions among the vector-bosons acts in the
non-Abelian sector. Indeed one would not normally expect the
hypercharge alone to become strong. However, we know that there is a strong violation 
of the custodial symmetry in the form of the top-quark mass. Actually the 
top-mass almost saturates the existing corrections to the $\rho$~parameter,
leaving no room for violations of the custodial symmetry in the anomalous
vector-boson couplings. Therefore, we conclude: {\it If there really are strong
vector-boson interactions, the mechanism for mass generation is unlikely 
to be the same for bosons and fermions}.

Eliminating the custodial symmetry violating interactions, we are left with the
simplified Lagrangian, containing ${\cal L}_1$, ${\cal L}_4$, ${\cal L}_5$.
Besides the vertices, there are 
also propagator corrections, in principle. We take the two-point functions
without explicit $T$.
Specifically, we add to the theory \cite{kastening1}
\beq
\label{laghctr}
\ts\Lag_{hc,tr}
=-\frac{1}{2\La_W^2} {\rm Tr}[(D_\al F_{\mu\nu})(D^\al F^{\mu\nu})]
+\frac{1}{2\La_B^2} {\rm Tr}[(\p_\al B_{\mu\nu})(\p^\al B^{\mu\nu})]
\eeq
for the transverse degrees of freedom of the gauge-fields, and
\beq
\label{laghclg}
\ts\Lag_{hc,lg}
=-\frac{g^2v^2}{4\La_V^2} {\rm Tr}[(D^\al V^\mu)(D_\al V_\mu)]
\eeq
for the longitudinal ones, where the $\La_X$ parametrise
the quadratic divergences and are expected to represent the
scales where new physics comes in. In phenomenological applications,
these contributions give rise to form factors in the propagators
\cite{kastening1, vdbij}. Introducing such cut-off dependent
propagators in the analysis of the vector-boson pair production is
similar to having $s$-dependent triple vector-boson couplings,
which is the way the data are usually analysed. 

This effective Lagrangian is very similar to the one in pion-physics.
Indeed, if one takes the limit vacuum expectation value (vev) 
fixed and gauge-couplings to zero,
one finds the standard pion Lagrangian. As it stands, one can use the LEP1 data
to put a limit on the terms in the two point vertices. Using a naive analysis
one finds \cite{kastening1} $1/\Lambda_B^2=0$. For the other two cut-offs
one has:

\begin{center}
{ A. The case $\Lambda_V^2>0,\Lambda_W^2<0$}:
\hspace{2cm} $\Lambda_V>0.49$ TeV, $|\Lambda_W|>1.3$ TeV.

{ B. The case $\Lambda_V^2<0,\Lambda_W^2>0$}:
\hspace{2cm} $|\Lambda_V|>0.74$ TeV, $\Lambda_W>1.5$ TeV. 
\end{center}
 
This information is important for further limits at high-energy colliders, 
as it tells us, how one has to cut off off-shell propagators. We notice
that the limits on the form factors are different for the transverse, 
longitudinal
and hypercharge form factors. The precise limits are somewhat qualitative
and should be taken as such. 
The current data show that $\Lambda=0.5$ TeV, which thus has to be considered
as a minimal possible value as long as a dipole form factor is used.
Further information comes from the 
direct measurements
of the three-point couplings at LEP2, which tell us that they are small. 
Similar limits at the Tevatron have to be taken with some care, as there is 
a cut-off dependence. As there is no known model that can give large 
three-point
interactions, we assume for the further analysis of the four-point vertices,
that the three-point anomalous couplings are absent. 
Two more constraints can be put on the remaining two 
four-point vertices . The first comes from 
consistency of chiral perturbation theory \cite{pelaez}. Not every 
effective chiral
Lagrangian can be generated from a physical underlying theory. 
 
A second condition comes from the $\rho$~parameter. Even the existing violation
of the custodial symmetry, though indirect via the minimal coupling to 
hypercharge,
gives a contribution to the $\rho$~parameter. It constrains the 
combination $5g_4 + 2g_5$. The remaining combination 
$2 {\cal L}_4 - 5 {\cal L}_5$ 
is fully unconstrained by experiment and  in principle gives a possibility
for very strong interactions to be present. However, this particular 
combination does
not seem to have any natural interpretation from underlying dynamics. 
Therefore,
one can  conclude presumably that both couplings $ g_4,g_5$ are small.
There is a loophole to this conclusion, namely when the anomalous couplings
are so large that the one-loop approximation, used to arrive at the limits,
is not consistent and resummation has to be performed everywhere. This is a
somewhat exotic possibility that could lead to very low-lying resonances and
which ought to be easy to discover at the LHC \cite{vanderBij:1999fp}.

\subsection{LHC processes}

Given the situation described above, one  has to ask oneself, what the
LHC can do and in which way the data should be analysed. There are essentially
three processes that can be used to study vector-boson vertices:
vector-boson pair production, vector-boson scattering, triple vector-boson production.
About the first two we have only a few remarks to make. 
They are discussed more fully
in other contributions to the workshop.

\subsubsection{Vector-boson pair production}

Vector-boson pair production can be studied in a relatively straightforward
way. The reason is that here the Higgs boson does not play a role in the
Standard Model, as we take the incoming quarks to be massless. Therefore
naive violations of unitarity can be compensated by the introduction of
smooth form-factors.

 One produces two vector-bosons via normal Standard Model processes
with an anomalous vertex added. The extra anomalous coupling leads to
unitarity-violating cross sections at high energy. As a total energy
of 14~TeV is available this is a serious problem, in principle. 
It is cured by introducing a form factor for the incoming off-shell
line connected to the anomalous vertex. Naively this leads to a
form-factor dependent limit on the anomalous coupling in question.
The LEP1 data gives a lower limit on the cut-off to be used inside
the propagator. When one wants an overall limit on the anomalous coupling,
one should use this value. This is particularly
relevant for the Tevatron. Here one typically takes
a cut-off of 2~TeV. This might give too strict a limit, as the LEP1 data
indicate that the cut-off can be as low as 500~GeV. For practical
purposes the analysis at the Tevatron should give limits on anomalous
couplings for different values of the cut-off form factors, including
low values of the cut-off. For the analysis at the LHC, one has much larger 
statistics. This means that one can do better and measure limits on the
anomalous couplings as a function of the invariant mass of the produced system.
This way one measures the anomalous form factor completely.

\subsubsection{Vector-boson scattering}

Here the situation is more complicated than in vector-boson pair production.
The reason is that within the Standard Model the process cannot be 
considered without intermediate Higgs contribution. This would violate 
unitarity.
However the incoming vector-bosons are basically on-shell and this allows the 
use of unitarisation methods, as are commonly used in chiral perturbation theory
in pion physics. These methods tend to give rise to resonances in longitudinal
vector-boson scattering. The precise details depend on the coupling constants.
The unitarisation methods are not unique, but generically give rise
to large $I=J=0$ and/or $I=J=1$ cross section enhancements. The literature
is quite extensive:  a good introduction is \cite{hikasa}; a recent review
is \cite{dominici}. 

\subsubsection{Triple vector-boson production}

In this case it is not clear how one should consistently approach an analysis 
of anomalous vector-boson couplings. Within the Standard Model the presence
of the Higgs boson is essential in this channel. Leaving it out,
one has to study
the unitarisation. This unitarisation has to take place not only
on the two-to-two scattering subgraphs, as in vector-boson scattering, but
also on the incoming off-shell vector-boson, decaying into three
real ones. The analysis here becomes too arbitrary to derive very meaningful
results. One cannot 
calculate confidently 
anything here without a fully known underlying model of new strong 
interactions. Also measurable cross sections tend to be small, so that the
triple vector-boson production is best used as corroboration of results in
vector-boson scattering. Deviations of Standard Model cross sections
could be seen, but the vector-boson scattering would be needed for
interpretation.

One therefore needs the Standard Model results. The total number of events 
with three vector-bosons in the final state is given in Table~\ref{jochum1}. 
We used an integrated luminosity
of 100 fb$^{-1}$ and an energy of 14 TeV throughout.

\begin{table}[htb]
\begin{center}
\caption{Number of events: before cuts and all decays
($\sqrt{s}=14$~TeV, 100~fb$^{-1}$).\label{jochum1}}
\vskip0.2cm
\begin{tabular}{lcccc}\hline
$M_{\rm Higgs}$ (GeV)  &200 &400  &600  &800    \\ \hline
$W^+W^-W^-$  &11675  &5084   &4780  &4800    \\ 
$W^+W^+W^-$  &20250  &9243   &8684  &8768    \\ 
$W^+W^-Z$    &20915  &11167   &10638  &10685    \\ 
$W^-ZZ$     &2294  &1181   &1113  &1113    \\ 
$W^+ZZ$     &4084  & 2243  &2108  &2165    \\ 
$ZZZ$       &4883  & 1332  &1087  &1085    \\ \hline
\end{tabular}
\end{center}
\end{table}

One sees from this table that a large part of the events comes from associated
Higgs production, when the Higgs is light. However for the study of
anomalous vector-boson couplings, the heavier Higgs results are arguably
more relevant.
Not all the events can be used for the analysis. If we limit ourselves
to events, containing only electrons,  
muons and neutrinos, assuming just acceptance cuts we find the results shown in
Table~\ref{jochum2}.

\begin{table}[htb]
\begin{center}
\caption{Number of events containing only leptonic decays. 
Cuts on leptons: $\vert \eta \vert < 3$, $p_T > 20$ GeV; no cuts on
missing energy ($\sqrt{s}=14$~TeV, 100~fb$^{-1}$).}
\label{jochum2}
\vskip 0.2cm
\begin{tabular}{lcccc}\hline
$M_{\rm Higgs}$ (GeV)  &200 &400  &600  &800    \\ \hline
$W^+W^-W^-$  &68  & 28  & 25 & 25  \\
$W^+W^+W^-$  & 112 &49   &44  & 44   \\ 
$W^+W^-Z$    & 32 & 17 &15  & 15   \\ 
$W^-ZZ$     &1.0  &0.51   &0.46  &0.45    \\ 
$W^+ZZ$     &1.7  & 0.88  &0.79  & 0.79   \\ 
$ZZZ$       &0.62  &0.18   &0.13  & 0.12   \\ \hline
\end{tabular}
\end{center}
\end{table}

We see that very little is left, in particular in the processes with at least
 two $Z$ bosons, where the events can be fully reconstructed. In order to see 
how sensitive we are to anomalous couplings, we assumed a 4$Z$ coupling
with a form factor cut-off at 2~TeV. We make here no correction
for efficiencies {\it etc.}. Using the triple $Z$ boson production, assuming
no events are seen in 100~fb$^{-1}$,
we find a limit $\vert g_4 + g_5 \vert < 0.09$ at the 95\% CL,
where $g_4$ and $g_5$ are the coefficients multiplying the operators
${\cal L}_4$ and ${\cal L}_5$.
This is to be compared with $-0.15 < 5 g_4 + 2 g_5 < 0.14$ \cite{valencia}
 or $-0.066 < (5 g_4 + 2 g_5) \Lambda^2({\rm TeV}) 
< 0.026$ \cite{kastening1,vdbij}.
So the sensitivity is not better than present indirect limits. Better limits
exist in vector-boson scattering \cite{eboli} or at a linear collider
\cite{ghinculov,bib:belanger,bib:boos}.

In the following tables we present numbers for observable cross sections
in different decay modes of the vector-bosons. We used the following cuts.
\begin{eqnarray}
& &  \vert \eta \vert_{lepton} < 3, \quad \quad
  \vert \eta \vert_{jet} < 2.5, \nonumber \\
& & \vert p_T \vert_{lepton} > 20\, {\rm GeV}, \quad \quad
  \vert p_T \vert_{jet} > 40\, {\rm GeV}, \quad
  \vert p_T \vert_{2 \nu} > 50\, {\rm GeV},  \nonumber \\ 
& & \Delta R_{jet,lepton} > 0.3, \quad \quad
  \Delta R_{jet,jet} > 0.5 \, .  \nonumber
\end{eqnarray}
States with more than two neutrinos are not very useful because of the
background from two vector-boson production. We did not consider final
states containing $\tau$-leptons.

With the given cuts, the total number of events to be expected is rather small.
In particular, this is the case because 
we did not consider the reduction in events due to
experimental inefficiencies, which may be relatively large because
of the large number of particles in the final state. For the
processes containing jets in the final state, there will be large
backgrounds due to QCD processes. A final conclusion on the significance
of the triple vector-boson production for constraining the four
vector-boson couplings will need more work, involving detector Monte 
Carlo calculations. 

However it is probably fair to say from the above
results, that no very strong constraints will be found from this process
at the LHC, but it is useful as a cross-check with other processes.
It may provide complementary information if non-zero anomalous couplings are
found.

\begin{table}[htb]
\begin{center}
\caption{Number of events from $ZZZ$ production in different decay modes
($\sqrt{s}=14$~TeV, 100~fb$^{-1}$).}
\vskip 0.2cm
\begin{tabular}{lccccc}\hline
$M_{\rm Higgs}$ (GeV)           &200  &300  &400  &500  &600  \\ \hline
$6 l$              &0.62 &0.29 &0.18 &0.14 &0.13    \\ 
$4 l, 2 \nu$       &5.1  &2.5  &1.5  &1.2  &1.1    \\ 
$4 l, 2 j$         &6.6  &3.8  &2.2  &1.7  &1.4    \\ 
$2 l, 2 j, 2 \nu$  & 34  &20   &12   &9.0  &7.7    \\ 
$2 l, 4 j$         & 24  &19   &11   &7.6  &6.0    \\ 
$2 \nu, 4j$           & 37  &34   &21   &15   &11    \\ 
$6 j$                 & 25  &31   &19   &12   &8.7    \\ \hline
\end{tabular}
\end{center}
\end{table}

\begin{table}[htb]
\begin{center}
\caption{Number of events from $WWZ$ production in different decay modes
($\sqrt{s}=14$~TeV, 100~fb$^{-1}$). }
\vskip 0.2cm
\begin{tabular}{lccccc}\hline
$M_{\rm Higgs}$ (GeV)           &200  &300  &400  &500  &600   \\ \hline
$4 l, 2 \nu$       &31   &20   &17   &16   &15    \\ 
$3 l, 2 j, 1 \nu$  &51   &40   &31   &28   &26    \\ 
$2 l, 4 j$         &19   &22   &17   &14   &13    \\ 
$2 \nu, 4 j$          &63   &74   &60   &51   &48    \\ 
$2 l, 2 j, 2 \nu$  &102  &68   &54   &49   &48    \\ 
$1 l, 4 j, 1 \nu$  &262  &196  &140  &127  &127   \\ 
$6 j$                 &86   &104  &78   &62   &56    \\ \hline
\end{tabular}
\end{center}
\end{table}

\begin{table}[htb]
\begin{center}
\caption[]{Number of events from 
$ZZW^-$(upper) and $ZZW^+$(lower) production in
different decay modes
($\sqrt{s}=14$~TeV, 100~fb$^{-1}$).}
\vskip 0.2cm
\begin{tabular}{lccccc}\hline
$M_{\rm Higgs}$ (GeV)           &200  &300  &400  &500  &600   \\ \hline
$5 l, 1 \nu$ &0.45&1.04&0.63&0.52&0.47        \\ 
          &0.80&1.69&1.08&0.91&0.81              \\ 
$3 l, 2 j, 1 \nu$  &3.37&6.89&5.36&4.18&3.73    \\ 
                   &5.9&11.5&9.3&7.4&6.5         \\ 
$1 l, 4 j, 1\nu$ &7.6&11.5&12.4&10.0&8.4            \\ 
                    &13.3&20.0&21.6&18&15            \\ 
$4 l, 2 j$  &0.29&1.0&0.54&0.38&0.32            \\ 
               &0.49&1.6&0.91&0.65&0.54             \\ 
$2 l, 2 j, 2 \nu$ &2.0&6.5&3.5&2.5&2.2     \\ 
                     &3.4&10.7&6.1&4.4&3.7      \\ 
$2 l, 4 j$    &2.5&7.4&5.4&3.6&2.9 \\ 
                  &4.7&9.5&9.5&6.9&5.6       \\ 
$4 j, 2\nu$ &8.9&27&18&12.6&10.4        \\ 
             &195.&54&38&28&23           \\ 
$6 j$     &5.3&12.3&13.3&8.8&7.4                 \\ 
         &9.1&20.7&23&16&12.5                  \\ \hline
\end{tabular}
\end{center}
\end{table}

\begin{table}[htb]
\begin{center}
\caption{Number of events from $W^-W^+W^+$ production in different decay modes
($\sqrt{s}=14$~TeV, 100~fb$^{-1}$).}
\vskip 0.2cm
\begin{tabular}{lccccc}\hline
$M_{\rm Higgs}$ (GeV)           &200  &300  &400  &500  &600   \\ \hline
$3 l, 3 \nu$       &66&44&37&35&33                \\  
$l^+ l^+, 2 j, 2 \nu$ &57&43&31&26&24          \\ 
$l^+ l^-, 2 j, 2\nu$   &13&7.9&5.3&4.4&4.0     \\
$l^+, 4 j, 1 \nu$         &148&129&86&66&58   \\ 
$l^-, 4 j, 1 \nu$    &99&61&36&26&23  \\ 
$6 j$                  &50&74&46&32&25   \\ \hline
\end{tabular}
\end{center}
\end{table}

\begin{table}[htb]
\begin{center}
\caption{Number of events from $W^+W^-W^-$ production in different decay modes
($\sqrt{s}=14$~TeV, 100~fb$^{-1}$).}
\vskip 0.2cm 
\begin{tabular}{lccccc}\hline
$M_{\rm Higgs}$ (GeV)           &200  &300  &400  &500  &600   \\ \hline
$3 l, 3 \nu$      &40&26&22&21&20                \\  
$l^- l^-, 2 j, 2 \nu$ &34&25&17&14&13          \\ 
$l^+ l^-, 2 j, 2\nu$  &78&45&30&25&23      \\ 
$l^-, 4 j, 1 \nu$       &90&76&49&37&33   \\ 
$l^+, 4 j, 1 \nu$   &59&35&20&15&13   \\ 
$6 j$                  &29&43&26&18&14  \\ \hline
\end{tabular}
\end{center}
\end{table}


%
%


%
%

%
%
\newcommand{\txt}[1]{{\mbox{\tiny #1}}} 
\newcommand{\pT}[1]{p_T(#1)}      
\newcommand{\Minv}[1]{M_\txt{inv}(#1)}  
\newcommand{\Lmachine}{\Lambda_{\txt{machine}}}
\newcommand{\LFF}{\Lambda_\txt{FF}}
%
%
%
%

\subsection{Unitarity limits and form factors \label{sec:tgc_formfact}}

Unitarity in the Standard Model depends directly on its gauge-structure. 
Departure from this structure can violate unitarity at
relatively low energies and so protection is provided in the effective
Lagrangian for triple gauge-boson vertices by expressing the
anomalous couplings as energy dependent form factors.
For experimental results at a given subprocess energy $\hat{s}$ ({\it i.e.}\ 
$e^+e^-$ colliders), the choice of form factor parametrisation is not
important since one can unambiguously translate between
parametrisations.  However, when results are integrated over a range
of $\hat{s}$ as they will be at the LHC, no simple translation is
possible and results depend crucially on the choice of the form
factors.  The form factor behaviour of anomalous couplings should not
be neglected, particularly in regions of $\hat{s}$ near to unitarity
limits.  Any measurement of anomalous couplings over integrated
energies carries with it {\it assumptions on the parametrisation of
  the form factor}.

This section outlines the considerations which influence the choice of
form factor and suggests a method for measuring energy dependent
anomalous couplings.



\subsubsection{Form factor parametrisation}

Triple gauge-boson vertices in di-boson production arise in the
$J=1$ partial wave amplitude only ($s$-channel exchange of a 
gauge-boson coupled to massless fermions).
$S$-matrix unitarity implies a constant bound to any partial wave
amplitude. This means unitarity is violated at asymptotically high
energies if constant anomalous couplings are assumed.
Unambiguous and model-independent constant unitarity constraints for
$WV$ production have been derived\footnote{
  Cancellations may occur if more than one anomalous coupling is allowed 
  non-zero at a time, which weakens the unitarity limits somewhat.
}~\cite{baur88unitarity}.

To conserve unitarity at arbitrary energies, anomalous couplings must
be introduced as form factors. Thus, an arbitrary anomalous coupling
$\tilde{A}=\tilde{A}_0\times{\cal~F}(q_1^2,q_2^2,P^2)$ vanishes when
$q_1^2,~q_2^2,$ or $P^2$ becomes large, where $q_1^2$ and $q_2^2$ are
the invariant masses squared of the production bosons and
$P^2=\hat{s}$ is the virtual exchange boson invariant mass squared. We
refer to $\tilde{A}_0$ as the ``bare coupling'' and $\tilde{A}$ as the
form factor
($\tilde{A}~\epsilon~\lambda^V,\Delta\kappa^V,h^V_i,$\ldots).  For
di-boson production, the final state bosons are nearly on-shell
$q_1^2,q_2^2\simeq M_V^2$ even when finite width effects are taken
into account, though large virtual exchange boson masses
$\sqrt{\hat{s}}$ will be probed at the LHC.

The choice of parametrisation for the form factors is arbitrary
provided unitarity is conserved at all energies for a sufficiently
small value of anomalous coupling.
A step function operating at a cutoff scale $\LFF$ is sufficient\footnote{ 
  {\it i.e.}\ assuming a step
  function form factor operating at 2~TeV, the $\lambda^\gamma$
  coupling conserves unitarity for $\lambda^\gamma <
  0.99$~\cite[Equation~23]{baur88unitarity}.  
}
though discontinuous and thus unphysical.  More common in the
literature is a generalised dipole form factor which is motivated by
the well known nucleon form factors and has further appeal because it
enters the Lagrangian in a form similar to that of a propagator of
mass $\LFF$. The parametrisation is
\begin{equation}
  \tilde{A} ~=~ \frac{ \tilde{A}_0 }
                     { (1+\frac{\hat{s}}{\LFF^2})^n }
\end{equation}
where $n>1/2~(n>1)$ is sufficient for the $WWV$ vertex anomalous
couplings $\Delta\kappa^V~(\lambda^V,~\Delta g_1^V)$ which grow like
$\hat{s}^{1/2},~(\hat{s})$.  For the $ZV\gamma$ vertex $n>3/2~(n>5/2)$
is sufficient for anomalous couplings $h_{1,3}^V,~(h_{2,4}^V)$ which
grow like $\hat{s}^{3/2},~(\hat{s}^{5/2})$.  The usual assumptions are
$n=2$ for $g_1^V,~\lambda^V,~\kappa^V$
~\cite{BHOWZ,BHOWg,BHOWW} and $n=3~(n=4)$ for
$h_{1,3}^V,~(h_{2,4}^V)$~\cite{baur93D47}.  Unitarity limits for
generalised dipole form factors have been
enumerated~\cite[Equations~22-26]{snowmass95}.

The form factor scale $\LFF$ can be regarded as a
regularisation scale. It is related to (but not necessarily identical
to) the energy scale at which new physics becomes important in the
weak boson sector. 


\subsubsection{Impact of form factor on $\hat{s}$ dependent
  distributions \label{sec:tgc_ffonshat}}

The impact of the form factor parametrisation on $\hat{s}$ dependent
distributions is illustrated in Figure~\ref{ff_impact} where the
reconstructed~\footnote{ Reconstructing $\Minv{WZ}$ requires knowledge
  of the neutrino longitudinal momentum which is obtained up to a
  two-fold ambiguity using the $W$ mass
  constraint. Each solution is given half weight
  in the $\Minv{WZ}$ spectrum.  
}
$\Minv{WZ}$ and $\pT{Z}$ spectra are plotted for LHC $W^+Z$ production
with leptonic decays at $O(\alpha_s)$. The Standard Model expectation
is compared to scenarios with a modest $\lambda^Z_0=0.05$ coupling for
various generalised dipole form factor parametrisations.

\begin{figure}
\begin{center}
\includegraphics[height=4in,width=5in]{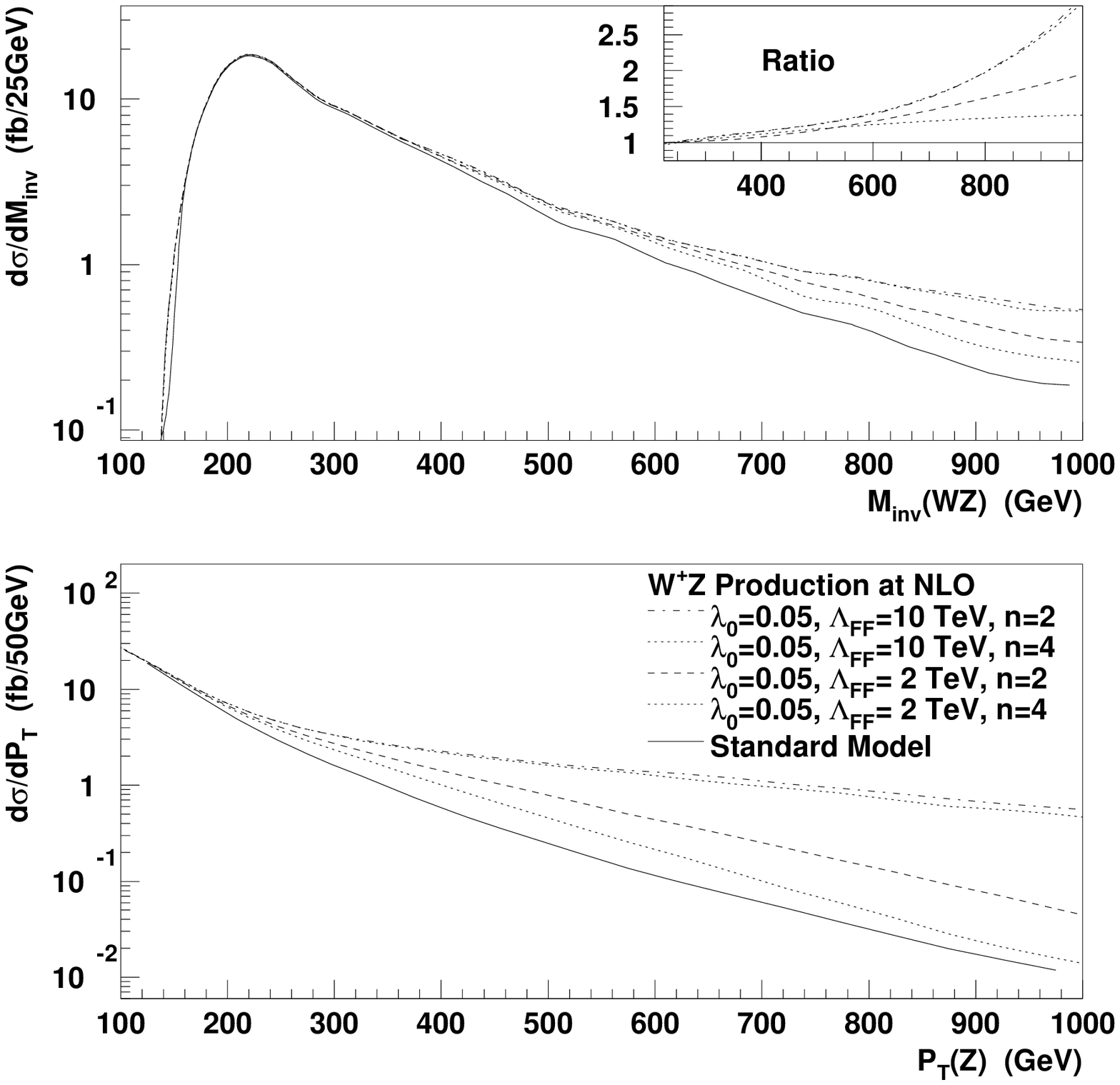}
\caption{\label{ff_impact} Reconstructed $\Minv{WZ}$ and $\pT{Z}$ 
  spectra are plotted for LHC $W^+Z$ production with leptonic decays
  at $O(\alpha_s)$ for the Standard Model and various choices of the
  generalised dipole form factor parametrisation with bare coupling
  $\lambda_0=0.05$.  }
\end{center}
\end{figure}


For the region of low invariant mass where $\sqrt{\hat{s}} \ll \LFF$,
the form factors remain essentially constant and distributions with
the same bare coupling agree well. As the form factor scale $\LFF$ is
approached, the distributions begin to be pushed back to the SM
expectation (visible at about $\Minv{WZ}=500$~GeV for the $\LFF$=2~TeV
case). For $\sqrt{\hat{s}}>\LFF$ the distribution returns to the SM
expectation. The exponent of the form factor $n$ dictates how fast the
``pushing'' occurs as $\LFF$ is approached.
Thus distributions sensitive to the $ZV\gamma$ vertex (for which
$n=$3~or~4 is the usual choice) exhibit a more pronounced form factor
behaviour than distributions sensitive to the $WWV$ vertex (for which
$n=2$ is usual).

Since distributions are constrained to the SM expectation at invariant
masses above the form factor scale, great care should be taken when
fitting to a form factor parametrised model in a region with data
where $\sqrt{\hat{s}}\geq\LFF$.  Effectively, since the anomalous
couplings are constrained near zero above $\LFF$ by the
parametrisation model, {\it there are no free parameters for the fit}
in this $\hat{s}$ region. For the case of observable non-zero
anomalous couplings, an analysis assuming a parametrisation of the
form factor with fixed $\LFF$ smaller than that provided by nature but
within the $\hat{s}$ accessible by the machine would overestimate the
anomalous coupling. This is because large bare coupling fit values are
necessary in the $\sqrt{\hat{s}}\geq\LFF$ region to counter the
(artificially imposed) form factor behaviour.


\subsubsection{Impact of form factor scale on sensitivity limits}

If triple gauge-coupling (TGC) 
measurements are consistent with the SM and confidence
limits are to be derived, it is impossible to avoid form factor
parametrisation assumptions.

The dependence of anomalous coupling limits on the form factor scale
$\LFF$ is illustrated in Figure~\ref{ff_limits} where the 95\%
confidence limits for $WW\gamma$ vertex anomalous
$\lambda^\gamma_0,\Delta\kappa^\gamma_0$ couplings in $W\gamma$
production with $W\rightarrow e\nu_e,\mu\nu_\mu$ are presented as a
function of $\LFF$ for a dipole form factor with $n=2$.  The limits
are for illustrative purposes only and have been derived at NLO
generator level using a binned maximum likelihood fit to the
$\pT{\gamma}$ distribution. No detector simulation has been applied
and the specific choice of cuts are unimportant.

\begin{figure}
\begin{center}
\includegraphics[height=4in]{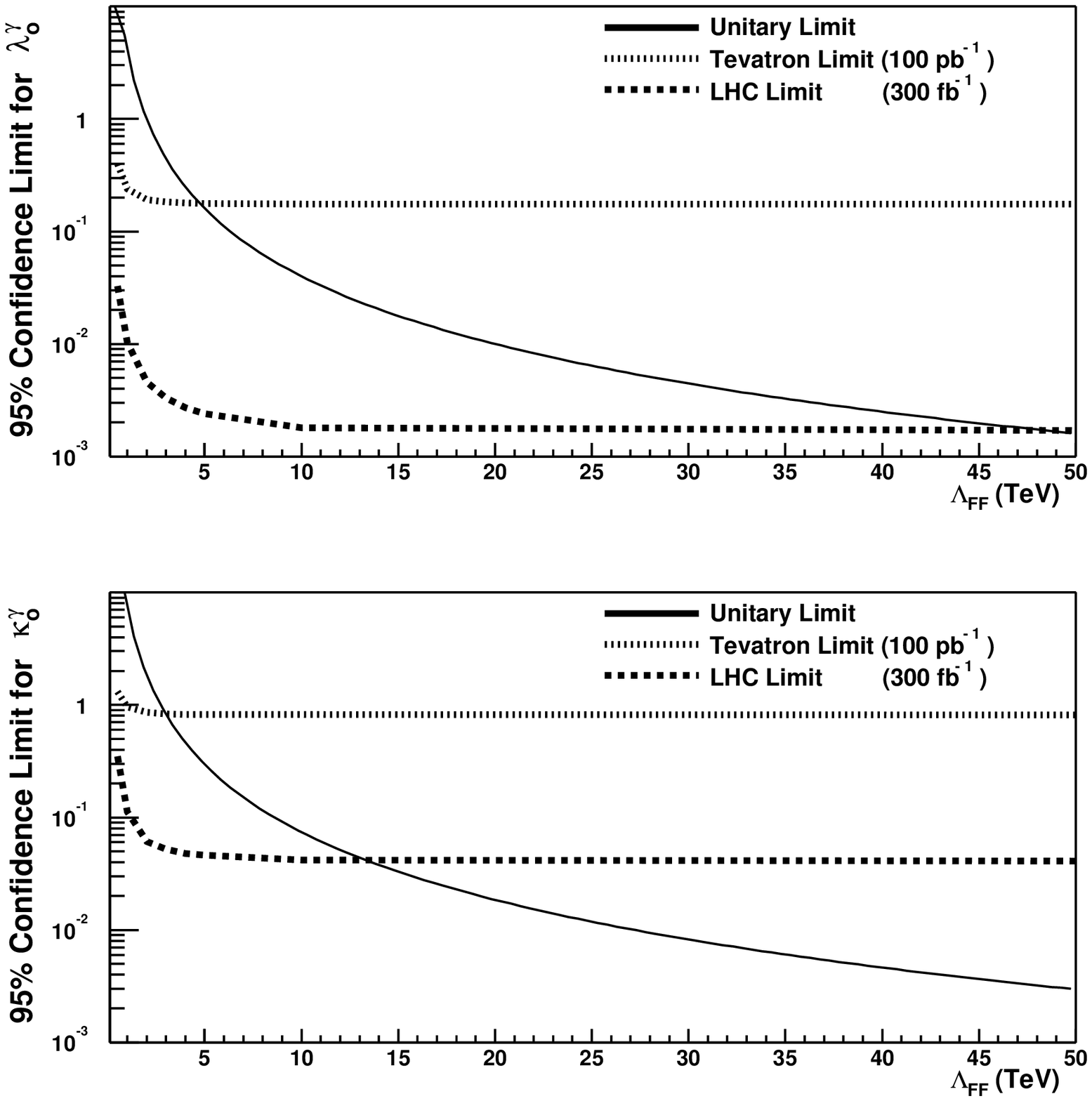}
\caption{\label{ff_limits} Limits for $WW\gamma$ vertex anomalous
  couplings at the 95\% confidence level as a function of
  $\LFF$ for a $n=2$ dipole form factor parametrisation are
  presented. The limits are derived at NLO generator level for the
  $W\gamma\rightarrow e\nu_e\gamma,\mu\nu_\mu\gamma$ channel using a
  binned maximum likelihood fit to the $\pT{\gamma}$ distribution. 
  The limits are for illustrative purposes only. Further details are
  provided in the text.
}
\end{center}
\end{figure}

The unitarity limit curve is superimposed. The region above this is
non-physical (violates unitarity). The curve is independent of
experiment and analysis but depends on the form factor
parametrisation. It goes asymptotically to zero for large $\LFF$
indicating TGC couplings are restricted to SM values at extreme
energies.

Simulated experimental limits for the Tevatron (2~TeV $p\bar{p}$ collisions,
${\cal{L}}=100$~pb$^{-1}$) and the LHC (14~TeV $pp$ collisions,
${\cal{L}}=300$~fb$^{-1}$) are presented. The limits depend on the
analysis and machine parameters. The restricted $\hat{s}$ accessible
by the machines result in an asymptotic behaviour wherein an optimal
limit for anomalous couplings is reached. We refer to the scale at
which this occurs as $\Lmachine$. A measurement with this scale
reflects the maximal discovery potential for anomalous couplings for a
given machine (since the full spectra in $\hat{s}$ contributes to the
limit).
It occurs at about 2~TeV for the Tevatron and about 5-10~TeV for the
LHC for $\lambda^\gamma,~\Delta\kappa^\gamma$ and lies below the
unitarity limit in both cases. 
The experimental limits are not sensitive to changes in
$\LFF$ for $\LFF\geq
\Lmachine$. Indeed, in this region the distributions
behave exactly as if the form factors were constants 
$\tilde{A}\equiv \tilde{A}_0$.
There is no contradiction with unitarity in approximating them as
such, provided we consider sufficiently small anomalous couplings so
as to remain far from the unitary limit {\it at the energy regimes
  accessible by the machines}.  
This is consistent with the basic assumption ($\Lambda \gg
\sqrt{\hat{s}}$) which allows for the effective Lagrangian
parametrisation of the TGC vertex keeping only the lowest dimensions:
it is sufficient to assume the form factor behaviour commences above
the observable scale so as to regulate the distributions before the
unitarity limit.

There is also a region on the extreme left side of the plots in
Figure~\ref{ff_limits} (although not indicated) which is excluded
by direct experimental searches. This is the region where physics is
believed to be well described by the SM.

Experimentally it is desirable to report confidence limits as a
function of $\LFF$. A result using $\LFF=\Lmachine$ should be included
(so long as $\Lmachine$ lies below the unitarity limit) as it is
motivated by machine parameters and provides a reasonable point of
reference for comparisons between different experiments.
Other scales (particularly those of theoretical
interest) should not be neglected\footnote{ 
  It should be noted that
  particularly for small choices of $\LFF$, a change in the
  analysis strategy may be necessary to increase sensitivity to the
  relevant regions of $\hat{s}$. }.


\subsubsection{Measuring form factors}

For a machine of sufficient luminosity such as the LHC, it is possible
to measure the energy dependence of anomalous
couplings\footnote{ The suggestion of making such a measurement is
  not new \cite{GLR99} but has received little attention
  in the literature.  }
by grouping the data into bins of invariant mass and extracting
constant anomalous couplings within these restricted domains. Such
a measurement does not carry any assumptions about the form factor
(until a fit to a given parametrisation is performed). It
is a viable method for measuring form factors, but
due to the restricted number of events in each bin, will not produce
competitive limits. The method is best employed in the
case where non-zero anomalous couplings have been observed. 

\begin{figure}
\begin{center}
\includegraphics[height=3in]{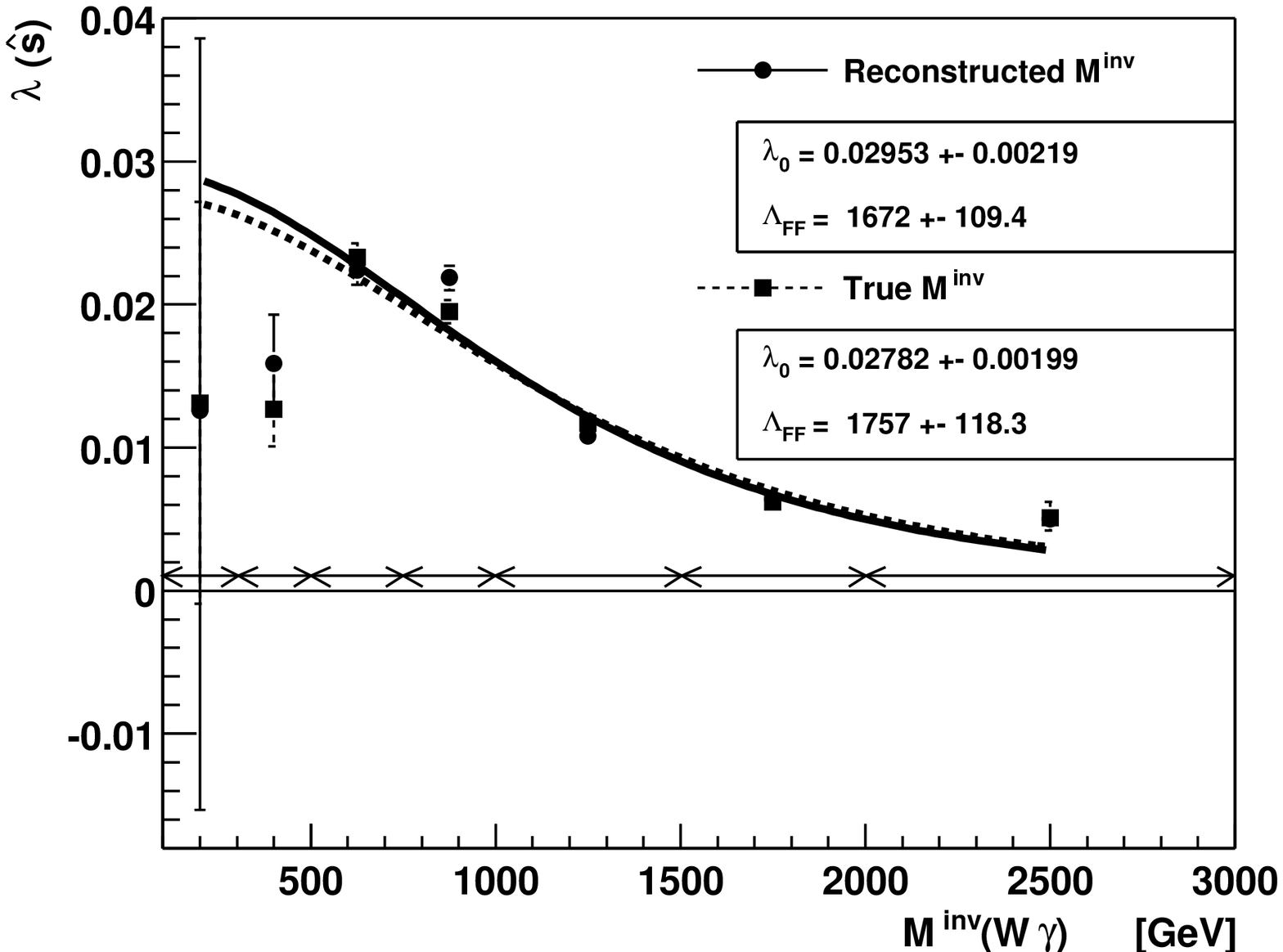}
\caption{\label{measure_ff} The $\lambda^\gamma$ form factor 
  is extracted in restricted invariant mass domains for 300~fb$^{-1}$
  of LHC data in the $W\gamma$ channel with $W\rightarrow
  e\nu_e,\mu\nu_\mu$ assuming nature provides an anomalous
  $\lambda_0^\gamma=0.025$ coupling described by an $n=2$ dipole form
  factor with $\LFF=$2~TeV.  A fit to a $n=2$ dipole form factor is
  performed to reconstruct the bare coupling and form factor scale.
  Arrows along the $x$-axis denote bin widths.  Further details are
  provided in the text.  }
\end{center}
\end{figure}

The method is illustrated in Figure~\ref{measure_ff} for the case of
the $W\gamma$ channel with $W\rightarrow e\nu_e,\mu\nu_\mu$ assuming
nature provides an anomalous $\lambda_0^\gamma=0.025$ coupling
described by an $n=2$ dipole form factor with $\LFF=$2~TeV.  Three
years of high luminosity (300~fb$^{-1}$) LHC events generated at NLO
are binned according to the reconstructed $\Minv{W\gamma}$.
The corresponding points derived using the generated (unobservable)
$\Minv{W\gamma}$ are superimposed for comparison. Bin widths (denoted
by arrows along the x-axis) are chosen so as to ensure sufficient data
in each $\Minv{W\gamma}$ domain. A measurement of the anomalous
coupling (assumed constant) is performed within each domain using a
binned maximum likelihood fit to the $\pT{\gamma}$ distribution.  No
detector simulation has been applied and the specific choice of cuts
is unimportant for this illustration. The results of the likelihood
fits are plotted as a function of $\Minv{W\gamma}$ and a fit to an
$n=2$ dipole form factor is performed. With this simple illustration,
the bare coupling and form factor scale are reconstructed as
$\lambda_0^\gamma=0.029$ and $\LFF=1.67$~TeV.  Sensitivity to the
anomalous coupling increases in the larger invariant mass domains,
reflecting the $\hat{s}$ growth of the $\lambda_0^\gamma$ coupling
(indeed the measurement in the first bin is consistent with zero).
Systematic effects related to the fit method (such as the non-uniform
distribution of events within the bins) have not been accounted for in
this illustration.

\newcommand{\order}{{\cal O}}
%
%
%
%

\subsection{Partonic simulation tools for di-boson production
\label{sec:tgc_nlo}}

Several Monte Carlo programs for hadronic di-boson event simulation
are in common use. General purpose programs such as
{\tt PYTHIA}~\cite{pythia} evaluate the matrix element at leading
order (LO) with no spin correlations for boson decay products. Limited
or no anomalous couplings are included.
In the past decade, programs have been implemented to calculate
di-boson production with leptonic decays to next-to-leading order (NLO)
in QCD.  The diagrams contributing to $O(\alpha_s)$ are: the squared
Born (LO) graphs, the interference of the Born with the virtual one-loop
graphs, and the squared real emission graphs.

The NLO generators by Baur, Han, and
Ohnemus~\cite{BHOWg,BHOWZ,BHOWW,BHOZg} (BHO) have been
available for several years. They employ the phase space slicing
method~\cite{Baer:1989jg} and the calculation is performed in
the narrow width approximation for the leptonically decaying
gauge-bosons. Non-standard TGC couplings are included.  Spin
correlations in the leptonic decays are included everywhere except in
the virtual contribution. The authors expect a negligible overall
effect from neglecting the spin correlations in the virtual
corrections as compared to the uncertainty from parton distribution
functions and the choice of factorisation scale. 
%
%
More recently Dixon, Kunszt, and Signer~\cite{Dixon:1999di} (DKS) have
implemented a program
with full lepton decay spin correlations (helicity amplitudes are
presented in~\cite{Dixon:1998py}). The
subtraction method~\cite{FKS,Ellis:1981wv} is employed in
the narrow width approximation including non-standard TGC couplings.
%
%
A third Monte Carlo program, \verb!MCFM!, by Campbell and
Ellis~\cite{Campbell:1999ah} exists. It does
not assume the narrow width approximation and includes singly resonant
diagrams but does not allow for non-standard TGC couplings.  The
effects of these improvements in \verb!MCFM! are largest in
off-resonant regions - such as near di-boson production thresholds.
The regions are of importance to studies of SM backgrounds to new
physics but contribute negligibly to the cross section
in TGC studies for typical choices of kinematic cuts~\cite{BHOWW}.

A common feature of the NLO generators is the inability to produce
unweighted events.  Both the phase space slicing and subtraction
methods produce events for which the weight may be either positive or
negative - thus it is only the integrated cross section over a region
of phase space ({\it i.e.}\ histogram bin) which is physical.  This makes
traditional Monte Carlo techniques for unweighting events (such as
hit-and-miss) difficult to apply, and we are aware of no universally
satisfactory technique for producing unweighted events using the NLO
generators\footnote{ One method involves reweighting events from a LO
  generator using a ``look-up table'' constructed at NLO.
}. Computationally this can render analyses very slow, since a large
fraction of CPU time can be spent processing events with
near-vanishing cross sections.

%
%

\subsubsection{Comparison of NLO particle level generators}

In this section, we present a comparison of the predictions from the
BHO and DKS generators, for which no published consistency check
exists, restricting ourselves to $W^+Z$ and $WW$ production for
simplicity.  The DKS and \verb!MCFM! packages have been found to be in
good agreement~\cite{Dixon:1999di}.

The comparison is performed at LHC energy ($14$~TeV $pp$
collisions) using CTEQ4M~\cite{cteq4} structure
functions\footnote{
  The choice of parton distribution function has an $\order(5\%)$ effect on
  the cross section.}. 
Input parameters are taken as
$\alpha_{EM}=\frac{1}{128}$,
$\sin^2\theta_W=0.23$,
$\alpha_s(M_Z)=0.116$,
$M_W=80.396$~GeV,
$M_Z=91.187$~GeV,
factorisation scale $Q^2=M_W^2$,
and Cabibbo angle $\cos\theta_{\rm C} =0.975$ with 
no 3rd generation mixing.
Branching ratios are taken as $BR(Z\rightarrow l^+ l^-)$ = 3.36\%,
$BR(W^\pm\rightarrow l^\pm\nu)$ = 10.8\%.
The $b$ quark contribution to parton
distributions has been taken as zero ($b\bar{b}\rightarrow W^+W^-$
contributes $\order(2\%)$ at LHC~\cite{Dixon:1999di}).
Kinematic cuts motivated by TGC analyses are chosen. The transverse
momentum of all leptons must exceed 25~GeV and the rapidity of all
leptons must be less than 3. Missing transverse momentum must be greater
than 25~GeV. A jet is defined when the transverse momentum of a parton
exceeds 30~GeV in the pseudorapidity interval $|\eta|<3$.

For $W^+Z$ production, the transverse momentum distribution of the $Z$ boson
$p_T(Z)$, the distribution of rapidity separation between the $W^+$ decay
lepton and the $Z$ boson $y(l)-y(Z)$, and total cross section are
compared at LO, inclusive NLO, and NLO with a jet veto.  Branching ratios to
$e,\mu$-type leptons are applied. 
For $WW$ production, the transverse momentum distribution of the lepton
pair from the $W^\pm$ decays $|\vec{p_T}(e^-)+\vec{p_T}(e^+)|$, the
distribution of rapidity separation between the $W$ decay leptons
$y(e^-)-y(e^+)$, the angle between the $W$ decay leptons in the
transverse plane $\cos\Phi(e^-,e^+)$, and the total cross section are
compared at LO, inclusive NLO, and NLO with a jet veto. Branching
ratios to one lepton flavour are applied.

The cross section results are presented in
Table~\ref{mc_comparison_table} and the distributions in
Figure~\ref{mc_comparison}.  
Consistency between generators is at the 1\% level for $WZ$ production
and 3-4\% level for $WW$ production.
Qualitative agreement is observed in the distribution shapes. 

\begin{table}[htbp]
\begin{center}
\caption{ \label{mc_comparison_table} 
  $W^+Z$ and $WW$ cross section predictions are tabulated for the BHO and DKS
  generators at LO, inclusive NLO, and NLO with a jet veto. 
  A jet is defined for $p_T(\mbox{jet})>$30~GeV, $\eta(\mbox{jet})<3$. 
  Statistical precision is ${\cal O}(\mbox{1~fb})$.}
\vskip0.2cm
  \begin{tabular}{lccc} \hline
    \multicolumn{4}{c}{\bf \boldmath $W^+Z$ \unboldmath Production } \\ 
    & Baur/Han/Ohnemus  & Dixon/Kunszt/Signer & \% diff. \\ \hline 
    \multicolumn{4}{c}{ Standard Model } \\ \hline
    $\sigma_{\txt{NLO inclusive}}$ & 127.9 fb & 129.8 fb & 1.4\% \\
    $\sigma_{\txt{NLO 0jet}}$      &  74.7 fb & 75.1 fb  & 0.5\% \\
    $\sigma_{\txt{Born}}$          &  70.5 fb & 70.9 fb  & 0.5\% \\ 
    \hline 
    \multicolumn{4}{c}{$ \Delta g^1_Z=0, \Delta\kappa_Z=0.5, \lambda_Z=0.1
      \hspace{1cm} (\Lambda=2~TeV)$} \\ \hline
    $\sigma_{\txt{NLO inclusive}}$ & 198.5 fb & 199.9 fb & 0.7\% \\
    $\sigma_{\txt{NLO 0jet}}$      & 107.5 fb & 106.8 fb & 0.7\% \\
    $\sigma_{\txt{Born}}$          & 119.7 fb & 119.9 fb & 0.2\% \\ 
    \hline 
\end{tabular}

\vspace{0.5cm}

  \begin{tabular}{lccc} \hline
    \multicolumn{4}{c}{\bf \boldmath $WW$ \unboldmath Production } \\ 
    & Baur/Han/Ohnemus  & Dixon/Kunszt/Signer & \% diff. \\ \hline 
    \multicolumn{4}{c}{ Standard Model } \\ \hline
    $\sigma_{\txt{NLO inclusive}}$ & 500.5 fb & 483.2 fb & 3.5\% \\
    $\sigma_{\txt{NLO 0jet}}$      & 321.0 fb & 309.6 fb & 3.6\% \\
    $\sigma_{\txt{Born}}$          & 294.0 fb & 295.5 fb & 0.5\% \\ 
    \hline 
    \multicolumn{4}{c}{$ \Delta g^1_Z=0.25, 
      \Delta\kappa_Z=\Delta\kappa_\gamma=0.1, \lambda_Z=\lambda_\gamma=0.1
      \hspace{1cm} (\Lambda=2~TeV) $} \\ \hline
    $\sigma_{\txt{NLO inclusive}}$ & 594.2 fb & 575.0 fb & 3.3\% \\
    $\sigma_{\txt{NLO 0jet}}$      & 363.0 fb & 349.6 fb & 3.8\% \\
    $\sigma_{\txt{Born}}$          & 351.6 fb & 353.7 fb & 0.6\% \\ 
    \hline 
\end{tabular}
\end{center}
\end{table}

\begin{figure}[htbp]
\begin{center}
\includegraphics[height=3in]{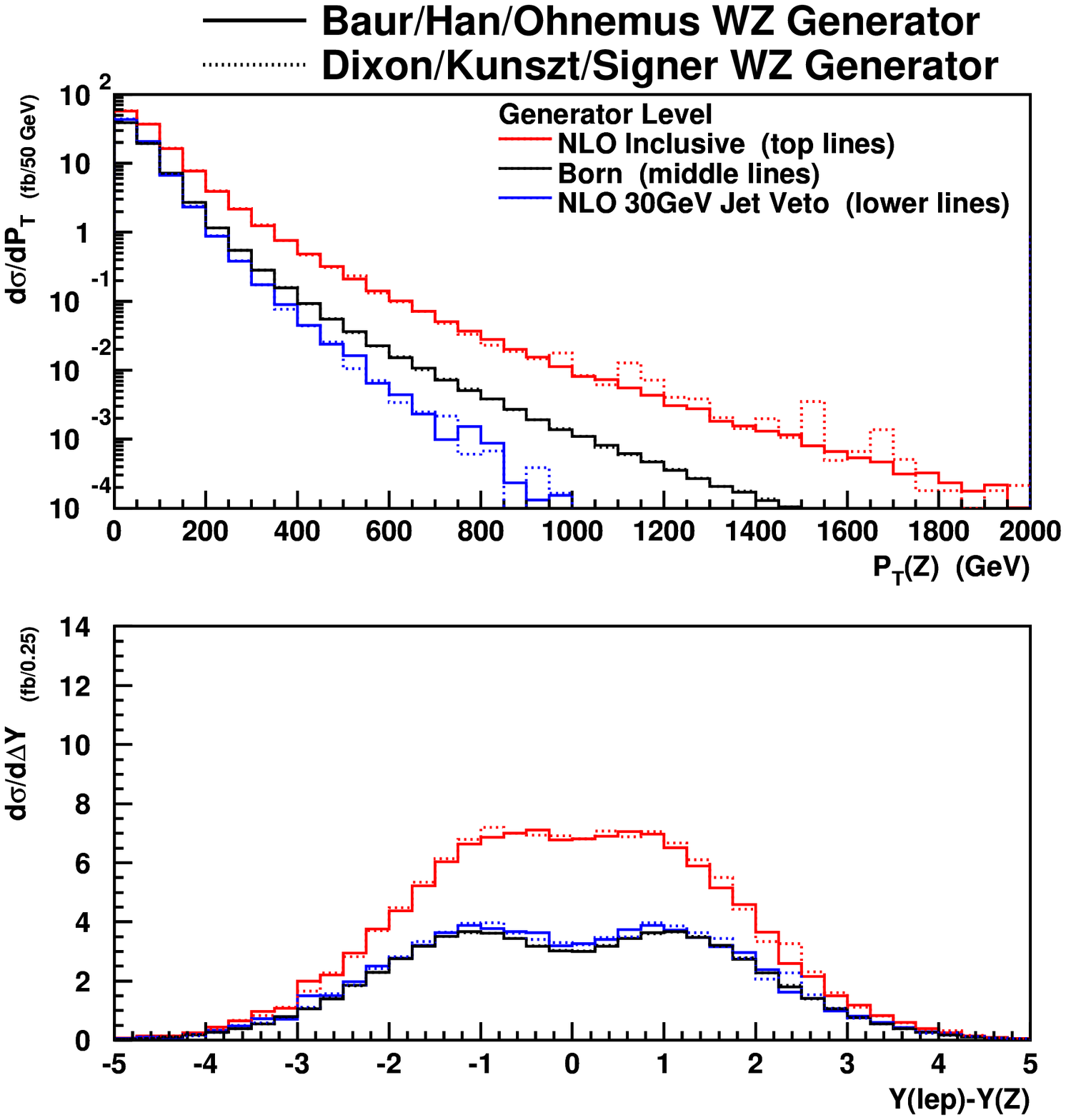}
\includegraphics[height=3in]{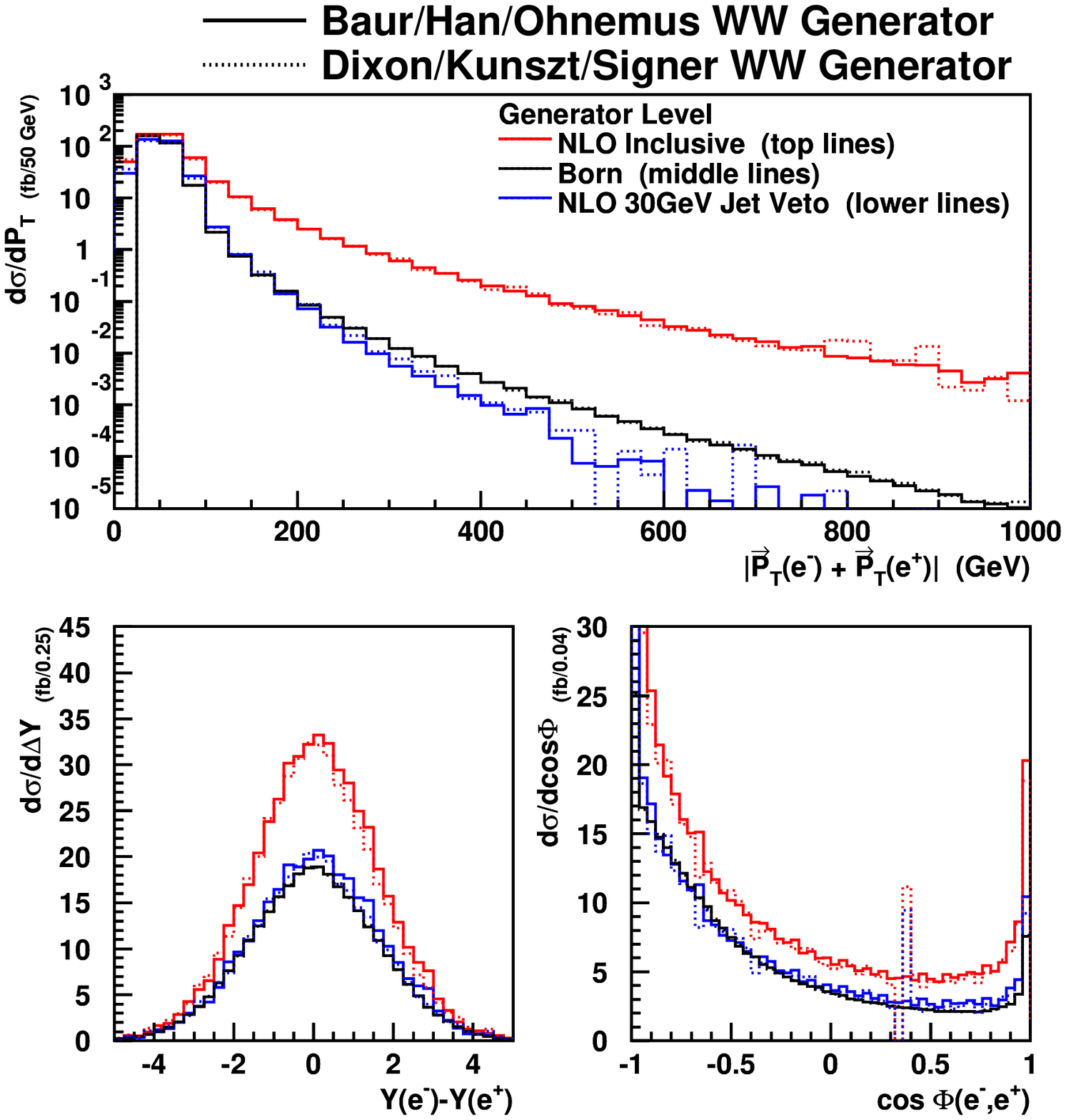}
\caption{\label{mc_comparison} 
  Distributions for $W^+Z$ production (left) and $WW$ production
  (right) from the Baur/Han/Ohnemus and Dixon/Kunszt/Signer generators
  are superimposed at Born level, inclusive NLO, and NLO with a jet
  veto (defined as $\pT{\mbox{jet}}>$30~GeV, $|\eta(\mbox{jet})|<3$).
  }
\end{center}
\end{figure}

%
%

\subsubsection{Effects of NLO corrections}

NLO corrections in hadronic di-boson production are large at LHC
energies, particularly in the region of high transverse momentum and
small rapidity separation (see Figure~\ref{mc_comparison}) which is
the same region of maximum sensitivity to anomalous TGCs.  The
corrections can amount to more than an order of magnitude.
The high quark-gluon luminosity at the LHC and a logarithmic
enhancement at high transverse momentum in the $qg$ and $\bar{q}g$
real emissions subprocesses are primarily
responsible~\cite{BHOWg,BHOWZ,BHOWW}.  In the channels which
exhibit radiation zero behaviour ({\it i.e.}\ $W\gamma$ and $WZ$
), the Born contribution is suppressed and NLO corrections are even
larger~\cite{BHOWg,BHOWZ}.  Since the $O(\alpha_s)$ subprocesses
responsible for the enhancement at large transverse momentum do not
involve TGCs, the overall effect of NLO corrections is a spoiling of
sensitivity to anomalous TGCs.

\paragraph{Jet veto}

Distributions obtained by vetoing hard jets in the central rapidity
region for one possible choice of jet definition
($\pT{\mbox{jet}}>30$~GeV, $|\eta(\mbox{jet})|<3$) are shown in
Figure~\ref{mc_comparison}.  The jet veto is effective in recovering
the qualitative shape of the LO distributions including the
approximate radiation zero in $WZ$ production
(Figure~\ref{mc_comparison}, bottom left). The jet veto serves to
recover anomalous TGC sensitivity which is otherwise lost when
introducing NLO corrections.  A 10-30\%
improvement in anomalous TGC coupling sensitivity limits in $WZ$
production can be achieved~\cite{BHOWZ} when a jet veto is applied as compared to the inclusive NLO
case. These limits are often close to those obtained at LO.
In general results derived at LO can be considered approximate zero
jet results and their conclusions remain interesting.
A jet veto also reduces the scale dependence of NLO
results~\cite{BHOWg,BHOWZ,BHOWW,Dixon:1999di}. 







\subsection{Determination of TGCs \label{sec:tgc_analysis}}
At the LHC the measurement of TGCs will benefit from both the large
statistics and the high centre-of-mass energy. The large available
statistics will allow the use of multi-dimensional distributions to
increase the sensitivity to the TGCs. 

This section discusses the experimental observables sensitive to TGCs
and describes the analysis methods employed to measure the TGCs.

\subsubsection{Experimental observables}

The experimental sensitivity to the TGCs comes from the increase of
the production cross section and the modification of differential
distributions with non-standard TGCs. The sensitivity is
enhanced at high centre-of-mass energies of the hard scattering
process, more significantly for $\lambda$-type TGCs than for $\kappa$-type 
TGCs in the case of $W\gamma$ and $WZ$ production. As an example,
the increase in the number of events
with large di-boson invariant masses is a clear signature of
non-standard TGCs as illustrated in Figure~\ref{fig:Mwg}, where the
invariant mass of the hard scattering is shown for $W\gamma$ events,
simulated with a parametric description of the ATLAS detector, for the
Standard Model and non-standard TGCs. A form factor of 10~TeV was used. 

\begin{figure}[h]
\begin{center}
\includegraphics[width=7cm]{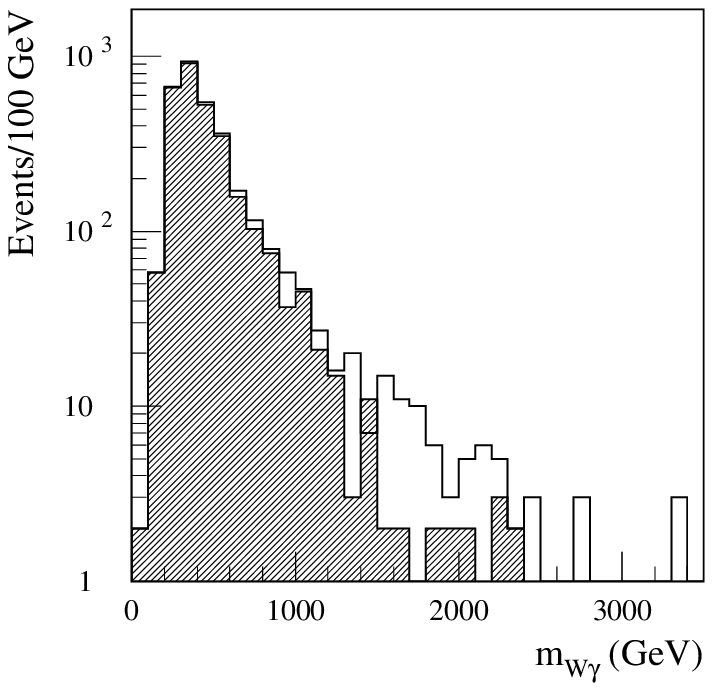}
\caption{The distribution of the invariant mass of the $W\gamma$
  system from $pp\rightarrow W\gamma$. Standard Model data (shaded histogram) and a non-standard
  value of 0.01 for $\lambda_{\gamma}$ (white histogram) are shown. Both charges of $W$ were generated
  using a parameterised Monte Carlo and summed. The number of
  events corresponds to an integrated luminosity of 30~fb$^{-1}$.}
\label{fig:Mwg}
\end{center}
\end{figure}

For the event generation employing non-standard values of the
TGCs, leading order (LO) \cite{ref:baur-zeppenfeld} as well as next to 
leading order (NLO) \cite{BHOWg,BHOZg} calculations 
have been used (see Section~\ref{sec:tgc_nlo}).
Limits on the
TGCs can be obtained from event counting in the high invariant mass
region. The disadvantage of such an approach alone is that the
behaviour of the cross section as function of the TGCs makes it
difficult to disentangle the contributions from different TGCs and
even their sign (with respect to SM). It is therefore advantageous to
combine it with information from angular distributions of the bosons
and possibly their decay angles; this improves the sensitivity and
improves the separation of contributions from different non-standard
TGCs.

In general it is possible experimentally to reconstruct up to four (six)
angular variables in the di-boson rest-frame describing an $W\gamma$
or $Z\gamma$ ($WZ$) event: 
\begin{itemize}
\item Boson production angles, $\Theta$ and $\Phi$, of the di-boson
  system with respect to the beam-axis in the di-boson rest-frame.
\item Decay angles of bosons, $\theta^*_{1(2)}$ and $\phi^*_{1(2)}$,
  in the rest-frame of the decaying bosons.
\end{itemize}

The azimuthal boson production angle, $\Phi$, has no sensitivity to
the TGCs.
In case of $W\gamma/WZ$, $\Theta$ is the most sensitive kinematical 
variable. 
The enhanced sensitivity to the TGCs in $WV$ production is due to the
vanishing of 
helicity amplitudes in the Standard Model prediction at $\cos\Theta
\sim 1/3$, affecting the small $|\eta|$ region
\cite{ref:baur-zeppenfeld}. Non-standard TGCs may partially eliminate
the radiation zero, although the zero radiation prediction
is less significant when including NLO corrections 
\cite{BHOWg}.
In $Z\gamma$ production, no radiation amplitude zero is present. 

In contrast, the
sensitivity to the TGCs from the decay angles is weak; the decay angles
primarily serve as projectors of different helicity components,
enhancing the sensitivity of other variables.

\begin{figure}[ht]
\begin{center}
\includegraphics[width=6cm]{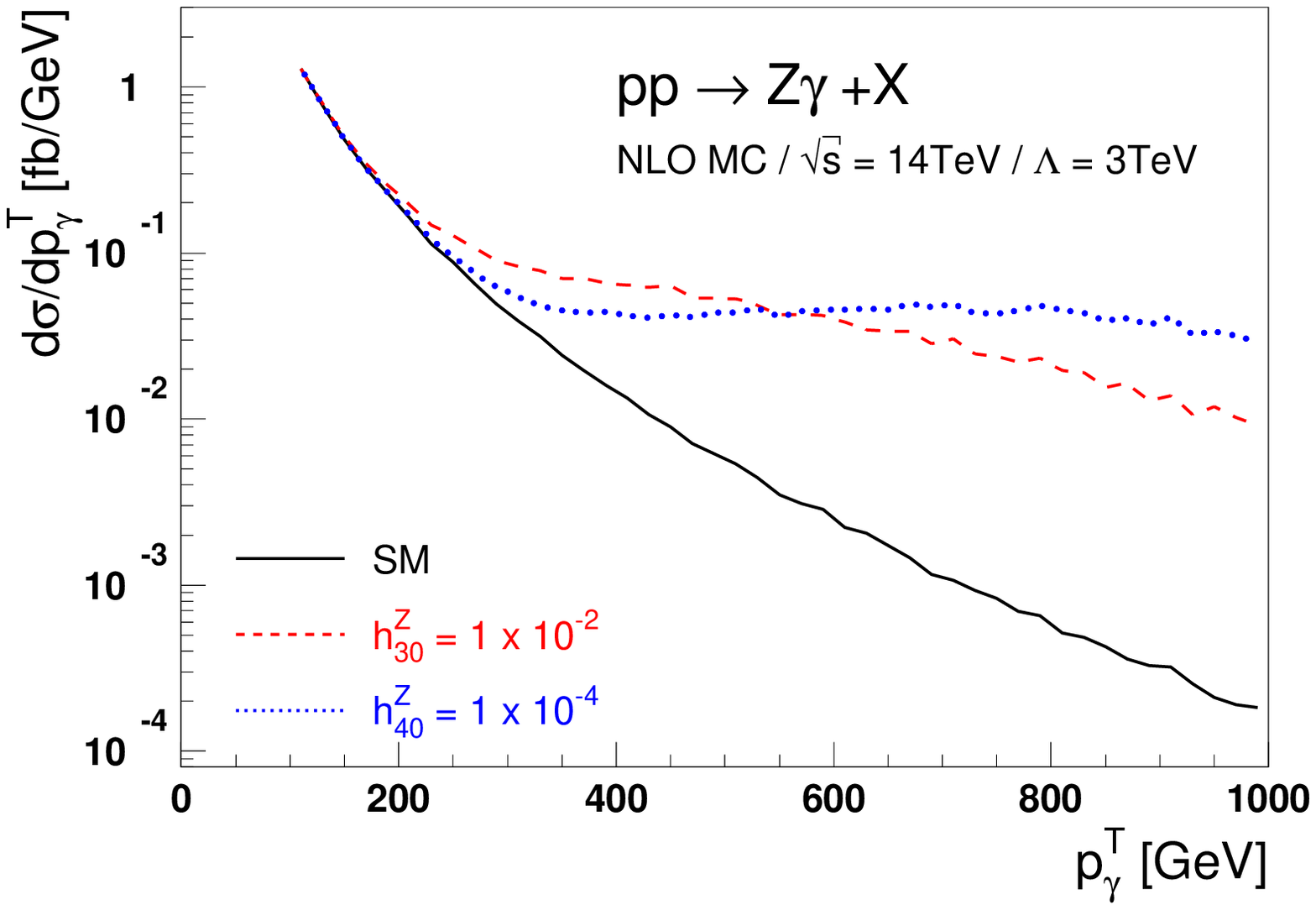}
\end{center}
\caption{
Differential cross section for $Z\gamma$ production versus $p_T^{\gamma}$ 
for Standard Model (solid line) and two different non-standard couplings 
(dashed and dotted lines) at LHC.}
\label{fig:pTg_Zg}
\end{figure}

In the study presented here, several experimentally derived observables
and combinations thereof have been studied to assess the possible
sensitivity to the TGCs. For both ($W\gamma$, $WZ$) and ($Z\gamma$,
$ZZ$) events 
the observables are very similar; for $WZ$, the $Z$ takes the role of
the $\gamma$. The actual behaviour of the observables as function of the
couplings and the energy is different between the processes, due to
the different masses of the involved bosons.

One observable, the transverse momentum, $p_T$, of the $\gamma$ or $Z$
(depending on the di-boson process), which has traditionally been used
at hadron 
colliders, has sensitivity from a combination of high mass event
counting and the $\Theta$ angular distribution. 
Figure~\ref{fig:pTg_Zg} shows the enhancement of di-boson production cross 
section for large values of the photon transverse momentum in presence of 
non-standard couplings.

The distribution of
$p_T^{\gamma,\; Z}$ assuming an integrated luminosity of 30\,fb$^{-1}$ is 
shown in Figure~\ref{fig:pT} for $W\gamma$ and
$WZ$ events, simulated with a parametric detector simulation
program, for the Standard Model and non-standard TGCs.
The enhancement for non-standard TGCs at high $p_T^{\gamma,\;
  Z}$ is clearly visible and, furthermore, the qualitative behaviour
is the same for different TGCs.
\begin{figure}[ht]
\begin{center}
\includegraphics[width=5.5cm]{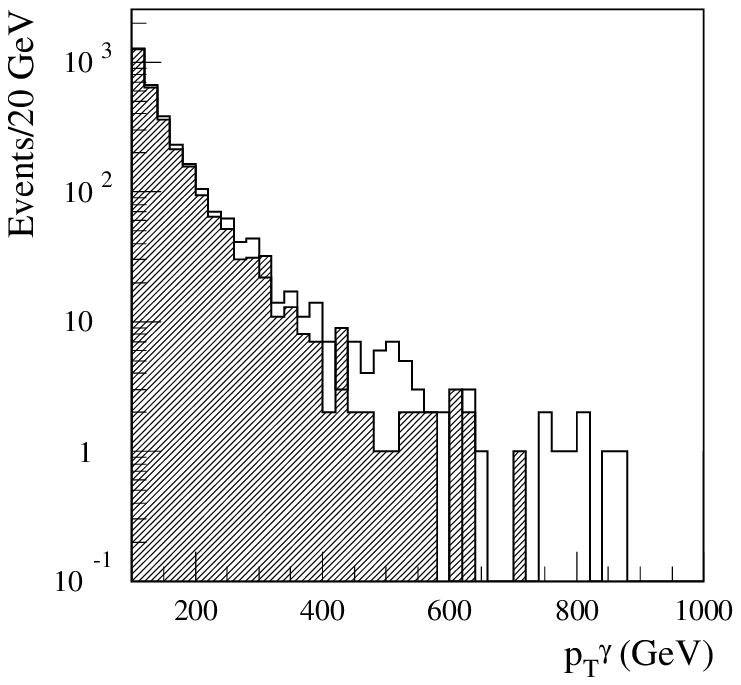}
\includegraphics[width=5.5cm]{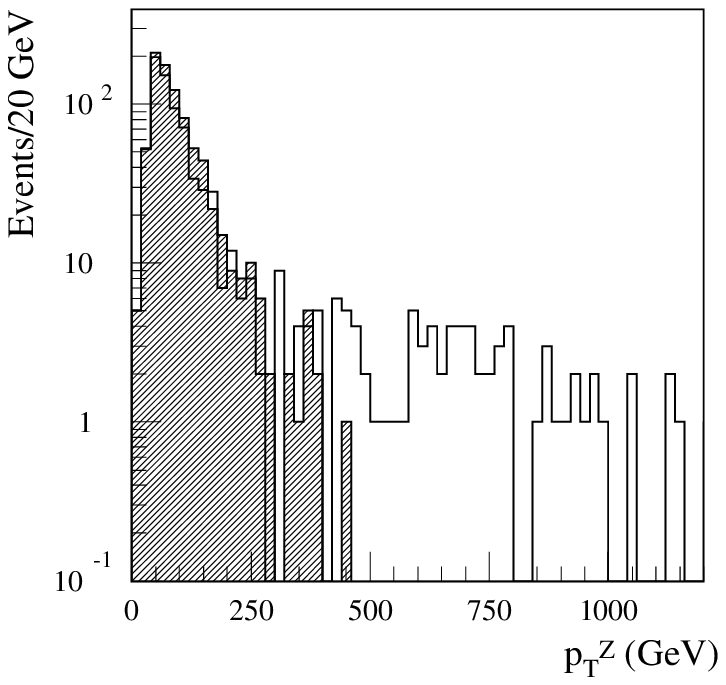}
\end{center}
\caption{
Distribution of $p_T^{\gamma,\; Z}$ for $W\gamma$ (left) and $WZ$ (right)
events for an integrated luminosity of 30~fb$^{-1}$. Distributions are
shown for the Standard Model (shaded histograms) and for non-standard
values (white histograms) $\lambda_{\gamma} = 0.01$ (left) and $\Delta
g_1^Z = 0.05$ (right).}
\label{fig:pT}
\end{figure}

For the statistics expected at the LHC, even after 3 years running at
low luminosity, one may enhance the experimental sensitivity
further by separating the different types of information in
multi-dimensional distributions. For $W\gamma$ and $WZ$ di-boson
production, two sets of variables have been studied (and the
equivalent set for $WZ$): $(m_{W\gamma},\; |\eta^*_{\gamma}|)$, and 
$(p_T^{\gamma},\; \theta^*)$, where $|\eta^*_{\gamma}|$ is the rapidity
of $\gamma$ with respect to the beam direction in the $W\gamma$
system (equivalent to $\Theta$), and $\theta^*$ is the polar decay
angle of the charged lepton in the $W$ rest-frame. Both sets consist of
one variable sensitive to the energy behaviour and one sensitive to
the angular information. For $|\eta^*_{\gamma}|$ and $\theta^*$, a complete
reconstruction of the $W$ is necessary. The momentum of the $W$ can be
reconstructed by using the $W$ mass as a constraint and assuming that
the missing transverse energy is carried away by the neutrino. This
leads to a two-fold ambiguity in the reconstruction. Alternatively,
$|\eta^*_{\gamma}|$, may be approximated by the rapidity difference
between the lepton from the $W$ and the $\gamma$. Distributions of
$|\eta^*_{\gamma}|$ and 
$\theta^*$ are shown in Figure~\ref{fig:eta},
for both the standard model expectation and different non-standard
TGCs. The high sensitivity to the TGCs from $|\eta^*_{\gamma}|$ is due
to the characteristic ``zero radiation'' gap. In contrast, the
sensitivity to the TGCs from the decay polar angle, $\theta^*$, is
weak.
\begin{figure}[ht]
\begin{center}
\includegraphics[width=5.5cm]{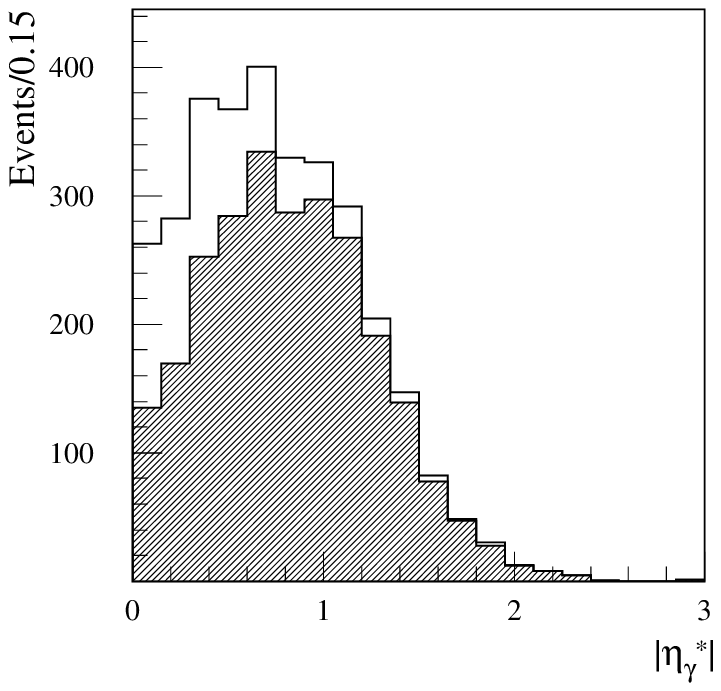}
\includegraphics[width=5.5cm]{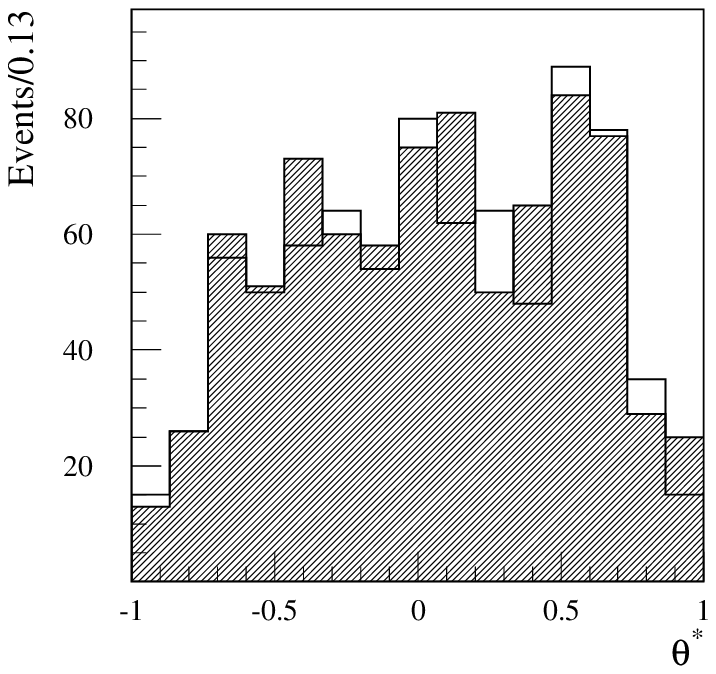}
\end{center}
\caption{
Distribution of $|\eta^*_{\gamma}|$ (left) and $\theta^*$ (right) from
$W\gamma$ and $WZ$ events, respectively, for an
integrated luminosity of 30~fb$^{-1}$. Distributions are shown for the
Standard Model (shaded histograms) and for non-standard values (white
histograms) $\Delta\kappa_{\gamma} = 0.2$ (left) and $\Delta\kappa_Z =
0.2$ (right).}
\label{fig:eta}
\end{figure}

\subsubsection{Analysis techniques for TGC determination}

Depending on the available statistics and the dimensionality of the
experimental distributions, different extraction techniques can be
used in the determination of the TGCs.

One approach employed in this study determines the couplings by a
binned maximum-likelihood fit to distributions of the observables,
combined with the total cross section information. The likelihood
function is constructed by comparing the fitted histogram with a
reference histogram using Poisson probabilities. The reference
distributions can be obtained for different values of the couplings by 
reweighting Monte Carlo events at generator level or equivalently
using several Monte Carlo event samples generated for different values
of the TGCs.

Although the expected number of events at the LHC will allow binning
in two dimensions, a general multidimensional binned fit using all the
TGC sensitive information will not be possible. In the latter case, an
unbinned maximum likelihood fit to the observed information can be
used, where the probability distribution functions can be constructed
by Monte Carlo techniques. In the case of many dimensions, this
approach can be time-consuming, but it may be advantageously combined
with the reweighting technique. The information from the absolute
prediction of the cross section can be included by the so-called
``extended maximum likelihood'' method \cite{ref:eml}.

%
%


\subsection{Sensitivities at LHC}

\newcommand{\intLumi}{\ensuremath{\int\mathcal{L}\,dt}}
\newcommand{\ifb}{\mbox{\,fb$^{-1}$}}
\newcommand{\LambdaFF}{\ensuremath{\Lambda_{FF}}}
\newcommand{\dk}{\ensuremath{\Delta \kappa}}
\newcommand{\lm}{\ensuremath{\lambda}}
\newcommand{\hZthreez}{\ensuremath{h^Z_{30}}}
\newcommand{\hZfourz}{\ensuremath{h^Z_{40}}}

Sensitivity limits have been derived for the triple gauge-couplings 
$WW\gamma$ (ATLAS, CMS), $WWZ$ (ATLAS) and $ZZ\gamma$ (CMS). 
The analysis techniques used by ATLAS and CMS are described in 
Section~\ref{sec:tgc_analysis}. 
The ATLAS studies assume an integrated luminosity of $\intLumi=30\ifb$, 
corresponding to three  years of LHC low luminosity operation. CMS 
assumes $100\ifb$, which is the expectation for one year of LHC high 
luminosity running. 

%
%

\begin{figure}[htbp]
\begin{center}
\includegraphics[width=7cm]{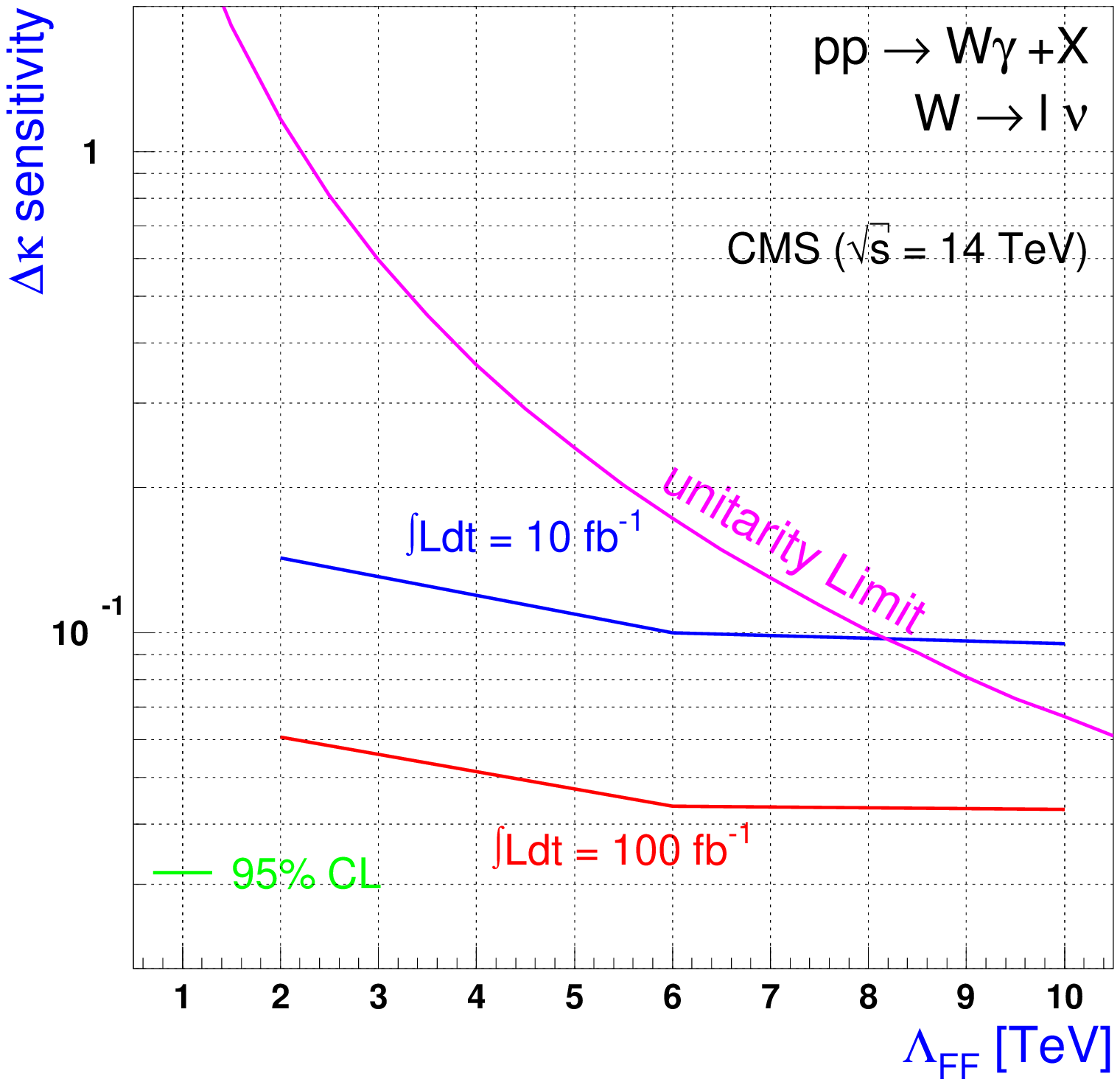}
\includegraphics[width=7cm]{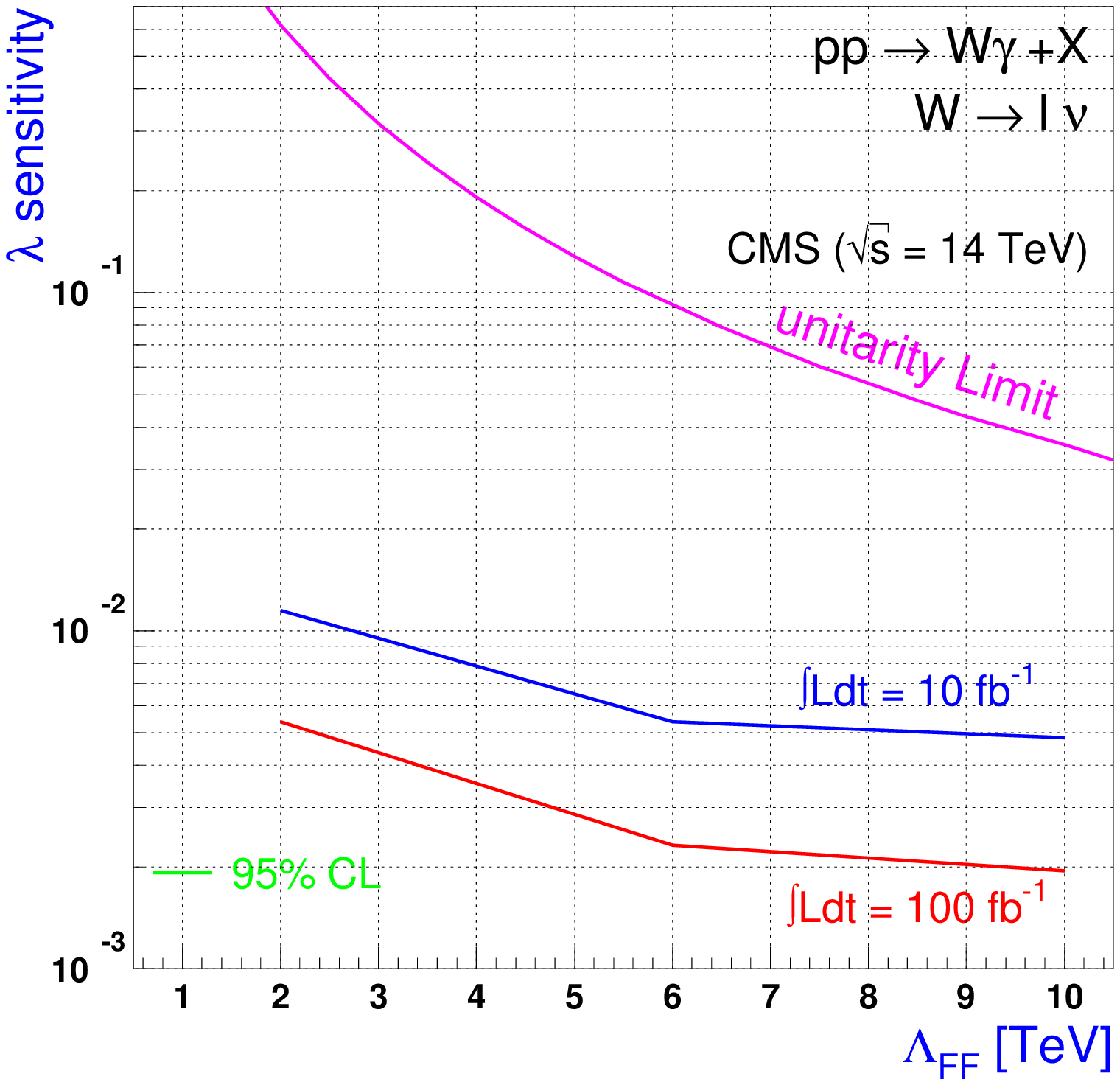} \\
\includegraphics[width=7cm]{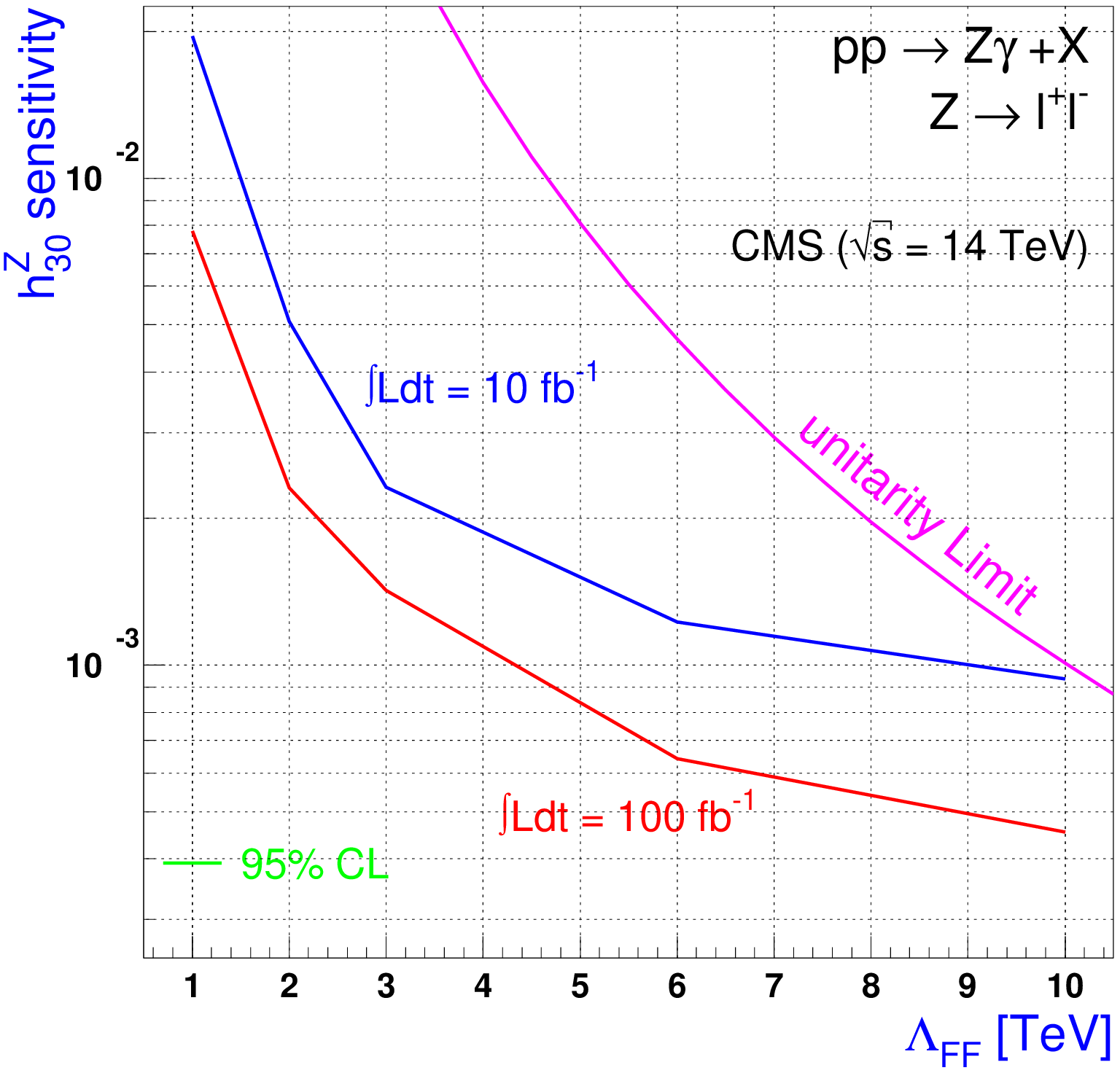}
\includegraphics[width=7cm]{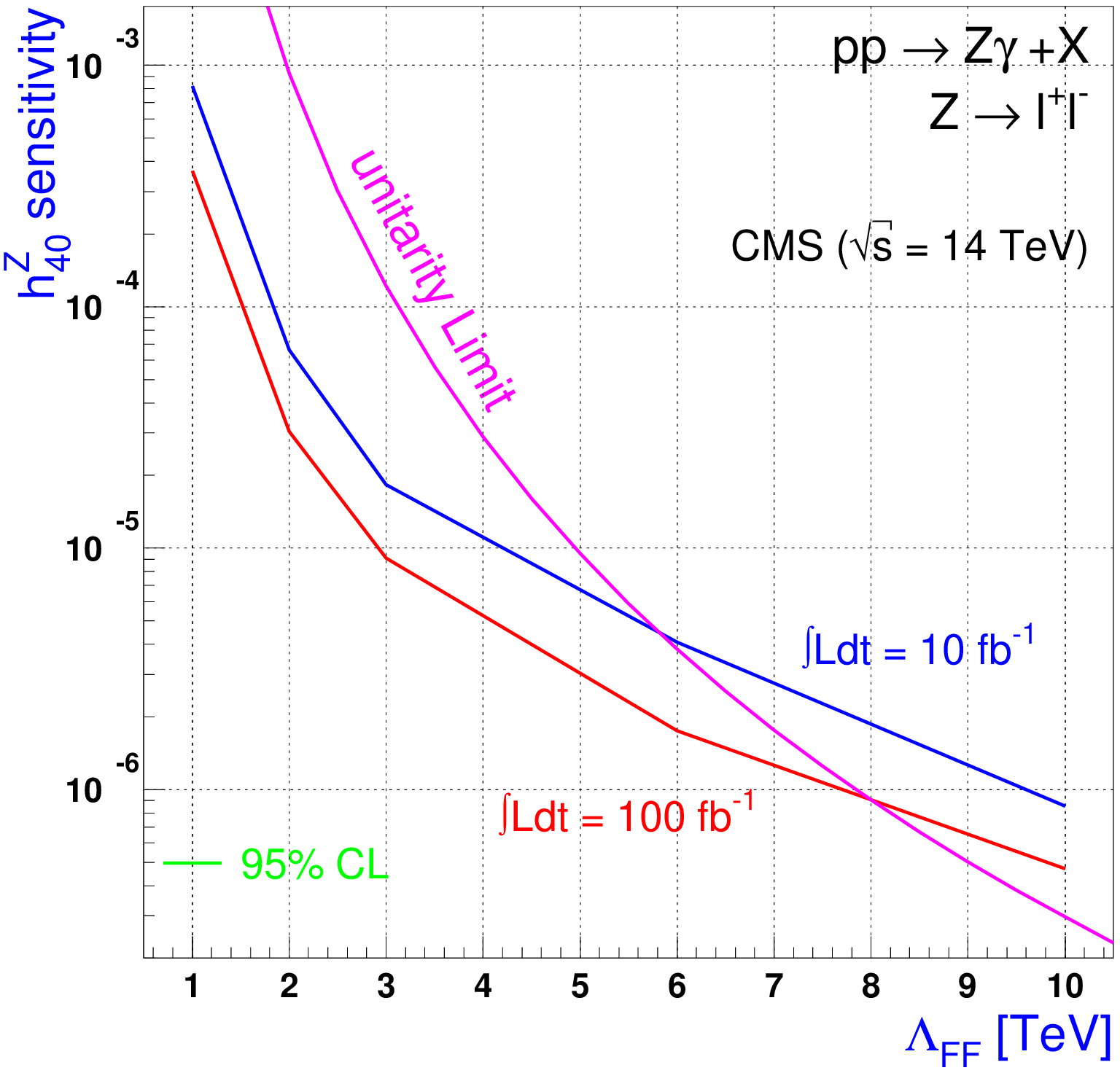}
\end{center}
\caption{
Sensitivity limits on the $WW\gamma$ (top) and $ZZ\gamma$ (bottom) coupling 
parameters from a two-dimensional likelihood fit as a function of the form 
factor scale \LambdaFF .
\label{fig:results_LambdaScanCMS}}
\end{figure}

CMS has performed its studies for a range of different form factor scales 
\LambdaFF , as motivated in Section~\ref{sec:tgc_formfact}.
The plots in Figure~\ref{fig:results_LambdaScanCMS} show the 
expected 95\% CL limits on the anomalous $WW\gamma$ and $ZZ\gamma$ 
coupling parameters together with the corresponding unitarity limits. 
Only the displayed coupling is considered to deviate from the Standard Model. 
The points where the experimental curves turn asymptotic with respect to 
\LambdaFF\ - or are crossed by the unitarity limit - give an 
indication on the range of form factor scales accessible by 
the experiments. While the current Tevatron measurements probe the 
triple gauge-couplings up to form factors of $\LambdaFF = 0.75$\,TeV and 
around 2\,TeV for $ZZ\gamma$ and ($WW\gamma, WWZ$), respectively 
\cite{EllisonWudka}, the LHC experiments will be able to study far smaller 
structures with scales up to 10~TeV, assuming an integrated 
luminosity of $100\ifb$.

\begin{figure}[ht]
\begin{center}
\includegraphics[width=7cm]{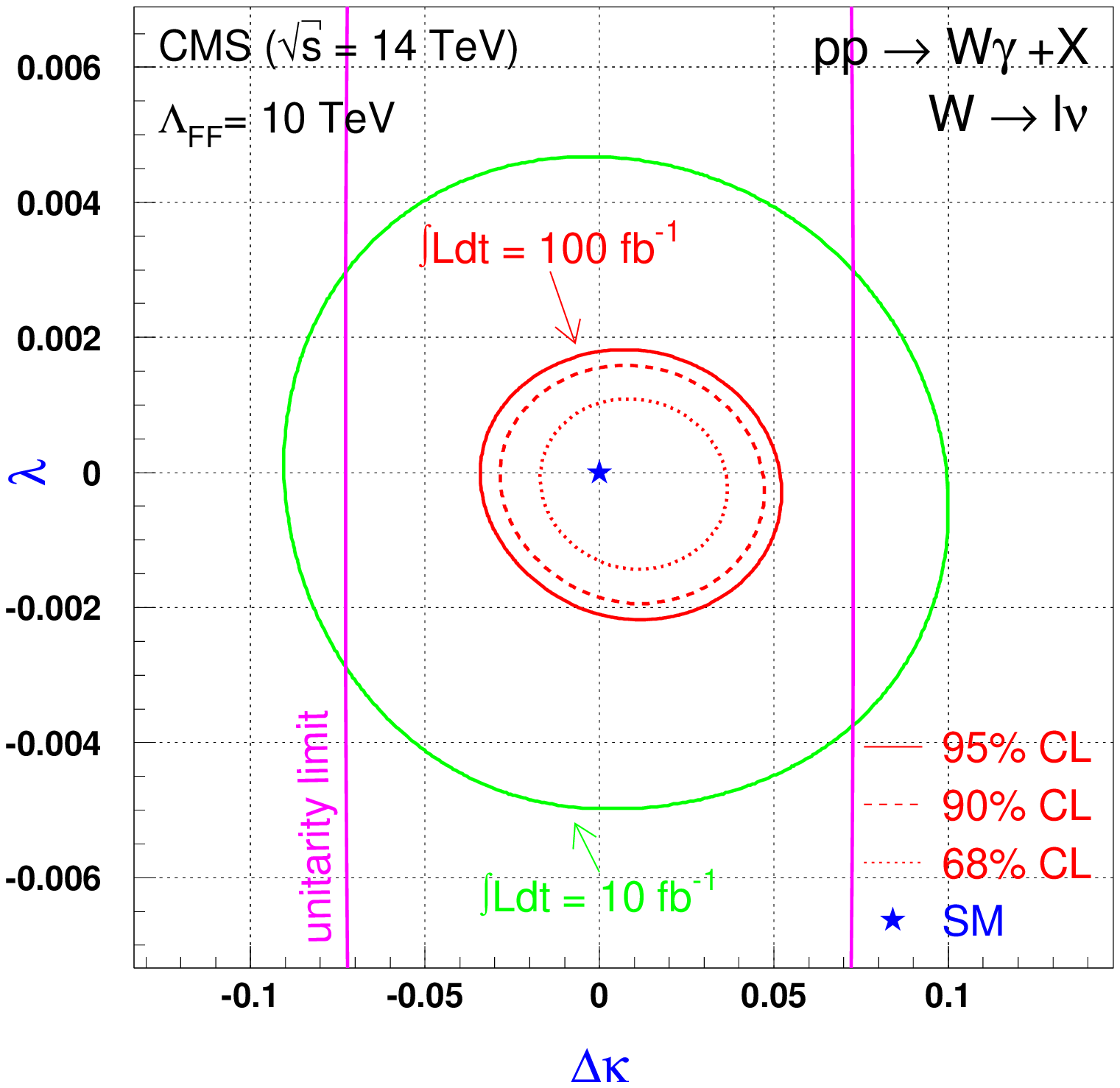}
\end{center}
\caption{
Sensitivity contours in the CP-conserving $W\gamma$ coupling 
space for integrated luminosities of 10\ifb\ and 100\ifb .}
\label{fig:Wg_contour}
\end{figure}

Multi-dimensional fits where several couplings are allowed to vary 
have also been performed \cite{CMS_tgcNote}. Here, the sensitivity limits 
extracted from the log likelihood curves form an ellipse for a particular 
confidence level. 
Figure~\ref{fig:Wg_contour} shows the typical $WW\gamma$ sensitivity 
contours in the two-dimensional CP-conserving 
$(\kappa \times \lambda)$ coupling space for a form factor scale of $10$\,TeV.

\begin{table}[ht]
\begin{center}
%
\caption
{Sensitivity limits (95\% CL), 
assuming integrated luminosities of $30\ifb$ and $100\ifb$, 
respectively. The form factor scale is $\LambdaFF = 10$\,TeV for 
$WW\gamma, WWZ$ and 6\,TeV for $ZZ\gamma$.\label{tab:results_ATLASvsCMS}}
\vskip0.2cm
\begin{tabular}{lcccc}
\hline
Vertex & Coupling & $(m_{W\gamma},|\eta^\ast|)$ & 
                    $(p_T^{\gamma}, \theta^\ast)$ & $p_T^{\gamma}$  \\
\hline
$WW\gamma$ & $\Delta\kappa_\gamma$ 
           & {$0.035$} & {$0.046$} 
           & {$0.043$} \\
{ $\Lambda_{FF}=10$\,TeV} & $\lambda_\gamma$ 
           & {$0.0025$} & {$0.0027$} 
           & {$0.0020$} \\
\hline
$WWZ$      & $\Delta g_1^Z$ 
           & {$0.0078$} & {$0.0089$} & ---  \\
           & $\Delta\kappa_Z$ 
           & {$0.069$} & {$0.100$} & ---  \\
{ $\Lambda_{FF}=10$\,TeV} & $\lambda_Z$ 
           & {$0.0058$} & {$0.0071$} & --- \\
\hline
$ZZ\gamma$ & $h^Z_{30}$ 
           & --- & --- & {$6.4\times10^{-4}$} \\
{ $\Lambda_{FF}=6$\,TeV} & $h^Z_{40}$ 
           & --- & --- & {$1.8\times10^{-6}$} \\
\hline
\end{tabular}
\end{center}
\end{table}

Table~\ref{tab:results_ATLASvsCMS} summarises the sensitivity limits obtained 
by ATLAS and CMS as reported in \cite{atlas-phystdr2,CMS_tgcNote}. In 
addition, ATLAS has performed a fit using the complete generator level 
phase space information \cite{atlas-phystdr2}. The results for this 
{\em ideal case} show that, as the high energy tails of the $p^T_\gamma$ 
distributions exhibit a very strong sensitivity to the $\lambda$-like 
anomalous couplings, the additional information does not improve the limits 
on this type of couplings considerably. However, the $\kappa$-type couplings 
may profit from a more sophisticated data analysis.

From the numbers in Table~\ref{tab:results_ATLASvsCMS}, we expect an 
improvement in sensitivity by up to two (four) orders of magnitude for 
anomalous $WW\gamma/WWZ$ ($ZZ\gamma$) couplings, with respect to the current 
Tevatron limits. The strong increase in sensitivity is 
due to the pronounced high $\hat{s}$ enhancement at the LHC, most prominently 
for $ZZ\gamma$ (see Section~\ref{sec:tgc_ffonshat}). 
A smaller choice of the form factor scale would cut 
off this enhancement and diminish the sensitivity considerably, as shown in 
the lower plots in Figure~\ref{fig:results_LambdaScanCMS}.

\subsection{Backgrounds to \boldmath $W \gamma$}

The $W \gamma$ signal has a very small cross section, compared to $W+$jet production for example, and can contain a significant amount of background. The dominant background to the $W\gamma$ signal is from $W$+jet production where the jet is misidentified as a photon, resulting in a fake signal. Radiative $W$ decay also contributes when the electron from the $W$ decay radiates a photon, and both $t\bar{t}\gamma$ and $b\bar{b}\gamma$ quark-gluon fusion processes can also produce a fake signal contributing to the background. $Z\gamma$ production and $W$($\tau \nu$)$\gamma$ also make a small contribution to the backgrounds.

Previous studies \cite{ATLASTP, Fouchez, Mackay, Reichel} have shown that the 
$W\gamma$ signal will be observable at the LHC provided that the backgrounds can be suppressed. 
All the backgrounds were generated with {\tt PYTHIA 5.7} \cite{pythia} in conjunction with the {\tt CMSJET} 
\cite{Cmsjet} fast detector simulation for the CMS experiment.

\subsubsection{$W+$ jet and $W \rightarrow l \nu \gamma$ backgrounds}

The dominant background to the process $pp \rightarrow W(e\nu)\gamma$ arises from $W+$jet events where the jet decays electromagnetically and is reconstructed in the calorimeter as a photon. The probability for the jet to fluctuate into an isolated electromagnetic shower is small, but the large number of jets above 10 GeV in the $W$ sample guarantees that some jets will look identical to photons. Even if the jet is not misidentified as a photon, it is possible for a radiative decay of the $W$ to produce the same signature as the signal. If the lepton from the $W$ decay radiates a photon, an event signature of $\gamma,l,\nu$ may be observed. Cuts must therefore be applied to reduce this background.

\paragraph{\boldmath $W+$\unboldmath jet}

Figure~\ref{ptgam} shows the $p_T$($\gamma$) spectrum for misidentified photon from the $W+$jet background and the real photon from the $W\gamma$ signal. 
A photon isolation cut has been applied to both data sets. A rejection power of nearly 7 can be obtained with an efficiency loss of less than 5$\%$, by using an isolation area of $\Delta R$ = 0.25 and a $p_T$ threshold of 2 GeV \cite{CMS-ECAL-TDR}. A greater rejection power with a  much smaller efficiency loss is available at low luminosity. Therefore an event is selected if the photon meets the isolation criteria and if it is within  $\eta$ = $\pm$2.5.
The isolation cut clearly makes it possible to observe the signal, especially at high $p_T$, however a cut at $p_T$($\gamma$) = 100 GeV further reduces the background. This would not harm the sensitivity to anomalous couplings greatly as the anomalies only manifest themselves at high $p_T$.

\begin{figure}[hbtp]
  \begin{center}
\includegraphics[width=9cm]{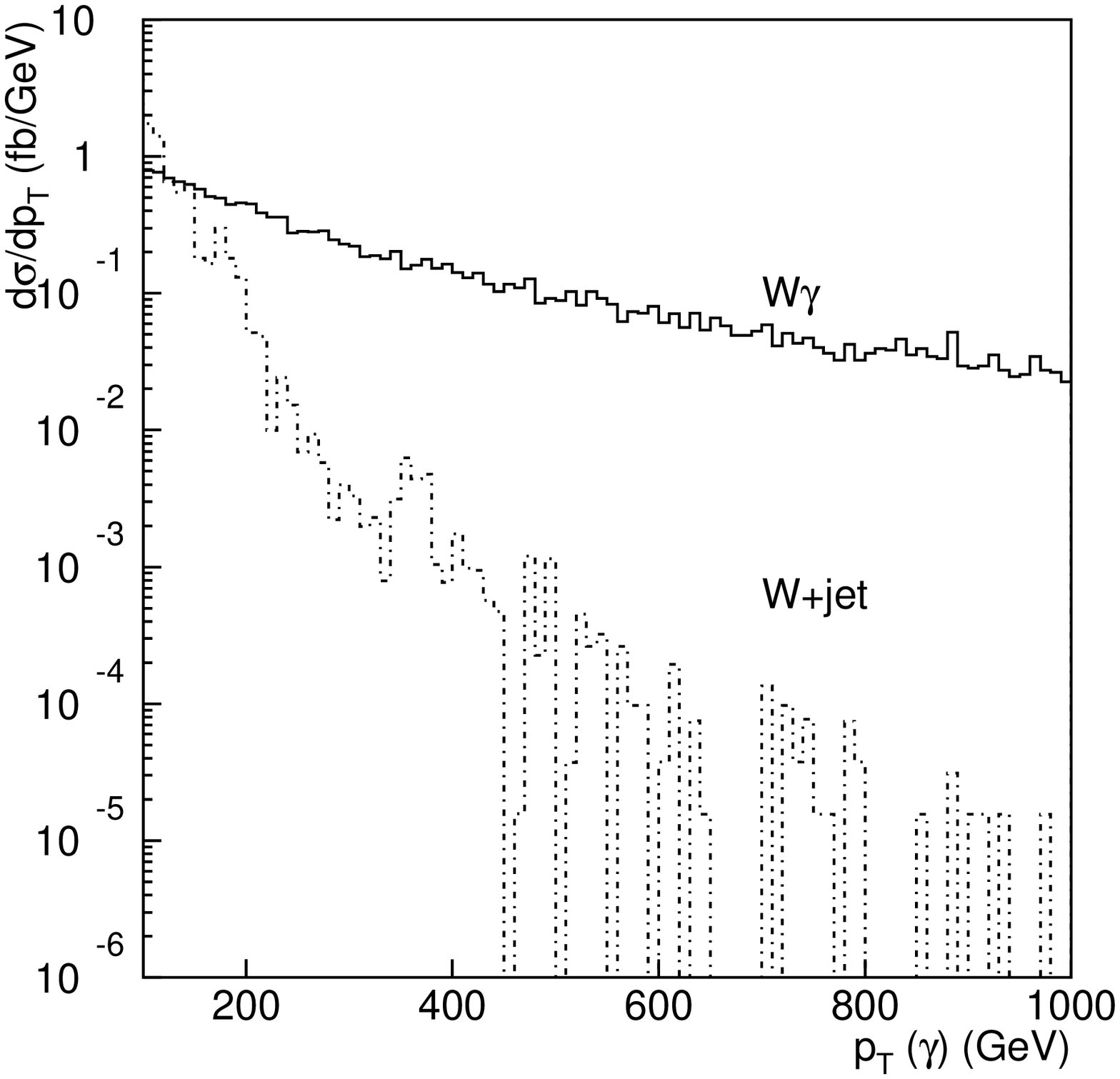}
    \caption{ p$_T$($\gamma$) distribution for the $W\gamma$ signal and the 
  $W+$jet background where the jet is misidentified as a photon.}
    \label{ptgam}
  \end{center}
\end{figure}

\paragraph{Radiative \boldmath $W$}

One method of reducing the background of radiative $W$ decays is to make a cut on the invariant mass of the $\gamma l \nu$ system.  For the $W\gamma$ signal, $M(\gamma l \nu)$ is always larger than $M_W$ if finite $W$ width effects are ignored.

However, the $M(\gamma l \nu)$ cannot be determined unambiguously as the four-momentum of the neutrino is unknown: even if the transverse momentum is correctly determined from the missing momentum in the event, there is no measurement of the missing longitudinal momentum. Therefore the cluster transverse mass, or minimum invariant mass, may be used instead \cite{clust}. 
The transverse mass is independent of the longitudinal momenta of the parent particle and its decay products.

For $W \rightarrow \gamma l \nu$ the cluster transverse mass sharply peaks at $M_W$ \cite{miminm} and drops rapidly above the $W$ mass. Thus $\gamma l \nu$ events originating from $W\gamma$ production and radiative $W$ decays can be distinguished if $M_T(\gamma l \nu)$ is cut slightly above $M_W$ \cite{radcut}. Hence a cut at  $M_T(\gamma l \nu) >$ 90~GeV should take into account the finite width of the $W$ whilst not significantly affecting the signal.

The $W\gamma$ signal produces the lepton and photon almost back-to-back. Ensuring that they are well separated will further reduce the radiative $W$ background. This can be done using the quantity $\Delta R = \sqrt{(\Delta \phi ^2 + \Delta \eta ^2)}$.
Leading order analysis of the signal and radiative background enabled a study of the optimum value of $\Delta R$ to use for separation. Typically a cut at $\Delta R > 0.5$ is used to ensure separation, but increasing the separation to $\Delta R > 0.7$ makes little difference to the signal whilst greatly reducing the background.

In order to suppress the radiative $W$ background events,
cuts of $\Delta R(\gamma,l)> 0.7$ and $M_T(\gamma l \nu) > 90$~GeV are used.

\subsubsection{Quark-Gluon fusion background}

Quark-gluon fusion is important at the LHC because the rate is extremely high. There are lots of available gluons in the proton at relatively high $x$, and because the $WW\gamma$ reaction is suppressed in some regions of phase space.

\paragraph{\boldmath $b\bar{b}\gamma$}

At the LHC 10$^{12}$ $b\bar{b}$ events \cite{bphys} are expected for a years running at high luminosity. Although the $b\bar{b}\gamma$ events are not kinematically similar to the signal, the expected number of events is so large that the background will be a problem unless it is reduced by cuts.

The $b\bar{b}\gamma$ background was generated using the processes: $q\bar{q} \rightarrow g\gamma$, and $q\bar{q} \rightarrow Z \gamma$. Events were generated from $\hat{p}_T$ = 500 GeV with a cross section of 1.055 pb. This parton-level requirement was for computational efficiency as only the very highest $p_T$ events
contribute to the background. A cut on missing $p_T$ can be made at 50 GeV in order to reduce the $b\bar{b} \gamma$ background. 

\paragraph{\boldmath $t\bar{t}\gamma$}

Since the $M_t > M_W+M_b$, $t\bar{t}$ events represent an irreducible background to $W\gamma$ pair production. $t\bar{t}\gamma$ production is a copious source of high $p_T$ photons in association with hard leptons and without cuts has a cross section, $\sigma \sim$ 300~pb, of at least 3 orders of magnitude more than the $W\gamma$ signal \cite{moretti}. The subsequent decay of top quarks into a $W$ boson and a $b$ quark and also the $W$ decay into a $f\bar{f}$ pair provide the same event signature as the $W\gamma$ signal. Therefore, due to the very large top quark production cross section at LHC energies, the process $pp \rightarrow t\bar{t}\gamma \rightarrow W\gamma+X$ represents a potentially significant background.

Events were generated by the process $q\bar{q} \rightarrow g\gamma$  and looking for $t\bar{t}$ production. This method is very inefficient, 4 million events were generated and 489~$t\bar{t}\gamma$ events were produced, with 10 events passing all of the cuts. 
The $t\bar{t}\gamma$ events were generated from $\hat{p}_T$~=~500 GeV (for the same reasons as $b\bar{b}$), with a cross section of 1.049~pb. The large cross section means that although only a few events pass the cuts, this background is a potential problem.

Studies for the SSC \cite{ssc} showed that the background can be reduced to a manageable level by requiring the photon to be isolated from the hadrons in the event, and by imposing a jet veto ({\it i.e.} by considering the exclusive reaction $pp \rightarrow W\gamma + 0$~jets).

Since the top quark decays predominantly into a $Wb$ final state, $t\bar{t}\gamma$ events are characterised by a large hadronic activity which frequently results in one or several high-$p_T$ jets. If the second $W$ boson decays hadronically, up to four jets are possible. This observation suggests that the $t\bar{t}\gamma$ background may be suppressed by vetoing high-$p_T$ jets. Such a ``zero jets'' requirement has been demonstrated to be very useful in reducing the size of the NLO QCD corrections in $pp \rightarrow W\gamma + X$ at SSC energies \cite{BHOWg}. If the second $W$ in the $t\bar{t}\gamma$ events decays hadronically, the number of jets in $pp \rightarrow t\bar{t}\gamma \rightarrow W\gamma + X$ is generally larger than for leptonic $W$ decays, and the jet veto is more efficient.

Unfortunately the jet veto also drastically reduces the number of signal events. Only 10\% of the signal survives the jet veto cut alone and only 4\% survive all the cuts and the jet veto. This suggests that an alternative method for reducing this background needs to be found for the LHC.

ATLAS \cite{Fouchez} studied the possibility of exploiting the number of jets in the $t\bar{t}\gamma$ events by imposing a cut on the second jet in the event. The $W\gamma$ signal will not have a 2nd jet, or if it does, it is a misidentified jet and will be of very low $p_T$. The $t\bar{t}\gamma$ events will have up to four high $p_T$ jets in each event. By cutting all events where the $p_T$ of the second jet is greater that 25~GeV, the majority of the $t\bar{t}\gamma$ events will be eliminated without greatly affecting the signal.
 
\subsubsection{$Z\gamma$ background}

There is a small background to $e\nu \gamma$ that comes from $Z(ee)\gamma$ events in which one of the electrons gives rise to significant missing energy (generally by entering a gap in the detector). As CMS is hermetic and the crystals of the ECAL are off-pointing with respect to the interaction point, this background is very small. ATLAS \cite{Fouchez} calculate this background to be $\sim$ 25 times smaller than the signal before any cuts are imposed. Thus the $Z\gamma$ background is assumed to be negligible.

\subsubsection{$W(\tau \nu)\gamma$ background}

The final background to $pp\rightarrow W(e \nu,\mu \nu)\gamma$ is $pp\rightarrow W(\tau \nu)\gamma$ where the $\tau$ lepton decays into an electron or muon. The background is very small because the decay of the tau lepton results in electrons or muons with significantly reduced $p_T$ and the kinematical threshold for an electron is 25~GeV. Previous studies at Fermilab have shown this background to be negligible \cite{Kelly}.

\subsubsection{Summary of backgrounds}

Table~\ref{'allcuts'} shows a list of all the cuts proposed to reduce the backgrounds to the $W\gamma$ signal. Having chosen each cut to reduce an individual background, it is important to understand how each cut effects both the signal and the other backgrounds.  \\  


\begin{table}[htb]
\caption{Proposed cuts to reduce the backgrounds to the $W\gamma$ signal.}
    \label{'allcuts'}
\vskip0.2cm
\begin{center}
\begin{tabular}{lccccccc}
\hline 
 Quantity & $|\eta(\gamma, l, jet)|$ & $p_T(\gamma)$ & $p_T(l)$ & 
$M_T(\gamma,l,\nu)$ & $\Delta R(\gamma,l)$ & $p_T(\nu)$ & 2nd jet\\ 
 & & (GeV)  & (GeV) & (GeV) &  & (GeV) & (GeV) \\
\hline 
Cut value & $<$ 2.5  & $>$ 100  & $>$ 25  & $>$ 90  &
 $>$ 0.7 & $>$ 50  & $<$ 25 \\ \hline
\end{tabular}
\end{center}
\end{table}


\nopagebreak
Table~\ref{'effic'} shows the efficiency of the individual cuts on the signal and the backgrounds. The $W+$jet and radiative $W$ backgrounds are treated together. \\
\nopagebreak
\begin{table}[htb]
\caption{Efficiency of individual cuts on the signal and backgrounds, errors are statistical.}
    \label{'effic'}
\begin{center}
\begin{tabular}{lcccc}
\hline 
{Cut}                   & { Signal (\%)} & \multicolumn{3}{c}{Background (\%)} \\ 
                            &      & { $W+$jet/Rad.$W$} & { $t\bar{t}\gamma$} & { $b\bar{b}\gamma$}  \\ \hline
    $p_T(\gamma)$         &67$\pm$0.49&0.06$\pm$0.008&72$\pm$5.33&84$\pm$0.22\\ 
    $p_T(l)$           &84$\pm$0.52&62$\pm$0.25&5$\pm$1.02& 0.2$\pm$0.001  \\ 
    $M_T(\gamma,l,\nu)$&85$\pm$0.52&19$\pm$0.14&87$\pm$4.2& 0.3$\pm$0.0115  \\ 
    $\Delta R(\gamma,l)$&95$\pm$0.55&94$\pm$0.3&95$\pm$4.4&94$\pm$0.23  \\ 
    $p_T(\nu)$            &86$\pm$0.53&60$\pm$0.25&43$\pm$2.9 &28$\pm$0.124 \\ 
    2nd jet         &89$\pm$0.54&42$\pm$0.2&0+0.2&34$\pm$0.14  \\ \hline
    All Cuts                &55$\pm$0.42&0.033$\pm$0.018&0+0.2 & 0.006$\pm$0.0019  \\ \hline

\end{tabular}
\end{center}
\end{table}

\begin{figure}[hbtp]
  \begin{center}
\includegraphics[width=9cm]{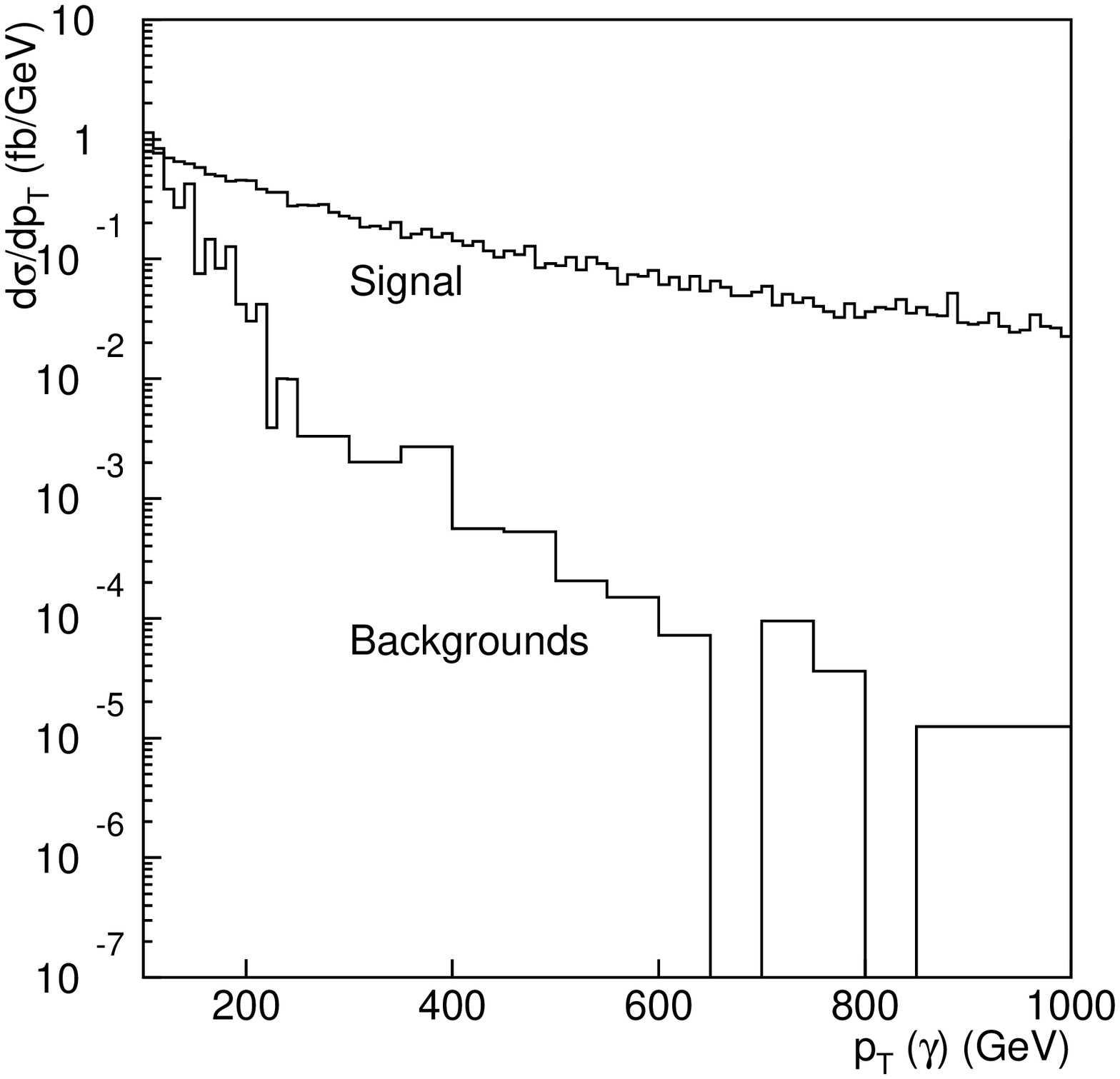}
    \caption{ $p_T(\gamma)$ distribution for the $W\gamma$ signal and the backgrounds.}
    \label{fig:cuts}
  \end{center}
\end{figure}

%

\subsubsection{Conclusion}

The backgrounds to the $W\gamma$ signal have been studied and cuts have been made in order to reduce the backgrounds to at least an order of magnitude less than the signal for $p_T(\gamma) > 200$~GeV. The $W+$jet and radiative $W$ backgrounds have been well studied and understood and the cuts made reduce these significantly. The quark-gluon fusion backgrounds are not so well understood in this work since a less than optimal generator for $t\bar{t}\gamma$ was used. However, the cuts studied for
this channel work well for the 
low statistic samples presented here. Further study of this background would be interesting.

Backgrounds to $WZ$ production have been  studied briefly and are similar, within statistical errors, to those in the $W\gamma$ channel presented here.

\section{VECTOR-BOSON FUSION AND SCATTERING
         \protect\footnote{Section coordinators: 
         Z.~Kunszt, R.~Mazini, D.~Rainwater}
         \label{sec:vbfs}}

\subsection{Searching for \boldmath $VV \to H \to \tau\tau$ \unboldmath 
\label{sec:vvhtt}}

\newcommand{\sla}[1]{/\!\!\!#1}

\subsubsection{Introduction}

The search for the Higgs boson and, hence, for the origin of electroweak 
symmetry breaking and fermion mass generation, remains one of the premier tasks 
of present and future high energy physics experiments. Fits to precision 
electroweak (EW) data have for some time suggested a relatively small Higgs 
boson mass, of order 100~GeV~\cite{EWfits,EWfits2}, hence we have studied an 
intermediate-mass Higgs, with mass in the $110-150$~GeV range, beyond the reach 
of LEP at CERN and perhaps of the Fermilab Tevatron. Observation of the 
$H\to\tau\tau$ decay channel in weak boson fusion events at the  Large 
Hadron Collider (LHC) is quite promising, both in the Standard Model (SM) and 
Minimal Supersymmetric Standard Model (MSSM). This channel has lower QCD 
backgrounds compared to the dominant $H\to b\bar{b}$ mode, thus offering the 
best prospects for a direct measurement of a $Hf\bar{f}$ coupling.

At the LHC, despite the fact that the cross section for Higgs production
by weak-boson fusion is significantly lower 
than that from gluon fusion (by almost one order of magnitude), 
it has the advantage of additional 
information in the event other than the decay products' transverse momentum and 
their invariant mass resonance: namely, 
the observable quark jets. Thus one can exploit 
techniques like forward jet 
tagging~\cite{Cahn,Cahn2,Cahn3,BCHP,BCHP2,BCHP3,BCHP4,DGOV,DGOV2} to reduce the 
backgrounds. Another advantage is the different colour structure of the signal vs 
the background. Additional soft jet activity (minijets) in the central region, 
which occurs much more frequently for the colour-exchange processes of the QCD 
backgrounds~\cite{bpz_minijet,bpz_minijet2}, are suppressed via a central jet veto.

We have performed first analyses of intermediate-mass SM $H\to\tau\tau$ and of 
the main physics and reducible backgrounds at the LHC,  considering separately
the decay modes $\tau\tau\to h^\pm l^\mp \sla{p}_T, e^\pm\mu^\mp \sla{p}_T$.
These modes 
demonstrate the feasibility of Higgs boson detection in this channel with 
modest luminosity~\cite{HRZ_tau,PRZ_tau}. We demonstrated that forward jet 
tagging, $\tau$ identification and reconstruction criteria alone yield a 
signal-to-background ($S/B$) ratio of approximately 1/1 or better. Additional large 
background suppression factors can be obtained with the minijet veto, achieving 
final $S/B$ ratios as good as 6/1, depending on the Higgs mass.

In the MSSM, strategies to identify the structure of the Higgs sector are much 
less clear. For large $\tan\beta$, the light neutral Higgs bosons may couple much 
more strongly to the $T_3 = -1/2$ members of the weak isospin doublets than its 
SM analogue. As a result, the total width can increase significantly compared to 
a SM Higgs of the same mass. This comes at the expense of the branching ratio 
$BR(h\to\gamma\gamma)$, the cleanest Higgs discovery mode, possibly rendering it 
unobservable over much of MSSM parameter space and forcing consideration of 
other observational channels. Instead, since $BR(h\to\tau\tau)$ is 
enhanced slightly, 
we have examined the $\tau$ mode as an alternative~\cite{PRZ_tau,PRZ}.


\subsubsection{Simulations of signal and backgrounds}

The analyses used full tree-level matrix elements for the weak boson fusion 
Higgs signal and the various backgrounds. Extra minijet activity was simulated 
by adding the emission of one extra parton to the basic signal and background 
processes, with the soft singularities regulated via a truncated shower 
approximation (TSA)~\cite{TSA,TSA2}. 

We simulated $pp$ collisions at the  LHC, $\protect\sqrt{s} = 14$~TeV. For 
all QCD effects, the running of the strong-coupling constant was evaluated at 
one-loop order, with $\alpha_s(M_Z) = 0.118$. We employed CTEQ4L parton 
distribution functions~\cite{cteq4} throughout. The factorisation scale was 
chosen as $\mu_f =$ min($p_T$) of the defined jets, and the renormalisation 
scale $\mu_r$ was fixed by $(\alpha_s)^n = \prod_{i=1}^n \alpha_s(p_{T_i})$. 
Detector effects were considered by including Gaussian smearing for partons and 
leptons according to ATLAS expectations~\cite{ATLASTP,CMSTP}.

At lowest order, the signal is described by two single-Feynman-diagram 
processes, $qq \to$ $qq(WW,$ $ZZ)$ $\to$ $qqH$, {\it i.e.} 
$WW$ and $ZZ$ fusion where the weak 
bosons are emitted from the incoming quarks \cite{qqHorig}. From a previous 
study of $H\to\gamma\gamma$ decays in weak boson fusion~\cite{RZ_gamma}, 
we know 
several features of the signal which we could  exploit directly here: the 
centrally produced Higgs boson tends to yield central decay products (in this 
case $\tau^+\tau^-$), and the two quarks enter the detector at large rapidity 
compared to the $\tau$'s and with transverse momenta in the 20-80 GeV range, 
thus leading to two observable forward tagging jets.

We considered separately the cases of one $\tau$ decaying leptonically 
($e$,$\mu$) and the other decaying hadronically (with a combined branching 
fraction of $45\%$), and both decaying leptonically but with different flavour 
($e\mu$ or $\mu e$, with a combined branching fraction of $6.3\%$). Our analyses 
critically employed transverse momentum cuts on the charged $\tau$-decay 
products and, hence, some care was taken to ensure realistic momentum 
distributions. Because of its small mass, we simulated $\tau$ decays in the 
collinear and narrow-width approximations and with decay distributions to 
$\pi$,$\rho$,$a_1$~\cite{HMZ}, adding the various hadronic decay modes according 
to their branching ratios. We took into account the anti-correlation of the 
$\tau^\pm$ polarisations in the decay of the Higgs.

\paragraph{Lepton-hadron mode}

Positive identification of the hadronic $\tau^\pm\to h^\pm X$ decay requires
severe cuts on the charged hadron isolation. We based our simulations on the 
possible strategies analysed by Cavalli {\it et al.}~\cite{Cavalli}. Considering 
hadronic jets of $E_T>40$~GeV in the ATLAS detector, they found non-tau 
rejection factors of 400 or more while true hadronic $\tau$ decays are retained 
with an identification efficiency of $26\%$.

Given the $H$ decay signature, the main physics background to the 
$\tau^+ \tau^- jj$ events of the signal arises from real emission QCD 
corrections to the Drell-Yan process $q\bar{q} \to (Z,\gamma) \to \tau^+\tau^-$, 
dominated by $t$-channel gluon exchange. All interference effects between 
virtual photon and $Z$-exchange were included, as was the correlation of 
$\tau^\pm$ polarisations. The $Z$ component dominates, so we call these 
processes collectively the ``QCD $Zjj$'' background.

An additional physics ``EW $Zjj$'' background arises from $Z$ and $\gamma$ 
bremsstrahlung in (anti)quark scattering via $t$-channel electroweak boson 
exchange, with subsequent decay $Z,\gamma\to \tau^+\tau^-$. Naively, this EW 
background may be thought of as suppressed compared to the analogous QCD 
process. However, the EW background includes electroweak boson fusion, 
$VV \to \tau^+\tau^-$, which has a momentum and colour structure identical to 
the signal and thus cannot easily be suppressed via cuts.

Finally, we considered reducible backgrounds, {\it i.e.} 
any event that can mimic the 
$Hjj$ signature of a hard, isolated lepton and missing $p_T$, a hard, narrow 
$\tau$-like jet, and two forward tagging jets. Thus we examined $W+jets$, where 
the $W$ decays leptonically ($e$,$\mu$) and one jet fakes a hadronic $\tau$, and 
$b\bar b+jets$, where one $b$ decays leptonically and either a light quark or 
$b$ jet fakes a hadronic $\tau$. We neglected other sources like $t\bar{t}$ 
events which had previously been shown to give substantially smaller 
backgrounds~\cite{Cavalli}. 

Fluctuations of a parton into a narrow $\tau$-like jet are considered with 
probability $0.25\%$ for gluons and light-quark jets and $0.15\%$ for $b$ jets 
(which may be considered an upper bound)~\cite{Cavalli}.

In the case of $b\bar b+jj$, we simulated the semileptonic decay $b\to l\nu c$ 
by multiplying the $b\bar{b}jj$ cross section by a branching factor of 0.395 and 
implementing a three-body phase space distribution for the decay momenta to 
estimate the effects of lepton isolation cuts. We normalised our resulting cross 
section to reproduce the same factor 100 reduction found in~\cite{Cavalli}.

\paragraph{Dual lepton mode}

For the dilepton mode, we consider decay only to $e,\mu$ pairs to completely 
eliminate the backgrounds from real $Z$ production decaying directly to $ee$ 
or $\mu\mu$. Tau decays were performed in the same manner as in the 
lepton-hadron channel. We again considered QCD and EW $Zjj;Z\to\tau\tau$ 
production as the physics backgrounds.

We calculated the primary contributions from reducible backgrounds by 
considering all significant sources of two $W$'s, which decay leptonically to 
form the signature $e,\mu$, and two forward jets. This consists of 
$t\bar{t} + jets$, as well as both QCD and EW $WWjj$ production. As with the 
EW $Zjj$ case, EW $WWjj$ processes contain an electroweak boson fusion component 
kinematically similar to the signal, and so cannot be ignored.

We also considered $b\bar{b}jj$ production, with each $b$ decaying 
semileptonically simulated by implementing the $V - A$ decay distributions of 
the $b$-quarks in the collinear limit, and multiplying the resultant cross 
section by a branching fraction 0.0218 (for the $e,\mu$ or $\mu,e$ final 
states).

Finally, we considered the overlapping contribution from the signal itself in 
the decay mode $H\to WW \to e\mu\sla{p}_T$, which can be significant above 
$M_H \geq \sim130$~GeV.


\subsubsection{Standard Model analysis}

The basic acceptance requirements must ensure that the two jets and two $\tau$'s 
are observed inside the detector (within the hadronic and electromagnetic 
calorimeters, respectively), and are well-separated from each other:
\begin{eqnarray}
\label{eq:tag1}
& p_{T_{j}} \geq 20~{\rm GeV} \, , \qquad |\eta_j| \leq 5.0 \, ,\qquad 
\Delta R_{jj} \geq 0.7 \, , & \nonumber\\
& |\eta_{\tau}| \leq 2.5 \, , \qquad \Delta R_{j\tau} \geq 0.7 \, .
\end{eqnarray}
Tau-tau separation and tau decay product $p_T$ requirements are slightly 
different for the two signatures and are discussed separately below.

The $Hjj$ signal is characterised by two forward jets with large invariant mass, 
and central $\tau$ decay products. The QCD backgrounds have a large 
gluon-initiated component and thus prefer lower invariant tagging jet masses. 
Also, their $\tau$ and $W$ decay products tend to be less central. Thus, to 
reduce the backgrounds to the level of the signal, we required tagging jets with 
a combination of large invariant mass, far forward rapidity, and high $p_T$, as 
well as $\tau$ decay products central with respect to the tagging 
jets~\cite{RZ_gamma}:

\begin{eqnarray}
\label{eq:tag2}
\eta_{j,min} + 0.7 < \eta_{\tau_{1,2}} < \eta_{j,max} - 0.7 \, , \qquad
\eta_{j_1} \cdot \eta_{j_2} < 0 \, , \nonumber\\
\Delta \eta_{tags} = |\eta_{j_1}-\eta_{j_2}| \geq 4.4 \, , \qquad
m_{jj} > m_{jj_{min}} \, ,
\end{eqnarray}

where $m_{jj_{min}}$ is chosen slightly differently for the two scenarios, as 
discussed below.

\paragraph{Lepton-hadron mode}

Here we required two additional cuts to form the tagging jet signature:

\begin{equation}
\label{eq:tag3}
p_{T_j} > 40,20 \, {\rm GeV} \, , \qquad \Delta R_{\tau\tau} \geq 0.7 \, .
\end{equation}

That is, the $p_T$ requirement on the tagging jets is staggered, and as one tau 
decay is hadronic, it must have a large separation from the leptonic tau.

Triggering the event via the isolated $\tau$-decay lepton and identifying the 
hadronic $\tau$ decay as discussed in~\cite{Cavalli} requires sizable 
transverse momenta for the observable $\tau$ decay products: 
$p_{T_{\tau,lep}} > 20~{\rm GeV}$ and $p_{T_{\tau,had}} > 40~{\rm GeV}$. It is 
possible to reconstruct the $\tau$-pair invariant mass from the observable 
$\tau$ decay products and the missing transverse momentum vector of the 
event~\cite{tautaumass}. The $\tau$ mass was neglected and collinear decays 
assumed, a condition easily satisfied because of the high $\tau$ transverse 
momenta required. The $\tau$ momenta were reconstructed from the charged decay 
products' $p_T$ and missing $p_T$ vectors. We imposed a cut on the angle 
between the $\tau$ decay products to satisfy the collinear decay assumption, 
$\cos\theta_{l h} > -0.9$, and demanded a physicality condition for the 
reconstructed $\tau$ momenta (unphysical solutions arise from smearing effects); 
that is, the fractional momentum $x_{\tau}$ a charged decay observables takes 
from its parent $\tau$ cannot be negative. Additionally, the $x_{\tau_{l}}$ 
distribution of the leptonically decaying $\tau$-candidate is softer for real 
$\tau$'s than for the reducible backgrounds, because the charged lepton shares 
the parent $\tau$ energy with two neutrinos. Cuts $x_{\tau_l} < 0.75$ and
$x_{\tau_h} < 1$ proved very effective in suppressing the reducible backgrounds.

Our Monte Carlo predicted a $\tau$-pair mass resolution of 10 GeV or better, so 
we chose $\pm 10$~GeV mass bins for analysing the cross sections. To further 
reduce the QCD backgrounds, which prefer low invariant masses for the tagging 
jets, we required  $m_{jj} > 1$~TeV. Additionally, the $Wj+jj$ background 
exhibits a Jacobian peak in its $m_T$ distribution~\cite{Cavalli}; hence a cut 
$m_T(l,\sla{p_T}) < 30$~GeV largely eliminates this background.

Finally, to compensate for overall rate loss based on ATLAS and CMS expected 
detector ID efficiencies, we apply a factor 0.86 to the cross section for each 
tagging jet, and a factor 0.95 for the charged lepton. 

Using all these cuts together, although not in a highly optimised combination, 
we expect already a signal to background ratio of 2/1 with a signal cross 
section of 0.4 fb for $M_H = 120$~GeV.

A probability for vetoing additional central hadronic radiation was obtained by 
measuring the fraction of events that have additional radiation in the central 
region, between the tagging jets, with $p_T$ above 20 GeV, using the matrix 
elements for additional parton emission. This minijet veto reduces the signal by 
about $15\%$, but eliminates typically $70\%$ of the QCD backgrounds; the EW 
$Zjj$ background is reduced by about $20\%$, indicating the presence of both 
boson bremsstrahlung and weak boson fusion effects. Because the veto probability 
for QCD backgrounds is found to be process independent, we applied the same 
value to the $bb+jj$ background.

Table~\ref{tab:lhcuts} summarises the signal and various background cross sections 
at progressive levels of the cuts, ID efficiencies and minijet veto as described 
above, for the case $M_H = 120$~GeV. Table~\ref{tab:lhsum} gives the expected 
numbers of events for 60 fb$^{-1}$ integrated luminosity (low luminosity 
running) at the LHC.

\begin{table}[htb]
\begin{center}
\caption{Signal and background cross sections $\sigma \cdot BR$ 
(fb) for $M_H = 120$~GeV 
$Hjj$ events in the lepton-hadron channel. Results are given for successive
cuts, as discussed in the text. The last column gives the ratio of the 
signal to the background cross sections listed in the previous columns.}
\label{tab:lhcuts}
\vspace{0.2cm}
\begin{tabular}{lcccccc}
\hline
Cuts & $Hjj$ & QCD $Zjj$ & EW $Zjj$ & $Wj+jj$ & $b\bar{b}+jj$ & $S/B$ \\
\hline
forward tagging &
                      68.4 & 1680 & 91   &      &      &       \\
$\tau$ identification &
                      1.99 & 20.0 & 1.45 & 26.4 & 7.6  & 1/28  \\
$110 < m_{\tau\tau} < 130 {\rm GeV}$ &
                      1.31 & 0.95 & 0.07 & 1.77 & 0.59 & 1/2.6 \\
$m_{jj}>1$~TeV, $m_T(l,\sla p_T) < 30$~GeV & 
                      0.69 & 0.16 & 0.04 & 0.11 & 0.15 & 1.5/1 \\
$x_{\tau_l} < 0.75$, $x_{\tau_h} < 1.0$ &
                      0.54 & 0.15 & 0.03 & 0.03 & 0.05 & 2.1/1 \\
ID efficiency ($\epsilon = 0.70$) &
                      0.38 & 0.10 & 0.03 & 0.03 & 0.05 & 2.1/1 \\
$P_{surv,20}$  & ${\it\times 0.87}$ & ${\it\times 0.28}$ 
               & ${\it\times 0.80}$ & ${\it\times 0.28}$ 
               & ${\it\times 0.28}$ & - \\
minijet veto &
                      0.33 & 0.03 & 0.02 & 0.004 & 0.011 & 5.2/1 \\
\hline
\end{tabular}
\end{center}
\end{table}

\begin{table}[htb]
\begin{center}
\caption{Number of expected events in the lepton-hadron channel for the 
signal and backgrounds, for 60~fb$^{-1}$ at low luminosity running; cuts, ID 
efficiency ($\epsilon = 0.70$) and minijet veto as in the last line of 
Table~\protect\ref{tab:lhcuts}; for a range of Higgs boson masses. Mass bins of 
$\pm 10$~GeV around a given central value are assumed. As a measure of the 
Poisson probability of the background to fluctuate up to the signal level, 
the last row gives $\sigma_{Gauss}$, the number of Gaussian equivalent 
standard deviations.}
\label{tab:lhsum}
\vspace{0.2cm}
\begin{tabular}{lccccc}
\hline
$M_H$ (GeV)                      & 110  & 120  & 130  & 140  & 150  \\
\hline
$\epsilon\cdot\sigma_{sig}$ (fb) & 0.38 & 0.33 & 0.25 & 0.16 & 0.08 \\
$N_S$                            & 22.9 & 19.6 & 15.2 &  9.5 &  4.6 \\
$N_B$                            & 10.2 &  3.8 &  2.4 &  1.8 &  1.5 \\
$S/B$                              &  2.2 &  5.2 &  6.4 &  5.2 &  3.1 \\
$\sigma_{Gauss}$                 &  5.6 &  6.6 &  6.3 &  4.7 &  2.6 \\
\hline
\end{tabular}
\end{center}
\end{table}

\begin{figure}[htbp]
\begin{center}
\includegraphics[angle=90,width=7cm]{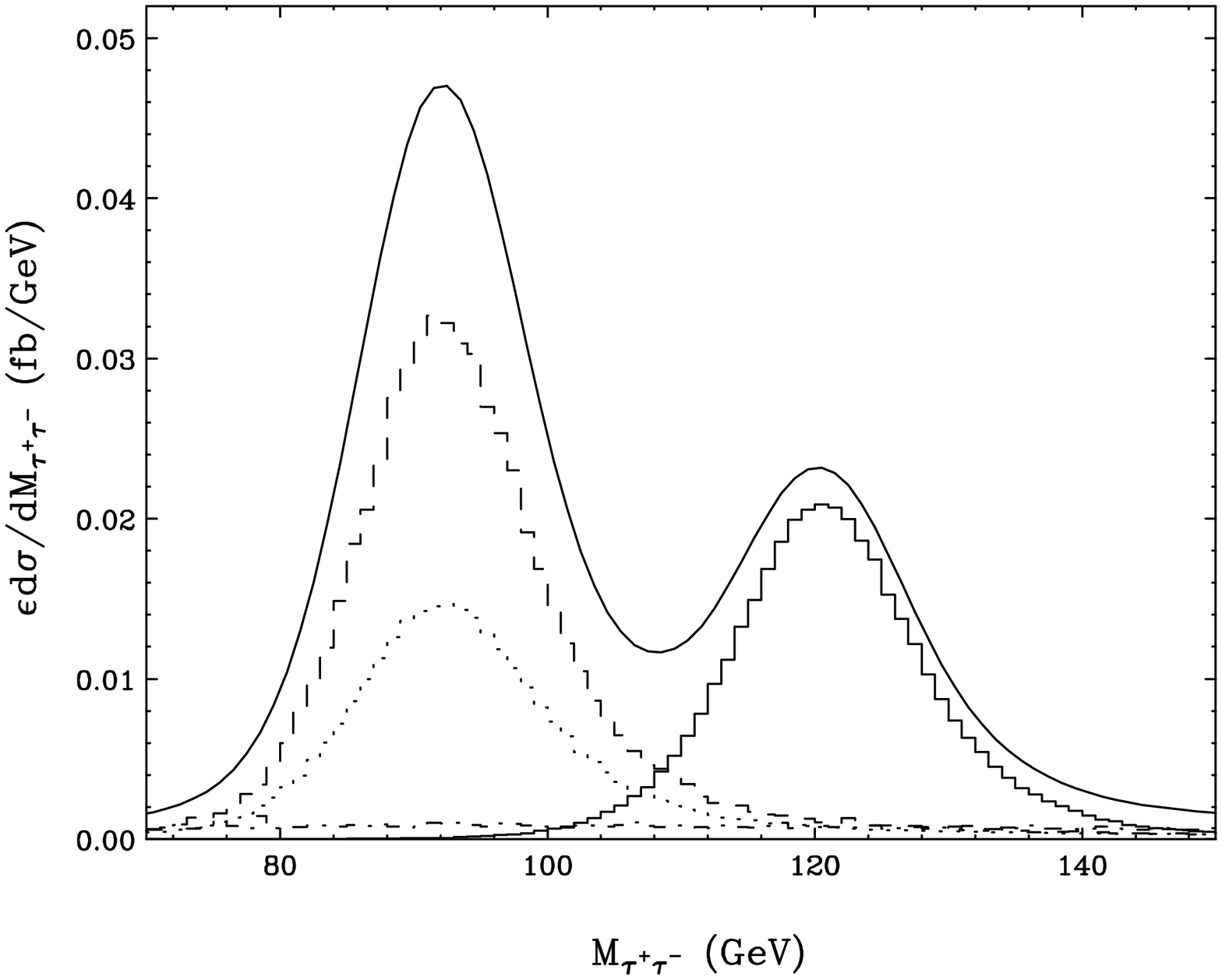}
\end{center}
\caption{Reconstructed $\tau$ pair invariant mass distribution for the signal 
(lepton-hadron channel) and backgrounds after all cuts and multiplication by 
the expected survival probabilities. The solid line represents the sum of the 
signal and all backgrounds. Individual components are shown as histograms: 
the $Hjj$ signal (solid), the irreducible QCD $Zjj$ 
background (dashed), the irreducible EW $Zjj$ background (dotted), and the 
combined $Wj+jj$ and $b\bar{b}jj$ reducible backgrounds (dash-dotted).}
\label{fig:Mtaulh}
\end{figure}

It is possible to isolate a virtually background-free $qq\to qqH\to jj\tau\tau$ 
signal at the LHC, leading to a $5\sigma$ observation of a SM Higgs boson with 
a mere 60 fb$^{-1}$ of data. The expected purity of the signal is demonstrated 
in Figure~\ref{fig:Mtaulh} showing the reconstructed $\tau\tau$ invariant mass 
for a SM Higgs of 120 GeV after all cuts, particle ID efficiency factors and a 
minijet veto have been applied. While the reducible $Wj+jj$ and $b\bar b+jj$ 
backgrounds are the most complicated and do require further study, they appear 
to be easily manageable. 

\paragraph{Dual lepton mode}

For this signature, we simulated tau decays as before, but with both decaying to 
final-state leptons. As this would form a different final state in experiment, 
to form the basic tagging jet signature we require the cuts of 
Equations~\ref{eq:tag1} and~\ref{eq:tag2} as before, but additionally a minimum 
separation of the charged leptons somewhat less than for the lepton-hadron 
scenario, $\Delta R_{\tau\tau} \geq 0.4$. To be able to trigger on the 
leptons, we require them to have minimum transverse momentum 
$p_{T_l} > 10$~GeV. In the LHC experiments, 
this may be slightly higher for electrons 
and slightly lower for muons, but we do not make the distinction here.

Both the $t\bar{t} + jets$ and $b\bar{b}jj$ backgrounds are about three orders 
of magnitude larger than the signal, but the contribution from $b\bar{b}jj$ may 
be reduced by a cut on missing transverse energy, $\sla{p}_T > 30$~GeV, and 
that from $t\bar{t} + jets$ may be severely restricted by vetoing additional 
jets in the central region between the tagging jets, which even before 
considering additional gluon radiation (minijets) may come from the decays of 
central final-state $b$-quarks. We veto all events with a central $b$ with 
$p_T > 20$~GeV. This provides approximately a factor 17 in reduction of the 
top quark background, which may be substantially improved to even lower $p_T$ 
threshold via a $b$-tag, which we cannot simulate.

As the dual lepton final state has a lower overall branching ratio than the 
lepton-hadron case, we retained more overall rate by making a looser cut on the 
tagging jet invariant mass, $m_{jj} > 800$~GeV. This cut was still necessary to 
reduce the QCD backgrounds.

Our Monte Carlo again predicted an excellent $\tau$-pair mass resolution, so we 
retain the mass binning of $\pm 10$~GeV. We also rejected non-tau's as in the 
lepton-hadron case, although our exact cut was somewhat differently defined: 
\begin{eqnarray}\nonumber
x_{\tau_1} , \; x_{\tau_2} > 0 \; , \qquad
x_{\tau_1}^2 + x_{\tau_2}^2 < 1 \; .
\end{eqnarray}

Finally, we found that a cut on the maximal separation of the two charged 
leptons is very useful in reducing the heavy quark backgrounds: 
$\Delta{R}_{e\mu} < 2.6$.

Efficiency factors for detection are the same as in the previous case, although 
with two final-state leptons an extra factor 0.95 was taken into account. A 
minijet veto was applied as before, although other analyses we have performed 
suggest the survival probabilities change slightly due to the lower hardness of 
the event, which is strongly correlated with $m_{jj}$ (see Table~\ref{tab:llcuts}).

Table~\ref{tab:llcuts} outlines the cross sections of signal and background for 
progressive levels of cuts as described above, for the case $M_H = 120$~GeV. 
Table~\ref{tab:llsum} gives the expected numbers of events for 60~fb$^{-1}$ 
integrated luminosity (low luminosity running) at the LHC.

\begin{table}[htb]
\begin{center}
\caption{Signal rates $\sigma\cdot BR(H\to\tau\tau\to e^\pm\mu^\mp\sla{p_T})$ for 
a SM Higgs of $M_H = 120$~GeV and progressive levels of cuts as discussed in the 
text. All rates are given in fb. Note: the fifth line, non-tau 
rejection, also includes a cut 90~GeV~$< m_{\tau\tau} < 160$~GeV.}
\label{tab:llcuts}
\vspace{0.2cm}
{\footnotesize
\begin{tabular}{lccccccccc}
\hline
     & $H\to\tau\tau$ & $H\to WW$ & QCD & EW & & & QCD & EW & \\
Cuts & signal & bkgd & $\tau\tau jj$ & $\tau\tau jj$ 
     & $t\bar{t} + jets$ & $b\bar{b}jj$ & $WWjj$ & $WWjj$ & $S/B$ \\
\hline
forward tags
& 2.2  &       & 57    & 2.3   & 1230  & 1050  & 4.9   & 3.3   & 1/1100 \\
$b$ veto
&      &       &       &       & 72    &       &       &       & 1/550  \\
$\sla{p}_T > 30$~GeV
& 1.73 &       & 29    & 1.57  & 62    & 29    & 4.1   & 2.9   & 1/74   \\
$M_{jj} > 800$~GeV
& 1.34 &       & 10.3  & 1.35  & 16.3  & 10.4  & 1.60  & 2.6   & 1/32   \\
non-$\tau$ reject.
& 1.15 &       & 5.2   & 0.63  & 0.31  & 0.42  & 0.032 & 0.042 & 1/5.8  \\
$\pm 10$~GeV mass bins
& 0.87 &       & 0.58  & 0.10  & 0.09  & 0.10  & 0.009 & 0.012 & 1/1   \\
$\Delta R_{e\mu} < 2.6$
& 0.84 & 0.023 & 0.52  & 0.086 & 0.087 & 0.028 & 0.009 & 0.011 & 1.1/1 \\
ID effic. (${\it\times 0.67}$)
& 0.56 & 0.015 & 0.34  & 0.058 & 0.058 & 0.019 & 0.006 & 0.008 & 1.1/1 \\
$P_{surv,20}$ & ${\it\times 0.89}$ & ${\it\times 0.89}$ 
              & ${\it\times 0.29}$ & ${\it\times 0.75}$ 
              & ${\it\times 0.29}$ & ${\it\times 0.29}$ 
              & ${\it\times 0.29}$ & ${\it\times 0.75}$ & - \\
minijet veto
& 0.50 & 0.014 & 0.100 & 0.043 & 0.017 & 0.006 & 0.002 & 0.006 & 2.7/1 \\
\hline
\end{tabular}
}
\end{center}
\end{table}

\begin{table}[htb]
\begin{center}
\caption{Number of expected events for a SM $Hjj$ signal in the 
$H\to\tau\tau\to e^\pm\mu^\mp\sla{p_T}$ channel, for a range of Higgs boson
masses. Results are given for 60~${\rm fb}^{-1}$ of data at low luminosity 
running, and application of all efficiency factors and cuts, including a minijet 
veto. As a measure of the Poisson probability of the background to fluctuate up 
to the signal level, the last line gives $\sigma_{Gauss}$, the number of 
Gaussian equivalent standard deviations.}
\label{tab:llsum}
\vspace{0.2cm}
\begin{tabular}{lccccccccccc}
\hline
$M_H$ & 100  & 105  & 110  & 115  & 120  & 125  & 130  & 135  & 140  & 145  & 150  \\
\hline
$\epsilon\cdot\sigma_{sig}$ (fb)
      & 0.62 & 0.61 & 0.58 & 0.55 & 0.50 & 0.44 & 0.37 & 0.30 & 0.23 & 0.16 & 0.11 \\
$N_S$ & 37.4 & 36.5 & 35.0 & 32.8 & 30.0 & 26.3 & 22.3 & 18.0 & 13.7 &  9.9 &  6.5 \\
$N_B$ & 67.7 & 45.4 & 27.4 & 16.8 & 11.2 &  8.4 &  7.1 &  6.4 &  6.1 &  5.9 &  5.7 \\
$S/B$   &  0.6 &  0.8 &  1.3 &  2.0 &  2.7 &  3.2 &  3.1 &  2.8 &  2.2 &  1.7 &  1.1 \\
$\sigma_{Gauss}$ & 4.1 & 4.8 & 5.6 & 6.4 & 6.8 & 6.7 & 6.1 & 5.3 & 4.3 & 3.2 & 2.2 \\
\hline
\end{tabular}
\end{center}
\end{table}

\begin{figure}[htbp]
\begin{center}
\includegraphics[angle=90,width=7cm]{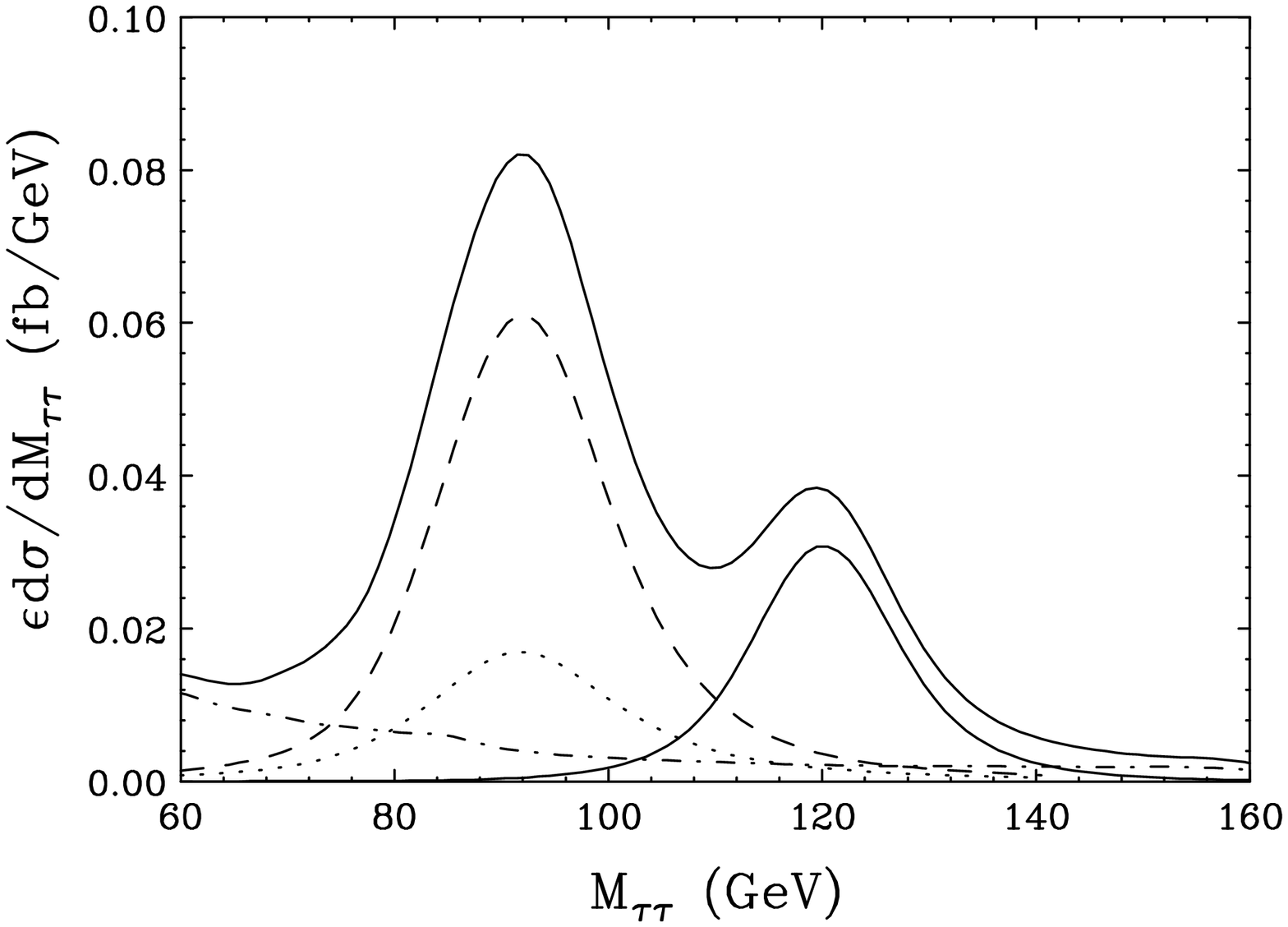}
\end{center}
\caption{Reconstructed $\tau$ pair invariant mass distribution for a SM 
$H\to\tau\tau\to e^\pm \mu^\mp \sla{p}_T$ ($M_H=120$~GeV) signal and backgrounds 
after all cuts, particle ID efficiencies and minijet veto. The double-peaked 
solid line represents the sum of the signal and all backgrounds. Individual 
components are: the $Hjj$ signal (solid), the irreducible QCD $Zjj$ background 
(dashed), the irreducible EW $Zjj$ background (dotted), and the combined 
reducible backgrounds from QCD + EW + Higgs $WWjj$ events and $t\bar{t} + jets$ 
and $b\bar{b}jj$ production (dash-dotted).}
\label{fig:Mtaull}
\end{figure}

Although the dual lepton channel does not appear to be able to achieve quite as 
high an $S/B$ ratio as the lepton-hadron channel, it is still better than 1/1 over 
much of the mass range of interest, which is also clearly evident in the tau 
pair invariant mass plot of Figure~\ref{fig:Mtaull}. Furthermore, the independent 
statistical significance of this channel is as good as that found for the 
lepton-hadron case.


\subsubsection{MSSM analysis}

\begin{figure}[htbp]
\begin{center}
\includegraphics[width=7cm]{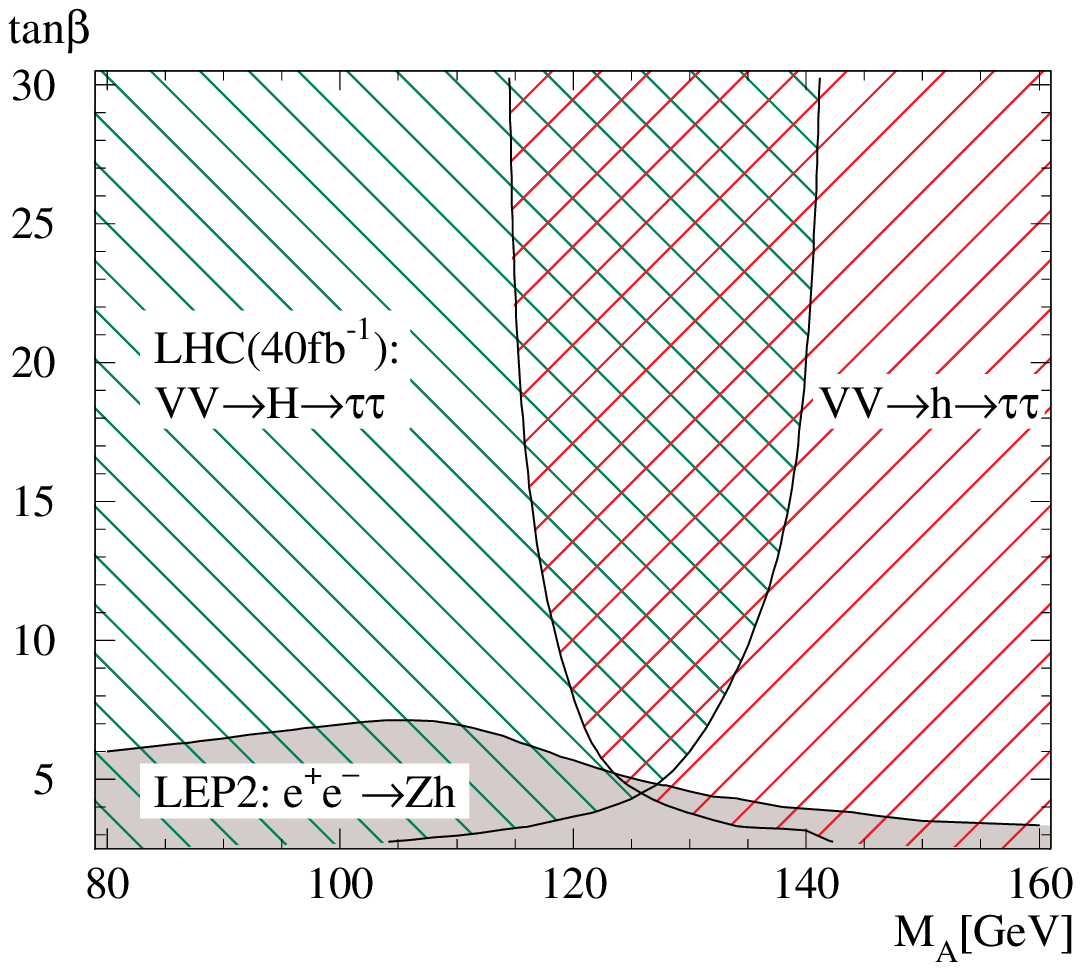}
\includegraphics[width=7cm]{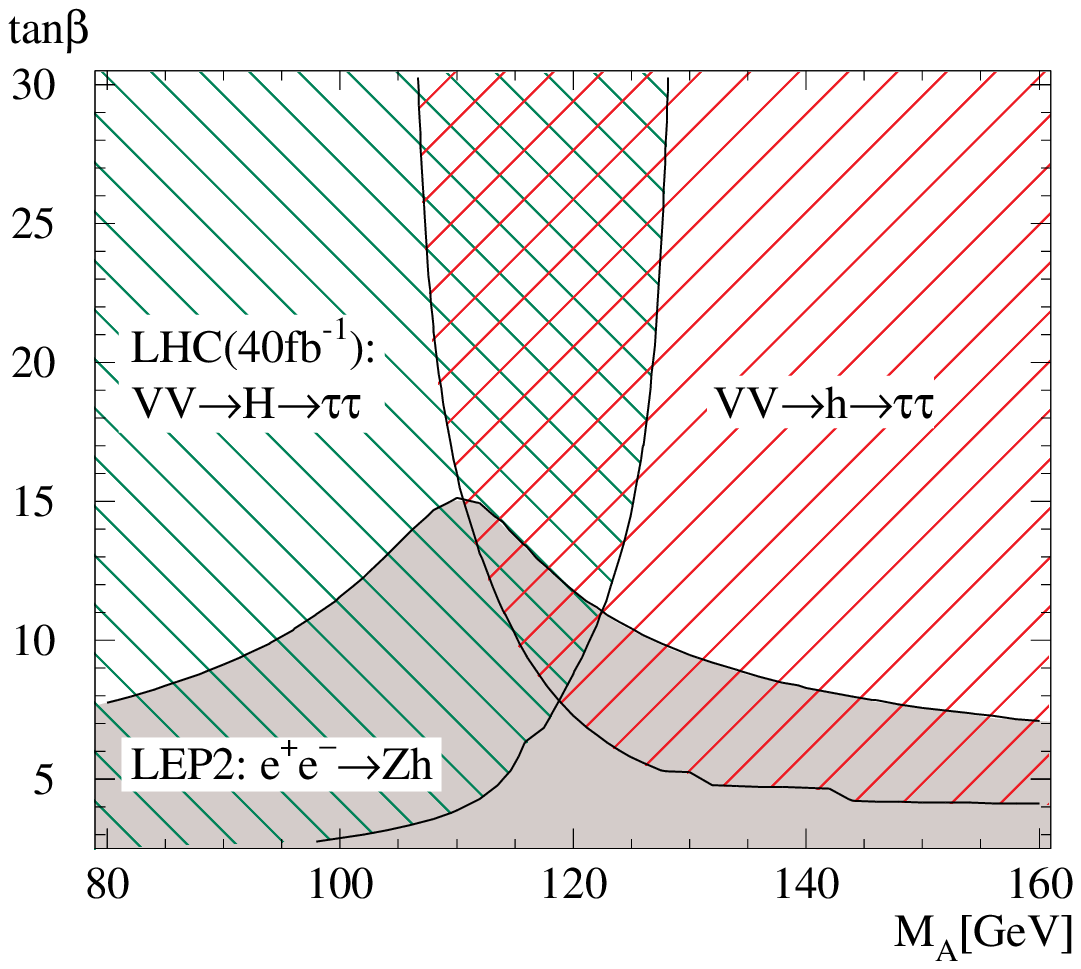}
\end{center}
\caption{$5\sigma$ discovery contours for $h\to\tau\tau$ and $H\to\tau\tau$ in
weak boson fusion at the LHC, with 40~fb$^{-1}$. Also shown are the
projected LEP2 exclusion limits. Results are shown for maximal mixing 
(left) and no mixing (right). From~\cite{PRZ_tau,PRZ}.}
\label{fig:MSSM}
\end{figure}

The production of CP even Higgs bosons in weak boson fusion is governed by the 
$hWW,HWW$ couplings, which are suppressed by factors 
$\sin(\beta-\alpha),\cos(\beta-\alpha)$, respectively~\cite{mssm}, compared to the SM case. 
Their branching 
ratios are modified with slightly more complicated factors. One can simply 
multiply SM cross section results from our analysis by these factors to 
determine the observability of $H \to \tau\tau$ in MSSM parameter space. We used 
a renormalisation group improved next-to-leading order calculation, which allows 
a light Higgs mass up to $\sim 125$~GeV, and examined two trilinear term mixing 
cases, no mixing and maximal mixing~\cite{PRZ_tau,PRZ}.

Varying the pseudoscalar Higgs boson mass $M_A$, one finds that $M_h$, 
$M_H$ each 
approach a plateau for the case $M_A \to \infty,0$, respectively. Below 
$M_A \sim 120$~GeV, the light Higgs mass will fall off linearly with $M_A$, 
while the heavy Higgs will approach $M_H \sim 125$~GeV, whereas above 
$M_A \sim 120$~GeV, the light Higgs will approach $M_h \sim 125$~GeV and the 
heavy Higgs mass will rise linearly with $M_A$. The transition region behaviour 
is very abrupt for large $\tan\beta$, such that the plateau state will go to 
$\sim 125$~GeV almost immediately, while for small $\tan\beta$ the transition is 
much softer and the plateau state reaches the limiting value via a more gradual 
asymptotic approach. 

With reasonable integrated luminosity and combination of the lepton-hadron and 
dual-lepton channels, 40 fb$^{-1}$ in the worst case, it will be possible to 
observe at the $5\sigma$ level either $h$ or $H$ decays to $\tau$ pairs when 
they are in their respective plateau region, with the possibility of some 
overlap in a small region of $M_A$, as shown in Figure~\ref{fig:MSSM}. Very low values 
of $\tan\beta$ would be unobservable, but already excluded by LEP2; there 
should be considerable overlap between this mode at the LHC and the LEP2 
excluded region. Furthermore, a parton shower Monte Carlo with full detector 
simulation should be able to optimise the analysis so that much less data is 
required to observe or exclude the MSSM Higgs.


\subsubsection{Conclusions}

The production of a neutral, CP even Higgs via weak boson fusion and decay 
$H\to\tau\tau$ at the LHC has been studied for the Standard Model and MSSM, 
utilising parton level Monte Carlo analyses. Each of the decay channels 
$\tau\tau\to h^\pm l^\mp \sla{p}_T, e^\pm\mu^\mp \sla{p}_T$ independently 
allows a $5\sigma$ observation of a Standard Model Higgs with an integrated 
luminosity of about 60~fb$^{-1}$ or less, and provides a direct measurement of 
the $H\tau\tau$ coupling. For the MSSM case, a highly significant signal for 
at 
least one of the Higgs bosons with reasonable luminosity is possible over the 
entire physical parameter space which will be left unexplored by LEP2. Only 
40~fb$^{-1}$ of data is required after combining the two channels.
We conclude that this mode provides 
a {\bf no-lose strategy} for seeing at least one of the CP even neutral MSSM Higgs 
bosons.

\subsection{Searching for \boldmath $VV \to H \to WW$}

In the previous section, vector-boson fusion forming a Higgs which then
decays to two $\tau$'s was identified as a valuable process by which to 
find a Higgs boson in the mass range 110 to 150~GeV.
Rainwater and Zeppenfeld have shown that a heavier Higgs in the 
range 130 to 200~GeV could be found by looking for the 
process $VV \to H \to WW \to e^{\pm} \mu^{\mp} \not{\!p}_T$ \cite{bib:vvhvv}.
As for the lighter Higgs, the forward jet tagging is a powerful tool for
removing background ($W$ pairs, $t\bar{t}$ and $Z \rightarrow \tau \tau$
accompanied by jets).
This approach appears more promising than the a search for an inclusive 
$H \to WW \to e^{\pm} \mu^{\mp} \not{\!p}_T$ signal, yielding a significant 
result with $\sim 5$~fb$^{-1}$.

Work has started in the context of the Workshop to investigate this with 
fast detector simulation, but has not yet been completed.



\def\lsim{\mathrel{\lower2.5pt\vbox{\lineskip=0pt\baselineskip=0pt
\hbox{$<$}\hbox{$\sim$}}}}
\def\gsim{\mathrel{\lower2.5pt\vbox{\lineskip=0pt\baselineskip=0pt
\hbox{$>$}\hbox{$\sim$}}}}
\def\wh{{\cal W}}
\def\bh{{\cal B}}
\def\wtu{\wh^{\mu \nu}}
\def\wtd{\wh_{\mu \nu}}
\def\btu{\bh^{\mu \nu}}
\def\btd{\bh_{\mu \nu}}
\def\gs{SU(2)_{\rm L} \times U(1)_{\rm Y}}
\def\mz{{m}_{\scriptscriptstyle Z}^{2}}
\def\w{{\tilde{W}^{\pm}}}
\def\h{{\tilde{H}^{\pm}}}
\def\sw{s_{{\scriptscriptstyle W}}}
\def\cw{c_{{\scriptscriptstyle W}}}
\def\Tr{{\rm Tr}} 
\newcommand{\slas}[1]{\rlap/ #1}
\newcommand{\diag}{{\rm diag}}
%



\subsection{The strongly interacting symmetry breaking sector}


One possible scenario for the spontaneous breaking
of the electroweak (EW) symmetry is 
a strongly interacting symmetry breaking sector (SBS), which generically
is formed by new particles with strong interactions at the TeV scale. 
This sector should provide a global $SU(2)_L\times SU(2)_R$ spontaneous
symmetry breaking down to the custodial $SU(2)_{L+R}$ subgroup, thus
triggering the Standard Model spontaneous breaking from the $SU(2)_L\times U(1)_Y$ gauge-symmetry down to $U(1)_{\rm em}$. This is the minimal symmetry 
pattern ensuring that $\rho\simeq 1+O(g^2)$.
 
By assuming that the new states appear at the TeV scale,
we are only left, at low energies, with  the 
three massless Goldstone Bosons (GB) 
associated to the global symmetry breaking.
We will refer to  this scenario as the minimal strongly interacting
symmetry breaking sector (MSISBS). In this case,
the low-energy EW interactions can be well described 
with the Electroweak Chiral Lagrangian (EChL) \cite{ApBe80,longhitano:81},
which is an $SU(2)\times U(1)$ gauge-invariant
effective field theory that couples the GB  
to the gauge-bosons and fermions, 
without any further assumptions than those just described. The EChL,
inspired in Chiral Perturbation Theory \cite{ChPT},
is organised as a derivative (momentum) expansion, 
with a set of effective 
operators of increasing dimension.
Although the lowest-order Lagrangian  is common to
all models satisfying the minimal assumptions, at higher orders
each effective operator has a coefficient, whose different
values will account for different underlying symmetry breaking mechanisms. 
Within this approach it is possible, not only to calculate at tree level,
but to include loops whose divergences will be absorbed 
in the coefficients of operators of higher dimension, thus yielding finite 
results order by order in the calculations.
The values of these renormalised parameters are expected in the 
$10^{-3}$ to $10^{-2}$ range. 

As far as physics at the LHC is concerned, the most characteristic feature
of a strong SBS is the enhanced  
production of longitudinal gauge-boson
pairs.  We will review the EChL 
amplitudes for these processes.
However, the  EChL perturbative 
predictions can only describe EW physics at low
energies, well below the mass of the heavy states. 
Indeed, any amplitude calculated with the EChL is obtained 
as a truncated series in powers of the external momenta. 
Hence, it will always
violate unitarity bounds at high enough energies. In addition,
it cannot reproduce any pole associated to new resonant
states. Consequently, in order to apply this formalism 
to study strong SBS phenomenology at the LHC, we have several 
ways to proceed: 
\begin{enumerate}
\item
Perform studies strictly within
the EChL, but restricted to subprocess
energies below 1.5 TeV
and to very small chiral parameters.
\item
Enlarge the EChL 
introducing explicitly the heavy resonances of each 
particular model, but this adds new unknown parameters, namely the mass and 
the width of each resonance.
\item
Follow a more model-independent approach,
by unitarising the EChL amplitudes 
and generating heavy resonances 
from the information contained in the chiral coefficients.
\end{enumerate}
 
In the last approach, it is possible to describe the
different resonant scenarios with just two chiral parameters.
Finally we present
a study of the LHC sensitivity reach within this parameter space,
using the signal of the cleanest leptonic decays of $ZZ$ and $WZ$  pairs.

\subsubsection{Effective Chiral Lagrangian description of electroweak 
interactions}

The EChL~\cite{ApBe80,longhitano:81} provides a phenomenological description 
of EW interactions when the SBS
is strongly-interacting.  The only degrees of freedom 
at low energies are the GBs associated to 
the $SU(2)_L\times SU(2)_R\rightarrow SU(2)_{L+R}$ global symmetry
breaking, which are coupled to the EW gauge and fermion
fields in an $SU(2)_L \times U(1)_L$ invariant way.
Customarily, the GBs, $\omega^a$ with $a=1,2,3$,
are gathered in an $SU(2)$ matrix
$U=\exp\left(i\omega^a\tau^a/v\right)$,
where $\tau^a$ are the Pauli matrices and  $v=246\;\hbox{GeV}$.
The C and P invariant 
effective bosonic operators up to dimension four are
(see the appendix for other notations)  
 \begin{eqnarray} 
 {\cal L}_{\rm EChL} &=& \frac{v^2}{4} \Tr (D_{\mu}U (D^{\mu}U)^\dagger) 
                 + a_0\frac{g'^2v^2}{4}[\Tr(TV_{\nu})]^2 
                 + a_1\frac{i g g'}{2} \btd \Tr (T \wtu ) 
\nonumber\\
                 &+& a_2\frac{ i g'}{2} \btd \Tr (T[V^{\mu},V^{\nu}])
                 + a_3 g \Tr(\wtd [V^{\mu},V^{\nu}])  
                 + a_4 [\Tr(V_{\mu}V_{\nu})]^2 
                               \nonumber \\
                 &+& a_5 [\Tr(V_{\mu}V^{\mu})]^2   
                 + a_6 \Tr(V_{\mu}V_{\nu})\Tr(TV^{\mu})\Tr(TV^{\nu}) 
                 + a_7 \Tr(V_{\mu}V^{\mu}) [\Tr(TV^{\nu})]^2 
\nonumber\\
                 &+& a_8\frac{g^2 }{4} [\Tr(T \wtd )]^2 
                 + a_9 \frac{g}{2}\Tr(T \wtd )\Tr(T[V^{\mu},V^{\nu}]) 
                 + a_{10}[\Tr(TV_{\mu})\Tr(TV_{\nu})]^2 
\nonumber \\                  
                 &+& \hbox{{e.o.m.} terms }+ \hbox{standard YM terms} 
\label{lag}               
 \end{eqnarray} 
where  we have defined 
$T \equiv U \tau^3 U^\dagger$ and $V_\mu \equiv (D_{\mu}U)U^{\dagger}$,
as well as 
\begin{eqnarray}
 D_\mu U &\equiv &\partial_\mu
U - g \wh_\mu U + g' U \bh_\mu , \quad  \quad  \quad 
\wh_\mu  \equiv \frac{ -i}{2}\; \vec{W}_\mu \cdot \vec{\tau} ,  \quad
\bh_\mu \equiv  \frac{ -i}{2} \; B_\mu \;
\tau^3, \nonumber\\
\wtd  & \equiv & \partial_\mu \wh_\nu - \partial_\nu \wh_\mu -
g [ \wh_\mu, \wh_\nu ], \quad 
\btd \equiv  \partial_\mu \bh_\nu - \partial_\nu \bh_\mu .
\end{eqnarray}
The ``e.o.m.'' terms  refer to operators that can be 
removed using the equations of motion and the ``standard YM terms'' 
are the usual Yang Mills Lagrangian together 
with the gauge-fixing and Faddeev-Popov terms.

The first operator in Equation~\ref{lag}, which
provides the $W$ and $Z$ masses, has dimension two and 
has the form of a gauged non-linear sigma model (NL$\sigma$M). 
Note that it is universal, since it only depends on $v$ -
that is why its predictions for longitudinal gauge-boson scattering amplitudes
are called ``Low Energy Theorems''.
In contrast, the  $a_i$ couplings  will have
different values depending on the underlying theory.

The gauge-boson observables are obtained from ${\cal L}_{\rm EChL}$ 
 as a double expansion in $p^n/(4\pi v)^n$, 
$p$ being an external momentum, and in the gauge-couplings 
$g$ and $g'$. The lowest-order predictions are given by the tree level
NL$\sigma$M, whereas the next order corrections are obtained
with a one-loop calculation using the NL$\sigma$M vertices
plus the tree level contributions of the other operators.
The  $a_i$ coefficients not only provide
a model independent parametrisation of the unknown dynamics,
but also some of them are used to absorb 
all the one-loop NL$\sigma$M divergences. This procedure could be carried
out to any desired order, adding higher dimensional operators, thus yielding finite
results order by order in the expansion.

In principle, the $a_i$ values for
a particular scenario can be obtained
by integrating out the heavy degrees of freedom.
In fact, they have been determined for the particular cases of
the SM with a heavy Higgs \cite{HeRu,HeRu2} and  
for technicolor theories in the large $N_{TC}$ limit \cite{aesenTC}.
In both cases, these couplings lie in the range $10^{-2}$ to 
$10^{-3}$, with either sign. They all have a constant contribution,
but those needed in the renormalisation also have a 
logarithmic term.

\subsubsection{Present bounds on the chiral parameters}

Let us now look at the present experimental constraints
on the EChL parameters $a_i$ from low energy EW data.
The best constraints come from the oblique radiative corrections, 
giving bounds on the $a_0$, $a_1$ and $a_8$ parameters that
contribute to the gauge-bosons two-point functions up to order $q^2$. 
The EChL calculation of the $S$, $T$ and $U$ \cite{PeTa} self-energy 
combinations give \cite{adobado}
\begin{eqnarray*}
S=16\pi\left[-a_1(\mu)+ \hbox{EChL loops}(\mu)\right],\quad
T=\frac{8\pi}{c_W^2}\left[a_0(\mu)+\hbox{EChL loops}(\mu)\right],\\
U=16 \pi \left[a_8(\mu)+\hbox{EChL loops}(\mu)\right]\hspace{3cm}
\end{eqnarray*}
Note that the $a_i$ have been renormalised to 
absorb the one-loop divergences from the NL$\sigma$M chiral loops, 
so that $S$, $T$ and $U$ are scale independent.
Using the $a_i$ values for a heavy Higgs 
boson \cite{HeRu,HeRu2}, the deviations of EW observables from
the SM predictions at a reference value of the Higgs mass $M_H$ 
are
\begin{eqnarray*}
\Delta S \equiv S - S_{\rm SM}(M_H) =
16\pi\left[-a_1(\mu)+ \frac{1}{12} \frac{5/6 -
\log M_H^2/\mu^2}{16 \pi^2} \right],\hspace{5.1cm}\\
\Delta T \equiv T - T_{\rm SM}(M_H) =
\frac{8\pi}{c_W^2}\left[a_0(\mu)- \frac{3}{8} \frac{
5/6 - \log M_H^2/\mu^2 }{16 \pi^2}  \right],\quad
 \Delta U \equiv U - U_{\rm SM}(M_H) = 16 \pi a_8.
\end{eqnarray*}
A global fit 
with $M_H=300$ GeV and $m_t=175$ GeV  to the low energy EW data gives
\cite{PDG} 
\begin{eqnarray*}
\Delta S=-0.26\pm0.14\quad,\quad
\Delta T=-0.11\pm0.16\quad,\quad
\Delta U=\;\;\;0.26\pm0.24
\end{eqnarray*}
which imply 
the following bounds for the three chiral couplings 
\begin{eqnarray*}
a_1(1 \hbox{TeV})=(6.8\pm2.8)\times10^{-3}, 
a_0(1 \hbox{TeV})=(4.3\pm4.9)\times10^{-3}, 
a_8(1 \hbox{TeV})=(4.9\pm4.7)\times10^{-3}. \label{afit}
\end{eqnarray*}
Other studies  agree with these values \cite{BFS}.
These data already disfavour the SM with a heavy Higgs boson and
set strong constraints in models with a dominance of vector 
resonances \cite{PeTa} (like technicolor). 
With further assumptions on
the underlying SBS dynamics, the latter 
give a negative contribution  to $a_1$.
However, the precision EW measurements leave room  for 
an strong SBS \cite{BFS}. 

Further constraints  come from the three-point functions,
whose  anomalous electroweak effective couplings were traditionally 
parametrised in terms of 
$g_1^{\gamma}, g_1^{Z}, \kappa_{\gamma}, \kappa_Z, 
\lambda_{\gamma}$ and $\lambda_Z$. A one-loop EChL calculation of these 
vertices \cite{Espriu} gives
\begin{eqnarray*}
g_1^\gamma-1&=&0+\hbox{EChL loops},\hspace{5.5cm}
g_1^Z-1=\frac{-g^2}{c_W^2}a_3+\hbox{EChL loops}(\mu)\\
\kappa_\gamma-1&=&g^2(a_2-a_3-a_1+a_8-a_9)+\hbox{EChL loops},\hspace{3cm}\lambda_\gamma=0\\
\kappa_Z-1&=&g^2(a_8-a_3-a_9)+g'^2(a_1-a_2)+\hbox{EChL loops}(\mu),
\hspace{1.7cm}\lambda_Z=0
\end{eqnarray*}
There are several analyses \cite{tril,vanderBij:1999fp} that 
constrain these chiral couplings
from LEP and Tevatron data. Ignoring 
the loops from the NL$\sigma$M, we get the
following values from present LEP data 
(the Tevatron precision is comparable)
$\lambda_\gamma=-0.037 {\scriptstyle{+0.035 \atop -0.036}}$,
\begin{eqnarray*}
\kappa_\gamma-1&=&0.038{\textstyle{+0.079 \atop -0.075}}
,\quad \longrightarrow 
\quad a_2-a_3-a_1+a_8-a_9=0.088{\textstyle{+0.184 \atop -0.174}},\\
g_1^Z-1&=&-0.010\pm0.033\quad \longrightarrow \quad a_3=0.018\pm0.059.
\end{eqnarray*}

Finally, some  indirect bounds on quartic couplings   
have also been found \cite{quartic,quartic2}. 
These indirect estimates come from loops 
containing $a_i$ vertices, but do not include 2-loop diagrams from
the NL$\sigma$M. They find bounds on $a_i$ for $i=4,5,6,7,10$
ranging from $10^{-1}$ to $10^{-2}$.

In summary, the present data on  the oblique EW corrections already 
sets significant bounds on the $a_0,a_1$ 
and $a_8$ chiral parameters, but  
there is not much 
sensitivity yet to those chiral parameters that contribute
to the three or four-point functions.
We will see next how, at the LHC, 
the situation will improve significantly.

\subsubsection{The Effective Chiral description at the LHC}

At the next generation of colliders, we 
will be probing the $W$ and $Z$ interactions at TeV energies. 
As long as we are only considering the GBs
 and no other fundamental fields 
up to the TeV scale, we expect the self-interactions
of longitudinal gauge-bosons, $V_L$, to become strong at LHC energies.
This can be easily understood since, intuitively,
longitudinal gauge-bosons are nothing but the GBs,
which interact strongly.
This  intuitive statement is rigorously given 
in terms of on-shell amplitudes and is
known as the Equivalence Theorem (ET),
\begin{equation}
A(V^a_L,V^b_L,V^c_L...\hbox{Other fields})\simeq A(\omega^a
\omega^b\omega^c...\hbox{Other fields})+O\left(M_W^2/\sqrt{s}\right),
\label{ET}
\end{equation}
which holds for any spontaneously broken
non-Abelian theory. Indeed, it was first derived for the
SM \cite{bib:ET,bib:ET2,bib:ET3}. Its usefulness is twofold:
it relates the pure SBS fields
with the observables, but also the calculations
can now be performed in terms of scalars instead of gauge-bosons,
at least in the high energy limit $s>>M_W^2$.
At first sight it may seem that the ET is incompatible with the
use of the EChL, since an effective theory is a low energy limit.
Nevertheless, the ET can still be applied with
the EChL, {\em only at leading 
order in $g$ and $g'$}, if we only consider energies below 
1.5 TeV and small chiral parameters \cite{ETnosotros,ETnosotros2,ETnosotros3}.

Hence, in a first approximation, we will simplify 
the high energy description of the strong SBS 
by neglecting EW corrections. Thus, due to our assumption
that $SU(2)_{L+R}$ is preserved in the SBS, only the
operators that respect custodial symmetry once the
gauge-symmetries are switched off will be relevant in this
regime. These are the universal term and the 
operators with $a_i$ couplings for $i=3,4,5$. 

At the LHC, the two most relevant processes of $V_L V_L$ production
are the scattering of two longitudinal vector-bosons in fusion reactions 
and the $V_L$ pair production from $q\bar{q}$ annihilation.
Through the ET, they are identified with GB elastic scattering
and $q\bar{q}\rightarrow\omega\omega$, respectively.
Customarily, GB elastic scattering is described in terms of partial wave
amplitudes of definite angular momentum, $J$,
and isospin, $I$, associated to the custodial $SU(2)_{L+R}$ group. 
With the EChL, these partial waves, $t_{IJ}$ are obtained as 
\begin{equation} 
t_{IJ}(s)=t^{(2)}_{IJ}(s)+t^{(4)}_{IJ}(s)+...,
\label{expansion}
\end{equation}
where the superscript refers to the corresponding power of momenta. 
They are given by \cite{ChPT,antoniomariajo,antoniomariajo2}
\begin{eqnarray}
 t^{(2)}_{00}&=&\frac{s}{16\,\pi v^2}, \quad\quad
 t^{(4)}_{00}=\frac{s^2}{64\,\pi v^4}
\left[\frac{16(11a_5+7a_4)}{3}
+\frac{101/9-50 \log(s/\mu^2)/9+4\, i\,\pi}{16\,\pi^2}
\right],\nonumber\\
 t^{(2)}_{11}&=&\frac{s}{96\,\pi v^2},  \quad\quad
 t^{(4)}_{11}=\frac{s^2}{96\,\pi v^4}\left[4(a_4-2a_5)
+\frac{1}{16\,\pi^2}\left(\frac{1}{9}+\frac{i\,\pi}{6}\right)
\right],\nonumber\\
 t^{(2)}_{20}&=&\frac{-s}{32\,\pi v^2}, \quad\quad
 t^{(4)}_{20}=\frac{s^2}{64\,\pi v^4}\left[\frac{32(a_5+2a_4)}{3} 
+\frac{273/54-20\log(s/\mu^2)/9
+i\,\pi}{16\,\pi^2}
\right].
\label{pertamplis}
\end{eqnarray}
Note that, within our approximations,
the above amplitudes only depend on $a_4$ and $a_5$.
The projection in angular momentum has been defined,
from the definite $I$ amplitude $T_I$, as
\begin{equation}
  t_{IJ}=\frac{1}{64\,\pi}\int_{-1}^1\,d(\cos\theta)
\,P_J(\cos\theta)\,T_I(s,t)\;.
\end{equation}

The $V_LV_L$ production from $q\bar{q}$ annihilation,
is very important since vector resonances can also
couple to this channel. By means of the ET, 
we are thus interested in $q\bar{q}\rightarrow\omega \omega$. 
As far as  GBs couple to quarks proportionally to
their mass,
the only relevant contribution comes from the
$s$-channel annihilation through a vector-boson. 
In practice, for the $WZ$ final state, the
$W\rightarrow\omega z$
interaction is described as $g\, F_V(s)$,
by means of a vector form factor, $F_V(s)$, which
is obtained from the EChL as
\begin{equation}
  \label{Fpert}
  F_V(s)=1+ F_V^{(2)}(s)+...\quad\quad \hbox{with}\quad\quad
 F^{(2)}_V(s)=\frac{s}{ (4 \, \pi \, v)^2} \left[ 64 \pi^2 a_3(\mu) 
 -  \frac{1}{6} \log{s \over \mu^2} + \frac{4}{9} + i \,
\frac{\pi}{6}
\right] 
\label{F2}
\end{equation}
Let us then review the studies of the LHC sensitivity to 
the chiral parameters via these two
processes.

\subsubsection{Non-resonant studies for LHC}

The EChL formalism has been applied to study the LHC sensitivity
to different non-resonant SBS sectors in  
\cite{noreson,noreson2,CMSTP,noresonnosotros2,noresonnosotros3,
Belyaev}. 
We  summarise in Table~\ref{tab:nos}
the results from \cite{CMSTP,noresonnosotros2,noresonnosotros3}
where the expected number of gold-plated $ZZ$ and $WZ$
from $VV$-fusion and $q \bar q$-annihilation 
was calculated for values of the custodial
preserving $a_3, a_4$ and $a_5$ parameters  in the 
10$^{-2}$ to 10$^{-3}$ range.
Since for values of $a_4$ or $a_5\geq 5\times10^{-3}$
\emph{unitarity violations cannot be ignored at energies beyond} 1.5 TeV,
these studies only include events in the region of low invariant mass
$V_L V_L$ pair, {\it i.e.} $M_{VV}\leq 1.5$ TeV. The rest 
of kinematical cuts are similar
to those given in Equation~\ref{cuts}. To illustrate the agreement
between these kinds of studies, we give in Table~\ref{tab:nos} 
other estimates \cite{Belyaev} of
the $a_i$ bounds attainable at the LHC.

\begin{table}[htbp]
\begin{center}
\caption{ Expected number of signal and  total (signal+background) 
gold-plated $WZ$ and $ZZ$ 
events~\cite{CMSTP,noresonnosotros2,noresonnosotros3}. 
The statistical significance 
is defined as $r = (N(a_i)-N(0))/\sqrt{N(0)}$
where $N(a_i)$ is the expected number of events for a given $a_i$. 
On the bottom right, expected limits on the chiral parameters attainable
at the LHC~\cite{Belyaev} are shown.\label{tab:nos}}
\vskip 0.2cm
\begin{tabular}{l|cccc|cccc}   \hline
 & \multicolumn{4}{c|}{$a_4$} 
& \multicolumn{4}{c}{$a_5$} 
\\  
${\cal L}= 100$ fb$^{-1}$ & $10^{-2}$ & $-10^{-2}$ &  $5 \times 10^{-3}$ & 
$-5 \times 10^{-3}$& $10^{-2}$ & $-10^{-2}$ &  $5 \times 10^{-3}$ & 
$-5 \times 10^{-3}$ 
\\ \hline

$W^{\pm}Z  \rightarrow W^{\pm}Z$ & 36 & 80 & 27 & 47 
& 22 & 58 & 23 & 41\\ 

total $W^{\pm}Z$ & 118 & 162 & 109 & 129&  104 &  139 & 105 & 122 \\ 
$r_{W\,Z}$ & 0.7 & 4.8 & 0.2 & 1.7& 0.7 & 2.6
 & 0.6 & 1.0 \\ 
$r_{WZ\,tagging}$ & 1.0 & 7.5 & 0.3 & 2.7
& 1.0 & 4.2 & 0.9 & 1.7\\ \hline

$W^+W^-\rightarrow  ZZ$ &  12  & 7 &  9 &  7 
&  21  & 7 &  13 &  6\\ 

$ZZ  \rightarrow  ZZ$ &  6 & 6 & 1 & 1 & 6 & 6 & 1 & 1\\ 

total $ZZ$ & 37  & 32 & 30 & 27& 46 & 32 & 33 & 26 \\ 
$r_{ZZ}$ & 1.9 & 0.9 & 0.5 & $\simeq$0 
& 3.8 & 0.9 & 1.2 & 0.1\\ 
$r_{ZZ\; tagging}$ & 3.5 & 1.8 & 0.9 & 0.1
& 6.6 & 1.8 & 2.3 & 0.2 \\ 
\hline 
\end{tabular}

\vspace{.3cm}

\begin{minipage}{6cm}
\begin{tabular}{l|cc}
\hline
 & \multicolumn{2}{c}{$a_3$}  \\  
${\cal L}= 100$ fb$^{-1}$ & $10^{-2}$ & $-10^{-2}$ \\ \hline
$qq'\rightarrow W^{\pm}Z$&96&139\\ 
$r_{WZ\,{tagging}}$ &1.4 &2.7 \\ \hline
\end{tabular}
\end{minipage}
\begin{minipage}{7cm}
\begin{tabular}{lc}
\hline
LHC Limits (90\% CL) & Process\\ \hline
$-0.0035 \leq a_4 \leq 0.015$
&{\normalsize  $W^\pm W^\pm,WZ,ZZ$}\\ 
$-0.0072 \leq a_5 \leq 0.013$
&{\normalsize  $W^\pm W^\pm ,WZ,ZZ$}\\ 
$-0.013 \leq a_6 \leq 0.013$
&{\normalsize  $WZ,ZZ$}\\ 
$-0.013 \leq a_7 \leq 0.011$
&{\normalsize  $WZ,ZZ$}\\ 
$-0.029 \leq a_{10} \leq 0.029$
&{\normalsize  $ZZ$}\\
\hline
\end{tabular}
\end{minipage}
\end{center}
\end{table}

It will be very difficult to detect these non-resonant signals over the
continuum background, since they just give small enhancements
in the high energy region of the $M_{VV}$ and $p_T$ distributions.
There is a general agreement that, although the
present bounds could be significantly improved, with
these non-resonant studies, the LHC would be 
hardly sensitive to values of the chiral
parameters down to the $10^{-3}$ level.
Like-sign $W^\pm W^\pm$ production may be better in these channels
\cite{BCHP,BCHPCK2}.

Obviously, these studies do not describe one of the most characteristic 
features of strong interactions: resonances.
Moreover, they are limited to moderate energies due to the 
unitarity violations mentioned already. These caveats can be overcome
by means of unitarisation procedures which we explain next.

\subsubsection{Unitarisation and resonances in the SBS}

In terms of the partial waves defined in Equation~\ref{pertamplis},
the elastic $V_LV_L$ scattering
unitarity condition, (basically, the Optical Theorem)
{\em for physical values of $s$}, is
\begin{equation}
\hbox{Im}\, t_{IJ}(s) =\mid t_{IJ}(s)  \mid ^2\quad \Rightarrow\quad
\hbox{Im}\,\frac{1}{t_{IJ}(s)}=-1,\quad \Rightarrow\quad
t_{IJ}(s)=\frac{1}{\hbox{Re}\,t^{-1}_{IJ}(s) - i}.
\label{tunit}
\end{equation}
Hence we only have to use the EChL
to approximate 
\begin{equation}
\hbox{Re}\,t^{-1}_{IJ}= (t^{(2)}_{IJ})^{-1}[1-\hbox{Re}\, 
t_{IJ}^{(4)}/t^{(2)}_{IJ}+...\,].  
\end{equation}
But since the EChL amplitudes satisfy
elastic unitarity {\em perturbatively}, {\it i.e.} 
\begin{equation}
\hbox{Im}\, t_{IJ}^{(4)}(s) = \mid t_{IJ}^{(2)}(s)  \mid ^2
\quad \Rightarrow\quad
 \frac{\hbox{Im}\, t_{IJ}^{(4)}(s)}{\mid t_{IJ}^{(2)}(s)  \mid ^2} = -1,
\label{pertunit}
\end{equation}
we find
\begin{equation}
t_{IJ}(s)=\frac{t^{(2)}_{IJ}}{1-t_{IJ}^{(4)}/t^{(2)}_{IJ}} 
\label{IAM}
\end{equation}
This is the $O(p^4)$ Inverse Amplitude Method (IAM), which
 has given remarkable results describing meson interactions, which 
have a symmetry breaking pattern almost identical to our present case
\cite{IAMpiones,IAMpiones2,IAMpiones3,IAMpiones4}. 
Note that it respects strict elastic unitarity, while keeping the
correct EChL low energy expansion. Furthermore, the extension of
Equation~\ref{IAM} to the complex plane can be justified using
dispersion theory \cite{IAMpiones,IAMpiones2,IAMpiones3,IAMpiones4}. 
In particular, it
has the proper analytical structure and, eventually,
poles in the second Riemann sheet 
for certain $a_4$ and $a_5$ values, that can
 be interpreted as resonances.
Thus, EChL+IAM formalism can 
describe resonances without increasing 
the number of parameters and respecting chiral symmetry
and unitarity.

The EChL+IAM has already been applied to the SBS \cite{Truong,Terron} 
to study some specific choices of $a_4$ and $a_5$ that mimic models
with vector or scalar resonances. The LHC sensitivity to resonances 
parametrised with  $a_4$ and $a_5$ was  first studied in \cite{Terron}
and \cite{JPR}, and more recently in \cite{jose}.
A map of 
these resonances in the $(a_4,a_5)$ space
was first obtained  in \cite{elnene}. We show in Figure~\ref{fig:chiral1}
the vector and scalar neutral resonances expected in the 
$(a_4,a_5)$ parameter space. As far as we expect $a_4$ and $a_5$ to lie  
between $10^{-2}$ and $10^{-3}$, we scan only that range.
Furthermore, the poles of the IAM amplitudes will give us the
positions and widths of the resonances. Note that, from 
Equation~\ref{pertamplis}
within our approximations, the $I=J=1$ and  $I=J=0$ channels only depend
on the $a_4-2a_5$ and $7a_4+11a_5$ combinations, respectively.
Thus the straight lines that keep these combinations constant
have the same physics in the corresponding channel.
We give several examples in the tables within the figure.
The fact that each IAM amplitude depends only on one combination
of $a_i$ implies that their mass and width are related by the 
KSFR relation \cite{KSFR,KSFR2}.
In addition, we locate five points that we will use later as illustrative examples.
The white area means that no resonances or saturation 
of unitarity is reached below $4\pi\,v\simeq3\,\hbox{TeV}$, which
we expect to be the region of applicability for our approach. 

\begin{figure}[hbtp]
\hspace{-.3cm}    \includegraphics[width=0.33\textwidth,clip]{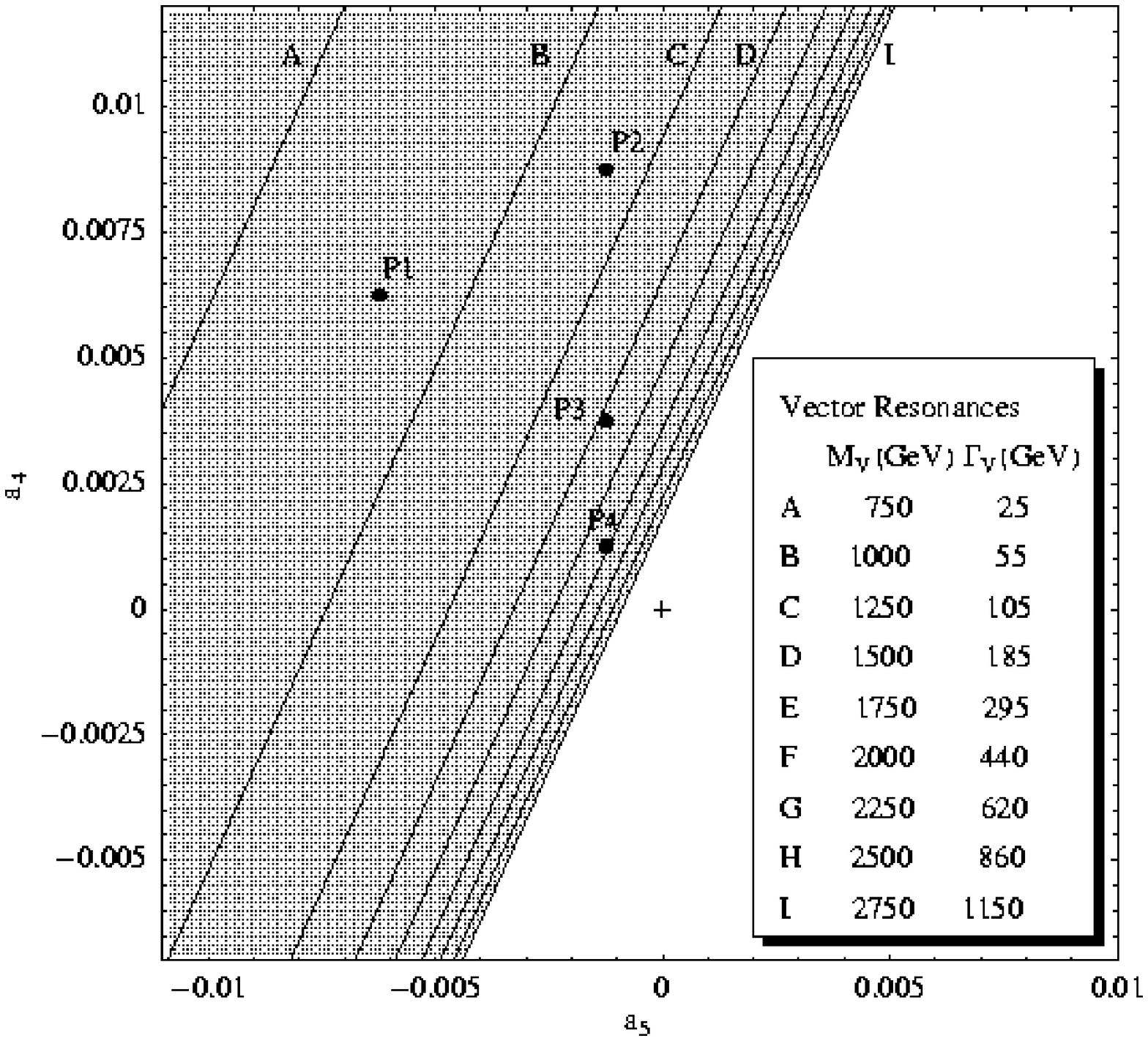}
\hspace{-.3cm}    \includegraphics[width=0.33\textwidth,clip]{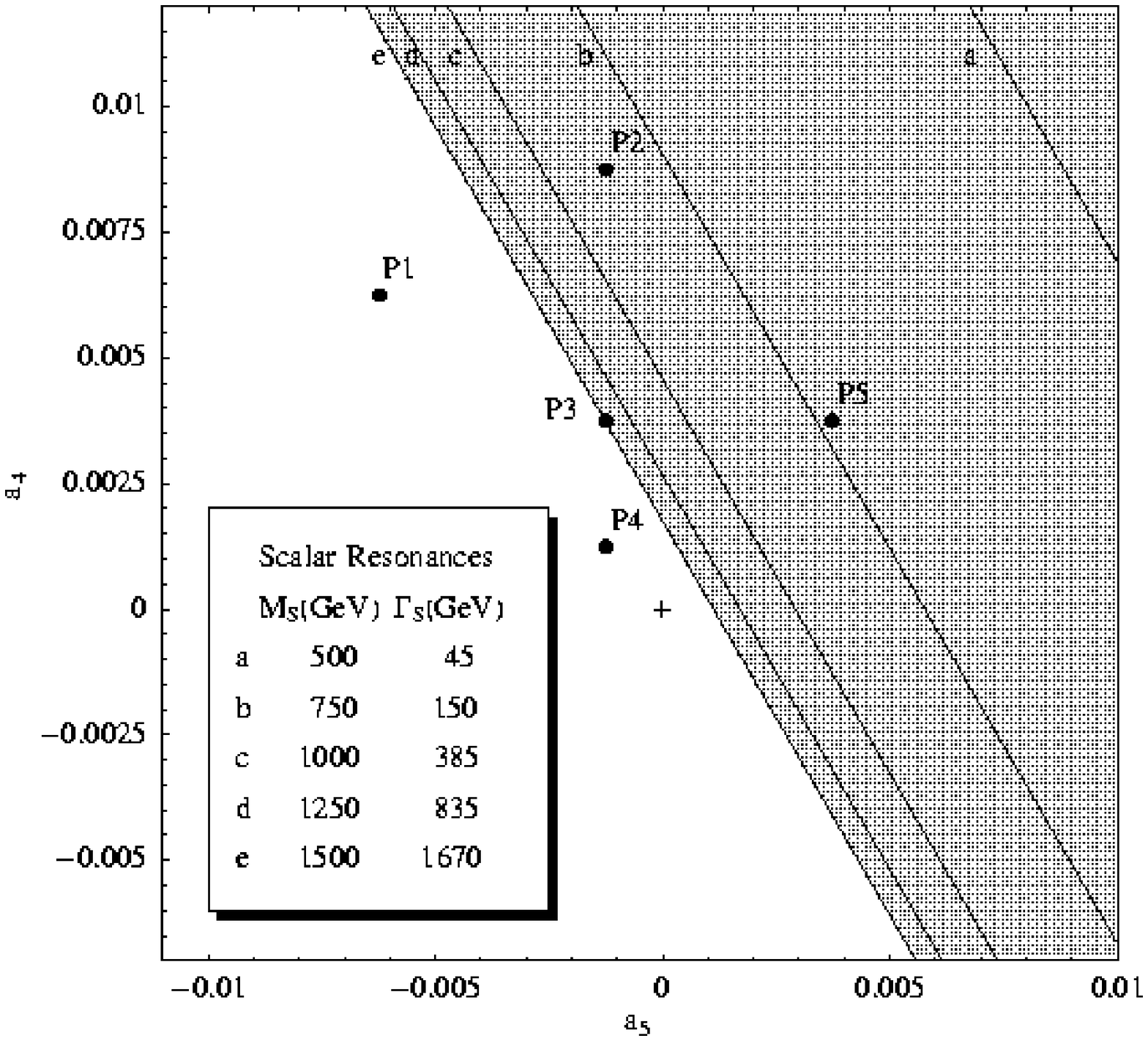}
\hspace{-.2cm}    \includegraphics[width=0.33\textwidth,clip]{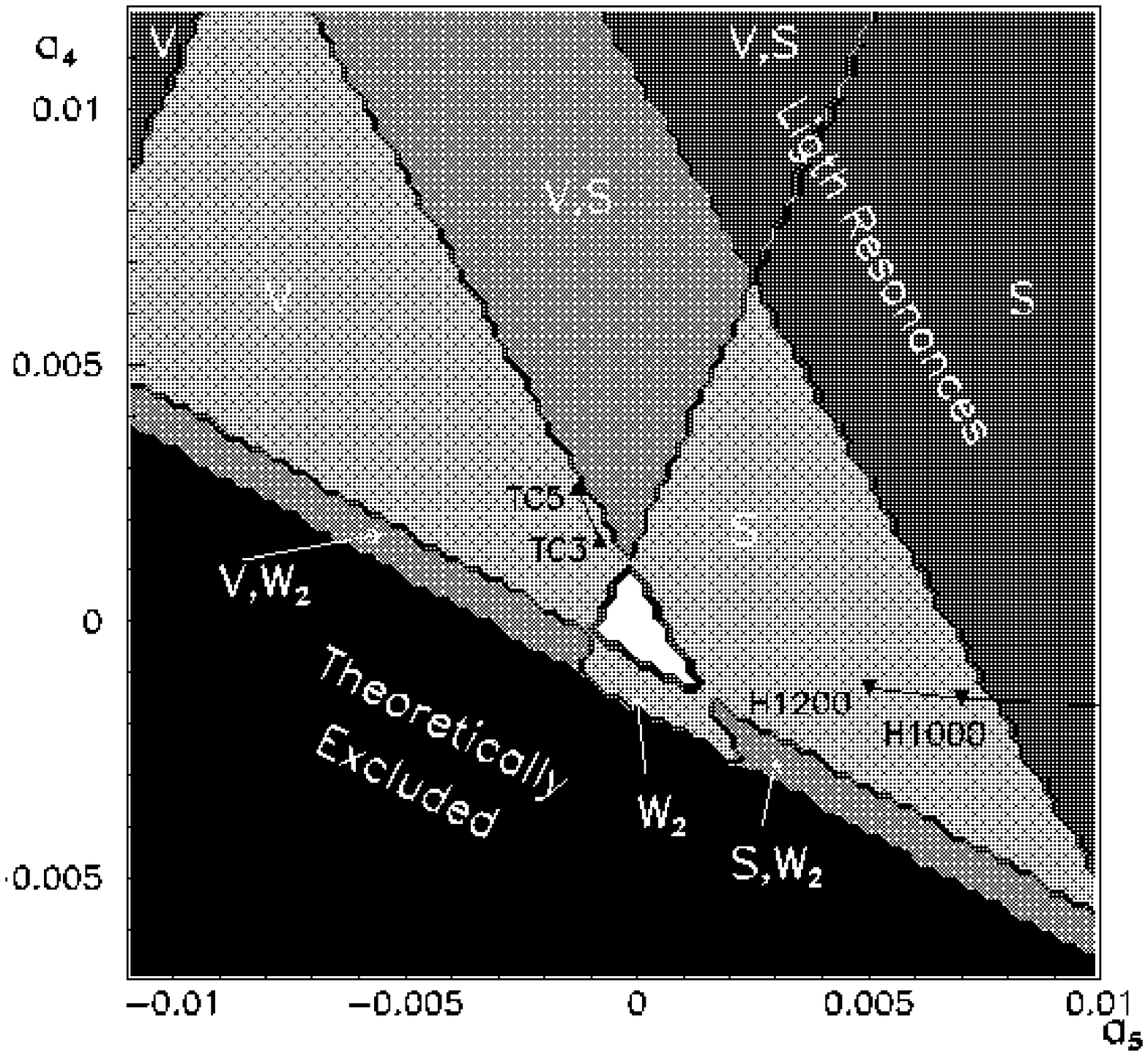}
\caption{Resonances in the $(a_4,a_5)$ space \cite{elnene}. 
In the tables
we give the resonance parameters for several lines. a) Left: Vector
resonances. The points with the same $a_4-2a_5$
have the same physics in the $I=J=1$ channel. b) Middle:
Scalar neutral neutral resonances. Those points with constant
$7a_4+11a_5$ have the same physics in this channel. c) Right:
General Resonance Spectrum of the strong SBS.
$V$ stands for vector resonances, $S$ for neutral scalar resonances 
and $W_2$, for wide
structures that  saturate the doubly charged ($I=2$) channel. 
For illustration, we have also located several simple and 
familiar models explained in the text.  \label{fig:chiral1}}
\end{figure}

We do not give results for the $I=2, J=0$ channel since 
we do not expect any heavy resonance 
with our minimal assumptions. Intuitively this occurs 
because the $I=2, J=0$ channel is
repulsive.

The general resonance spectrum of the MSISBS is gathered in
the last plot of Figure~\ref{fig:chiral1}~\cite{elnene}. Depending 
on $a_4$ and $a_5$, we find one scalar resonance ($S$), one vector
resonance ($V$), two resonances ($S,V$), a resonance and 
a doubly charged wide saturation effect ($W_2$) or even no 
resonances below 3 TeV (white area). For illustration, we have included  
points for some simple and familiar scenarios: 
minimal technicolor models with 3 and 5 technicolors
($TC3$ and $TC5$), and the heavy Higgs 
SM case, with a tree
level mass of 1000 and 1200 GeV ($H1000$ and $H1200$).
The black region is excluded
by the constraints on the $I=2, J=0$ wave \cite{elnene}.
In the dark ``Light Resonances'' areas
(lighter than 700 GeV), our 
results should be interpreted cautiously. Outside these areas,
we estimate that the predictions of Figure~\ref{fig:chiral1} are 
reliable within $\sim$~20\%  \cite{jose}.

Once we have the general spectrum, our aim is to study to what 
extent the LHC is
sensitive to different resonant scenarios via $V_LV_L$ production.
For that purpose, we cannot forget the
unitarisation of $q\bar{q}\rightarrow V_LV_L$, since we expect the
final state to re-scatter strongly, in particular
when there is a resonance in the $I=J=1$ elastic channel. This effect
can be parametrised in terms of a vector form factor, $F_V$.  Again, 
the $F_V$ obtained from the EChL
does not satisfy exactly its unitarity condition
\begin{equation}
\hbox{Im}\,F_V (s) = F_V(s) t_{11}^*(s),
    \label{Funit}
\end{equation}
which implies that 
the phases of $F_V$ and $t_{11}$ should be the same
(Watson's Final State Theorem). Moreover, the poles
of $F_V$ should be those of $t$. Hence,
we can relate the combination of $a_i$ that appears
in the perturbative expansion of $F_V$ (Equation~\ref{F2})
with $a_4-2\,a_5$.  All in all,
it is possible to unitarise $F_V$
using only the $t_{11}$ EChL result, as follows \cite{jose}:
\begin{equation}
  \label{Fradikal}
F_V\simeq\frac{1}{1-t^{(4)}_{11}/t^{(2)}_{11}}.
\end{equation}
In summary, $F_V$ is determined just by $a_4-2\,a_5$, 
and we can still use the map of resonances in Figure~\ref{fig:chiral1}.

\subsubsection{Study of the LHC sensitivity to the resonance spectrum
of the strong SBS}

We will restrict the study to 
$ZZ$ and $WZ$ production, assuming that their gold-plated decays, 
$ZZ \to 4l$ and $WZ \to l\nu \ ll$ (with $l= e,\mu$)
can be identified and reconstructed with a 100\% efficiency. 
We do not consider like-sign $W^\pm W^\pm$ production,
since, as we have seen, we do not expect $I=2$ resonances.

To evaluate $VV$ fusion processes, we use the  leading-order 
Effective-$W$ Approximation (EWA) \cite{EWA}.
 Non-fusion diagrams are not
included since they are expected to
be small in our kinematic region.
We also use the CTEQ4 \cite{CTEQ} parton distribution functions 
at $Q^2= M_W^2$ for $VV$ fusion  
and at $Q^2 = s$ for $q \bar q$ annihilation
and $gg$ fusion, with $\sqrt s$ being the 
centre of mass energy of the parton pair.
More detail can be found in  \cite{jose}.

Since we do not consider final $W$ and $Z$ decays, 
the cuts are set directly on the gauge-boson variables. 
A first criterion to enhance
the strong $V_L V_L$ signal over the background is to require
high invariant mass $M_{VV}$ and small rapidities.
We have applied the following set of minimal cuts:
\begin{equation}
500 \  {\rm  GeV} \leq \ M_{V_1 V_2} \  \leq 10 \ {\rm TeV},\quad
|y_{\rm lab}(V_1)|, \ |y_{\rm lab}(V_2)|  \ \leq \  2.5,\quad
p_T(V_1), \ p_T(V_2) \  \geq   \ 200  \ {\rm GeV},
\label{cuts}
\end{equation}
which are also required by our approximations,
mainly by the ET.
An additional invariant mass cut around 
each resonance will be imposed later.

The $ZZ$ production signal occurs through the 
$W^+_L W^-_L \to Z_L Z_L$ and
$Z_L Z_L \to Z_L Z_L$
fusion  processes.
In addition, we have included the following backgrounds
\begin{eqnarray*}
q \bar q \to Z Z,\;    (61\%),\quad
W^+ W^- \to Z Z,\;   (18\%),\quad
g g \to  Z Z,\;    (21\%)
\end{eqnarray*}
where we also give their relative contribution to the total background
with the minimal cuts.
The continuum from $q \bar q$ annihilation has only tree level SM
formulae, which is probably too optimistic
since 
the NLO QCD corrections \cite{WWOhn,WZOhn,WZit,WWit}
can  enhance significantly the tree level cross sections. 
The second  background is calculated in the SM at tree level,
with at least one transverse weak boson. 
Finally, the one-loop $g g \to  Z Z$ 
amplitude has been taken from~\cite{GG}. 

For $W^\pm Z$ final states, two processes contribute to the signal:
$W^\pm_L Z_L \to W^\pm_L Z_L$ and 
$q \bar q' \to  W^\pm_L Z_L$, whereas 
the backgrounds, calculated at tree level within the SM, are
\begin{eqnarray*}
W^\pm  Z \to W^\pm Z,  (18\%),\quad  
\gamma  Z \to W^\pm Z,  (15\%),\quad
q \bar q' \to  W^\pm Z,   (67\%).
\end{eqnarray*}
The $W^\pm  Z \to W^\pm Z$ amplitudes have
at least one transverse boson and exclude the Higgs contribution.
In the $q \bar q' \to  W^\pm Z$ background, we have
excluded the amplitude with a $V_LV_L$ pair, 
which is part of the signal.
The QCD corrections to $q \bar q'$
annihilation would give an enhancement
in both the signal and the background, so we expect that they 
will not modify considerably our estimates of the statistical significance 
of vector resonance searches.
We have not studied the 
$t \bar t$ background since it can be efficiently 
suppressed after imposing kinematic constraints and isolation cuts to
high-$p_T$ leptons~\cite{ATLASTP,CMSTP,atlas-phystdr2}.

\begin{figure}
    \includegraphics[width=0.5\textwidth,clip]{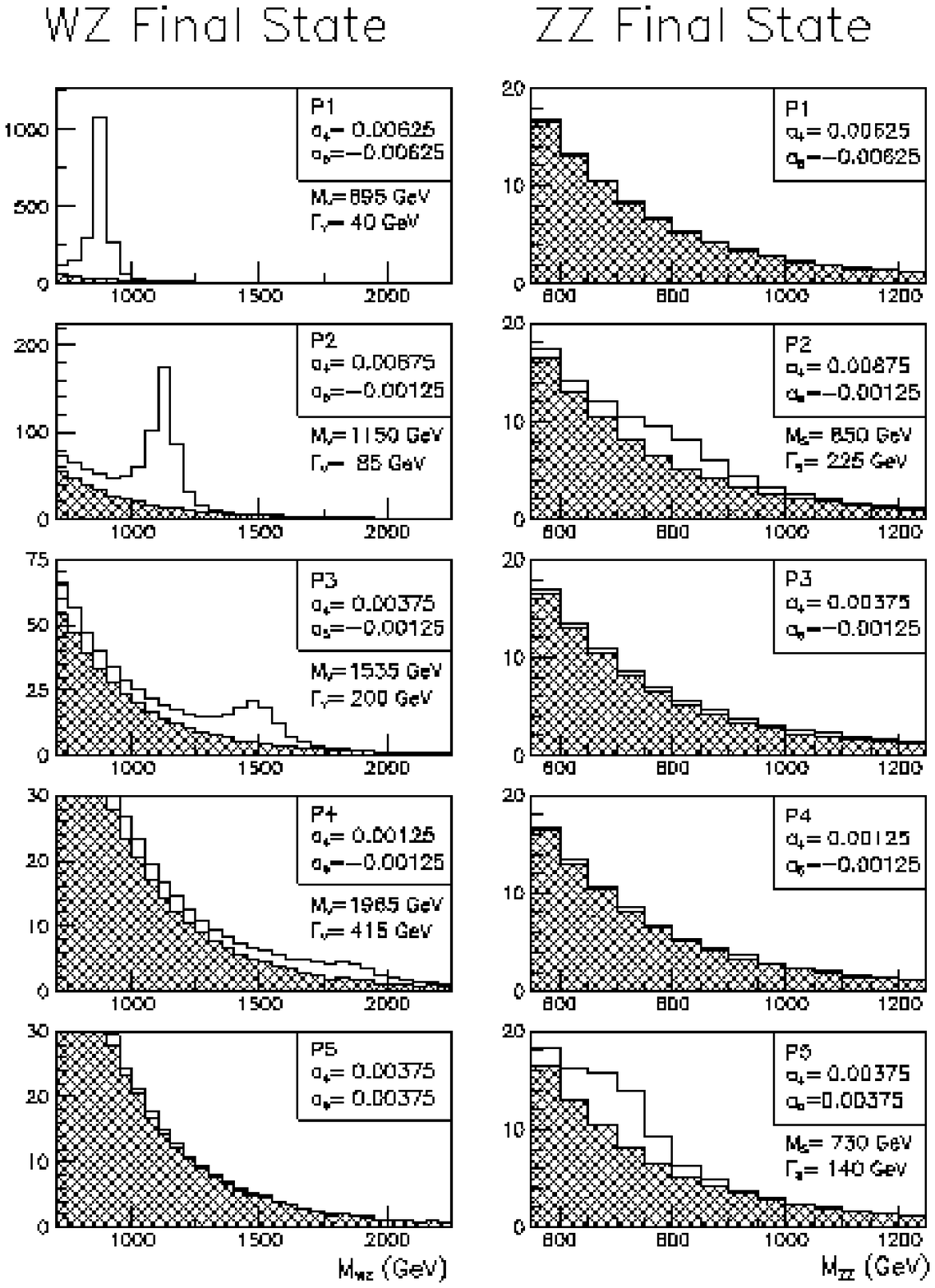}
     \includegraphics[width=0.5\textwidth,clip]{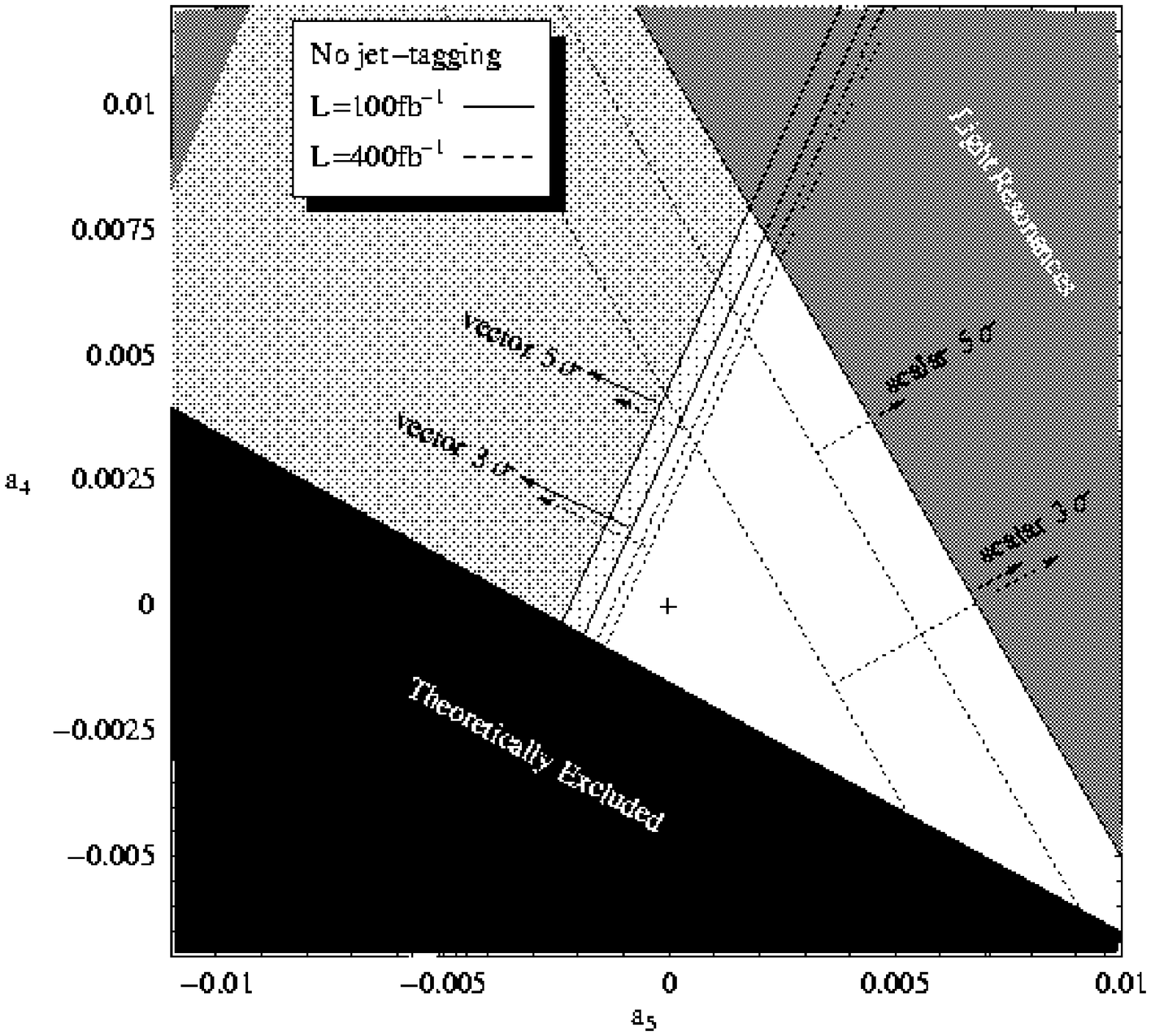}
   \caption{ a) Left: Distribution of gold-plated 
events from $WZ$ and $ZZ$ production \cite{jose}.
The shaded histogram corresponds to 
the background as described in the text.
On top of it we have plotted the signal as a white histogram. The points 
labelled $P1$ to $P5$ correspond to those in Figure~\ref{fig:chiral1} and 
are representative of cases which, from top to bottom, present:
one narrow vector resonance, a vector
and a scalar resonance, an intermediate vector resonance, a very wide
vector resonance and, finally, a ``narrow'' scalar resonance. 
b) Right: Sensitivity of the LHC to the resonance spectrum of the strong SBS
with $WZ$ and $ZZ$ gold plated events \cite{jose}.
In the $(a_4,a_5)$ parameter space, we show the $3\sigma$ and 
$5\sigma$ reach with an integrated luminosity of 100 fb$^{-1}$ (solid 
lines limiting the shaded areas)
and 400 fb$^{-1}$ (dashed lines), both for scalar and vector resonances.
    \label{fig4}}
  \end{figure}

For illustrative purposes, let us first concentrate
on the five representative
points given in Figure~\ref{fig:chiral1}.
Points~1, 3~and~4 represent models containing 
a $J=I=1$ resonance with masses in the range 900-2000~GeV.
Point~5 represents a model with a scalar resonance with  mass
730~GeV and a width of 140~GeV. Finally, point~2 
represents both a scalar and a vector resonance.
The $M_{VV}$ distributions for these five models are shown
in Figure~\ref{fig4}, where we have plotted the signal on top
of the background for gold-plated $ZZ$ and $WZ$ events, assuming
an integrated luminosity of 100 fb$^{-1}$.
The vector resonances in points 1 to 4 can be seen as peaks in
the distribution of final $WZ$ pairs.
The scalar resonances in points 2 and 5 give small enhancements
of $ZZ$ pairs. Note that as both $a_4$ and $a_5$ tend to 0,
the resonances become heavier and broader, yielding a
less significant signal. It seems evident 
that it will be much harder to detect scalar than
vector resonances. The reasons are that 
scalars are wider, they are not
 produced with a significant rate from $q \bar q$ annihilation,
and there is a smaller rate of $ZZ$ production from $VV$~fusion.
Furthermore, the  $ZZ$ branching ratio to leptons is smaller 
that that of $WZ$.

The contributions to signal and background 
for $WZ$ and $ZZ$ production at these representative points
are given in Table~\ref{tab:tab1}.
In order to enhance the signal to background ratio,
we have optimised the $M_{VV}$ cut, keeping events
within approximately one resonance width around
the resonance mass (see the second column of these tables).
From the $WZ$ results, it is clear that
the LHC will have a very good sensitivity to
light vector resonances, due to the $q \bar q'$-annihilation, which 
dominates by far the $VV$-fusion process.
As the vector resonance mass increases, the $q\bar q$ contribution  
is damped faster than that of $VV$ fusion, and both 
signals become comparable for vector masses around 2 TeV.
Let us remark that, in $ZZ$ production, there is only strong interaction
signal in $VV$ fusion, and therefore to tag forward jets
is always convenient in this final state in order to reject non-fusion 
processes. This is not the case, however, for vector resonance searches
since it is mostly due  to $q \bar q$ annihilation.
In these tables, we have also estimated the statistical significance, 
$\hbox{Signal}/\sqrt{\hbox{Bkgd}}$, assuming integrated luminosities
of 100 and 400~fb$^{-1}$. In $ZZ$ final states, we also give the 
significance assuming perfect forward jet-tagging.

\begin{table}[h]
\vspace{-0.3cm}
\caption{Expected number of signal and background gold-plated  $VV$
events at the LHC with ${\cal L} = 100 fb^{-1}$.
a) Top: For $W^\pm Z$ final state and four different
$(a_4,a_5)$ values representing vector resonances. 
b) Bottom:  For $ZZ$ and two representative
$(a_4,a_5)$ values with scalar resonances.
The statistical significance is also given for ideal forward jet-tagging.
\label{tab:tab1}}
\begin{center}
{\footnotesize
\begin{tabular}{lccccccccc}
\hline
$M_V$, $\Gamma_V$ (GeV) & Cuts: & 
Signal & Signal & Signal  &  Bkgd & Bkgd & Bkgd &
$S/\sqrt{B}$ & $S/\sqrt{B}$  \\
$\hspace{4mm} (a_4, a_5)\times 10^3$  & $(M_{VV}^{min}, M_{VV}^{max})$ & 
Fusion & $q \bar q$ & Total  &  Fusion & $q \bar q$ & Total  &
100 fb$^{-1}$  & 400 fb$^{-1}$ \\ \hline
\begin{tabular}{r}
$P1$:  894, 39 \\
 (-6.25,6.25) \end{tabular}
& (700,1000) & 123 & 1630 & 1743 & 74 & 150 & 224 & 116 & 232 \\
\begin{tabular}{r}
$P2$:  1150, 85 \\
 (-1.25,8.75) \end{tabular}
& (900, 1300) & 65 & 369 & 434 & 50 & 84 & 134 & 37 & 75 \\
\begin{tabular}{r}
$P3$:  1535 , 200 \\
 (-1.25,3.75) \end{tabular}
& (1250, 1700) & 24 & 56 & 80 & 21 & 27 & 48 & 11 & 23 \\
\begin{tabular}{r}
$P4$:  1963 , 416 \\
 (-1.25,1.25) \end{tabular}
& (1500, 2350 ) & 10 & 12 & 22 & 14 & 16 & 30 & 4 & 8\\
\hline
\end{tabular}
}

\vspace{.1cm}

{\footnotesize \setlength{\tabcolsep}{1.83mm}
\begin{tabular}{lccccccccc}
\hline
$M_S$, $\Gamma_S$ (GeV) & Cuts: & 
Signal &  Bkgd & Bkgd & Bkgd & Bkgd &
$S/\sqrt{B}$ & $S/\sqrt{B}$  & $S/\sqrt{B}$  \\
$\hspace{4mm} (a_4, a_5)\times 10^3$  & $(M_{VV}^{min}, M_{VV}^{max})$ & 
Fusion &  Fusion &  $ gg $ & $q \bar q$ & Total  &
100 fb$^{-1}$  & jet-tagging & 400 fb$^{-1}$ \\ \hline
\begin{tabular}{r}
$P2$:  850, 225 \\
 (-1.25,8.75) \end{tabular}
& (600, 1050) & 15 & 10 & 11 & 34 & 55 & 2 & 5 & 4 \\
\begin{tabular}{r}
$P5$:  750 , 140 \\
 (3.25,3.75) \end{tabular}
& (550, 900) & 21 & 10 & 14 & 39 & 63 & 3 & 6 & 5 \\
\hline
\end{tabular}
}
\end{center}
\end{table}

Finally, we also show in Figure~\ref{fig4} the regions of the 
$(a_4,a_5)$ space accessible at the LHC,
giving 3 and 5$\sigma$ contours and assuming integrated luminosities of
100 and 400 fb$^{-1}$. In terms of resonance mass
reach limits, we find that with 100 fb$^{-1}$,
scalar resonances could be discovered (5$\sigma$)
in gold-plated $ZZ$ events up to a mass of 800 GeV 
with forward jet-tagging. Vector resonances could be discovered
using gold-plated $WZ$ events up to a mass of 1800 GeV. 
These numbers are in good agreement with more 
realistic studies \cite{ATLASTP,CMSTP,atlas-phystdr2} of particular cases.
We can also see that there is a central region in the
$(a_4, a_5)$ space that does not give significant
signals in gold-plated $ZZ$ and $WZ$ events. This region corresponds
to models in which either the resonances are too heavy or there 
are no resonances in the SBS and the scattering
amplitudes are unitarised smoothly.
It is a key issue as to whether this type of non-resonant 
$V_LV_L$ signal could be probed at the LHC.
It has been argued that doubly-charged $WW$ production
could be relevant to test this non-resonant region.
But non-resonant $VV$ distributions would only have slight 
enhancements at high energies, and a very accurate
knowledge of the backgrounds and the detector performance
would be necessary in order to establish their existence.


\subsubsection{Appendix}

\begin{table}[h]
\vspace{-1cm}

\caption{Relation between different notations in the literature. 
 \label{tab:notations}}

\vspace{.2cm}

{\footnotesize
\begin{tabular}{lccccccccccc}
\hline
Ours \cite{HeRu,HeRu2}&$a_0$ &$a_1$ &$a_2$ &$a_3$ &$a_4$ &$a_5$&$a_6$&$a_7$ &$a_8$ &$a_9$ &$a_{10}$ \\
App.\& Longh. \cite{ApBe80,longhitano:81}& $\frac{g^2}{g'^2}\beta_1$ & $\frac{g}{g'}\alpha_1$ 
& $\frac{g}{g'}\alpha_2$ & $-\alpha_3$&
 $\alpha_4$ & $\alpha_5$ & $\alpha_6$ & $\alpha_7$ 
& $-\alpha_8$ & $-\alpha_9$ & $\frac{1}{2}\alpha_{10}$ \\
 S.Alam \cite{tril,vanderBij:1999fp}& $\frac{1}{g'^2}\beta_1$ &  $\alpha_1$ &  $\alpha_2$  & $-\alpha_3$ &  $\alpha_4$ & 
 $\alpha_5$ &  $\alpha_6$ &  $\alpha_7$ &  $-\alpha_8$ &  $-\alpha_9$ &  $\frac{1}{2}\alpha_{10}$\\
He {\it et al.} \cite{noreson,noreson2}&$\frac{l_0}{16\pi^2\,g'^2}$ &$\frac{l_1}{16\pi^2}$ &$\frac{l_2}{16\pi^2}$ &
$\frac{l_3}{16\pi^2}$ &$\frac{l_4}{16\pi^2}$ &$\frac{l_5}{16\pi^2}$ &$\frac{l_6}{16\pi^2}$ &
$\frac{l_7}{16\pi^2}$ & $\frac{-l_8}{16\pi^2}$ &$\frac{-l_9}{16\pi^2}$ &
$\frac{-l_{10}}{32\pi^2}$ \\
Vertex & 2 & 2,3 & 3 &  3,4 & 4 & 4 & 4 &4 & 2,3,4 & 3,4 & 4 \\
$SU(2)_{L+R}$& no &no&no&yes&yes &yes &no &no& no&no&no\\
\hline
\end{tabular}
}
\end{table}

\subsection{Vector-boson scattering
}

\def\pt{p_T}
\def\emiss{E_T^{miss}}
\newcommand{\WpWp} {\mbox{$W_L^+W_L^+$}}
\newcommand{\WmWm} {\mbox{$W_L^-W_L^-$}}
\def\rhot{$\rho_T$}
\def\etal{{\it et al.},}

The search for a fundamental scalar particle which would be 
responsible for electroweak symmetry breaking has so far proven 
unsuccessful. While the existence of a light Standard Model (SM) Higgs 
alone would be consistent 
with all precision electroweak measurements, the well known hierarchy 
problems~\cite{SMDEF} make the theory unsatisfactory. The model makes 
{\it ad hoc} assumptions about the shape of the potential, 
responsible for electroweak symmetry breaking, and provides no 
explanation for the values of the parameters. Although supersymmetry 
is an appealing alternative, no indication exists, yet, of its 
validity. Therefore, in the absence of a low mass Higgs particle, a strongly 
coupled theory must be considered. The study of electroweak 
symmetry breaking will require measurements of the production rate of 
pairs of longitudinal gauge-bosons, since they are the Goldstone 
bosons of the symmetry breaking process. It will also be essential to 
search for the presence of resonances which regularise the vector-boson 
scattering cross-section. Scalar resonances occur in models 
with a heavy SM Higgs boson, and vector resonances, in charged or 
neutral channels, are also predicted in dynamical theories, such as 
technicolor. 

In this section, 
different channels for scattering of high energy gauge-bosons 
at the LHC are considered
These include heavy Higgs production and resonant $WZ$ as 
well as non-resonant $WZ$ and $W^+W^+$ production in the Chiral 
Lagrangian model. High mass gauge-boson pair production in a 
multi-scale technicolor model is also examined. The possibility of 
making such measurements at the LHC is evaluated. 
  
\subsubsection{Heavy Higgs signal \label{sec:vv_heavyhiggs}}

It is now generally believed that a SM Higgs should be 
light, its mass being bound by requirements of vacuum stability and 
by the validity of the SM to high scales 
in perturbative calculations~\cite{QUIROS}.
The parameters of the Higgs used in this study were calculated 
at tree level.
One should note that in NNLO, the resonance saturates
\cite{19-3}. Nevertheless, the search for such a resonance 
at the LHC can serve as a testing ground for the measurement of the 
production of high mass longitudinal gauge-boson pairs or for the 
search of a generic resonance.
The $H\rightarrow WW \rightarrow l\nu jj$ channel is 
presented in this section as an example of a typical analysis of a 
heavy Higgs signal.
In fact, $V_LV_L$ fusion is also detectable in the case of a heavy 
Higgs resonance, through the processes $H\rightarrow ZZ$, up to 
$M_H \sim 800$~GeV. Simultaneous detection of a heavy Higgs in other 
signals would not only confirm the discovery but also provide 
additional information on the Higgs couplings, which are essential 
for determining the nature of the resonance.

\paragraph{\boldmath $H\rightarrow WW \rightarrow l\nu jj$}

In the vector-boson fusion process of Higgs production, $qq 
\rightarrow qqH$, the rate for this channel is sufficient to be 
observed at low luminosity with a very distinctive signature 
\cite{19-58,19-61,bib:savard}:
\begin{itemize}
\item{A high-$p_T$ central lepton ($|\eta_l| <$2).}
\item{A large $\emiss$.}
\item{Two high-$p_T$ jets from the $W\rightarrow jj$ decay in the 
central region and close-by in space ($\Delta R \sim 0.4$) arising from 
the large boost of the $W$~boson.}
\item{Two tag jets in the forward regions ($|\eta_j| > 2$).}
\item{No extra jet in the central region (central jet veto).}
\end{itemize}
The main backgrounds are:
\begin{itemize}
\item{$W$+jet which gives the largest contribution but also suffers 
from significant theoretical uncertainties due to higher-order 
corrections \cite{19-59}.}
\item{$t\bar{t} \rightarrow l\nu b~jj\bar{b}$, with the presence 
of a real $W\rightarrow jj$ decay, but also additional hadronic 
activity from the $b$-jets in the central region.}
\item{$WW\rightarrow l\nu jj$ continuum production, which has a 
much lower rate but is irreducible in the central region.}      
\end{itemize}
In addition to central jet veto and forward tag jets cuts, other cuts 
(high-$p_T$ cuts) have been used to optimise the statistical 
significance of the signal. They are:
\begin{itemize}
\item{Lepton cuts: $p_T^l$, $\emiss >$~100~GeV, $p_T^{W 
\rightarrow l\nu}>$~350~GeV.}
\item{Jet cuts: two jets reconstructed within $\Delta R = 0.2$ with 
$p_T>$~50~GeV and  $p_T^{W \rightarrow jj}>$~350~GeV.}
\item{$W$ mass window: $m_{jj} = m_W\pm 2\sigma$, where $\sigma$ is the 
resolution on $m_{jj}$.}
\end{itemize}
Table~\ref{tab:HWWACC} shows the number of events resulting from this 
selection, for an integrated luminosity of 30~fb$^{-1}$, for 
$M_H=$~1~TeV and $M_H=$~800~GeV as  evaluated with the ATLAS fast 
simulation program ({\tt ATLFAST}, \cite{richter-98}).
A significant signal remains above 
background. Variation of the 
$E_{tag}$ cut provides the possibility to compare the shape and cross 
section of the resonance production to the expected parameters of the 
Higgs signal (see Figure~\ref{fig:HWWM}). 
\begin{table}[htbp]
\begin{center}
\caption{$H\rightarrow WW \rightarrow l \nu jj$ with $M_H=$~1~TeV 
and $M_H=$~800~GeV and $\cal L =$30~fb$^{-1}$. Accepted signal and 
background events after high-$p_T$ cuts, central jet veto and a 
double forward tag with E$_{tag}>$~300~GeV.}
\label{tab:HWWACC}
\vskip0.2cm
\begin{tabular}{lccccc}
\hline
                    & Higgs  & $t\bar{t}$    & $W$+jets         &  $WW$   
         & $S/\sqrt{B}$  \\
                    & signal &($p_T>$~300~GeV) &($p_T>$~250~GeV) & 
($p_T>$~50~GeV) &               \\
\hline
$M_H=$~1~TeV        & 37.9   &  3.3           &  9.2           &   
1.0          &  10.3         \\
$M_H=$~800~GeV      & 43.5   &  3.3           &  9.2           &   
1.0          &  11.8         \\
\hline
\end{tabular}
\end{center}
\end{table}   
\begin{figure}
\begin{center}
\includegraphics[width=6cm]{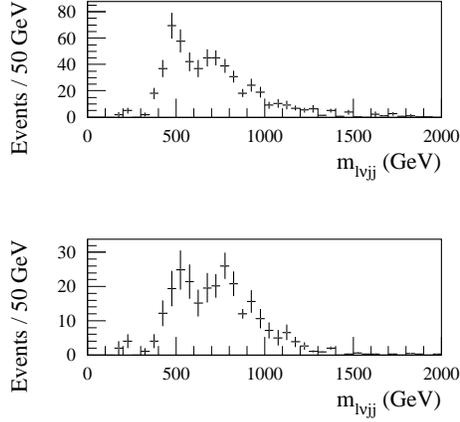} 
\caption{$m_{l\nu jj}$ distribution for the summed signal+background 
obtained with $M_H=$~800~GeV and $\cal L =$~30~fb$^{-1}$ after 
requiring two tag jets with  E$_{tag}>$~200~GeV (top) and 
E$_{tag}>$~400~GeV (bottom) \cite{bib:savard}.}
\label{fig:HWWM}
\end{center}
\end{figure}

The $H\rightarrow ZZ\rightarrow ll\nu\nu$ and  
$H\rightarrow ZZ\rightarrow lljj$ channels in ATLAS have also 
been studied \cite{19-55,19-56,19-61} over most of the mass range 
from 300~Gev to 1~TeV. It has been shown that forward jet tagging ($2 
< | \eta_j | < 5$),  is a powerful method for rejecting background 
and selecting $qq\rightarrow qqH$ production, {\it i.e.} the vector-boson 
fusion process.   

\subsubsection{Strong vector-boson scattering}

\paragraph{Chiral Lagrangian model}
In the  Chiral Lagrangian model~\cite{Pelaez},  the  form of  the 
Lagrangian is only constrained by symmetry considerations which  are 
common to any strong electroweak symmetry  breaking  sector.  
Differences among  underlying theories appear  through the  values of 
 the parameters of  the Chiral Lagrangian.  Within the  chiral 
approach, the low-energy  Lagrangian is built   as an expansion   in   
derivatives  of  the  Goldstone  boson fields. There is  only one 
possible  term  with two derivatives  which respects $SU(2)_{L+R}$ 
symmetry: 
   \[ {\cal L}^{(2)} = \frac{v^2}{4}  {\rm Tr}(D_{\mu}U D^{\mu}U^{\dagger}) 
\]
where $D_{\mu} U$ = $\partial_{\mu} U$ - $W_{\mu} U + U B_{\mu}$, $W_{\mu} = 
- i g \sigma^{a} W_{\mu}^{a} /2$, $B_{\mu} =  i g \sigma^{3} B_{\mu} 
/2$.

The dependence on the different models appears at next order through 
two phenomenological parameters $L_1$ and $L_2$:
  \[  {\cal L}^{(4)} = L_{1} ({\rm Tr}(D_{\mu}U D^{\mu}U^{\dagger}))^{2} + 
L_{2} ({\rm Tr}(D_{\mu}U D^{\nu}U^{\dagger}))^{2} \] 

 The $SU(2)_{L+R}$ symmetry allows us to define a weak isospin $I$. The 
$W_{L}W_{L}$ scattering can then be written in terms of isospin 
amplitudes, exactly as in low energy hadron physics. We assign 
isospin indices as follows:
\[    W_{L}^{a} W_{L}^{b} \rightarrow W_{L}^{c} W_{L}^{d} \]
where $W_{L}$ denotes either $W_{L}^{\pm}$ or $Z_{L}$, where 
$W_{L}^{\pm} = (1/\sqrt{2})$ $(W_{L}^{1} \mp i W_{L}^{2})$ and $Z_{L} 
= W_{L}^{3}$. The scattering amplitude is given by:
  \[ {\cal M}(W_{L}^{a} W_{L}^{b} \rightarrow W_{L}^{c} W_{L}^{d}) 
\equiv A(s,t,u) \delta^{ab} 
\delta^{cd}+A(t,s,u)\delta^{ac}\delta^{bd}+A(u,t,s)\delta^{ad}\delta^{
bc} \]
where $a,b,c,d$ =1,2,3 and $s,t,u$ are the usual Mandelstam  
kinematical variables.

 In this approach it is possible to compute the function $A(s,t,u)$ 
in ${\cal O}(p^{4})$ \cite{21-5,21-5_2}:
\begin{eqnarray*}
A(s,t,u) &  = & \frac{s}{v^{2}} + \frac{1}{4 \pi v^{4}} (2 L_{1} 
s^{2}  + L_{2} (t^{2}+u^{2})) \\
         &    &  + \frac{1}{16\pi^{2}v^{4}} \left( -\frac{t}{6} 
(s+2t)\log(-\frac{t}{\mu^{2}}) 
-\frac{u}{6}(s+2u)\log(-\frac{u}{\mu^{2}}) 
-\frac{s^{2}}{2}\log(-\frac{s}{\mu^{2}}) \right) 
\end{eqnarray*}
 The values of $L_{1}$ and $L_{2}$ depend on the model, but are 
expected  to be in the range $10^{-2}$ to $10^{-3}$.  

The usual Chiral Lagrangian approach does not respect unitarity at 
high energies. 
The Inverse Amplitude Method (IAM) \cite{IAMpiones,IAMpiones2,Pelaez}, which 
is based on the assumption that the inverse of the amplitude has the 
same analytic properties as the amplitude itself, has been very 
successful at describing low energy hadron scattering. The most 
interesting feature of this approach is that it allows us to describe 
different reactions by using only the two parameters $L_{1}$ and 
$L_{2}$.

In analogy to $\pi\pi$ scattering, there are three possible isospin 
channels $I$ = 0,1,2. At low energies, the states of lowest momentum $J$ 
are the most important, and thus only the $a_{00}$, $a_{11}$ and 
$a_{20}$ partial waves are considered. It is possible to reproduce, 
with the IAM model, the broad Higgs-like resonance in ($I,J$) = (0,0) 
channel as well as resonant and non-resonant scattering in the 
channel (1,1) by selecting appropriate values for $L_{1}$ and 
$L_{2}$. It has been shown \cite{elnene} that in the ($I=1, J=1$) 
channel there may exist narrow resonances up to 2500~GeV and this 
scattering only depends on the combination of ($L_2-2L_1$).
   
\paragraph{Resonant \boldmath $W_LZ_L \rightarrow W_LZ_L$ 
\unboldmath channel \label{subsection:WZRES}}

As a reference for the IAM model, the process $W_LZ_L \rightarrow 
W_LZ_L$, with $Z \rightarrow ll$ ($l=e, \mu$) and $W 
\rightarrow jj$ is used \cite{19-130}. A modified version of 
{\tt PYTHIA~5.7} was 
used to generate $V_LV_L$ scattering processes for each value of $L_1$ 
and $L_2$. The simulation was done for two values of ($L_2 - 2L_1$) = 
0.006 and 0.01, which yield $\sigma \times BR$ of 1.5~fb and 2.8~fb, 
with mass peaks at 1.5~TeV and 1.2~TeV respectively.

Irreducible background arises from continuum $WZ$ production and the 
main QCD background is from $Z$+jets production with two final state 
jets faking the $W$ decay if their invariant mass is close to $m_W$. 
$t\bar{t}$ production is potentially dangerous but is 
efficiently suppressed by a cut on the invariant mass of leptons from 
the $W$ decay \cite{19-130}. The following cuts were used for background 
rejection:
\begin{itemize}
\item{Two isolated leptons with the same flavour and opposite charges 
in the region $| \eta |<2.5$ and $p_T>100$~GeV. Their invariant mass 
was required to lie in the region $|m_{ll}-m_Z|<6$~GeV.}
\item{Jets were reconstructed in a cone of width $\Delta R=0.2$. Only two 
jets with $p_T>50$~GeV were allowed in the central region ($| \eta 
|<2$)  and  $|m_{jj}-m_W|<15$~GeV was required. Only $W$ and $Z$ with 
$p_T>200$~GeV were kept.}
\item{In the forward region ($2 <| \eta |<5$), jets were reconstructed 
in a cone of width $\Delta R=0.5$ and events were accepted only 
if jets with $p_T>30$~GeV and $E_{jet}>500$~GeV were present in each 
hemisphere.}
\end{itemize} 
The expected number of signal and background events after all cuts 
and for ${\cal L} = 100$~fb$^{-1}$ are presented in 
Table~\ref{tab:WZRES}. The mass spectra obtained after  all cuts 
(Figure~\ref{fig:WZRES}) shows a clear peak with 
a width of 75~GeV (100~GeV) 
for the 1.2~TeV (1.5~TeV) resonance and  14 (8) 
signal events in the window $|m_{WZ}-m_V|<2\sigma$. The contribution 
from irreducible backgrounds is negligible and is below 0.05 events 
inside the mass window. It is clear that such a narrow resonance 
could be detected easily after a few years of high luminosity. 

\begin{table}[htbp]
\begin{center}
\caption{Number of signal and background events after all cuts for 
${\cal L} = 100$~fb$^{-1}$ with ($L_2 - 2L_1$) = 0.01 and 0.006, 
corresponding to $m_V=1.2$~TeV and $m_V=1.5$~TeV respectively.}
\vskip0.2cm
\begin{tabular}{lcccc}
\hline
                 & \multicolumn{2}{c}{$M_V$=1.2~TeV}  &   
\multicolumn{2}{c}{$M_V$=1.5~TeV}  \\
   Cuts              &       $W_LZ_L$  &   $Z$+jets         &     $W_LZ_L$  
&   $Z$+jets             \\
\hline
Central jets cut &          284    &  2187            &     145       
&  1781                \\
$m_{jj} = m_W \pm 15$~GeV & 101    &  154             &     46        
&   82                 \\
Leptonic cuts    &          70     &  84              &     36        
& 47                   \\
Forward jet tagging&        14     &  3               &     8         
& 1.3                  \\
\hline
\end{tabular}
\label{tab:WZRES}
\end{center}
\end{table}
         
\begin{figure}[htbp]
\begin{center}
\includegraphics[width=8cm]{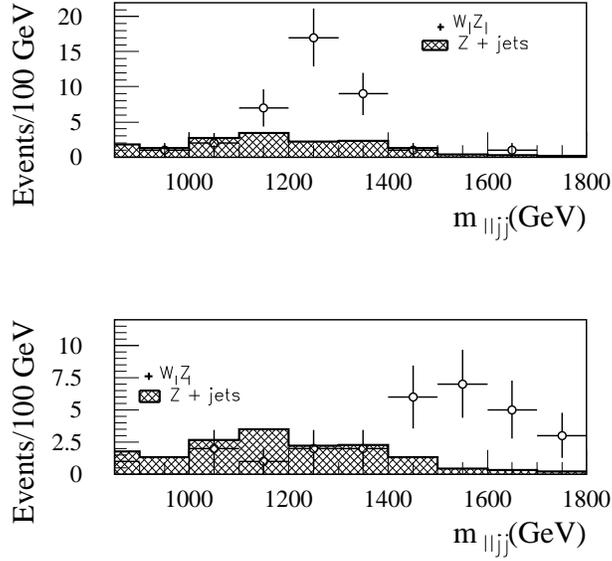} 
\caption{Reconstructed distribution of the $WZ$ system for 1.2~TeV and 
1.5~TeV resonances and $\cal L$~=~300~fb$^{-1}$.}
\label{fig:WZRES}
\end{center}
\end{figure}
\vspace{-0.5cm} 

\paragraph{Non-resonant channels \label{sec:vv_nonreschan}}

If nature does not provide resonances in $V_LV_L$ scattering, the 
measurement of cross sections at high mass for non-resonant channels 
becomes the only probe for the mechanism of regularisation of the 
cross section. It would then be essential to understand very well the 
magnitude and energy dependence of backgrounds. Those channels can be 
particularly important since it has been shown that a complementary 
relationship exits between resonant and non-resonant processes 
\cite{BCHPCK2,BCHP,19-133}. Both $W_LZ_L$ and $W_LW_L$ scattering 
have been studied within the ATLAS framework.
  
\subparagraph{\boldmath $W_LZ_L \rightarrow W_LZ_L$ \unboldmath}

The non-resonant $W_LZ_L \rightarrow W_LZ_L$ process, with 
$Z\rightarrow ll$ and $W\rightarrow l\nu$ ($l=e,\mu$),  was 
incorporated in {\tt PYTHIA} and used with two values of $L_1$: 0.003 and 
0.01, leading to $\sigma \times BR =$ 0.19~fb and 0.11~fb 
respectively. The main features of the signal are:
\begin{itemize}
\item{The presence of two high-$p_{T}$ leptons of same flavour and 
opposite charge in the barrel region, having an invariant mass  
consistent with the mass of the $Z$ boson.}
\item{One additional high-$p_{T}$ lepton in the barrel region.} 
\item{Significant missing momentum in the event due to the presence 
of a neutrino.}
\item{The presence of energetic jets in the forward region.}
\end{itemize}
The main irreducible background, coming from continuum $WZ$ production, 
was generated by {\tt PYTHIA} with $\sigma \times BR =$ 13.5~fb. The main 
reducible background is the QCD process $Zt\overline{t}$ where one of 
the $W$ bosons from a $t$-quark decays into a lepton and an anti-neutrino. 
The value of $\sigma \times BR$ of this process is 26.3~fb. A less 
important contribution comes from $ZZ$ production with $\sigma \times 
BR =$ 1.52~fb. These different backgrounds were rejected with a high 
efficiency by using the following cuts:
\begin{itemize}
\item{Two isolated leptons of same flavour and opposite charge were 
required in the central region with $p_T>30$~GeV and invariant mass 
satisfying $|m_{ll}-m_Z|<6$~GeV. One additional lepton was required.} 
\item{A missing momentum of at least 75~GeV.}
\item{At least one jet with $p_T>40$~GeV and $E_{jet}>500$~GeV should 
be present in the forward region.}        
\end{itemize}
In order to analyse $WZ$ scattering in the high-mass region, the 
transverse mass $M_T$ 
$$
M^2_T = \left[ \sqrt{M^2(lll) + p^2_T(lll)} + \mid \not{\!p}_T \mid 
\right]^2 - \left[ \vec{p}_T(lll) + \not{\!\vec{p}}_T \right]^2
$$
was used. $M(lll)$ and $p_T(lll)$ are the invariant mass and 
transverse momentum of the three charged leptons and $\not{\!p}_T$ is 
the missing momentum in the event. The transverse mass $M_{T}$ 
distribution for the $W_{L}Z_{L}$ scattering and for $Zt\overline{t}$ 
background, after the application of cuts, is shown in 
Figure~\ref{fig:WZ}. The number of signal and background events with 
the invariant mass of $WZ$ system larger then 600 GeV for an integrated 
luminosity of ${\cal L} =500$ fb$^{-1}$ and applying 
different cuts, are shown in Table~\ref{tab:WZ}. The $ZZ$ 
background is not shown since it is effectively removed by the 
requirement of missing transverse momentum.

\begin{figure}[htbp]
\begin{center}
\includegraphics[width=6cm]{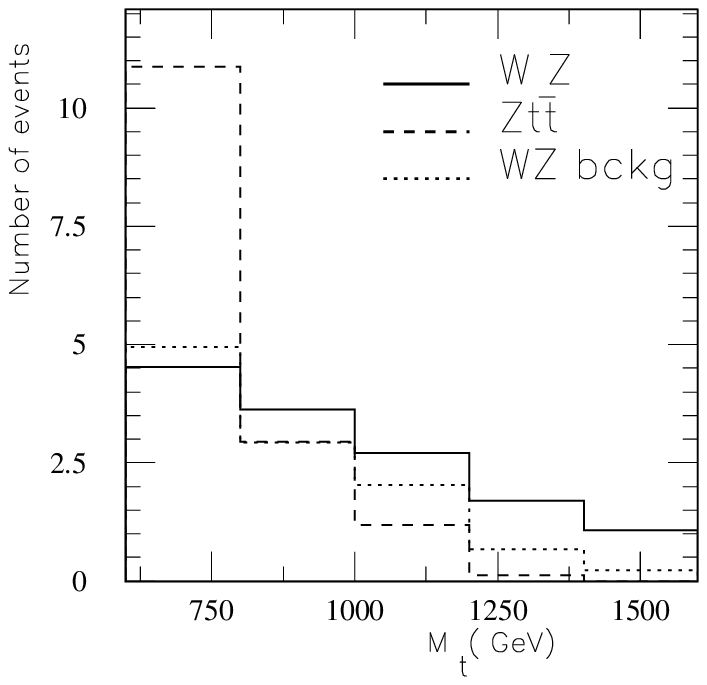} 
\caption{The transverse mass $M_{T}$ distribution for $ZW$ system (GeV) 
for $W_{L}^{\pm} Z_{L}$ scattering and for $Zt\bar{t}$.}
\label{fig:WZ}
\end{center}
\end{figure}
\begin{table}[htbp]
\begin{center}
\caption{Number of expected events for the $WZ$ signal and 
backgrounds with an integrated luminosity of 500 fb$^{-1}$.}
\label{tab:WZ}
\vskip0.2cm
\begin{tabular}{lcccccc} 
\hline
Cuts               & $L_{1}$=0.003 & $L_{1}$=0.01 & $Zt\overline{t}$ 
 & $WZ$            &  \multicolumn{2}{c}{$S/\sqrt{B}$} \\ 
 & & & &                  &  $L_{1}$=0.003 &  $L_{1}$=0.01     \\ 
\hline
Leptonic cuts       & 33.3          & 18.3         &  223.            
 & 762    &                &                   \\ 
Missing momentum    & 25.9          & 14.3         & 85.1             
 &  405   &                &                   \\ 
$p_{T}(Z) > M_{T}$/4  & 22.2          & 12.2         & 67.              
 & 300    &                &                   \\ 
Forward jet tagging & 14            & 7.3          & 15               
 & 10.8   &   2.7          &  1.43             \\ 
\hline
\end{tabular}
\end{center}
\end{table}

\subparagraph{ Like-sign \boldmath $W$ \unboldmath pair production}

$W^+_LW^+_L$ production has been extensively studied 
\cite{19-134}. As possible scenarios for this process by $W^+_LW^+_L$ 
scattering, the following are considered:
 \begin{itemize}
\item{A $t$-channel exchange of a Higgs with $M_H$ = 1 TeV, ($W_LW_L$ 
only), simulated with {\tt PYTHIA} with $\sigma \times BR$~=~1.33~fb (the 
same parameters of the resonance as in Section~\ref{sec:vv_heavyhiggs} 
were used).}
\item{The K-matrix unitarised amplitude \cite{BCHPCK2,19-135} 
$a_{IJ}^K =\frac{Re(a_{IJ})}{1-iRe(a_{IJ})}$, where $a_{IJ}$ is the 
low-energy theorem amplitude, proportional to $s$. This model is 
constructed to satisfy explicitly  elastic unitarity and would yield 
the maximum expected signal. The $\sigma\times BR$~=~1.12~fb.}
\item{A Chiral Lagrangian model, as in the $WZ$ resonant channel, 
with the same parameters: $L_1$~=~0, and $L_2$~=~0.006 or 0.01, 
leading to $\sigma\times$ BR = 0.484 and 0.379 fb, respectively.}
\end{itemize}
Backgrounds from continuum $WW$ bremsstrahlung produce mostly 
transverse $W$'s. Other backgrounds include processes involving 
non-Higgs exchange, as well as QCD processes of order 
$\alpha\alpha_s$ in amplitude, with gluon exchange and $W$ 
bremsstrahlung from interacting quarks. The effects of $Wt\bar{t}$ 
and $WZ$ backgrounds are also considered. The signal was generated 
with {\tt PYTHIA~6.2} and backgrounds were incorporated into 
{\tt PYTHIA} from a
Monte Carlo generator based on Barger's work \cite{19-121}, 
which takes into account all 
diagrams. The contribution from electroweak processes not involving 
the Higgs were estimated by assuming a low-mass Higgs ($M_H =$ 
100~GeV).

An analysis was performed using the fast ATLAS detector simulation 
({\tt ATLFAST}), with parameters set for high luminosity. The following 
leptonic cuts were first applied:
\def\labelenumi{L\arabic{enumi}.}
\begin{enumerate}
\item Two positively charged isolated leptons in the central region 
($p_T > 40$ GeV and 
$|\eta|<1.75$) must be identified. They will satisfy the trigger 
requirement.
\item The opening angle between the two leptons, in the transverse 
plane, must satisfy:
$\cos\Delta\phi < -0.5$. This cut selects preferentially events with 
longitudinal $W$'s which have high $p_T$.
The invariant mass of the two leptons was further required to satisfy 
$m_{ll} > 100$ GeV.
This latter cut eliminates few events in the low $m_{ll\nu\nu}$ region.
\end{enumerate}
At the jet level, backgrounds can be reduced by requiring that:
\def\labelenumi{J\arabic{enumi}.}
\begin{enumerate}
 \item No jet having $p_T > 50 $ GeV be present in the central 
region ($|\eta|<2$). 
This reduces significantly the background from the $Wt\bar t$ process.
\item Two jets must be present  in the forward and backward regions: 
$\eta>2$ and $\eta<-2$, with energies $>$ 300 GeV.
\item A lower $p_T$ was required for the forward jets: $p_T <$ 150 
GeV for the first and $p_T <$ 90 GeV for the second.
\end{enumerate} 

\begin{figure}[h]
\begin{center}
\includegraphics[width=6cm]{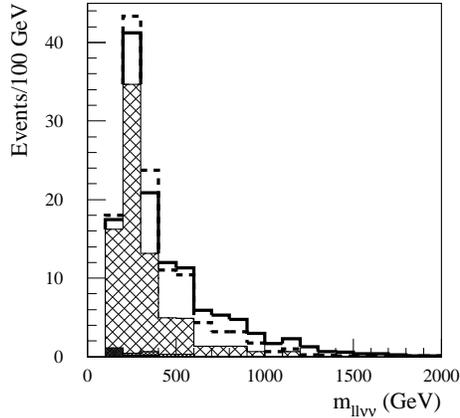} 
\caption{Distribution of invariant transverse mass of the two leptons 
with $E_T^{miss}$ in the $\WpWp \to l^+l^+\nu\nu$ process, after 
three years of high luminosity running.
Full line: K-matrix unitarisation; dashed
line: Higgs with $M_H = 1$~TeV, at tree level; hatched area: background
from transverse $W$'s. }
\label{fig:invmass}
\end{center}
\end{figure}

Figure~\ref{fig:invmass} shows expected mass distribution of the 
$ll\nu\nu$ system, for an integrated cross section of 300 fb$^{-1}$, 
after all cuts were applied, accounting only for transverse momentum. 
No correction was made for pile-up effects in jet tagging or central 
jet veto. If one counts only events with $m_{ll\nu\nu}>$ 400 GeV, a 
significant signal to background ratio is obtained (see 
Table~\ref{tab:wpwp3}). As expected, the K-matrix scenario gives the 
highest signal \cite{BCHPCK2} - this could be 
observable after a few years of high 
luminosity running.
By contrast, it was shown in Section~\ref{subsection:WZRES} 
that if the $\rho$ 
resonance is itself clearly observable in the resonant channel, 
then the signal will be very low.
The major remaining background, especially at low 
values of $m_{ll\nu\nu}$, is from continuum  transverse $W$ pairs. Note that 
only a \WpWp\  signal was searched for in this analysis. Combining the 
results with \WmWm\ would add approximately one-half to one-third of 
the signal and backgrounds. The Chiral Lagrangian model, with its 
parameters leading to a resonance in the $WZ$ system, would yield a 
very weak signal in the $W^+W^+$ channel, confirming the 
complementarity relationship between those two channels 
\cite{BCHPCK2,BCHP,19-133}.

\begin{table}[htbp]
\begin{center}
\caption{ Number of events expected for an integrated
luminosity of 300 fb$^{-1}$, after successive applications of cuts. 
The results are 
for $m_{ll\nu\nu} > $ 400 GeV.}
\vskip0.2cm
\begin{tabular}{lccccc}
\hline
                    & \multicolumn{2}{c}{Lepton cuts} & 
\multicolumn{3}{c}{Jet cuts}   \\ 
                    &   L1         &     L2            &   J1    &   
J2   &   J3         \\ 
\hline
$M_H$=1 TeV         &   59         &     56            &   43    &   
24   &   19.0       \\
K-matrix            &   90         &     86            &   69    &   
41   &   32         \\
Chiral Lagrangian $L_2$=0.006     &   22         &     21            &   15.8  &  
9.3   &   7.1        \\
Chiral Lagrangian $L_2$=0.01      &   15.1       &     14.1          &   10.4  &  
6.0   &   4.6        \\ \hline
$W_TW_T$            &  350         &   243             &   68    &  
54    &  14.0        \\
gluon   exchange         &   76         &    51             &   3.2   &  0 
    &   0          \\
$Wt\bar t$          &   93         &    71             &   2.0   &  0 
    &   0          \\
$WZ$                &   36         &    35             &  19.1   &  
0.5   &   0.3        \\ \hline
\end{tabular}
\label{tab:wpwp3}
\end{center}
\end{table}

\subsubsection{Technicolor}

Technicolor (TC) provides a framework
for dynamical electroweak symmetry breaking \cite{technicol,technicol_suss}. 
It assumes the existence of 
techni-fermions possessing a technicolor charge and interacting 
strongly at high scale. Chiral symmetry is broken by techni-quark 
condensates giving rise to Goldstone bosons, the techni-pions, which 
are the longitudinal degrees of freedom of the $W$ and $Z$ gauge-bosons. 
TC has been extended (extended TC, or ETC) to allow the generation of 
fermion masses \cite{21-2,21-2_2}. In order to account for the absence of 
FCNCs, the coupling constant is required to ``walk'', rather than 
``run''. To achieve a walking $\alpha_{TC}$, multi-scale TC models 
contain several representations of the fundamental family, and lead 
to the existence of techni-hadron resonances accessible at LHC 
energies. Such models \cite{21-3,21-4} are constrained by precision 
electroweak data \cite{21-5,21-5_2}, but not necessarily excluded 
\cite{21-6,21-7}. However, the constraints from those data make it 
unnatural to have a large top quark mass. In top-colour-assisted TC 
(TC2) models \cite{21-8,21-9}, the top quark arises in large part 
from a new strong top-colour interaction, which is a separate broken 
gauge-sector. 

The possible observation of TC resonances using the ATLAS detector 
is described in \cite{noteTC}. In particular, the search for a 
($I$=1, $J$=1) techni-rho resonance, a techni-pion and a techni-omega has 
been performed. Although certain models, with a given set of 
parameters, are used as reference, the signals studied can be 
considered generic in any model which predicts resonances. The model 
adopted here is that of multi-scale TC \cite{21-13,21-14}, with the TC 
group $SU(N_{TC})$ where $N_{TC}$~=~4 and two isotriplets of techni-pions. 
The longitudinal gauge-boson and the techni-pions mix 
$$|\Pi_T> = \sin\chi |W_L> + \cos\chi |\pi_T>$$
with a mixing angle which has a value
$\sin \chi=1/3$. The decay  constant of the mixed state is
$F_T=F_{\pi}\sin \chi =82$~GeV and the charge of the up-type 
(down-type) techni-fermion is $Q_U=1$ ($Q_D=0$). This model is 
incorporated in {\tt PYTHIA~6.1}. The decay channels of \rhot\ depend on 
the assumed masses of the techni-particles. Some mass scenarios have 
been considered to be representative of what one may expect to probe 
at the LHC and it is also assumed that the $\pi_T$ coupling to the top 
quark is very small, as may be expected in TC2 models. The following 
sections present an example showing a typical analysis for 
extracting TC signals. More channels and an 
extensive description can be found in \cite{noteTC}.   

\paragraph{\boldmath $\rho_T^\pm \to W^\pm Z \to l^\pm \nu 
l^+ l^-$ \unboldmath}

This decay could be the cleanest channel  for the techni-rho detection 
and complements the study shown in Section~\ref{sec:vv_nonreschan}. 
The  good efficiency 
of the ATLAS and CMS detectors  for lepton detection and missing   
transverse energy  measurement will provide good identification 
of the  $W$ and $Z$  bosons. Table~\ref{tab:rhotWZpar} shows  the 
parameters for the various sets of events which were generated.
For each set, $10^4$ events were generated and the signal was normalised
to three  years of low 
luminosity  running at the LHC (30~fb$^{-1}$). The  
branching ratios quoted include a preselection  on the transverse 
 mass ($\hat{m} >$ 150, 300, 600 GeV for    $m_{\rho_T^{\pm}}$=  220, 
  500 and    800  GeV respectively).

\begin{table}[htb]            
\begin{center}              
\caption{Signal parameters  for the    $\rho^{\pm}_T   
\rightarrow W^{\pm}   Z
  \rightarrow    l^{\pm}\nu  l^+l^-$. 
The last column gives the 
significance ($S/\sqrt{B}$) for three years of low luminosity running. }    
\label{tab:rhotWZpar}
\vskip0.2cm
\begin{tabular}{cccccc} \hline 
$m_{\rho_T}$ & $m_{\pi_T}$ &  $\Gamma_{\rho_T}$ & $BR$ & $\sigma \times BR$ 
& $S/\sqrt{B}$ \\
  (GeV)        &  (GeV)    & (GeV)  &       &  (pb)    &  \\ \hline
        220  & 110 (a)     & 0.93      & 0.13  & 0.16          & 31.6 \\ 
\hline
             &  110 (b)    &  67.1     & 0.014 & $1.0\times10^{-3}$& 0.7 \\

         500 &   300 (c)   & 4.47      &  0.21 & $1.3\times10^{-2}$ & 14.7 \\ 
             & 500 (d)     & 1.07      & 0.87  & $5.4\times10^{-2}$ & 64.2 \\
\hline  
             & 110 (e)     &  130.2    & 0.013 &  $1.5\times10^{-4}$& 0.3 \\
       800   & 300 (f)     & 52.4      & 0.032 &   $3.6\times10^{-4}$& 1.2 \\ 
             &  500 (g)    & 7.6       & 0.22  & $2.5\times10^{-3}$ & 10.9 \\ 
\hline  
\end{tabular}  
\end{center}
\end{table}  

The only background which needs to be considered is the continuum 
production of $WZ$ gauge-bosons, with $\sigma=21$~pb. The  cuts which
were applied are:
\begin{itemize}
\item{At least three charged  leptons were required (with 
$E_T>20$~GeV for electrons and $E_T>6$~GeV for muons), two  of which 
must  have the same flavour and opposite charge.}
\item{The invariant mass of the lepton  pair with the same flavour and 
opposite sign should be close to that of the $Z$: 
$|m_{l^+l^-} - m_Z| < 5$~GeV.}
\item{The longitudinal momentum  of the neutrino is calculated
(with a  2-fold ambiguity) from     the missing transverse energy 
and the momentum   of the    unpaired    lepton assuming  an 
invariant  mass $m_{l\nu} =  m_W$. Once the $W$  and  $Z$ were 
reconstructed, their transverse momentum was required  to be larger 
than  40~GeV.}
\item{Only   events  for which the   decay  angle with  respect to 
the direction   of   the $WZ$ system  ($\rho_T$)   in  its  rest   
frame was $|\cos\hat\theta| < 0.8$ were accepted.}
\end{itemize}


The significance ($S/\sqrt{B}$) of the signal ($S$) above the background ($B$)
is shown in 
Table~\ref{tab:rhotWZpar}. 
The number of signal 
and  background events was counted in mass  regions around the  
$\rho_T$ peak: 210 to 240,  460 to 560  and 740 to 870   for 
$m_{\rho_T} = $220, 500  and 800~GeV respectively.  No evident  signal 
can be observed for  cases (b), (e) and (f) (see 
Figure~\ref{fig:rhomass}),  principally because the $\rho_T$ resonance
is too wide.  

\begin{figure}[htb]
\begin{center}
\includegraphics[width=12cm]{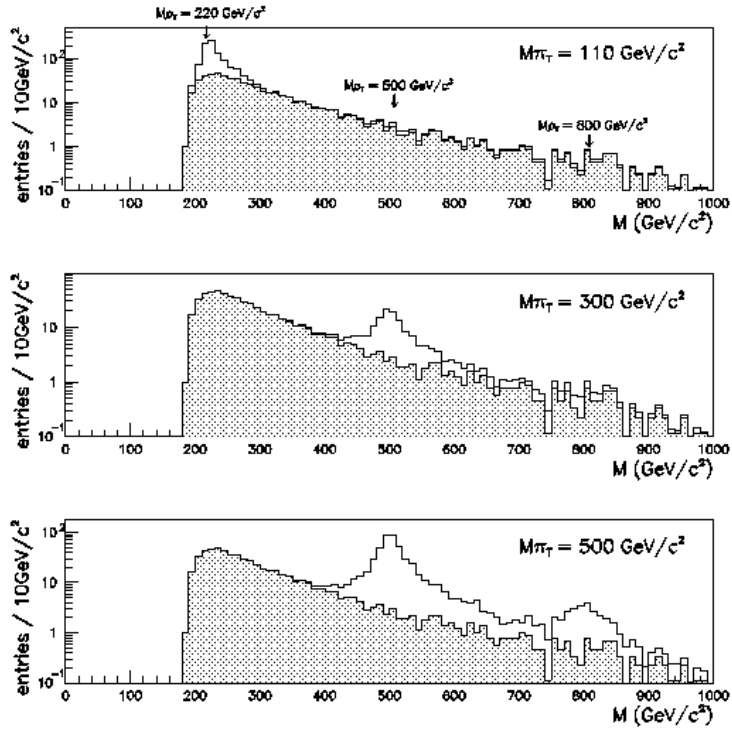} 
\caption{Reconstructed $W^{\pm}Z$  invariant mass.  The solid 
line is  for the    $\rho_T$  signal and    the filled   area for  
the   $WZ$ background. The three plots, each characterised by the
value of $m_{\pi_T}$, correspond to the cases (a,b,e), (c,f) and (d,g) 
defined in Table~\ref{tab:rhotWZpar}.}
\label{fig:rhomass}
\end{center}
\end{figure}

The Authors would like to thank M.~Chanowitz, K.~Lane, M.~Mangano, 
J.R.~Pel\'aez, S.R.~Slabospitsky and P.~Savard for their technical 
help with some Monte Carlo generators and for fruitful discussions.

\subsection{The degenerate BESS Model at the LHC \label{sec:bess}}

It is well known that na\"{\i}ve Dynamical Symmetry Breaking (DSB) models like standard
QCD-scaled technicolor generally tend to provide
large corrections to electroweak precision observables. New physics
effects are naturally small if decoupling holds. In fact in this case 
the corrections to electroweak observables are power suppressed in the 
limit in which the masses of the new particles are made large. It is thus a
natural question as to whether examples of DSB models with decoupling do
exist.

Here we will focus on a scheme of DSB, called degenerate BESS (D-BESS)
\cite{Casalbuoni:1996qt} in which decoupling is naturally satisfied in the low energy limit.
 The model predicts the existence of two triplets of new resonances
corresponding to the gauge-bosons of an additional gauge-symmetry $SU(2)_L
\otimes SU(2)_R$.
The global symmetry group of the theory is  $(SU(2)_L\otimes SU(2)_R)^3$
breaking down spontaneously to $SU(2)_D\otimes(SU(2)_L\otimes SU(2)_R)$ and
giving rise to nine Goldstone bosons. 
Six of these give mass to the new gauge-bosons,
which turn out to be degenerate. As soon as we perform the gauging of the
subgroup $SU(2)_L \otimes U(1)_Y $, the three remaining Goldstone bosons 
disappear
giving masses to the SM gauge-bosons.

What makes the model \cite{Casalbuoni:1996qt} so attractive is the fact that, due to the
degeneracy of the masses and couplings of the extra gauge-bosons
$(L^\pm,L_3,R^\pm,R_3)$, it decouples, so all the deviations in the low-energy
parameters from their SM values are strongly suppressed. Also, the degeneracy is
protected  by the additional ``custodial'' symmetry $(SU(2)_L\otimes SU(2)_R)$.
The deviations from the SM  predictions come from the mixing of $({\bf L}_\mu,
{\bf R}_\mu)$ with the standard gauge-bosons. In order to compare with the
experimental data, radiative corrections have to be taken into account. Since
the model is an effective parametrisation of a strongly interacting symmetry
breaking sector, one has to introduce a UV  cut-off $\Lambda$. We neglect the
new physics loop corrections and assume for D-BESS the same radiative
corrections as for the SM with $M_H=\Lambda=1$ TeV \cite{Casalbuoni:1996qt}. The 95\%~CL
bounds on the parameter space of the model coming from the precision electroweak
data can be expressed by the following approximated relation: $M$(TeV)$\ge 2.4
~g/g''$, where $M$ is the common mass of the new resonances, $g$~and $g''$~are the
standard $SU(2)_L$ and the new strong gauge-couplings respectively. Therefore
one  has a large allowed region available for the model even for the choice
$M_H=\Lambda=1$~TeV - a value highly disfavoured by the fit within the SM
\cite{nalep}. Also, the bounds on the D-BESS model from the direct search for new gauge
bosons performed at Tevatron are very loose \cite{Casalbuoni:1996qt}. This allows the
existence of a strong electroweak sector at relatively low energies such that it
may be accessible with accelerators designed for the near future. A peculiar
feature of this strong electroweak symmetry breaking model is the absence of $W
W$ enhancement due to the absence of direct couplings of the new resonances to
the longitudinal weak gauge-bosons. For this reason, the gold plated channels to
consider for discovering $({\bf L}_\mu, {\bf R}_\mu)$ are the fermionic ones.

\begin{figure}[htb]
\begin{center}
\begin{tabular}{ll}
\mbox{\psfig{file=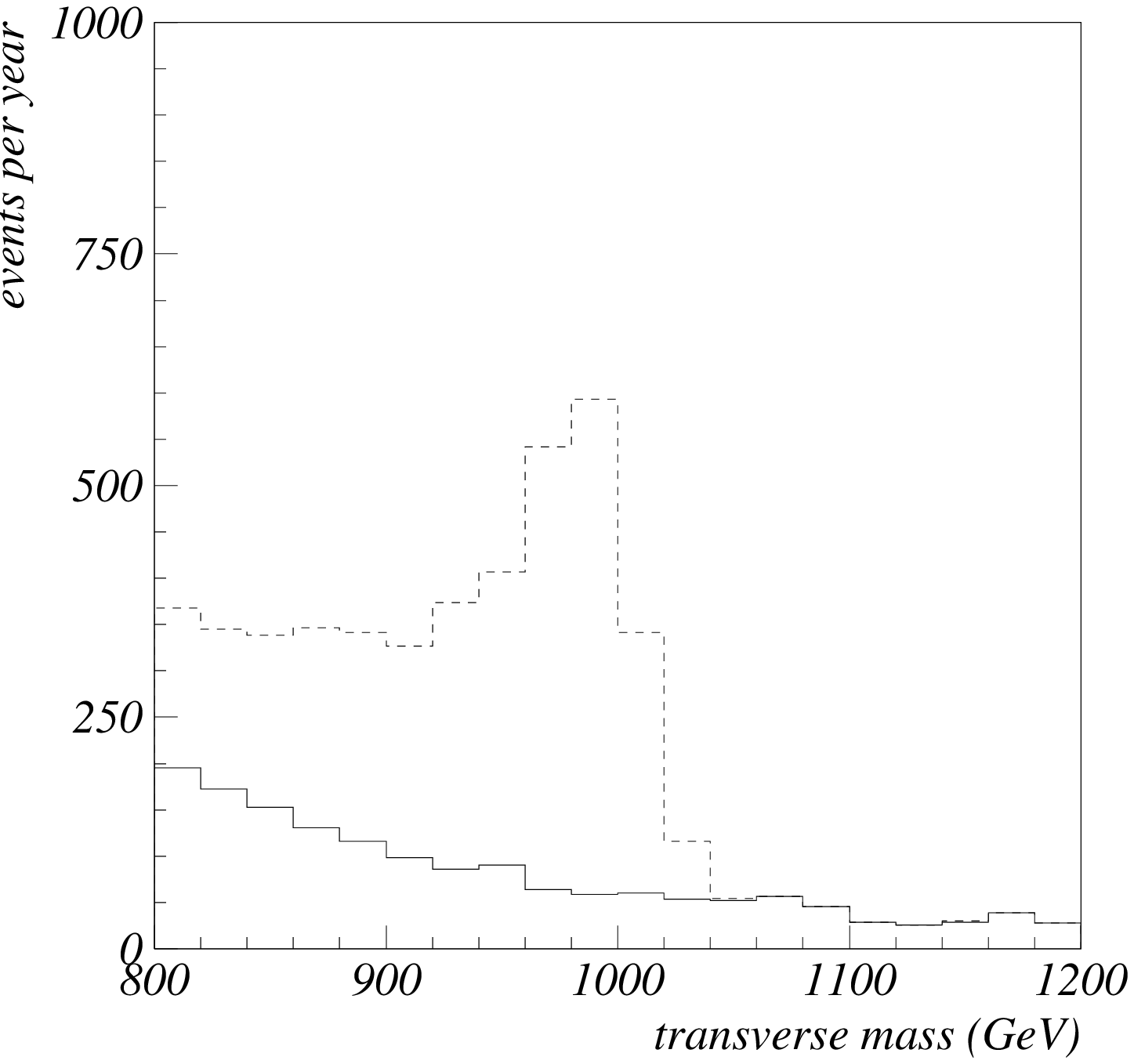,height=7 truecm}} &
\hspace{-1cm} \mbox{\psfig{file=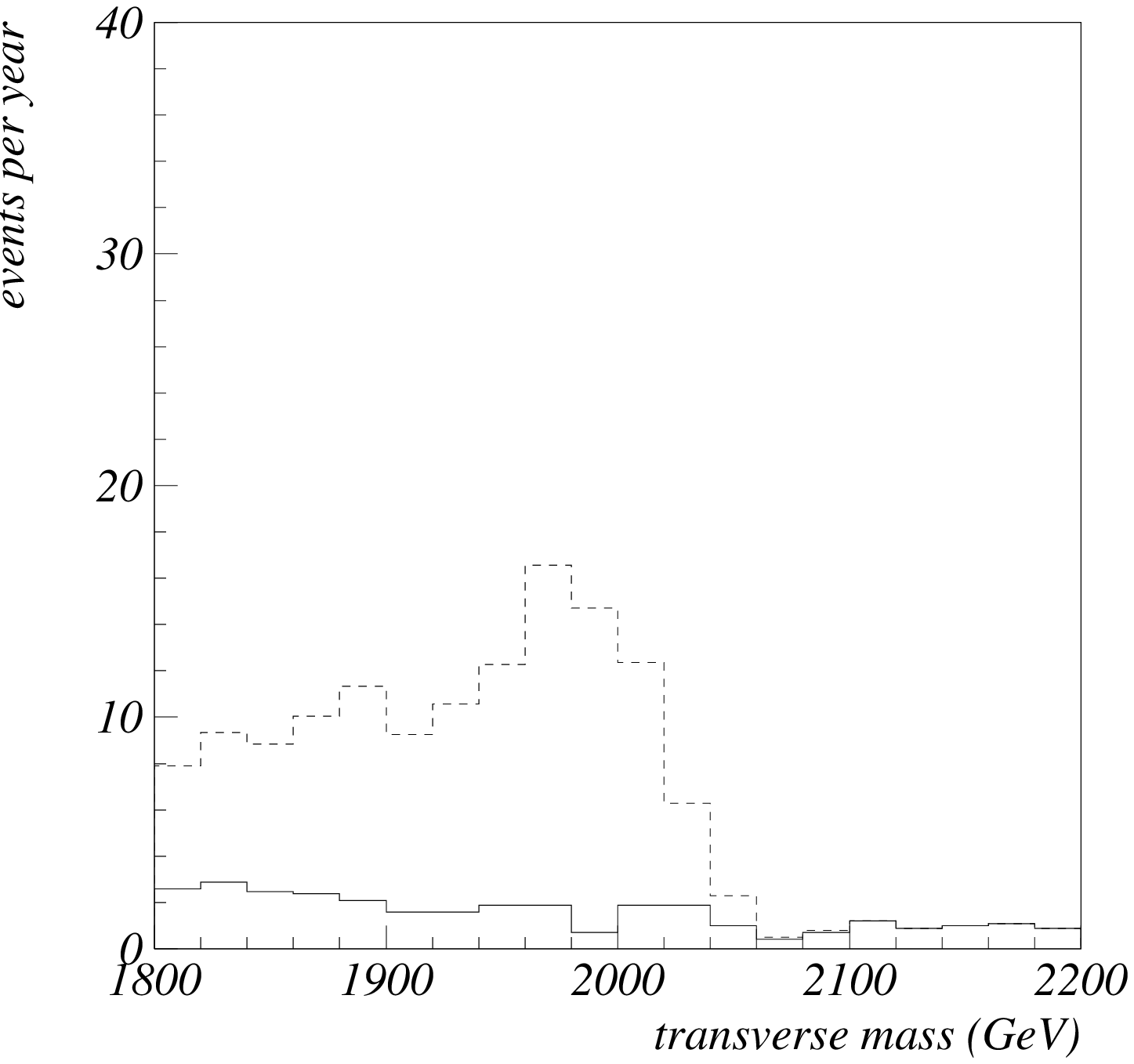,height=7 truecm}}\\
\end{tabular}
\end{center}
\caption{{Transverse mass differential distributions for
$pp\to L^\pm$,$W^\pm\to e\nu_e$ events at the LHC within the D-BESS model (dash
line) for $g/g''=0.1$ and $M=1$~TeV (left), $M=2$~TeV (right). The solid line is
the SM prediction.}\label{fig:bess1}}
\end{figure}

Here we have considered the production of these new resonances at the LHC for the
following configuration $\sqrt{s}=14$~TeV and ${\cal L}=10^{34}$~cm$^{-2}$sec$^{-1}$
and for the electron channel decay (the muon channel was studied in
\cite{pierre}). The events were generated using {\tt PYTHIA} Monte Carlo
(version 6.136)~\cite{pythia}.
Only the Drell-Yan mechanism for production was considered since it turns out to be
the dominant one. We have  analysed the production of the charged resonances in
$pp \to L^\pm, W^\pm\to e \nu_e$ ($R^\pm$ are completely decoupled) and neutral
ones in $pp\to L_3,R_3,Z,\gamma\to e^+e^-$. The signal events were compared with
the background from SM production.
 We have performed a rough simulation of the detector,  in particular, assuming a
$2\%$ smearing in the momenta of charged leptons and a resolution
 $\Delta E_T^{miss}=0.6 \sqrt{E_T^{miss}}$ in the missing transverse energy.
In the neutral channel, we have assumed an error of $2\%$ in the reconstruction
of the $e^+e^-$ invariant mass, which  includes bremsstrahlung 
effects~\cite{denegri}. We have considered several choices of the model parameters, in
the region allowed by the present bounds, and for each case we have selected
cuts to maximise the statistical significance of the signal. In Figure~\ref{fig:bess1} we show
the transverse mass distributions for the signal and for the SM background for
the case $M=1$~TeV (left)
 and $M=2$~TeV (right) and $g/g''=0.1$. The following cuts have been applied
for $M=1$~TeV: $|p_{T}^e|$ and $|p_{T}^{miss}|>0.3$~TeV and $M_T>0.8$~TeV. The
number of signal events per year is 3200, 
the corresponding background is of 1900 
events. The corresponding statistical significance $S/\sqrt{S+B}$ for one
year of running is 44. 
For $M=2$~TeV, the applied cuts are: $|p_{T}^e|$ and
$|p_{T}^{miss}|>0.7$~TeV and $M_T>1.8$~TeV, resulting in $S=108$, $B=46$ and
$S/\sqrt{S+B}=8.7$.

\begin{figure}[htb]
\begin{center}
\begin{tabular}{ll}
\mbox{\psfig{file=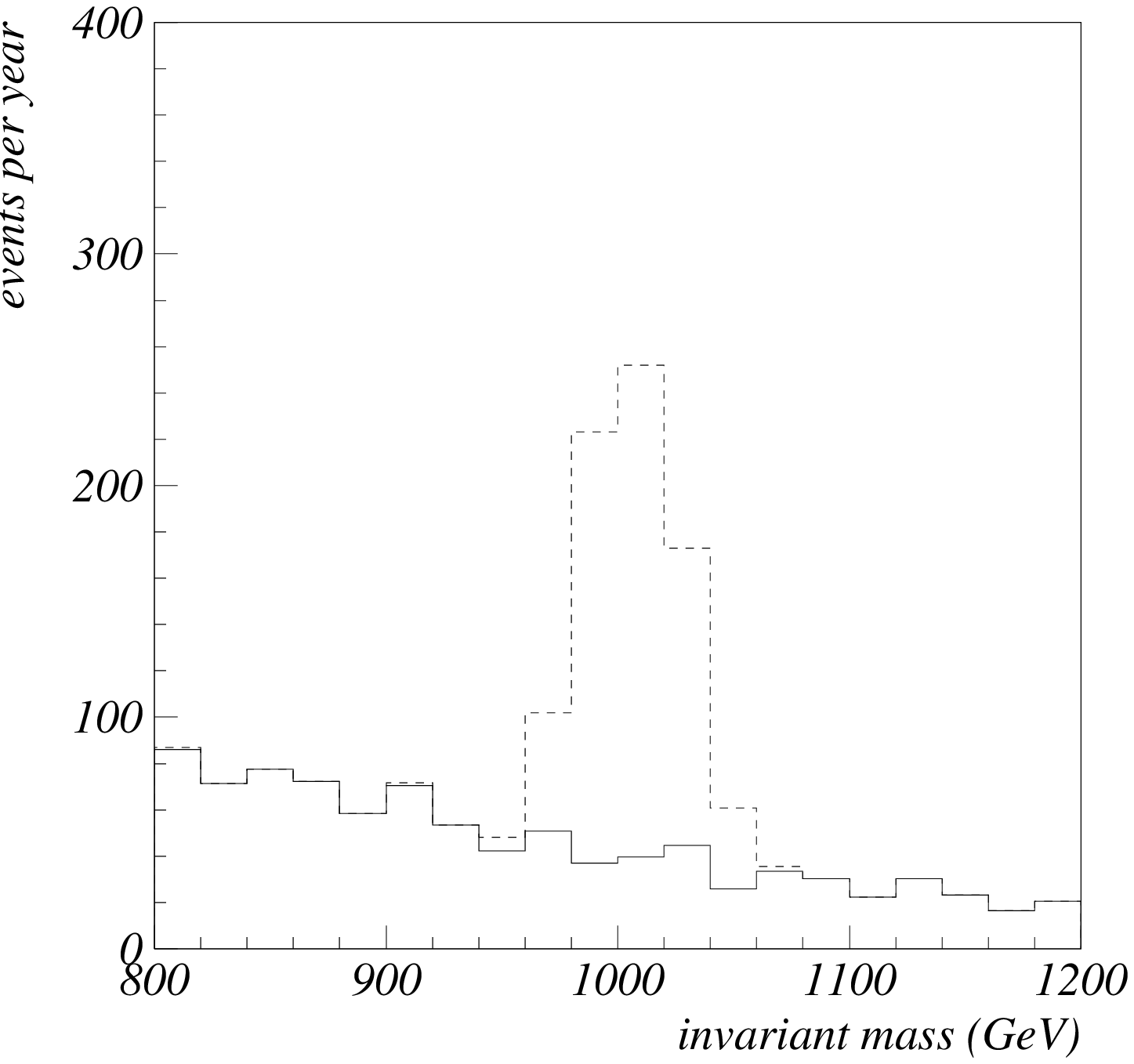,height=7 truecm}} &
\hspace{-1cm} \mbox{\psfig{file=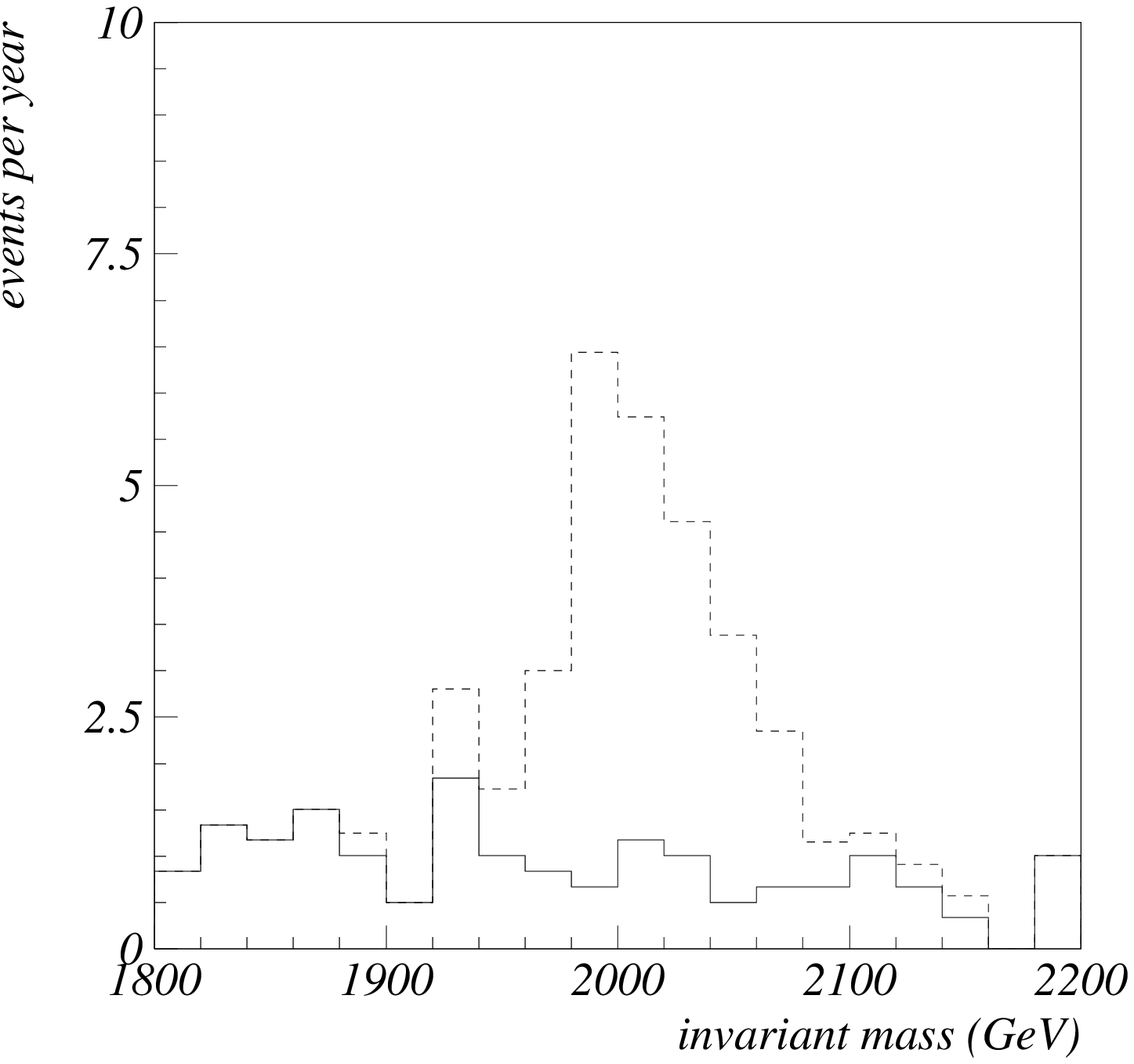,height=7 truecm}}\\
\end{tabular}
\end{center}
\caption{{Invariant mass differential distributions for
$pp\to L_3$,$R_3$,$Z$,$\gamma\to e^+e^-$ events at the LHC within the D-BESS model
(dash line) for $g/g''=0.1$ and $M=1$~TeV (left), $M=2$~TeV (right). The solid
line is the SM prediction.}\label{fig:bess2}}
\end{figure}

In Figure~\ref{fig:bess2}, we show the results of our simulation for the
   same choice of the parameters as in Figure~\ref{fig:bess1}  
for the neutral channel.  The
following cuts have been applied for $M=1$~TeV:
$|p_{T}^{{e^+}}|$ and $|p_{T}^{{e^-}} |>0.3$~TeV  and $M_{e^+e^-}>0.8$~TeV. The
number of signal events per year is 620, 
the background is of 1200 
events with a corresponding statistical significance of 15. 
For $M=2$~TeV, the cuts
are: $|p_{T}^{{e^+}}|$ and $|p_{T}^{{e^-}}|>0.7$~TeV and $M_{e^+e^-}>1.8$~TeV, 
resulting in
$S=24$, $B=30$ and $S/\sqrt{S+B}=3.3$. It turns out that the
cleanest signature is in the neutral channel, but the production rate is lower
than for the charged one. Also we observe that, due to the fact that the
D-BESS resonances are almost degenerate ($\Delta M/M\sim (g/g'')^2$), it will be
impossible to disentangle $L_3$ and $R_3$ which both contribute to the peak of
the signal in Figure~\ref{fig:bess2}.

Our conclusion is that the LHC will be able to discover a strong electroweak
resonant sector as described by the degenerate BESS model for masses up to 
2~TeV - in some cases with very significant numbers of events. Furthermore, if no
deviations from the SM predictions are seen within the statistical and
systematic errors, the LHC with $L=100$~fb$^{-1}$ will put a 95\%~CL bound
$g/g''<0.04-0.06$ for $0.5<M($TeV$)<2$ \cite{pierre}.



\end{document}